\documentclass{ws-ijmpd}
\usepackage[super,compress]{cite}
\usepackage{amssymb}
\usepackage{amsmath}
\usepackage{latexsym}
\usepackage{dcolumn}
\usepackage{graphicx}
\usepackage{epsfig}
\usepackage{pdfpages}

\usepackage{epstopdf}

\def\fnote#1#2{\begingroup\def\thefootnote{#1}\footnote{#2}\endgroup}
\input psfig.sty
\begin{document}
\markboth{Yu.\,L.\,Bolotin, A.\ Kostenko,O.\,A.\,Lemets,D.\,A.\,Yerokhin }
{Cosmological Evolution With Interaction Between Dark Energy And Dark Matter}

%
\catchline{}{}{}{}{}
%

\title{Cosmological Evolution With Interaction Between Dark Energy And Dark Matter}

\author{Yuri L.\, Bolotin}
\address{A.I.Akhiezer Institute for
Theoretical Physics,\\
 National Science Center "Kharkov Institute of
Physics and Technology",\\
 Akademicheskaya Str. 1, 61108 Kharkov,
Ukraine\\
\email{ybolotin@gmail.com}}

\author{Alexander Kostenko}
\address{A.I.Akhiezer Institute for
Theoretical Physics,\\
 National Science Center "Kharkov Institute of
Physics and Technology",\\
 Akademicheskaya Str. 1, 61108 Kharkov,
Ukraine\\
\email{alexander\_kostenko@yahoo.com}}

\author{Oleg\,A.\,Lemets}
\address{A.I.Akhiezer Institute for
Theoretical Physics,\\
 National Science Center "Kharkov Institute of
Physics and Technology",\\
 Akademicheskaya Str. 1, 61108 Kharkov,
Ukraine\\
\email{oleg.lemets@gmail.com}}
\author{Danylo\,A.\,Yerokhin}
\address{A.I.Akhiezer Institute for
Theoretical Physics,\\
 National Science Center "Kharkov Institute of
Physics and Technology",\\
 Akademicheskaya Str. 1, 61108 Kharkov,
Ukraine\\
\email{denyerokhin@gmail.com}}

\maketitle

\begin{abstract}
 In this review we consider in detail  different theoretical topics associated with interaction in the dark sector. We study linear and non-linear interactions which depend on the dark matter and dark energy densities. We consider a number of different models (including the holographic dark energy and dark energy in a fractal universe) with interacting dark energy (DE) and dark matter (DM), have done a thorough analysis of these models. The main task of this review was not only to give an idea about the modern set of different models of dark energy, but to show how much can be diverse dynamics of the universe in these models.   We find that the dynamics of a Universe that contains interaction in the dark sector can differ significantly from the Standard Cosmological Model (SCM).
\end{abstract}
\ccode{98.80.-k}
\maketitle\flushbottom

\tableofcontents
\section{INTRODUCTION}

The driving forces behind scientific progress are contradictions between entrenched thories and new observations. The Ptolemaic system was brought down not by the heliocentric theory, but by the first telescopes. The discovery of accelerated expansion of the Universe  \cite{Riessetal,Perlmutteretal} was one such contradiction, and led to the replacement of the Big Bang model with SCM. This model found great success. A vast number of observations can be explained and reconciled between eachother if we assume that we live in a planar Universe undergoing accelerated expansion,for which the values of relative density of dark energy ${\Omega_{de}}$ , dark matter ${\Omega_{dm}}$ ,baryonic matter ${\Omega_b}$ and radiation ${\Omega_r}$ are
\begin{equation}\label{int1}
\begin{array}{l}
{\Omega_{dm}} = 23\%  \pm 4\% ;\\
{\Omega_b} = 4\%  \pm 0.4\% ;\\
{\Omega_{de}} = 73\%  \pm 4\% ;\\
{\Omega_r} = 5 \times {10^{ - 5}}
\end{array}
\end{equation}
Unlike fundamental theories, physical models only reflect the current state of our understanding of the process or phenomenon that they were created to describe. A model's flexibility plays a big part in its success – a model must be able to modernize and evolve as new information comes in. For this reason, the evolution of any broadly applied model is accompanied by multiple generalizations that aim to resolve conceptual problems, as well to explain the ever-increasing array of observations\cite{UFN}. In the case of the SCM, one of the more promising directions of generalization is the replacement of the cosmological constant with a more complicated, dynamic form of dark energy \cite{CopelandSami,Amendola_Tsujikawa,Bamba_Capozziello,Yoo}, as well as the inclusion of interaction between the dark components \cite{Wetterich,Anderson,Carroll,Overduin,Amendola1,Amendola2,Billyard,Herrera_Pavon_Zimdahl,0411221,Holden,Zimdahl,Jai-chan_Hwang,Amendola3,Campo,Hoffman,Chimento,Amendola4,Guo,Zimdahl1,Jesus}. Typically, DE models are based on scalar fields minimally coupled to gravity, and do not implement the explicit coupling of the field to the background matter. However there is no fundamental reason for this assumption in the absence of an underlying symmetry which would suppress the coupling. Given that we do not know the true nature of either DE or DM, one cannot exclude that there exists a coupling between them. Whereas new forces between DE and normal matter particles are heavily constrained by observations (e.g. in the solar system and gravitational experiments on Earth), this is not the case for DM particles. In other words, it is possible that the dark components interact with each other, while not being coupled to standard model particles. In the absence of the aforementioned underlying symmetry, the study of the interaction of DE and DM is an important and promising research direction. Moreover, disregarding the potential existence of an interaction between dark components may result in misinterpretations of observational data. Since the gravitational effects of DE and of DM are opposite (i.e., gravitational repulsion versus gravitational attraction), even a small change of their relative concentrations can have an effect on cosmological dynamics. Models where the DM component of the Universe interacts with the DE field were originally proposed as solutions to the cosmic coincidence problem, since in the attractor regime, both DE and DM scale in the same way. It is remarkable that the scaling solutions in such models can lead to late-time acceleration, while this is not possible in the absence of coupling. It can also produce interesting new features in the large-scale structure. Therefore, the possibility of DE-DM interaction must be looked at with the utmost seriousness.

\section{PHYSICAL MECHANISM OF ENERGY EXCHANGE} \label{PHYSICAL_MECHANISM}

Models where DM interacts with the DE field can be realized if we only make an obvious assumption: the mass of the cold DM particles is a function of the DE field responsible for the present acceleration of the Universe. Variable-mass particles generally arise in models where the scalar (quintessence) field is coupled to the non-baryonic DM. Such a coupling represents a particularly simple and relatively general form of modified gravity. These particles appear, in fact, in scalar-tensor models and in simple versions of higher-order gravity theories in which the action is a function of the Ricci scalar. In the lagrangian, these couplings could be of the form $g(\varphi ){m_0}\bar \psi \psi $ or $h\left( \varphi  \right)m_0^2{\phi ^2}$ for a fermionic or bosonic dark matter represented by $\psi $  and $\phi $  respectively, where the functions $g$  and $h$  of the quintessence field $\varphi $   can, in principle, be arbitrary. Let's demonstrate the mechanism behind how the interaction appears by looking at a simple model \cite{Hoffman}. The dark matter particles in this model will be collisionless and nonrelativisic. Hence, the pressure of this fluid   and its energy density are
\begin{equation}\label{int2}
{p_{dm}} = 0,\;{\rho_{dm}} = nm,
\end{equation}
where $m$ is the rest mass and $n$ is the number density. We define
\begin{equation}\label{int3}
m = \lambda \varphi,
\end{equation}
where $\lambda $  is a dimensionless constant and $\varphi $   is a scalar field. The energy density associated with this fluid is thus
\begin{equation}\label{int3}
{\rho_{dm}} = \lambda n\varphi.
\end{equation}
We will assume that this species of particle froze out in the early Universe so that the comoving number density of dark matter particles is constant during the epochs of interest, i.e the particles are neither created nor destroyed. Thus, the number density is only a function of physical volume and $n = {n_0}{a^{ - 3}}$, where ${n_0}$  is the present number density of dark matter particles.
 The energy density and pressure associated with the scalar field in the potential $V\left( \varphi  \right)$  are:
\begin{equation}\label{int4}
{\rho_\varphi } = \frac{1}{2}{\dot \varphi ^2} + V\left( \varphi  \right),\quad {\rho_\varphi } = \frac{1}{2}{\dot \varphi ^2} - V\left( \varphi  \right).
\end{equation}
Since the energy density of the dark matter particles depends on $\varphi $ , the scalar field feels an additional effective  potential when it is in a ``bath'' of dark matter particles. Taking this effect into account, the equation of motion for the scalar field becomes
\begin{equation}
\begin{gathered}\label{int5}
  \ddot \varphi  + 3H\dot \varphi  + \frac{{d{V_{eff}}}}{{d\varphi }} = 0, \hfill \\
  {V_{eff}} = V(\varphi ) + \lambda n\varphi.  \hfill \\
\end{gathered}
\end{equation}
Consequently,
\begin{equation}\label{int6}
\ddot \varphi  + 3H\dot \varphi  + \frac{{d{V_{eff}}}}{{d\varphi }} =  - \lambda {n_0}{a^{ - 3}}.
\end{equation}
The only difference between this equation and that of a noninteracting dynamical dark energy model is the term on the right hand side, which accounts for the interaction.
Let's take a small detour. In the model which consists of a scalar $\varphi $  and a particle species $\psi $ (bosonic or fermionic),  the mass of $\psi $ is imagined to come from the vacuum expectation  value of $\left\langle \varphi  \right\rangle $ , with the constant of proportionality being some dimensionless parameter$\lambda $
\begin{equation}\label{int7}
{m_\psi } = \lambda \left\langle \varphi  \right\rangle.
\end{equation}
As an example \cite{Anderson}, let's look at the following potential:
\begin{equation}\label{int8}
V\left( \varphi  \right) = {u_0}{\varphi ^{ - p}}\;\left( {p > 0} \right).
\end{equation}
This model possesses no stable vacuum state; in empty space $\varphi $  tends to roll to infinity. We consider instead the behavior of $\varphi $  in a homogeneous background of $\psi $  with the number density ${n_\psi }$ . In that case, the dependence of the free energy on the value of $\varphi $  comes both from the potential $V\left( \varphi  \right)$  and the rest energy of the $\psi $ particles, which have a mass proportional to $\varphi $ . The equilibrium value of a homogeneous configuration  is therefore one which minimizes an effective  potential of the form
\begin{equation}\label{int9}
{V_{eff}}\left( \varphi  \right) = V\left( \varphi  \right) + \lambda {n_\psi }\varphi.
\end{equation}
The additional contribution to the effective potential is related to the fact that an increase of $\varphi $ leads, in this model, to the increase of the density of energy of   $\psi $-particles on account of an increase in their mass. The expectation value of $\varphi $ is
\begin{equation}\label{int10}
\left\langle \varphi  \right\rangle  = {\left( {\frac{{p{u_0}}}{{\lambda {n_\psi }}}} \right)^{1/1 + p}}.
\end{equation}	
The mechanism of the increase of the mass of the $\psi $ - particles is clear: ${m_\psi } \propto \left\langle \varphi  \right\rangle  \propto {\left( {{n_\psi }} \right)^{ - 1/1 + p}}$  , and in an expanding Universe, the density ${n_\psi }$ falls as time passes - ${n_\psi } \propto {a^{ - 3}}$ .

In order to derive an evolution equation for the dark matter energy density, we first consider the divergence of the stress-energy tensor for each dark component. Since neither dark component interacts directly with any other species, the divergence of the sum of their stress-energy tensors must vanish. However, due to the interaction, the divergence of each stress-energy tensor is not necessarily zero. The derivative operator is linear, so \fnote{\dag}{We use the signature $(+, -, -, -)$,
and definition $R^{\rho}_{~\sigma\mu\nu} = \partial_{\mu}\Gamma^{\rho}_{\nu\sigma}-
 \partial_{\nu}\Gamma^{\rho}_{\mu\sigma}+\Gamma^{\rho}_{\mu\lambda}\Gamma^{\lambda}_{\nu\sigma} - \Gamma^{\rho}_{\nu\lambda}\Gamma^{\lambda}_{\mu\sigma} $, ${ R}_{\nu \mu} = R_{\nu \beta
\mu}^ {~~~~~\beta}$, $R = g^{\mu \nu}R_{\mu \nu}.$}
\begin{equation}\label{int11}
{\nabla_\mu }\left( {{T_{(dm)}}_\nu ^\mu  + {T_{(\varphi )}}_\nu ^\mu } \right) = {\nabla_\mu }{T_{(dm)}}_\nu ^\mu  + {\nabla_\mu }{T_{(\varphi )}}_\nu ^\mu  = 0,
\end{equation}
which implies
\begin{equation}\label{int12}
{\nabla_\mu }{T_{(dm)}}_\nu ^\mu  =  - {\nabla_\mu }{T_{(\varphi )}}_\nu ^\mu.
\end{equation}
The stress-energy tensor for the nonrelativistic dark matter, ${T_{(dm)}}_\nu ^\mu $, is fairly simple. The only nonvanishing component is ${T_{(dm)}}_0^0 = {\rho_{dm}}$. For the scalar field, the stress-energy tensor is
\begin{equation}\label{int13}
	{T_{(\varphi )}}_\nu ^\mu  = {\partial ^\mu }\varphi \,{\partial_\nu }\varphi  - \delta_\nu ^\mu \left[ {\frac{1}{2}{\partial ^\alpha }\varphi \,{\partial_\alpha }\varphi  - V\left( \varphi  \right)} \right],
\end{equation}
and its divergence is
\begin{equation}\label{int14}
{\nabla_\mu }{T_{(\varphi )}}_\nu ^\mu  = {\partial_\mu }{T_{(\varphi )}}_\nu ^\mu  + \Gamma_{\mu \beta }^\mu {T_{(\varphi )}}_\nu ^\beta  - \Gamma_{\mu \nu }^\beta {T_{(\varphi )}}_\beta ^\mu  =  - \left( {\ddot \varphi  + 3H\dot \varphi  + \frac{{dV}}{{d\varphi }}} \right){\partial_\nu }\varphi.
\end{equation}
Using the equation of motion for the scalar field  \eqref{int6}, this expression simplifies to
\begin{equation}\label{int15}
{\nabla_\mu }{T_{(\varphi )}}_\nu ^\mu  = \lambda n{\partial_\nu }\varphi.
\end{equation}
The evolution equation for the dark matter energy density is then calculated by combining \eqref{int11}-\eqref{int15}:
\begin{equation}\label{int16}
{\nabla_\mu }{T_{\left( {dm} \right)}}_0^\mu  = {\dot \rho_{dm}} + 3H{\rho_{dm}} = \lambda n\dot \varphi.
\end{equation}
	
In this case, too, the inclusion of interaction leads to the appearance of a ``source'' in the right side of the ``conservation equation''. In the presence of a flux of energy between the dark components, this term must be taken with quotation marks.
Let's now look at alternative ways of introducing dark sector interaction. Dark energy represents the simplest explanation for the acceleration of the Universe within the $\Lambda CDM$ paradigm. Dark energy is generally associated with a cosmological constant, and can be thought of as being physically equivalent to vacuum energy.
We define \cite{Wands} a vacuum energy, $V,$ as having an energy-momentum tensor proportional to the metric
\begin{equation}\label{int17}
T_\nu ^\mu  = Vg_\nu ^\mu.
\end{equation} 	
By comparison with the energy-momentum tensor of a perfect fluid
\begin{equation}\label{int24}
T_\nu ^\mu  = \left( {\rho  + p} \right){u^\mu }{u_\nu } - pg_\nu ^\mu.
\end{equation} 	
we identify the vacuum energy density and pressure with $\rho  =  - p = V$.  A vacuum energy that is homogeneous throughout spacetime, ${\nabla_\mu }V = 0$ , is equivalent to a cosmological constant in Einstein gravity $\Lambda  = 8\pi GV$ .
We will consider the possibility of a time and/or space dependent vacuum energy. From Eq. \eqref{int17} we have
\begin{equation}\label{int19}
{\nabla_\mu }\hat T_\nu ^\mu  = {F_\nu },\quad {F_\mu } \equiv {\nabla_\mu }V.
\end{equation}
We can therefore identify an inhomogeneous vacuum, ${\nabla_\mu }V \ne 0$ , with an interacting vacuum, ${F_\mu } \ne 0$. The conservation of the total energy-momentum (including matter fields and the vacuum energy)

\begin{equation}\label{int20}
{\nabla_\mu }\left( {{T_{(de)}}_\nu ^\mu  + {T_{(dm)}}_\nu ^\mu } \right) = 0.
\end{equation} 	

implies that the vacuum ${T_{(de)}}_\nu ^\mu $ or dark energy transfers  energy-momentum to or from the matter fields ${T_{(dm)}}_\nu ^\mu $
\begin{equation}\label{int23}
{\nabla_\mu }{T_{(de)}}_\nu ^\mu  =  - {T_{(dm)}}_\nu ^\mu  = {F_\nu },
\end{equation} 	
where the ${F_\mu }$  is the 4-vector of interaction between dark components and its form is not known a priori.

We must now see how dark sector interaction affects the actual dynamics - the Friedmann equations - and therefore obtain general equations of motion for dark energy interacting with dark matter \cite{Solano}. We assume a Universe formed by only dark matter and dark energy. The equations of motion that describe the dynamics of the Universe as a whole are the Einstein field equations
\begin{equation}\label{int22}
{R_{\mu \nu }} - \frac{1}{2}R{g_{\mu \nu }} = 8\pi G\left( {{T_{(de)}}_{\mu \nu } + {T_{(dm)}}_{\mu \nu }} \right),
\end{equation}

			



Equations \eqref{int23}  can be projected on the time or on the space direction of the comoving observer.  We project these equations   in a part parallel to the velocity ${u^\mu }$
\begin{equation}\label{int25}
\begin{gathered}
  {u^\mu }{\nabla ^\nu }{T_{\left( {dm} \right)\mu \nu }} =  - {u^\mu }{F_\mu }, \hfill \\
  {u^\mu }{\nabla ^\nu }{T_{\left( {de} \right)\mu \nu }} = {u^\mu }{F_\mu }, \hfill \\
\end{gathered}
\end{equation}
and in other part orthogonal to the velocity using the projector ${h_{\beta \mu }} = {g_{\beta \mu }} - {u_\beta }{u_\mu }$

\begin{equation} \label{GrindEQ__2_21_}
\begin{array}{l} {\; \; h^{\mu \beta } \nabla ^{\nu } T_{\left(dm\right)\mu \nu } =-h^{\mu \beta } F_{\mu } ,} \\ {\; h^{\mu \beta } \nabla ^{\nu } \nabla ^{\nu } T_{\left(de\right)\mu \nu } =h^{\mu \beta } F_{\mu } }. \end{array}
\end{equation}
Using  \eqref{int24} ,  \eqref{int25} and \eqref{int24}    we obtain the Euler equations for each component,
\begin{equation} \label{GrindEQ__2_23_}
\begin{array}{l} {h^{\mu \beta } \nabla_{\mu } p_{dm} +\left(\rho_{dm} +p_{dm} \right)u^{\mu } \nabla_{\mu } u^{\beta } =-h^{\mu \beta } F_{\mu } ,} \\ {h^{\mu \beta } \nabla_{\mu } p_{de} +\left(\rho_{de} +p_{de} \right)u^{\mu } \nabla_{\mu } u^{\beta } =h^{\mu \beta } F_{\mu } } \end{array}.
\end{equation}

We assumed that the background metric is described by the flat FLRW metric . In the comoving coordinates we choose $u^{\mu } =\left(1,0,0,0\right)$. With this choice
\begin{equation} \label{GrindEQ__2_24_}
\begin{array}{l} {\nabla_{\mu } u^{\mu } =3H,} \\ {u^{\mu } \nabla_{\mu } u^{\nu } =0} \end{array}.
\end{equation}
Using the notation $u^{\mu } F_{\mu } =Q(a)$, we transform the equations \eqref{GrindEQ__2_23_} to their final form
\begin{equation} \label{GrindEQ__2_25_}
\begin{array}{l} {\dot{\rho }_{dm} +3H\rho_{dm} =Q,} \\ {\dot{\rho }_{de} +3H(\rho_{de} +p_{de} )=-Q} .\\ {} \end{array}
\end{equation}
The function  Q(a) is known as the interaction function, and depends on the scale factor. We note that the equations \eqref{GrindEQ__2_21_} are satisfied identically  (taking into account the condition $h^{\mu \nu } F_{\nu } =0$ )  and do not produce any new equations.

The presence of interaction between the dominant dark components can be interpreted in a different light. In accordance with current theories, the surrounding macroworld is controlled by electromagnetic and gravitational forces. Can we be sure that there are no other forces in nature besides the ones we know? This is a delicate question. One thing is certain - if these forces exist, they must be so weak on analysed spatial scales so as not to go outside of the margin of error of existing prescision measurements. The introduction of new forces requires the definition of objects between which they act, their intensity and radius of action.

The idea of a long-range ``fifth force'' (besides strong, weak, electromagnetic and gravitational forces) is a very popular and old one, although it's hard to incorporate it into a compelling working model. Hopes (and even announcements of discoveries of such forces) were quickly replaced with disappointment. With the discovery of dark sector interaction, however, many are looking at this field with renewed interest. Currently, the physics of the dark sector is effectively  unknown. Due to this, greater and greater popularity is being obtained by scenarios in which a purely dark sector interaction exists, resulting from a nonminimal coupling of dark matter to a scalar field, and this coupling in turn is interpreted as the fifth force.

Instead of coupling dark matter to dark energy, we can modify the coupling of dark matter particles with themselves. One class of models of this type \cite{Bean_Flanagan} involve an interaction between fermionic dark matter, $\psi $, and an ultra-light pseudo scalar boson, $\varphi $, that interacts with the dark matter through a Yukawa coupling with the strength g, described by the Lagrangian,
\begin{equation} \label{GrindEQ__2_26_}
L=i\bar{\psi }\gamma_{\mu } \nabla ^{\mu } \psi -m_{\psi } \bar{\psi }\psi -\frac{1}{2} \nabla_{\mu } \varphi \nabla ^{\mu } \varphi -\frac{1}{2} m_{\varphi } \varphi ^{2} +g\varphi \bar{\psi }\psi.
\end{equation}
For $g\ne 0$ , on scales smaller than $r_{s} =m_{\varphi }^{-1} $ , the Yukawa interaction acts like a long-range fifth force in addition to gravity. The effective  potential felt between two dark matter particles is
\begin{equation} \label{GrindEQ__2_27_}
V(r)=-\frac{Gm_{\psi }^{2} }{r} \left[1+\alpha \exp \left(-\frac{r}{r_{s} } \right)\right].
\end{equation}
with $\alpha =2g^{2} \frac{\bar{M}_{P}^{2} }{m_{\psi }^{2} } ,\bar{M}_{P} =\left(8\pi G\right)^{-1/2} $ . The cosmological implications of Yukawa-like interactions of dark matter particles have  been considered across a range of astrophysical scales.

The term "fifth force'' is usually brought up in a select few cases \cite{1206.1225v1}: couplings between dark energy and dark matter (coupled quintessence); couplings between dark energy and neutrinos; universal couplings with all species (scalar-tensor theories and $f(R)$). In all of these cosmologies the coupling produces a fifth force, complementary to standard gravitational attraction. The availability of a new force, generated by the DE scalar field (at times called the 'cosmon' \cite{Wetterich}, seen as the agent of cosmological interaction) can substantially change the growth of the cosmic structure \cite{Zhao1,Zhao2,Zhao3,Barrow_5th_force,Barrow_N_body}.

The scalar field, providing an additional degree of freedom (which can either indirectly interact with other types of matter via gravity or be directly related to the matter), generates a fifth force, which acts on the matter and violates the Weak Equivalence Principle (WEP).
The possibility of direct interaction between the scalar field and other matter fields is in agreement with the assumption that this type of interaction can help resolve the coincidence problem (see subsection \ref{CCP}).
In the case of direct interaction between the scalar field and baryons, the baryons must experience the action of the fifth force, which is severely constrained by observations, as long as there are no special mechanisms that suppress this effect. The resolution of this problem may lie in the assumption that the scalar field only interacts with dark matter.

The interaction of the scalar field with dark matter could affect cosmic structure
formation in different ways \cite{Barrow_5th_force,1206.1225v1}.
First of all, this interaction will change the rate of the background expansion of the Universe, which in turn affects the clusterization rate of matter particles; secondly, the interaction changes the effective mass of the dark matter particles, thereby changing the source term of the Poisson equation due to the added contribution of the density perturbations of the scalar field; third, this interaction will cause a fifth force to appear between the matter particles, which will lead to more intense clusterization of matter; finally, there will appear an additional, velocity-dependent force, that acts on the particles of matter and that can either be interpreted as a part of the fifth force, or as an additional force of friction, which will also lead to more intense clusterization of these particles. It must be stated that not all models prominently feature these effects - often, one or more of these effects can be ignored.

Let's look at how the main equations change, completely following \cite{1206.1225v1}.

In order to get the equations of motion that interest us, we start from a Lagrangian
\begin{equation}
\mathcal{L}=\frac{1}{2}\left[ \frac{R}{\kappa }-\nabla ^{a}\varphi \nabla
_{a}\varphi \right] +V(\varphi )-C(\varphi )\mathcal{L}_{\mathrm{DM}}+
\mathcal{L}_{\mathrm{S}},\ \
\end{equation}
where $R$ is the Ricci scalar, $\kappa =8\pi G$ with $G$ as the gravitational
constant, $\mathcal{L}_{\mathrm{DM}}$ and $\mathcal{L}_{\mathrm{S}}$ are
respectively the Lagrangian densities for dark matter and standard model
fields, $\varphi $ is the scalar field, and $V(\varphi )$ its potential; the
coupling function $C(\varphi )$ describes the coupling between $\varphi $
and dark matter. A model is fully specified when $V(\varphi )$ and $C(\varphi )$ are given.

Varying the total action with respect to the metric $g_{ab}$, the
following expression for the total energy-momentum tensor in this model can be obtained:
\begin{equation}\label{eq:emt_tot}
T_{ab} = \nabla_{a}\varphi \nabla_{b}\varphi -g_{ab}\left[ \frac{1}{2}\nabla^{c}\nabla _{c}\varphi -V(\varphi )\right]
+C(\varphi )T_{ab}^{\mathrm{DM}}+T_{ab}^{\mathrm{S}},
\end{equation}
where $T_{ab}^{\mathrm{DM}}$ and $T_{ab}^{\mathrm{S}}$ are the
energy-momentum tensors for (uncoupled) dark matter and standard model
fields.
Clearly, the existence of the scalar field and its interaction with matter fields changes the form of the energy-momentum tensor, and therefore changes the rate of the background expansion of the Universe, which in turn affects structure formation.

A nonminimally coupled scalar field generates a direct interaction (fifth force) by exchanging quanta of the scalar field with dark matter particles.
This can best be illustrated by tracing the changing geodesic equation for particles of dark matter.

\begin{equation}\label{eq:geodesic}
\frac{d^{2}\mathbf{r}}{dt^{2}}=-\vec{\nabla}\Phi -\frac{C_{\varphi }(\varphi
)}{C(\varphi )}\vec{\nabla}\varphi,
\end{equation}
where $\mathbf{r}$ is the position vector, $t$ is the (physical) time, $\Phi $
is the Newtonian potential and $\vec{\nabla}$ is the spatial derivative. $C_{\varphi }=dC/d\varphi $. The second term on the right-hand side is the
fifth force and only exists for coupled matter species (dark matter in our
model). As stated before, the fifth force also changes the dark matter's clusterization capability. Note also that on very large scales, the scalar field $\varphi$ must be uniform, and therefore, the fifth force must vanish.

As already mentioned, in terms of the Lagrangian, the coupling is introduced by allowing
the mass $m$ of matter fields to depend on a scalar field $\phi$ via the function
$m(\phi)$, which defines the interaction. As an example, let's look at an analogous action with a more concretely defined, Yukawa-like type of interaction:
\begin{equation}
  \label{mg:cde:action} {\cal S} = \int{d^4x \sqrt{-g} \left[\frac{1}{2}\partial^\mu
\phi \partial_\mu \phi + U(\phi) + m(\phi)\bar{\psi}\psi - {\cal
L}_{kin}[\psi]\right] },
\end{equation}
where $U(\phi)$ is the potential in which the scalar field $\phi$ rolls, $\psi$
describes matter fields, and $g$ is defined in the usual way as the determinant
of the metric tensor.
Using the standard relations (index $\alpha$ corresponds to all of the interacting components)
\begin{equation}
\nabla_{\nu}T_{(\alpha)\mu}^{\nu}=Q_{(\alpha)\mu}\,,\label{tensor_conserv_alpha}
\end{equation}
with the constraint
\begin{equation}
\sum_{\alpha}Q_{(\alpha)\mu}=0\label{Q_conserv_total} ,
\end{equation}
one can obtain the background conservation equations:
\begin{eqnarray}
 \label{cons_phi} \frac{d\rho_{\phi}}{d\eta} = -3 {\cal H} ( p_{\phi} +  \rho_{\phi})  +
\beta(\phi) \frac {d \phi}{d\eta} (1-3 w_{\alpha}) \rho_{\alpha} ~~~, \\
\label{cons_gr} \frac{ d \rho_{\alpha}}{d\eta} = -3 {\cal H} (p_{\phi} +  \rho_{\phi})  -
\beta(\phi) \frac{d\phi}{d\eta} (1-3 w_{\alpha}) \rho_{\alpha}.
\end{eqnarray}
The choice of the mass function $m(\phi)$ corresponds to the choice of $\beta(\phi)$ and, therefore, to the source of interaction $Q_{(\alpha)\mu}$, thereby defining the intensity of the interaction:
\begin{equation} \label{mass_def}
Q_{(\phi)\mu}=\frac{\partial\ln{m(\phi)}}{\partial\phi}
T_{\alpha}\,\partial_{\mu}\phi ~~~~~,~~~~~~
m_\alpha=\bar{m}_\alpha ~ e^{-{\beta(\phi)}{\phi}}.
\end{equation}

Like with the equations for the perturbations, the interaction can be included into the modified Euler equation
\begin{eqnarray}
& &\frac {d{\bf{{v}_\alpha}}}{d\eta} + \left({\cal H} - {\beta(\phi)} \frac{d \phi}{d\eta} \right) {\bf
{v}_\alpha} - {\bf{\nabla }} \left[\Phi_\alpha + \beta \phi \right] = 0
\,.
\end{eqnarray}
The Euler equation in terms of the cosmic time (${\rm d}t = a\, {\rm d}\tau$)  can also be
rewritten as the equation of motion of a particle with the coordinate ${\bf{r}}$:
\begin{equation}
\label{CQ_euler}
\dot{{\bf{v}}}_{\alpha} = -\tilde{H}{\bf{v}}_{\alpha} - {\bf{\nabla
}}\frac{\tilde{G}_{\alpha}{m}_{\alpha}}{r} \,.
\end{equation}
This last equation explicitly contains all of the main terms that are caused by interaction:
\begin{enumerate}
 \item a fifth force ${\bf{\nabla }} \left[\Phi_\alpha + \beta \phi
\right]$ with an effective $\tilde{G}_{\alpha} = G_{N}[1+2\beta^2(\phi)]$ ;
\item a velocity dependent term $\tilde{H}{\bf{v}}_{\alpha} \equiv H \left(1 -
{\beta(\phi)} \frac{\dot{\phi}}{H}\right) {\bf{v}}_{\alpha}$
\item a time-dependent mass for each particle $\alpha$, evolving according to
(\ref{mass_def}).
\end{enumerate}

Therefore, the scalar field interaction affects the growth of cosmic structure
chiefly owing to a velocity-dependent force(a so-called fifth force), but also due to an alteration of
the particle mass (or the source of the Poisson equation) and a modification of the rate of the background expansion of the Universe.

From this, it follows that  the fifth force, which is the most well-known consequence of a interaction between dark matter and a scalar field, is not the only one (and sometimes not even the most significant one) that affects the structure formation. Depending on the type of bond between the scalar field and the dark matter, other new effects are also brought
in, and could have important consequences.

\section{Phenomenology of Interacting Models} \label{Phenomenology_of_Interacting_Models}

We have already stated that since there is no fundamental theoretical approach that may specify the functional form of the coupling between DE and DM, presently coupling models are necessarily phenomenological. Of course, one can always provide arguments in favour of a certain type of correlation. However, until the creation of a microscopic theory of dark components, the effectiveness of any phenomenological model will be defined only by how well it corresponds to observations.

The interaction between dark matter and dark energy is described by following modified energy conservation equations
\begin{equation} \label{GrindEQ__3_1_}
\begin{array}{l} {\dot{\rho }_{dm} +3H\rho_{dm} =Q,} \\ {\dot{\rho }_{de} +3H\left(1+w_{de} \right)\rho_{de} =-Q}. \end{array}
\end{equation}
Here $Q$ is the rate of energy transfer and $w_{de}$ is the equation of state parameter (EoS) . The sign of $Q$ defines the direction of the energy flux:
\[Q\; \left\{\begin{array}{l} {>0} \\ {<0} \end{array}\right. \to energy\; transfer\; is\quad \left\{\begin{array}{l} {dark\; energy\to dark\; matter} \\ {dark\; matter\to dark\; energy} \end{array}\right. \]

We will focus our attention on DE in the form of a scalar field. In this case
\begin{equation} \label{GrindEQ__3_2_}
w_{de} =w_{\varphi } =\frac{p_{\varphi } }{\rho_{\varphi } } =\frac{\frac{1}{2} \dot{\varphi }^{2} -V\left(\varphi \right)}{\frac{1}{2} \dot{\varphi }^{2} +V\left(\varphi \right)}
\end{equation}
The modified (by interaction) Klein-Gordon equation is
\begin{equation} \label{GrindEQ__3_3_}
\ddot{\varphi }^{2} +3H\dot{\varphi }+\frac{dV}{d\varphi } =-\frac{Q}{\dot{\varphi }}.
\end{equation}

It is useful to note that the system \eqref{GrindEQ__3_1_}, which describes the interactiong dark components, can be transformed into the standard form that corresponds to non-interacting components by re-defining the parameters $w_{de} $ è $w_{dm} =0$ \cite{Bohmer}. If we write the equations \eqref{GrindEQ__3_1_} in the form
\begin{equation} \label{GrindEQ__3_4_}
\dot{\rho }_{i} +3H(1+w_{eff,i} )\rho_{i} =0,\; i=de,dm,
\end{equation}
then
\begin{equation} \label{GrindEQ__3_5_}
w_{eff,dm} =-\frac{Q}{3H\rho_{dm} } ,\quad w_{eff,de} =w_{de} +\frac{Q}{3H\rho_{de} },
\end{equation}
It follows that
\[\begin{array}{l} {Q>0\to \left\{\begin{array}{l} {w_{eff,dm} <0\quad dark\; matter\; redshifts\; slower\; than\; a^{-3} } \\ {w_{eff,de} >w_{de} \quad dark\; energy\; has\; les\; accelerating\; power} \end{array}\right. } \\ {Q<0\to \left\{\begin{array}{l} {w_{eff,dm} >0\quad dark\; matter\; redshifts\; faster\; than\; a^{-3} } \\ {w_{eff,de} <w_{de} \quad dark\; energy\; has\; more\; accelerating\; power} \end{array}\right. } \\ {} \\ {} \end{array}\]
If we "turn off" the interaction $(Q=0)$, we return to the original EoS parameters:$w_{eff,dm} =w_{dm} =0,\; w_{eff,de} =w_{de} \quad $

The equation \eqref{GrindEQ__3_5_} can be given an alternative interpretation. It is convenient to introduce the effective pressures $\Pi_{dm} $ and $\Pi_{de} $
\begin{equation} \label{GrindEQ__3_6_}
Q\equiv -3H\Pi_{dm} =+3H\Pi_{de},
\end{equation}
with the help of which
\begin{equation} \label{GrindEQ__3_7_}
\begin{array}{l} {\dot{\rho }_{dm} +3H(\rho_{dm} +\Pi_{dm} )=0,} \\ {\dot{\rho }_{de} +3H\left(\rho_{de} +p_{de} +\Pi_{de} \right)=0} \end{array}.
\end{equation}
In this case, the conservation equations formally look as those for two independent fluids. A coupling between them has been mapped into the relation $\Pi_{dm} =-\Pi_{de} $ .

In order to illustrate how interaction between the dark components acts on cosmological dynamics, consider the time evolution of the ratio $r\equiv \rho_{dm} /\rho_{de} $ ,
\begin{equation} \label{GrindEQ__3_8_}
\dot{r}=\frac{\rho_{dm} }{\rho_{de} } \left(\frac{\dot{\rho }_{m} }{\rho_{m} } -\frac{\dot{\rho }_{de} }{\rho_{de} } \right)=3Hr\left(w_{de} +\frac{1+r}{\rho_{dm} } \frac{Q}{3H} \right)
\end{equation}
Let's analyse the obtained expression \cite{Zimdahl_Pavon}  We take $r=r_{0} a^{-\xi } $ ($r_{0} $is the energy-density ratio at the present time and $\xi $ is a constant, non-negative parameter). In this case, for the interaction term, we obtain
\begin{equation} \label{GrindEQ__3_9_}
\frac{Q}{3H\rho_{dm} } =-\frac{w_{de} +\frac{\xi }{3} }{1+r}.
\end{equation}
Eq. \eqref{GrindEQ__3_9_} demonstrates that by choosing a suitable interaction between both components, we may produce any desired scaling behavior of the energy densities. The uncoupled case, corresponding  to $Q=0$ , is given by $\xi /3+w_{de} =0$ .  The SCM  model ( the special uncoupled case) corresponds to $w_{de} =-1,\; \xi =3$ .  Generally, interacting models are parameterized by deviations from $\xi =-3w_{de} $ . Any solution which deviates from $\xi =-3w_{de} $ represents a testable, non-standard cosmological model. For $\xi >0$, the interaction \eqref{GrindEQ__3_9_} becomes very small for $a\ll 1$ . Consequently, the interaction is not relevant at high redshifts. This guarantees the existence of an early matter-dominated epoch. Note also that energy transfer from DE to DM, i.e. $Q>0$ , requires $w_{de} +\frac{\xi }{3} <0$.

Let's now lay some ground rules for dynamical systems that are described by the Eqs \eqref{GrindEQ__3_1_} \cite{Zimdahl2}. To accomplish this, it is convenient to introduce an effective pressure $\Pi $ by $Q=-3H\Pi $ and to replace the derivatives with respect to cosmic time with derivatives with respect to $\ln a^{3} $ , denoted by a prime. Then, the dynamics of the two-component system are given by

\begin{equation} \label{GrindEQ__3_10_}
\begin{array}{l} {\frac{\rho '_{dm} }{\rho_{dm} } =-1-\frac{\Pi }{\rho_{dm} } ,} \\ {\frac{\rho '_{de} }{\rho_{de} } =-(1+w_{de} )+\frac{\Pi }{\rho_{de} } } \end{array}.
\end{equation}
or, alternatively, by
\begin{equation} \label{GrindEQ__3_11_}
\begin{array}{l} {\rho '=-\left(1+\frac{w_{de} }{1+r} \right)\rho ,} \\ {r'=r\left[w_{de} -\frac{\left(1+r\right)^{2} }{r\rho } \Pi \right]} \end{array}.
\end{equation}
In the interaction-free limit $\Pi =0$, the stationary point $r_{s} =0$  together with $w_{de} =-1$ corresponds   to the de Sitter space as the long-time limit of the SCM  model. This important result can be clarified in the following way. The system \eqref{GrindEQ__3_11_}, in the absence of interaction, is equivalent to the following system for the relative densities $\Omega_{dm} $ and $\Omega_{de} $
\begin{equation} \label{GrindEQ__3_12_}
\begin{array}{l} {\Omega '_{dm} =w_{de} \Omega_{dm} \Omega_{de} ,} \\ {\Omega '_{de} =-w_{de} \Omega_{dm} \Omega_{de} } \\ {} \end{array}
\end{equation}
The system has two stable points: $\Omega_{dm} =1,\; \Omega_{de} =0$ (Einstein-de Sitter Universe) and $\Omega_{dm} =0,\; \Omega_{de} =1$(de Sitter Universe), of which only the second one (for $w_{de} <0$ ) is stable.

The relevant critical points of the first equation of \eqref{GrindEQ__3_11_}  are given by

\begin{equation} \label{GrindEQ__3_13_}
r_{c} =-(1+w_{de} )
\end{equation}
Consequently, for positive values of $r$ , the existence of a critical point requires $w_{de} <-1$ , i.e., dark energy of the phantom type. This conclusion does not depend on the interaction. A non-zero stationary value for the ratio $r$  can be interpreted as an alleviation of the coincidence problem. The condition $r'=0$ together with \eqref{GrindEQ__3_13_}  provides us with

\begin{equation} \label{GrindEQ__3_14_}
\rho_{c} =-\frac{w_{de} }{1+w_{de} } \Pi_{c}.
\end{equation}
In general, $\Pi_{c} =\Pi_{c} \left(\rho_{c} ,r_{c} \right)$ . Therefore \eqref{GrindEQ__3_14_} is not an explicit relation for $\rho_{c} $ . Moreover, $\rho_{c} $ remains undetermined for a linear dependence of $\Pi $ on $\rho $ . This case is degenerate and does not admit a dynamical system analysis. On the other hand, for $\Pi \propto \rho $ the system of equations \eqref{GrindEQ__3_11_} breaks up into non-related equations, and can be subsequently solved.

Since $w<-1$ , a positive stationary energy density $\rho_{c} $  in \eqref{GrindEQ__3_14_} requires $\Pi_{c} <0$ , which is equivalent

to $Q_{c} >0$ . Regardless of the specific  interaction (excluding only a linear dependence $\left(\Pi \propto \rho \right)$), the existence of the critical points $r_{c} $  and $\rho_{c} $  requires a transfer from dark energy to dark matter . We emphasize that  the results for the critical points so far do not depend on the structure of interaction.

During comparisons of model dynamics with observational results, it is useful to analyse all dynamic variables as functions of redshift, not of time. Let's use the fact that
\[\frac{d}{dt} =\frac{d}{dz} \frac{dz}{da} \frac{da}{dt} =-(1+z)H(z)\frac{d}{dz} \]
and transform the base equations \eqref{GrindEQ__3_1_} to the form
\begin{equation} \label{GrindEQ__3_15_}
\begin{array}{l} {\frac{d\rho_{dm} }{dz} -\frac{3}{1+z} \rho_{dm} =-\frac{Q(z)}{(1+z)H(z)} ,} \\ {\frac{d\rho_{de} }{dz} -\frac{3}{1+z} (1+w_{de} )\rho_{de} =\frac{Q(z)}{(1+z)H(z)} } \end{array}.
\end{equation}
Also, let's introduce \cite{Solano} the dimensionless interaction function $I_{Q} (z)$ ,
\[I_{Q} (z)\equiv \frac{1}{\rho_{crit}^{0} (1+z)^{3} H(z)} Q\left(z\right)\]
Moving to relative densities, we finally get
\begin{equation} \label{GrindEQ__3_16_}
\begin{array}{l} {\frac{d\Omega_{dm} }{dz} -\frac{3}{1+z} \Omega_{dm} =-(1+z)^{2} I_{Q} (z),} \\ {\frac{d\Omega_{de} }{dz} -\frac{3}{1+z} (1+w_{de} )\Omega_{de} =(1+z)^{2} I_{Q} (z)} \end{array}.
\end{equation}
The function $I_{Q} (z)$ is useful during analysis of observational data \cite{Solano}.

\subsection{Simple Linear Models}\label{Delta_Q_a}

In general, the coupling term $Q$ can take any possible form $Q=Q\left(H,\rho_{dm} ,\rho_{de} ,t\right)$ . However, physically, it makes more sense that the coupling be time-independent. Among the time-independent options, preference is given to a factorized $H$ dependence $Q=Hq(\rho_{dm} ,\rho_{de} )$ . During this kind of factorization, the effects of the coupling on the dynamics of $\rho_{dm} $  and $\rho_{de} $  become effectively independent from the evolution of the Hubble scale $H$ .The latter is related to the fact that the time derivatives  that go into the conservation equation can be transformed in the following way: $d/dt\to Hd/d\ln a$. It is important to note, [29], that the decoupling of the dynamics of the two dark components from $H$ is valid in any theory of gravity, because it is based on the conservation equations. Any coupling of this type can be approximated at late times by a linear expansion
\begin{equation} \label{GrindEQ__3_17_}
q=q_{0}^{*} +q_{dm}^{*} \left(\rho_{dm} -\rho_{dm,0} \right)+q_{de}^{*} \left(\rho_{de} -\rho_{de,0} \right),
\end{equation}

the constants $q_{0}^{*} ,q_{dm}^{*} ,q_{de}^{*} $ can always be redefined in order to give the coupling $q$ the form
\begin{equation} \label{GrindEQ__3_18_}
q=q_{0} +q_{dm} \rho_{dm} +q_{de} \rho_{de}.
\end{equation}
Special cases of this general expression:
\begin{equation} \label{GrindEQ__3_19_}
\begin{array}{l} {q\propto \rho_{dm} ,\quad q_{0} =q_{de} =0;} \\ {q\propto \rho_{de} ,\quad q_{0} =q_{dm} =0;} \\ {q\propto \rho_{total,\quad } q_{0} =0,\; q_{dm} =q_{de} } \end{array}.
\end{equation}

Let's look at these special cases in greater detail. It can be shown \cite{Rosenfeld}, that the introduction of the coupling function $\delta $(a) between dark energy and dark matter  as
\begin{equation} \label{GrindEQ__3_20_}
\delta (a)=\frac{d\ln m_{\psi } (a)}{d\ln a}.
\end{equation}
(see Section 2) results in the following equation for the evolution of the DM energy density $\rho_{dm} $
\begin{equation} \label{GrindEQ__3_21_}
\dot{\rho }_{dm} +3H\rho_{dm} -\delta \left(a\right)H\rho_{dm} =0.
\end{equation}
The time dependence of the DM energy density is easily obtained as the solution of \eqref{GrindEQ__3_21_}
\begin{equation} \label{GrindEQ__3_22_}
\rho_{dm} (a)=\rho_{dm}^{(0)} a^{-3} \exp \left(-\int_{a}^{1}\delta (a')d\ln a' \right).
\end{equation}
This solution shows that the interaction causes $\rho_{dm} $ to deviate from the standard SCM scaling - $a^{-3} $.This is related to the fact that if the dark energy   is decaying into dark matter  particles, this  component will dilute more slowly compared to its conserved evolution. Consider the simple example of a constant coupling $\delta $ . In this case we obtain

\begin{equation} \label{GrindEQ__3_23_}
\rho_{dm} (a)=\rho_{dm,0} a^{-3+\delta }.
\end{equation}
The deviation from the standard evolution is characterized by a positive interaction constant $\delta $ .

The conservation of  the total energy density implies  that the dark energy density  should obey
\begin{equation} \label{GrindEQ__3_24_}
\dot{\rho }_{de} +3H\left(\rho_{de} +p_{de} \right)+\delta (a)H\rho_{dm} =0.
\end{equation}
The solution of this equation for a constant EoS parameter $w_{de} $  and constant coupling $\delta $is
\begin{equation} \label{GrindEQ__3_25_}
\rho_{de} (a)=\rho_{de,0} a^{-3\left(1+w_{de} \right)} +\frac{\delta \rho_{dm,0} }{\delta +3w_{de} } \left(a^{-3\left(1+w_{de} \right)} -a^{-3+\delta } \right).
\end{equation}
The first  term  is the usual evolution of dark energy at $\delta =0$. From this solution, it is easy to see that one must require a positive value of the coupling $\delta $ $>$ 0 in order to have a positive value of $\rho_{de} $  for earlier epochs of the Universe. For $w_{de} =-1,\; \delta \ne 0$ this expression reduced to
\begin{equation} \label{GrindEQ__3_26_}
\rho_{\Lambda } (a)=\rho_{\Lambda ,0} -\frac{\delta \rho_{dm,0} }{3-\delta } a^{-3+\delta }.
\end{equation}
This expression can be interpreted in terms of a time-dependent cosmological constant $\Lambda (t)$ (see section $\Lambda (t)$)

Before going further, let's also write, without any additional comments, the forms of the densities of energy $\rho_{de} (a)$ and $\rho_{dm} (a)$ for the case of $Q=\delta H\rho_{de} \; \left(\delta =const\right)$ :
\begin{equation} \label{GrindEQ__3_27_}
\begin{array}{l} {\rho_{de} (a)=\rho_{de0} a^{-\left[3\left(1+w_{de} \right)+\delta \right]} ,} \\ {\rho_{dm} \left(a\right)=\frac{-\delta \rho_{de0} }{3w_{de} +\delta } a^{-\left[3\left(1+w_{de} \right)+\delta \right]} +\left(\rho_{dm0} +\frac{\delta \rho_{de0} }{3w_{de} +\delta } \right)a^{-3} } \end{array}.
\end{equation}
Let's now look at a more general linear model for the expansion of a Universe that contains two fluids with the equations of state \cite{Chimento1,Barrow1}
\begin{equation} \label{GrindEQ__3_28_}
\begin{array}{l} {p_{1} =\left(\gamma_{1} -1\right)\rho_{1} ,} \\ {p_{2} =\left(\gamma_{2} -1\right)\rho_{2} } \end{array}.
\end{equation}
and energy exchange
\begin{equation} \label{GrindEQ__3_29_}
\begin{array}{l} {\dot{\rho }_{1} +3H\gamma_{1} \rho_{1} =-\beta H\rho_{1} +\alpha H\rho_{2} ,} \\ {\dot{\rho }_{2} +3H\gamma_{2} \rho_{2} =\beta H\rho_{1} -\alpha H\rho_{2} ,} \\ {} \end{array}
\end{equation}
Here $\alpha $ and $\beta $ are constants  describing  the energy exchanges between the two fluids. Using \eqref{GrindEQ__3_29_} and first Friedmann equation  we can eliminate the densities to obtain a single master  equation for $H(t)$ ,
\begin{equation} \label{GrindEQ__3_30_}
\begin{array}{l} {\ddot{H}+H\dot{H}\left(\alpha +\beta +3\gamma_{1} +3\gamma_{2} \right)+\frac{3}{2} H^{3} \left(\alpha \gamma_{1} +\beta \gamma_{2} +3\gamma_{1} \gamma_{2} \right)=} \\ {=\ddot{H}+AH\dot{H}+BH^{3} =0,} \\ {A\equiv \alpha +\beta +3\gamma_{1} +3\gamma_{2} ,\; B\equiv \frac{3}{2} \left(\alpha \gamma_{1} +\beta \gamma_{2} +3\gamma_{1} \gamma_{2} \right)} \end{array}
\end{equation}
The equation \eqref{GrindEQ__3_30_} has a simple solution
\begin{equation} \label{GrindEQ__3_31_}
H=\frac{h}{t} ,\; \; h\ne 0,
\end{equation}
as long as the following demand holds true
\begin{equation} \label{GrindEQ__3_32_}
Bh^{2} -Ah+2=0,
\end{equation}
Since the solution of this equation is
\begin{equation} \label{GrindEQ__3_33_}
h_{\pm } =\frac{A\pm \sqrt{A^{2} -8B} }{2B}
\end{equation}
real  power-law  solutions for $H(t)$  exist  if  $A^{2} \ge 8B$ . It can be shown [32], that for $\alpha ,\beta ,\gamma_{1} ,\gamma_{2} \ge 0$ and $\gamma_{1} \ne \gamma_{2} $ this inequality is always satisfied. For  $A^{2} >8B$ we find the solution
\begin{equation} \label{GrindEQ__3_34_}
H^{2} =a^{-A/2} \left(c_{1} a^{\sqrt{A^{2} -8B} /2} +c_{2} a^{-\sqrt{A^{2} -8B} /2} \right)
\end{equation}
where $c_{1} $ and $c_{2} $ are constants. As $a\to \infty $
\begin{equation} \label{GrindEQ__3_35_}
H^{2} \to a^{-\left(A-\sqrt{A^{2} -8B} \right)/2}
\end{equation}
and,  as $a\to 0$,
\begin{equation} \label{GrindEQ__3_36_}
H^{2} \to a^{-\left(A+\sqrt{A^{2} -8B} \right)/2}
\end{equation}
These two equations can be integrated to obtain
\begin{equation} \label{GrindEQ__3_37_}
a_{\pm } \propto t^{\left(A\pm \sqrt{A^{2} -8B} \right)/2B}.
\end{equation}
These asymptotics correspond to the solution of \eqref{GrindEQ__3_31_}, provided that \eqref{GrindEQ__3_33_} hold true. By integrating \eqref{GrindEQ__3_34_} we can show explicitly the existence of the above power-law attractors, and the smooth evolution of $a$  between them.

The conservation equations \eqref{GrindEQ__3_24_} can be used to construct the second-order differential equation
\begin{equation} \label{GrindEQ__3_38_}
\frac{\rho ''_{2} }{\rho_{2} } +A\frac{\rho '_{2} }{\rho_{2} } +2B=0
\end{equation}
where  primes denote derivative   with respect to the variable $N=\ln a$ . This equation can be solved for $\rho_{2} $ ,
\begin{equation} \label{GrindEQ__3_39_}
\rho_{2} =\rho_{20} a^{M},
\end{equation}
where $\rho_{20} $ is constant and $2M=-A\pm \sqrt{A^{2} -8B} $ . For the density of the second component, we find
\begin{equation} \label{GrindEQ__3_40_}
\rho_{1} =\rho_{10} a^{M},
\end{equation}
where $\rho_{10} =\frac{N+3\gamma +\alpha }{\beta } \rho_{20} $ is constant. It is immediately apparent that $\rho_{2} $  and $\rho_{1} $  evolve at the same rate and so the ratio $\rho_{2} /\rho_{1} $  is a constant quantity
\begin{equation} \label{GrindEQ__3_41_}
\frac{\rho_{2} }{\rho_{1} } =\frac{\beta }{N+3\gamma_{1} +\alpha }.
\end{equation}

It is this constant ratio of energy densities of two fluids  (during a period described by the power-law evolution \eqref{GrindEQ__3_37_}) with different  barotropic indices $\gamma_{1,2} $ that looks very promising from the point of view of the possible resolution of the coincidence problem (see section \ref{CCP}).

As an example of the effectiveness of the above analysis, let's look at the case of a decaying cosmological constant located in equilibrium with a radiation background. In this case $\gamma_{1} =0,\; \gamma_{2} =4/3,\; \alpha =0,\; \beta >0$ and therefore
\begin{equation} \label{GrindEQ__3_42_}
\begin{array}{l} {A=\beta +4,} \\ {B=2\beta ,} \\ {\delta =\frac{B}{A^{2} } =\frac{2\beta }{\left(\beta +4\right)^{2} } ,} \\ {h_{+} =\frac{1}{2} ,\; h_{-} =\frac{2}{\beta } } \end{array}
\end{equation}
The first  of these corresponds to the degenerate situation with pure radiation $\left(H=1/2t\right)$ . The second solution has $a\propto t^{2/\beta } $ and requires $\beta >3$  if the evolution of the Universe is to have a matter-dominated era following a radiation era. As the value $\beta $  increases, the dominance of the vacuum contribution slows the expansion whereas in the limit $\beta \to 0$  the expansion rate increases without bound and the dynamics approach the usual vacuum-energy dominated de Sitter expansion with $a\propto \exp \left(\sqrt{\rho_{1} } t/3\right)$ .

\subsection{Non-linear interaction in the dark sector} \label{non-linear}

We have already stated multiple times that our current lack of understanding of the structure of the dark components leaves us with only dimensional limitations on the choice of the form of interaction between them. Previously, we analysed linear interactions: the interaction term in the conservation equations of the individual components is proportional either to DM density, to DE density, or to a linear combination of both densities. However,  from a physical point of view, an interaction between two components should depend on the product of the abundances of the individual components, as, for instance, in chemical or nuclear reactions. Consequently, a product coupling, i.e., an interaction proportional to the product of DM density and DE density looks more appealing. An analysis of cosmological models with specific non-linear interactions was performed in \cite{Mangano,Baldi,Jian-Hua,Yin-Zhe}.

Following \cite{Arevalo}, we investigate, in a flat Universe, the dynamics of a simple two-component model $\left(de+dm\right)$ with a number of non-linear interactions . Motivated by the structure
\begin{equation} \label{GrindEQ__3_43_}
\begin{array}{l} {\rho_{dm} =\frac{r}{1+r} \rho ,\quad \rho_{de} =\frac{1}{1+r} \rho ,\quad r\equiv \frac{\rho_{dm} }{\rho_{de} } ,\quad \rho =\rho_{dm} +\rho_{de} } \\ {} \end{array},
\end{equation}
we consider  the ansatz concern to effective pressure $\left(Q=-3H\Pi \right)$
\begin{equation} \label{GrindEQ__3_44_}
\Pi =-\gamma \rho ^{m} r^{n} \left(1+r\right)^{s},
\end{equation}
where $\gamma $  is a positive coupling constant . The powers $m,n,s$  specify the interaction. For fixed values $m,n,s$the only free parameter is $\gamma $ .  A linear dependence of $\Pi $ on $\rho $  corresponds to $m=1$ . The effective interaction pressure $\Pi $ is proportional to powers of products of the densities of the components for the special cases $m=s$. Notice that, according to Friedmann's equations, ($\rho \propto H^{2} $). This implies that the interaction quantity $Q$  is not necessarily linear in the Hubble rate. For $m=s$ the ansatz \eqref{GrindEQ__3_23_} is equivalent to
\begin{equation} \label{GrindEQ__3_45_}
Q=3\gamma H\rho_{de}^{m-n} \rho_{dm}^{n} =3\gamma H\rho_{de}^{m} r^{n}.
\end{equation}
The ansatz \eqref{GrindEQ__3_23_}also includes the previously analysed linear cases. The combination $\left(m,n,s\right)=\left(1,1,-1\right)$ corresponds to $Q=3\gamma H\rho_{dm} $ , while $\left(m,n,s\right)=\left(1,0,-1\right)$reproduces $Q=3\gamma H\rho_{de} $. We can therefore state that the ansatz \eqref{GrindEQ__3_23_} contains a large variety of interactions, which have been studied in literature [38-43] as special cases.

We consider now three particular combinations of the parameters $\left(m,n,s\right)$ which give rise to analytically solvable models with non-linear interaction terms.

\subsubsection{Case $Q=3H\gamma \frac{\rho_{dm} \rho_{dx} }{\rho } ,\quad \left(m,n,s\right)=\left(1,1,-2\right)$ }.

For such an interaction, the system  \eqref{GrindEQ__3_11_} is reduced to
\begin{equation} \label{GrindEQ__3_46_}
\begin{array}{l}
{\rho '=-\left(1+\frac{w_{de} }{1+r} \right)\rho ,} \\ {r'=r\left(w_{de} +\gamma \right)}.
\end{array}
\end{equation}
The solutions of this system are
\begin{equation} \label{GrindEQ__3_47_}
\begin{array}{l}
 {r=r_{0} a^{3\left(w_{de} +\gamma \right)} ,} \\ {\rho =\rho_{0} a^{-3\left(1+w_{de} \right)} \left[\frac{1+r_{0} a^{3(w_{de} +\gamma )} }{1+r_{0} } \right]^{\frac{w_{de} }{w_{de} +\gamma } } ,} \\ {\rho_{dm} =\rho_{dm0} a^{-3(1-\gamma )} \left[\frac{1+r_{0} a^{3(w_{de} +\gamma )} }{1+r_{0} } \right]^{-\frac{\gamma }{w_{de} +\gamma } } ,} \\
 {\rho_{de} =\rho_{de0} a^{-3(1+w)} \left[\frac{1+r_{0} a^{3(w_{de} +\gamma )} }{1+r_{0} } \right]^{-\frac{\gamma }{w_{de} +\gamma } } }.
 \end{array}
\end{equation}
An ansatz $r=r_{0} a^{-\xi } $ for the energy density ratio corresponds to $\gamma =-\left(w_{de} +\frac{\xi }{3} \right)$ . For$a\ll 1$ (matter-dominated epoch), we obtain the correct behaviour of the density - $\rho \propto a^{-3} $ . The SCM model is recovered for $w_{de} =-1,\; \gamma =0\; \left(\xi =3\right)$ .

\subsubsection{Case $Q=3H\gamma \frac{\rho_{dm}^{2} }{\rho } ,\quad \left(m,n,s\right)=\left(1,2,-2\right)$}

The analytical solution in this case is
\begin{equation} \label{GrindEQ__3_48_}
\begin{array}{l} {r=r_{0} \frac{w_{de} }{\left(w_{de} +\gamma r_{0} \right)a^{-3w_{de} } -\gamma r_{0} } ,} \\ {\rho =\rho_{0} a^{-3\left(1-\frac{\gamma w_{de} }{w_{de} -\gamma } \right)} \left[\frac{\left(w_{de} +\gamma r_{0} \right)a^{-3w_{de} } +r_{0} \left(w_{de} -\gamma \right)}{w_{de} (1+r_{0} )} \right]} \end{array}.
\end{equation}
For $a\ll 1$  (the high redshift limit) the ratio $r$ becomes a constant,$r\to \left|w\right|/\gamma $ . In the opposite limit $(a\gg 1)$, $r\propto a^{-3} $, as in the SCM case.

\subsubsection{Case $Q=3H\gamma \frac{\rho_{de}^{2} }{\rho } ,\quad \left(m,n,s\right)=\left(1,0,-2\right)$   }

For $w_{de} <0,$ i.e. $w_{de} =-\left|w_{de} \right|$, the solutions are
\begin{equation} \label{GrindEQ__3_49_}
\begin{array}{l} {r=\left(r_{0} -\frac{\gamma }{\left|w_{de} \right|} \right)a^{-3\left|w_{de} \right|} +\frac{\gamma }{\left|w_{de} \right|} ,} \\ {\rho =\rho_{0} a^{-3\left(1-\frac{w_{de}^{2} }{\left|w_{de} \right|+\gamma } \right)} \left[\frac{\left|w_{de} \right|+\gamma +\left(\left|w_{de} \right|r_{0} -\gamma \right)a^{-3\left|w_{de} \right|} }{\left|w_{de} \right|\left(1+r_{0} \right)} \right]^{\frac{\left|w_{de} \right|}{\left|w_{de} \right|+\gamma } } } \end{array}.
\end{equation}
The ratio $r$ scales as $a^{-3\left|w_{de} \right|} $ for $a\ll 1$ . For $w_{de} =-1$, this coincides with the scaling of its SCM counterpart. In the opposite limit, $a\gg 1$ (far future), the $\rho $-solution corresponds to a matter dominated period, $\rho \propto a^{-3\left(1-\frac{w_{de}^{2} }{\left|w_{de} \right|+\gamma } \right)} $, which generally does not correspond to a de Sitter phase.

In conclusion of this section, let's try to solve the opposite problem. Instead of postulating  the form of the interaction, let's fixate the ratio
\begin{equation} \label{GrindEQ__3_50_}
r=\frac{\rho_{dm} }{\rho_{de} } =f(a).
\end{equation}
where $f(a)$  is any differentiable function of the scale factor. We then have
\begin{equation} \label{GrindEQ__3_51_}
\begin{array}{l} {\dot{\rho }_{dm} =\dot{\rho }_{de} f+\rho_{de} f'\, \dot{a};} \\ {\dot{\rho }_{de} =\frac{\dot{\rho }_{dm} }{f} -\frac{\rho_{dm} f'\dot{a}}{f^{2} } ,\quad f'=\frac{df}{da} } \end{array}
\end{equation}
From this, we find that
\begin{equation} \label{GrindEQ__3_52_}
Q=\frac{f}{1+f} \left(\frac{f'}{f} a-3w_{de} \right)H\rho_{de}.
\end{equation}
To see that the interaction is non-linear in nature, note that $f/\left(1+f\right)=\Omega_{dm} $ .Therefore,
\begin{equation} \label{GrindEQ__3_53_}
Q=\left(\frac{f'}{f} a-3w_{de} \right)H\rho_{de} \Omega_{dm}.
\end{equation}
We note that if $f=a^{\xi } $ , then
\begin{equation} \label{GrindEQ__3_54_}
Q=\left(\xi -3w_{de} \right)H\rho_{de} \Omega_{dm}.
\end{equation}
For SCM, $\xi =3,\; w_{de} =-1$, and we return to the obvious result - $Q=0$ .

\subsection{Cosmological models with a change of the direction of energy transfer}

In this section, we consider one more type of interaction, a $Q$ \cite{Wei}, whose sign (i.e., the direction of energy transfer) changes when the mode of decelerated expansion is replaced by the mode of accelerated expansion, and vice versa. Recently, publications have appeared \cite{Abdalla1,Abdalla2}, in which attempts, based on observational data, are made to determine not only the possibility itself of interaction existing in the dark sector, but also its concrete form and sign.  In this analysis, the whole set of redshifts $z$  is divided into intervals, in each of which the function $\delta (z)=Q/\left(3H\right)$  is considered to be constant. This analysis has shown that $\delta (z)$  most likely takes a zero value, $\delta =0$ , in the range of red shifts $0.45\le z\le 0.9$. It turns out that this remarkable result gives rise to new problems. Indeed, when an interaction is considered in literature  for a given model, the interaction is almost always either positive or negative, i.e. it cannot change sign. A change of sign is only possible in the case $Q\propto \gamma (t)\rho,$ where $\gamma (t)$  can change the sign of $Q$ , or in the case $Q=3H\left(\alpha \rho_{dm} +\beta \rho_{de} \right),$ where $\alpha $  and $\beta $  have different signs.

As noted in Ref. \cite{Abdalla1}, the solution to this problem requires the introduction of a new type of interaction, capable of changing its sign during the evolution of the Universe. In Ref. \cite{Wei}, one such type of interaction $Q$ was proposed, and its cosmological consequences were examined. It was noted  that the range of redshifts within which the function $\dagger$$\delta (z)$ must change sign includes the moment at which expansion of the Universe stopped decelerating and started accelerating .

Therefore, the simplest type of interaction that can explain the
above mentioned property is the case when the source $Q$ is
proportional to the deceleration parameter $q:$
\begin{equation}
\label{Q(q)}
  Q=q(\alpha\dot{\rho}+3\beta H\rho)\,
\end{equation}
where $\alpha$ and $\beta$ are dimensionless constants, and the sign
of $Q$ will change with the transition of Universe from the
decelerated expansion stage $(q>0)$ to the accelerated stage $(q<0).$
The authors also consider the cases
\begin{eqnarray}
&&Q=q(\alpha\dot{\rho}_m+3\beta H\rho_m),\label{q_eq5}\\
 &&Q=q(\alpha\dot{\rho}_{tot}+3\beta H\rho_{tot}),\label{q_eq6}\\
 &&Q=q(\alpha\dot{\rho}_{_{DE}}+3\beta H\rho_{_{DE}})\label{q_eq7}.
\end{eqnarray}
The paper \cite{HaoWei_Q(q)1} considers a model of the Universe with a
decaying cosmological constant
$$
 \dot{\rho}_\Lambda=-Q\,.
$$
The Friedmann and Raychaudhuri equations thus take the form
\begin{eqnarray}
   H^2 &=& \frac{\kappa^2}{3}\rho_{tot}=  \frac{\kappa^2}{3}(\rho_\Lambda+\rho_m)\,,\label{q-eq6}\\
   \dot{H} &=&-\frac{\kappa^2}{2}(\rho_{tot}+p_{tot})=
 -\frac{\kappa^2}{2}\rho_m,\label{q-eq7}
 \end{eqnarray}
where $\kappa^2\equiv 8\pi G.$ Following the paper
\cite{HaoWei_Q(q)1}, in the succeeding subsections we consider
cosmological models with interaction of the type
\eqref{q_eq5}-\eqref{q_eq7}.
\subsubsection{Case  $Q=q(\alpha\dot{\rho}_m+3\beta H\rho_m)$}
To start off, we consider the case when the interaction takes
the form \eqref{q_eq5} and insert this expression into the
conservation equation \eqref{GrindEQ__3_1_}, resulting in the following
 \begin{equation}\label{q-eq19}
 \dot{\rho}_m=\frac{\beta q-1}{1-\alpha q}\cdot 3H\rho_m\,.
\end{equation}
Substituting the obtained expression into the equation
\eqref{q_eq5}, we finally get
\begin{equation}\label{q-eq20}
 Q=\frac{\beta-\alpha}{1-\alpha q}\cdot 3qH\rho_m.
\end{equation}
From the equation \eqref{q-eq7}, one obtains
 \begin{equation}\label{q-eq21}
\rho_m=-\frac{2}{\kappa^2}\dot{H}.
\end{equation}
Inserting it into the equation \eqref{q-eq19}, one finds that
 \begin{equation}\label{q-eq22}
\ddot{H}=\frac{\beta q-1}{1-\alpha q}\cdot 3H\dot{H}\,,
\end{equation}
Thus we obtained a second order differential equation for the
function $H(t).$ Transforming from the time derivative to
differentiation with respect to the scale factor (denoted by the
prime ${}^\prime$), the equation \eqref{q-eq22} takes on the
form
 \begin{equation}\label{q-eq23}
 aH^{\prime\prime}+\frac{a}{H}H^{\prime\,2}+H^\prime=
 \frac{\beta q-1}{1-\alpha q}\cdot 3H^\prime\, .
\end{equation}
This expression represents a second order differential expression
for the function $H(a).$ Note that the deceleration parameter
\[
 q=-1-\frac{\dot{H}}{H^2}=-1-\frac{a}{H}H^\prime\,,
\]
is also a function of $H$ and $H^\prime$, except in the case of
 $\alpha\not=0,$ the equation has no exact solution and it
 represents a transcendental differential equation of the second order.
 That is why we consider solely the case of $\alpha=0.$ Thus, the interaction \eqref{q_eq5} takes the form
 \[
 Q=3\beta qH\rho_m.
\]
With $\alpha=0$, the solution \eqref{q-eq23} can be presented in the
form
  \begin{equation}\label{q-eq26}
 H(a)=C_{12}\left[\,3C_{11}(1+\beta)-(2+3\beta)
 \,a^{-3(1+\beta)}\,\right]^{1/(2+3\beta)},
 \end{equation}
where $C_{11}$ and $C_{12}$ are the integration constants determined
below. We find the relative density of dark matter as the following
 \begin{equation}\label{q-eq27}
 \Omega_m\equiv \frac{\kappa^2 \rho_m}{3H^2}
 =-\frac{2\dot{H}}{3H^2}=-\frac{2aH^\prime}{3H}\,.
  \end{equation}
Inserting the equation \eqref{q-eq26} into \eqref{q-eq27}, one gets
\begin{equation}\label{q-eq28}
 \Omega_m=\frac{2\left(1+\beta\right)}{2+3\beta
 -3C_{11}\left(1+\beta\right)\,a^{3\left(1+\beta\right)}}\,.
 \end{equation}
With the requirements $\Omega_m(a=1)=\Omega_{m0}$ and $H(a=1)=H_0,$
the integration constants take the form
 \begin{eqnarray}
   C_{11} &=& \frac{\Omega_{m0}(2+3\beta)
 -2(1+\beta)}{3\Omega_{m0}(1+\beta)}, \label{q-eq29}\\
   C_{12} &=& H_0\left[3C_{11}(1+\beta) -(2+3\beta)\right]^{-1/(2+3\beta)}. \label{q-eq30}
 \end{eqnarray}

Substitution of the expressions \eqref{q-eq29} and \eqref{q-eq30}
into the equation \eqref{q-eq26} finally gives the result
 \begin{equation}\label{q-eq31}
 E\equiv\frac{H}{H_0}=\left\{1-
 \frac{2+3\beta}{2(1+\beta)}\,\Omega_{m0}\left[1-
 (1+z)^{3(1+\beta)}\right]\right\}^{1/(2+3\beta)}.
 \end{equation}
The model contains two free parameters: $\Omega_{m0}$ and
 $\beta.$ We note that if $\beta=0,$ the equation \eqref{q-eq31} reduces to
 $E(z)=\left[\Omega_{m0}(1+z)^3+\left(1-\Omega_{m0}\right)
 \right]^{1/2},$ which is equivalent to the $\Lambda$CDM model.
Using the relation \[q(z)=
-\frac{(1+z)}{E(z)}\frac{d}{dz}\left(\frac{1}{E(z)}\right)-1,\] one
finds the dependency of the deceleration parameter on the redshift in
the considered model
\begin{equation}\label{Q(q)_1}
   q(z)= -1+\frac{3}{2}\Omega_{m0}\frac{(1+z)^{3(1+\beta)}}{E^{(2+3\beta)}}.
\end{equation}
The effective parameter of the equation of state is known to equal
\[w_{\rm eff}\equiv \frac{p_{tot}}{\rho_{tot}}=\frac{(2q-1)}{3}.\]
 \begin{center}
 \begin{figure}[tbhp]
 \centering
 \includegraphics[width=1.0\textwidth]{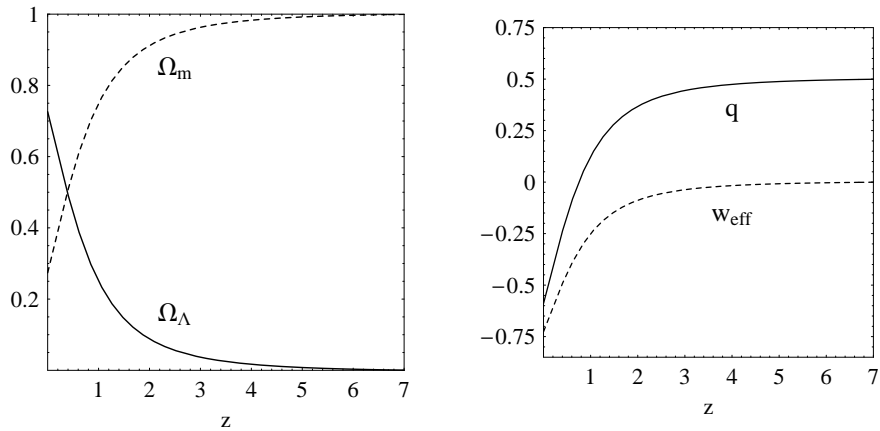}
 \caption{\label{q-fig2} $\Omega_m,$ $\Omega_\Lambda,$ $q$ and
 $w_{\rm eff}$ as functions of the redshift $z$ at $\Omega_{m0}=0.2738$ and $\beta=-0.010$ in the case $Q=3\beta qH\rho_m$\cite{HaoWei_Q(q)1}.}
 \end{figure}
 \end{center}
 The figure \ref{q-fig2} presents the plots for dependency of some
 cosmological parameters on the redshift $z.$ The free parameters $\Omega_{m0}$
 and $\beta$ were chosen to provide the best agreement with observations.
One can find that in the considered model, the transition from
decelerated expansion $(q>0)$ to accelerated expansion $(q<0)$ took
place at $z_t=0.7489,$ the parameter $\beta$ is negative and
therefore dark matter decays into dark energy when $z>z_t,$ and vice
versa at $z<z_t.$ The Universe lacks interaction in the dark sector at
$z_t.$
\subsubsection{Case
 $Q=q(\alpha\dot{\rho}_{tot}+3\beta H\rho_{tot})$}
Now we consider the case \eqref{q_eq6}, and proceeding completely
analogously to the above considered case, we obtain
\begin{equation}\label{q-eq32}
    Q=\frac{3qH^3}{\kappa^2}\left(2\alpha\frac{\dot{H}}{H^2}
 +3\beta\right).
\end{equation}
Inserting the equations \eqref{q-eq21} and \eqref{q-eq32} into
\eqref{GrindEQ__3_1_}, and transforming, as before, to differentiation with respect to
the scale factor, we obtain
 \begin{equation}\label{q-eq34}
 aH^{\prime\prime}+\frac{a}{H}H^{\prime\,2}+
 \left(4+3\alpha q\right)H^\prime+\frac{9\beta qH}{2a}=0\,.
\end{equation}
As in the previous case, we have once again obtained a differential equation
of the second order for the function $H(a).$ The exact solution exists only
in the case of $\alpha=0:$
 \begin{equation}\label{q-eq36}
 H(a)=C_{22}\cdot a^{-3(2-3\beta+r_1)/8}\cdot\left(a^{3r_1/2}+C_{21}\right)^{1/2},
 \end{equation}
where $C_{21},$ $C_{22}$ are integration constants and  $
r_1\equiv\sqrt{4+\beta\left(4+9\beta\right)}.$ Inserting
\eqref{q-eq36} into \eqref{q-eq27}, we get
 \begin{equation}\label{q-eq38}
 \Omega_m=\frac{1}{4}\left[2-3\beta
 +\left(\frac{2C_{21}}{a^{3r_1/2}+C_{21}}-1\right)r_1\right].
 \end{equation}
The integration constants are determined as usual from the condition
$\Omega_m(a=1)=\Omega_{m0},~H(a=1)=H_0:$
 \begin{equation}\label{q-eq39}
 C_{21}=-1+\frac{2\,r_1}{2-3\beta-4\Omega_{m0}+r_1},~~C_{22}=H_0\left(1+C_{21}\right)^{-1/2}.
 \end{equation}
We finally get
 \begin{equation}\label{q-eq41}
 E\equiv\frac{H}{H_0}=(1+z)^{3(2-3\beta+r_1)/8}\cdot
 \left[\frac{(1+z)^{-3r_1/2}+C_{21}}{1+C_{21}}\right]^{1/2}.
 \end{equation}
In the considered model there are also two free parameters
$\Omega_{m0}$ and $\beta.$ Using the condition $0\leq\Omega_m\leq 1$
with $a\to 0,$ from the equation \eqref{q-eq38} one gets $\beta\geq
0.$ The best agreement of the model under consideration with
observational data occurs at $\Omega_{m0}=0.2701$ and $\beta=0.0.$
This means that the considered interaction model is in worse agreement
with observations than $\Lambda CDM.$ A more detailed discussion can be found in the paper \cite{HaoWei_Q(q)1} by the
author of the considered model. The transition from the decelerated
expansion phase $(q>0)$ to the accelerated phase $(q<0)$ occurs at
$z_t=0.7549.$
 \begin{center}
 \begin{figure}[tbhp]
 \centering
 \includegraphics[width=1.0\textwidth]{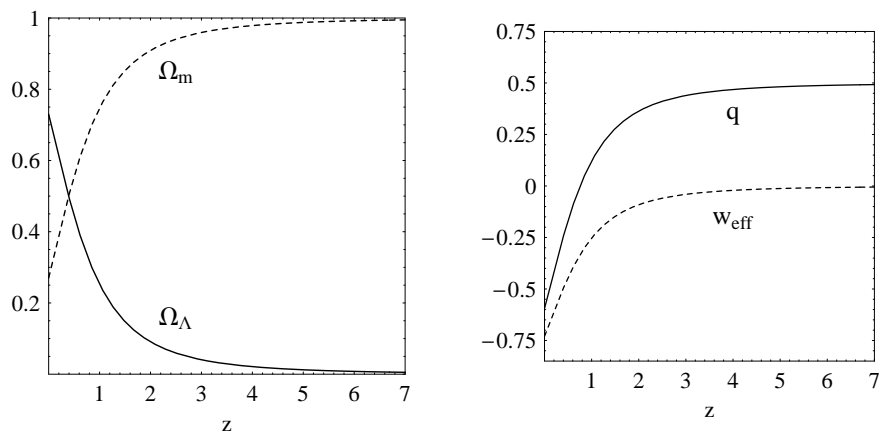}
 \caption{\label{q-fig4} Same as on Fig.~\ref{q-fig2}, but for the case of interaction of the form
 $Q=3\beta qH\rho_{tot}$ under the condition $\beta\geq 0$\cite{HaoWei_Q(q)1}.}
 \end{figure}
 \end{center}
\subsubsection{Case
 $Q=q(\alpha\dot{\rho}_\Lambda+3\beta H\rho_\Lambda)$}
For the conclusion we consider the case \eqref{q_eq7}. Following the
same procedure as in the two preceding cases, one obtains
 \begin{equation}\label{q-eq43}
 Q=\frac{3\beta qH\rho_\Lambda}{1+\alpha q}\,.
 \end{equation}
With the equation \eqref{q-eq21}, one has
 \begin{equation}\label{q-eq44}
 \rho_\Lambda=\frac{3}{\kappa^2}H^2-\rho_m=
 \frac{1}{\kappa^2}\left(3H^2+2\dot{H}\right).
\end{equation}
Therefore the equation for the Hubble parameter in terms of the
scale factor takes the form:
 \begin{equation}\label{q-eq46}
 aH^{\prime\prime}+\frac{a}{H}H^{\prime\,2}
 +\left(4+\frac{3\beta q}{1+\alpha q}\right)H^\prime+
 \frac{9\beta qH}{2a(1+\alpha q)}=0.
\end{equation}
The exact solution can be obtained in the case of $ Q=3\beta
qH\rho_\Lambda,$ namely
\begin{equation}\label{q-eq48}
 H(a)=C_{32}\cdot a^{-3(2-5\beta+r_2)/[4(2-3\beta)]}\cdot
 \left(a^{3r_2/2}+C_{31}\right)^{1/(2-3\beta)},
\end{equation}
where $C_{31},$ $C_{32}$ are the integration constants, and
$r_2\equiv\sqrt{\left(2-\beta\right)^2}=\left|\,2-\beta\,\right|\,.$
Inserting \eqref{q-eq48} into \eqref{q-eq27}, we get
\begin{equation}\label{q-eq50}
 \Omega_m=\frac{1}{2\left(2-3\beta\right)}\left[2-5\beta+
 \left(\frac{2C_{31}}{a^{3r_2/2}+C_{31}}-1\right)r_2\right]\,.
 \end{equation}
Assuming that $\Omega_m(a=1)=\Omega_{m0}$ and $H(a=1)=H_0,$ we can
write
 \begin{equation}\label{q-eq51}
 C_{31}=-1+\frac{2\,r_2}{2-5\beta+
 r_2+2\Omega_{m0}\left(3\beta-2\right)},~~C_{32}=H_0\left(1+C_{31}\right)^{1/(3\beta-2)},
  \end{equation}
and finally get
 \begin{equation}\label{q-eq53}
 E\equiv\frac{H}{H_0}=(1+z)^{3\left(2-5\beta+r_2\right)/\left[
 4\left(2-3\beta\right)\right]}\cdot\left[\frac{(1+z)^{-3r_2/2}
 +C_{31}}{1+C_{31}}\right]^{1/\left(2-3\beta\right)}.
\end{equation}
As before, the model has two free parameters : $\Omega_{m0}$ and $\beta.$
The maximum plausibility method for the free parameters of the
considered model gives the result \cite{HaoWei_Q(q)1}
$\Omega_{m0}=0.2717,~\beta=0.0136.$ Unlike the two preceding models,
the observational data analyzed in \cite{HaoWei_Q(q)1} give
evidence in favor of $\beta>0.$ A more detailed discussion can be found in the paper \cite{HaoWei_Q(q)1} by the
author of the model.
 \begin{center}
 \begin{figure}[tbp]
 \centering
 \includegraphics[width=1.0\textwidth]{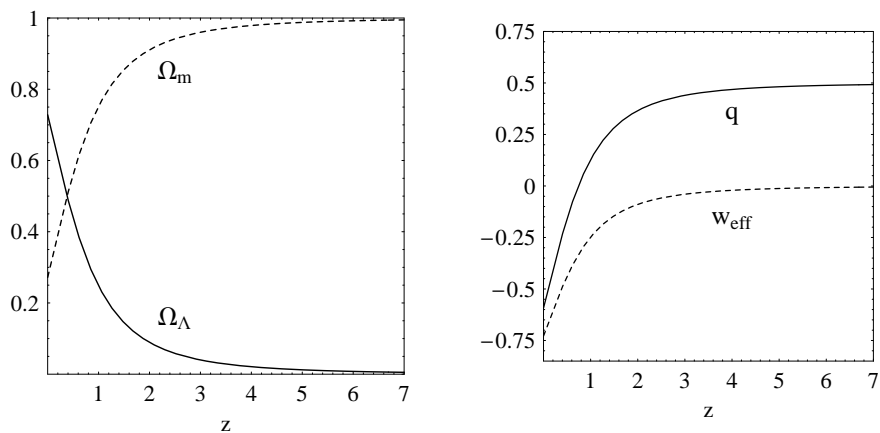}
 \caption{\label{fig8} Same as on Fig.~\ref{q-fig2}, but for the case of $Q=3\beta qH\rho_\Lambda$\cite{HaoWei_Q(q)1}.}
 \end{figure}
 \end{center}
The plot \ref{fig8} displays the dependencies of the deceleration
parameter and the effective equation of state parameter $w_{\rm
eff}\equiv p_{tot}/\rho_{tot}=(2q-1)/3$ as functions of the redshift
$z$, with the parameters obtained by the maximum plausibility method.
It is easy to show that the transition from the decelerated
expansion phase $(q>0)$ to the accelerated phase $(q<0)$ takes place
at $z_t=0.7398.$ As the parameter $\beta$ obtained from observations
satisfies $\beta>0,$ the dark energy decays into dark matter
($Q>0$) for $z>z_t,$ and vice versa $(Q<0)$ for $z<z_t.$

\subsection{Degeneration problem}

What are the root causes of degeneration? If space is uniform and isotropic, its metric is defined by one function - the scale factor $a(t)$ . There are two independent functions in the energy-momentum tensor, $\rho (t)$ and $p(t)$ . The Friedmann equations can only obtain the behaviour of one of them, usually taken to be $\rho $(t), while the pressure is defined with the help of the equation of state $p(t)=w(t)\rho (t)$ so that $w(t)$ is a function of time. Photons and baryonic matter are detected through their non-gravitational interactions, and their contribution to the energy-momentum tensor  can be measured directly. However,  if dark components are only detected gravitationally, we can only measure the total  energy-momentum tensor  $T_{(de)\,\mu \nu } +T_{(dm)\,\mu \nu } $. Hence there is a degeneracy between the dark energy equation of state $w(t)$ and the dark matter density parameter $\Omega_{dm} $ . Without additional assumptions, we cannot measure either of them. Any further freedom, like sub-dividing the dark EMT into dark matter and dark energy, or introducing couplings between the dark constituents, cannot be directly measured and will introduce degeneracy  \cite{Kunz}. As an example, let's look at a flat Universe composed of matter and dark energy with an unknown $w(z)$  and a given $H(z)$ .  In this case
\begin{equation} \label{GrindEQ__3_59_}
w(z)=\frac{H(z)^{2} -\frac{2}{3} H(z)H'(z)(1+z)}{H_{0}^{2} \Omega_{m} (1+z)^{3} -H^{2} (z)} ,\quad H'(z)=\frac{dH}{dz}.
\end{equation}
We see that for any choice of $\Omega_{m} $, there is a corresponding $w(z)$  which reproduces the measured expansion history of the Universe $H(z)$ .

Let's now look at the degeneration problem in models where the dark energy and the dark matter interact. The total energy momentum tensor for the dark components  has to be conserved. As long as $\left(T_{\left(dm\right)\mu \eta } +T_{\left(de\right)\mu \eta } \right)_{;\mu } =0$ holds true,  we can either keep it as a single unified  dark fluid model, we can divide it into a coupled dark matter -- dark energy system,  or we can also divide it into uncoupled dark matter and dark energy.

Let's analyse the simplest form of interaction between the dark components
\begin{equation} \label{GrindEQ__3_60_}
\begin{array}{l} {\dot{\rho }_{dm} +3H\rho_{dm} =Q(t),} \\ {\dot{\rho }_{de} +3H\left(1+w\right)\rho_{de} =-Q(t),} \\ {Q(t)=\gamma H\rho_{dm} ,\quad \gamma =const} \end{array}.
\end{equation}
The equations \eqref{GrindEQ__3_60_} are easily solved:
\begin{equation} \label{GrindEQ__3_61_}
\begin{array}{l} {\rho_{dm} =\rho_{dm0} (1+z)^{3-\gamma } ,} \\ {\rho_{de} =\left(\rho_{de0} +\rho_{dm0} \frac{\gamma }{\gamma +3w} \right)\left(1+z\right)^{3\left(1+w\right)} ,} \\ {H^{2} =H_{0}^{2} \left[\Omega_{dm0} \left(1-\frac{\gamma }{\gamma +3w} \right)\left(1+z\right)^{3-\gamma } +\left(1-\frac{3\Omega_{dm0} w}{\gamma +3w} \right)\left(1+z\right)^{3\left(1+w\right)} \right]} \end{array}.
\end{equation}
 Using $H(z)$ we can then derive a family of uncoupled models, using Eq. \eqref{GrindEQ__3_29_}, as well as families of models with other interactions.

\subsection{Duality invariance and dynamics of interacting components}

Regardless of the fact than an equation's symmetry does not always carry over into its solutions, it nevertheless significantly simplifies the process of finding these solutions, and also has an impact on their structure. Often, only symmetry-based ideas allow us to decrease the number of dynamical variables and to reach an understanding of complicated solutions. A classic example is the well known cosmological principle, which allows us to simplify the complicated, non-linear Einstein field equations to the relatively simple Friedmann equations. It is of great importance that symmetry can correlate solutions that correspond to different stages of evolution of the dynamical system.

The Hubble parameter is present in the first first  Friedmann equation quadratically. This gives rise to a useful symmetry within a class of FLRW models \cite{0609104}.  Because of this quadratic dependence, Friedmann's equation remains invariant under a transformation  $H\to -H$ for the spatially flat case. This means it describes both expanding and contracting solutions.  The transformation $H\to -H$ can be seen as a consequence of the change $a\to 1/a$ of the scale factor of the FLRW  metric.

If, instead of just the first Friedmann equation, we want to make the whole system of Universe-describing equations invariant relative to this transformation, we must expand the set of values that undergo symmetry transformations. Then, when we refer to a duality transformation, we have in mind the following set of transformations

\begin{equation} \label{duality_1_}
H\to \bar{H}=-H,\quad \rho \to \bar{\rho }=\rho ,\quad p\to \bar{p}=-2\rho -p.
\end{equation}
As a result of this transformation, the conservation equation $\dot{\rho }+3H\left(\rho +p\right)=0$    remains invariant due to $\rho +p\to -\left(\rho +p\right)$ . Consequently, if the weak energy condition is satisfied  in a given cosmological  model, i.e. $\rho +p\ge 0,\; \rho \ge 0$ , it is violated in its ``dual'', and vice versa. The transformation law $H\to \bar{H}=-H$   implies the transformation rule $a\to \bar{a}=1/a$  for the scale factor. Accordingly, if a certain configuration (say, the unbarred one) describes a phase of contraction, the barred one describes a phase of expansion. These cosmological solutions are said to be ``dual'' to each other. In particular, there is a duality between a final  contracting big crunch and a final expanding big rip. In general, the barotropic indices  $\gamma \equiv w+1$ will change as a result of a duality transformation according to
\begin{equation} \label{duality_2_}
\gamma =\frac{\rho +p}{\rho } \to \bar{\gamma }=\frac{\bar{\rho }+\bar{p}}{\bar{\rho }} =-\bar{\gamma }.
\end{equation}
The only invariant case is $p=-\rho $. This is related to the fact that the de Sitter Universe is free of singularities.

Let's now extend the technique of dual symmetry transformations that preserve the form of Einstein's equations to the case where the expansion of the Universe is dominated by two fluids (dark matter and dark energy) that interact with each other \cite{0505096}. Following this article, let us consider a homogeneous, isotropic and spatially flat  Universe filled with two fluids with the energy densities and pressures $\rho _{i} $  and $p_{i} $  (with $i=1,2$ ) respectively. Then, the Friedmann equation and the  conservation equation are
\begin{equation} \label{duality_3_}
\begin{array}{l} {3H^{2} =\rho _{1} +\rho _{2} ,} \\ {\dot{\rho }_{1} +\dot{\rho }_{2} +3H\left(\rho _{1} +\rho _{2} +p_{1} +p_{2} \right)=0} \end{array}.
\end{equation}
In this  general scenario there is a dual symmetry relating this cosmology to another one (with two fluids of energy densities  and pressures, $\bar{\rho }_{i} $and $\bar{p}_{i} $), generated by
\begin{equation} \label{duality_4_}
\begin{array}{l} {\bar{\rho }_{1} =\alpha \rho _{1} +\left(1-\beta \right)\rho _{2} ,} \\ {\bar{\rho }_{2} =\left(1-\alpha \right)\rho _{1} +\beta \rho _{2} ,} \\ {\bar{H}=-H} \end{array}.
\end{equation}
where the parameters of the transformation are
\begin{equation} \label{duality_5_}
\alpha =\frac{\bar{\gamma }_{2} +\gamma _{1} }{\bar{\gamma }_{1} +\bar{\gamma }_{2} } ,\quad \beta =-\frac{\gamma _{2} +\bar{\gamma }_{1} }{\bar{\gamma }_{1} +\bar{\gamma }_{2} }.
\end{equation}
and solely depend on the barotropic indexes of the fluids. As usual, these indexes are given by $\gamma _{i} =1+\frac{p_{i} }{\rho _{i} } $, and by analogous expressions for the $\bar{\gamma }_{i} $  of the other cosmology. We define the overall barotropic index $\gamma =\left(\gamma _{1} \rho _{1} +\gamma _{2} \rho _{2} \right)/\left(\rho _{1} +\rho _{2} \right)$ for the unbarred cosmology. An entirely analogously expression exists for $\bar{\gamma }$  in the other cosmology. Obviously the duality transformation connects these two indexes by $\bar{\gamma }=-\gamma $ . This means that $\rho _{1} +\rho _{2} +p_{1} +p_{2} \to -\left(\rho _{1} +\rho _{2} +p_{1} +p_{2} \right)$. Put another way, if the dominant energy condition $\left(\rho \ge 0,\; -\rho \le p\le \rho \right)$  is fulfilled in one cosmology, then it is violated in the other. The transformation law $H\to -H$, as in the one-component case,  implies $\bar{a}=1/a$. Accordingly, if one cosmology (say, the unbarred one) describes a phase of contraction, the barred one describes a phase of expansion, i.e., both cosmologies are dual to each other.

As we saw, (see \eqref{GrindEQ__3_7_}, in the case that is of interest to us, the system \eqref{duality_3_} transforms into
\begin{equation} \label{duality_6_}
\begin{array}{l} {3H^{2} =\rho _{1} +\rho _{2} ,} \\ {\dot{\rho }_{1} +3H\gamma _{1} \rho _{1} =-3H\Pi ,} \\ {\dot{\rho }_{2} +3H\gamma _{2} \rho _{2} =3H\Pi } \end{array}.
\end{equation}
Automatically, the above dual symmetry gets restricted to the following transformation: $\rho _{i} \to \rho _{i} ,\; H\to -H,\quad \gamma _{i} \to -\gamma _{i} ,\quad \Pi \to -\Pi $, with the overall barotropic index transforming as $\gamma \to -\gamma $. Therefore, there is a duality between the two cosmologies, driven by two interacting fluids through the set of equations \eqref{duality_3_}, that have the sign of the individual barotropic indexes reversed. As a consequence, superaccelerated expansion  (for example, phantom) can be obtained from decelerated ones an viceversa  without  affecting the field equations also in the case of interacting DM and DE.

\section{ Peculiarities of dynamics of scalar fields coupled to dark matter} \label{ISFMs}

\subsection{ Interacting quintessence model}

A vast number of cosmological observations have shown that the EoS (equation of state) parameter $w$ of dark energy  lies in a small interval near $w_{de} =-1$. The interval  $-1\le w_{de} <-1/3$ can be realized with the help of scalar fields with canonical Lagrangians. ìîæåò áûòü ðåàëèçîâàí ñ ïîìîùüþ ñêàëÿðíûõ ïîëåé ñ êàíîíè÷åñêèì ëàãðàíæèàíîì. The lower border, $w_{de} =-1$, corresponds to the cosmological constant, while the upper border, $w_{de} =-1/3$, is tied to the accelerated expansion of the Universe. These scalar fields are called quintessence. The quintessence EoS parameter is
\begin{equation} \label{GrindEQSF__1_}
w=\frac{p}{\rho } =\frac{\frac{1}{2} \dot{\varphi }^{2} -V\left(\varphi \right)}{\frac{1}{2} \dot{\varphi }^{2} +V\left(\varphi \right)}.
\end{equation}
We see that $w$ can take any value between $-1$ (if $\dot{\varphi }^{2} \ll V(\varphi )$ quintessence behaves as a cosmological constant $w\approx -1$ (slow-rolling regime))  and $w\approx +1$ (if $\dot{\varphi }^{2} \gg V(\varphi )$ (fast evolution regime)).

     Given that the quintessence field and the dark matter have unknown physical natures, there seem to be no a priori reasons to exclude a coupling between the two components.

    Let us consider a two-component system with  the energy density and pressure
\begin{equation} \label{GrindEQSF__2_}
\rho =\rho _{s} +\rho _{dm} ,\quad p=p_{s} +p_{dm}.
\end{equation}
The subscript $s$  refers to the scalar field component. If some interaction exists between the scalar field and the dark matter component,
\begin{equation} \label{GrindEQSF__3_}
\begin{array}{l} {\dot{\rho }_{dm} +3H\left(\rho _{dm} +p_{dm} \right)=Q,} \\ {\dot{\rho }_{s} +3H\left(\rho _{s} +p_{s} \right)=-Q} \end{array}.
\end{equation}
Using the effective pressures $\Pi _{s} $ and $\Pi _{dm} $
\begin{equation} \label{GrindEQSF__4_}
Q\equiv -3H\Pi _{dm} =3H\Pi _{s}.
\end{equation}
we  can rewrite \eqref{GrindEQSF__3_} in the form
\begin{equation} \label{GrindEQSF__5_}
\begin{array}{l} {\dot{\rho }_{dm} +3H\left(\rho _{dm} +p_{dm} +\Pi _{dm} \right)=0,} \\ {\dot{\rho }_{s} +3H\left(\rho _{s} +p_{s} +\Pi _{s} \right)=0} \end{array}.
\end{equation}
Consider now the time evolution of the very important (in terms of describing the dynamics of the Universe) ratio $r=\rho _{dm} /\rho _{s} $ . This evolution is described by the equation
\begin{equation} \label{GrindEQSF__6_}
\dot{r}=r\left(\frac{\dot{\rho }_{dm} }{\rho _{dm} } -\frac{\dot{\rho }_{s} }{\rho _{s} } \right).
\end{equation}
Moving to the barotropic  index $\gamma _{i} =w_{i} +1\quad \left(i=s,dm\right)$, we obtain
\begin{equation} \label{GrindEQSF__7_}
\dot{r}=-3Hr\left[\gamma _{dm} -\gamma _{s} +\frac{1+r}{r} \Pi _{dm} \right].
\end{equation}
The existence of a stationary solution $\dot{r}=0$ is guaranteed by the condition
\begin{equation} \label{GrindEQSF__8_}
\Pi _{dm} =\left(\gamma _{s} -\gamma _{dm} \right)\frac{r}{1+r}.
\end{equation}
For cold dark matter, $\gamma _{dm} \approx 1$, and for dark energy as quintessence, $\gamma _{s} =\frac{\dot{\varphi }^{2} }{\rho _{s} } $.  The ñoupling term $Q$ in this case is
\begin{equation} \label{GrindEQSF__9_}
Q=-3H\left(\gamma _{s} -1\right)\frac{r}{1+r} \rho _{s}.
\end{equation}
In a spatially flat Universe $H^{2} =\frac{1}{3} \rho $, and consequently
\begin{equation} \label{GrindEQSF__10_}
Q=-\sqrt{3} \left(\gamma _{s} -1\right)\frac{\rho _{s} \rho _{dm} }{\sqrt{\rho _{s} +\rho _{dm} } }.
\end{equation}
This important result shows that we can introduce an interaction between the cold dark matter and the scalar field (quintessence) that guarantees a constant ratio $r$  of the energy densities of the two components. As we will see shortly, this possibility makes the coincidence problem much easier to solve.

\subsection{ Interacting phantom}

What values of the parameter $w$ can we use? This is a difficult question to answer when dealing with a component of energy about which we know so little. In General Relativity, it is customary to limit the possible values of the components of the energy-momentum tensor with so-called "energy conditions". One of the simplest of these conditions is the so-called NDEC (Null Dominant Energy Condition) $\rho +p\ge 0$. The physical motivation behind this condition is the prevention of vacuum instability. When applied to the dynamics of the Universe, NDEC demands that the density of any allowable component of energy not rise as the Universe expands.

As stated previously, our lack of understanding regarding the dark components prevents us from completely discarding dark energy possibilities that violate NDEC (as well as other energy conditions) - the dark energies for which $w_{de} <-1$. These types of components are collectively called phantom energy.

The action of a phatom field $\varphi $, minimally coupled to gravity, differs from the canonical action of a scalar field in the sign of the kinetic term. In this case, the density of energy and pressure of the phantom field are defined through $\rho _{\varphi } =T_{00} =-\frac{1}{2} \dot{\varphi }^{2} +V\left(\varphi \right);\quad p_{\varphi } =T_{ii} =-\frac{1}{2} \dot{\varphi }^{2} -V\left(\varphi \right)$, while the EoS parameter is
\begin{equation} \label{GrindEQSF__12_}
w_{\varphi } =\frac{p_{\varphi } }{\rho _{\varphi } } =\frac{\dot{\varphi }^{2} +2V(\varphi )}{\dot{\varphi }^{2} -2V(\varphi )}.
\end{equation}

Let's say that the Universe contains only non-relativistic matter $(w_{m} =0)$ and a phantom field $(w_{\varphi } <-1)$. The densities of these components evolve separately: $\rho _{m} \propto a^{-3} $ and $\rho _{\varphi } \propto a^{-3\left(1+w_{\varphi } \right)} $. If matter domination ends at $t_{m} $, the solution for the scale factor at $t>t_{m} $ is
\begin{equation} \label{GrindEQSF__13_}
a(t)=a(t_{m} )\left[-w_{\varphi } +(1+w_{\varphi } )\left(\frac{t}{t_{m} } \right)\right]^{\frac{2}{3(1+w_{\varphi } )} }.
\end{equation}
From here, it immediately follows that for $w_{\varphi } <-1$ at the moment of time $t_{BR} =\frac{w_{\varphi } }{(1+w_{\varphi } )} t_{m} $, the scale factor, as well as a series of other cosmological characteristics of the Universe (like scalar curvature, density of energy of the phantom field) become infinite. This catastrophe has earned the name "Big Rip".

    One of the way to avoid the unwanted big rip singularity is to allow for a suitable interaction  between the phantom energy and the background  dark matter. Through a special choice of interaction, one can mitigate the rise of the phantom component and make it so that components decrease with time if there is a transfer of energy from the phantom field  to the dark matter.

    Let us consider \cite{Rong-Gen-Cai}  the  simplest possible interaction between the cold dark matter and the dark energy
\begin{equation} \label{GrindEQSF__14_}
\begin{array}{l} {\dot{\rho }_{dm} +3H\rho _{dm} =\delta H\rho _{dm} ,} \\ {\dot{\rho }_{de} +3H(1+w_{de} )\rho _{de} =-\delta H\rho _{dm} } \end{array}.
\end{equation}
where $\delta $ is a dimensionless coupling function. If $\delta $ depends on the scale factor only,
\begin{equation} \label{GrindEQSF__15_}
\rho _{dm} =\rho _{dm0} a^{-3} e^{\int \delta (a)d\log a }.
\end{equation}
We once again make the assumption that
\begin{equation} \label{GrindEQSF__16_}
r\equiv \frac{\rho _{dm} }{\rho _{de} } =\frac{\rho _{dm0} }{\rho _{de0} } a^{-\xi } =A^{-1} a^{-\xi } ,\quad A\equiv \frac{\rho _{de0} }{\rho _{dm0} } =\frac{\Omega _{de0} }{\Omega _{dm0} }.
\end{equation}
Let us consider the case with a constant parameter $w_{de} $ . From \eqref{GrindEQSF__16_} we have
\begin{equation} \label{GrindEQSF__17_}
\rho _{de} =\frac{Aa^{\xi } }{1+Aa^{\xi } } \rho _{tot} ,\quad \rho _{dm} =\frac{1}{1+Aa^{\xi } } \rho _{tot}.
\end{equation}
Then, the total energy density satisfies
\begin{equation} \label{GrindEQSF__18_}
\frac{d\rho _{tot} }{da} +\frac{3}{a} \frac{1+\left(1+w_{de} \right)Aa^{\xi } }{1+Aa^{\xi } } \rho _{tot} =0.
\end{equation}
Integrating \eqref{GrindEQSF__18_}, we obtain
\begin{equation} \label{GrindEQSF__19_}
\rho _{tot} =\rho _{tot0} a^{-3} \left[1-\Omega _{de0} \left(1-a^{\xi } \right)\right]^{-3w_{de} /\xi } ,\quad \rho _{tot0} =\rho _{de0} +\rho _{dm0}.
\end{equation}
Consequently, the first Friedmann equation can be written as
\begin{equation} \label{GrindEQSF__20_}
H^{2} =H_{0}^{2} a^{-3} \left[1-\Omega _{de0} \left(1-a^{\xi } \right)\right]^{-3w_{de} /\xi }.
\end{equation}
Using \eqref{GrindEQSF__14_}, one can get the coupling function
\begin{equation} \label{GrindEQSF__21_}
\delta =3+\frac{1}{H} \frac{\dot{\rho }_{dm} }{\rho _{dm} } =-\frac{\left(\xi +3w_{de} \right)Aa^{\xi } }{1+Aa^{\xi } } =-\left(\xi +3w_{de} \right)\frac{\rho _{de} }{\rho _{tot} }.
\end{equation}
This relation can be expressed  as
\begin{equation} \label{GrindEQSF__22_}
\delta =\frac{\delta _{0} }{\Omega _{de0} +\left(1-\Omega _{de0} \right)a^{-\xi } } ,\quad \delta _{0} \equiv -\Omega _{de0} \left(\xi +3w_{de} \right).
\end{equation}
Let's analyse this expression. The asymptotic of the interaction $\delta $ is a constant, $\delta (a\to \infty )=\delta _{0} /\Omega _{de0} $. Therefore, if the dynamics of the expansion are such that $\xi >-3w_{de} $ , then $\delta <0$, which implies that the energy flow is from the dark matter to the dark energy. On the contrary, when $0<\xi <3w_{de} $ , the energy flow is from the phantom dark energy to the dark matter. Furthermore, we can see from \eqref{GrindEQSF__12_} that there is no coupling between the dark energy and the dark matter at $\xi =-3w_{de} $ .Specifically, there is no coupling in SCM, for which $\xi =3,\; w_{de} =-1$ . In addition, we can see from \eqref{GrindEQSF__17_} that in this model, the Universe is dominated by dark matter at early times, and dominated by phantom dark energy at later times.

Let's now look at how coupling between the phantom dark energy and dark matter acts on the time of transition from a decelerated phase to an accelerated one. To do this, let's analyse the deceleration parameter
\begin{equation} \label{GrindEQSF__23_}
q=-\frac{\ddot{a}}{aH^{2} } =-1+\frac{\dot{H}}{H^{2} } =-1+\frac{3}{2} \frac{1-\Omega _{de0} +\left(1+w_{de} \right)\Omega _{de0} a^{\xi } }{1-\Omega _{de0} \left(1-a^{\xi } \right)}.
\end{equation}
Note that $q\left(a\to 1\right)$ and $q\left(a\to \infty \right)$ are negative, as is expected from the era of domination of dark energy.

\subsection{ Tachyonic interacting scalar field}
We consider a flat Friedmann Universe  filled with a spatially homogeneous tachyon field $T$  evolving according to the Lagrangian
\begin{equation} \label{GrindEQSF__24_}
L=-V(T)\sqrt{1-g_{00} \dot{T}^{2} }
\end{equation}
The energy density and the pressure of this field are, respectively

\begin{equation} \label{GrindEQSF__25_}
\rho _{T} =\frac{V(T)}{\sqrt{1-\dot{T}^{2} } }
\end{equation}
and
\begin{equation} \label{GrindEQSF__26_}
p_{T} =-V(T)\sqrt{1-\dot{T}^{2} },
\end{equation}
The equation of motion for the tachyon is
\begin{equation} \label{GrindEQSF__27_}
\frac{\ddot{T}}{1-\dot{T}^{2} } +3H\dot{T}+\frac{1}{V(T)} \frac{dV}{dT} =0.
\end{equation}
Using the approach described in sectionv4.1, let's build the interaction $Q$ between the tachyon field and the cold dark matter, which fulfils the condition $\dot{r}=0,\; r\equiv \rho _{dm} /\rho _{T} $ \cite{Herrera_Pavon_Zimdahl}. The equation \eqref{GrindEQSF__27_} can be written as
\begin{equation} \label{GrindEQSF__28_}
\dot{\rho }_{T} =-3H\dot{T}^{2} \rho _{T}
\end{equation}
Acting as one did during the derivation of \eqref{GrindEQSF__10_}, one obtains that the condition $\dot{r}=0$ is realized for the interaction
\begin{equation} \label{GrindEQSF__29_}
Q=3H\frac{r}{\left(r+1\right)^{2} } \left(1-\dot{T}^{2} \right)\left(\rho _{T} +\rho _{dm} \right)
\end{equation}
Since $\dot{T}^{2} <1$, we have $Q>0$. Therefore, the stationary solution $\left(\dot{r}=0\right)$ exists only when the energy of the tachyon field is transferred to the dark matter.  A stability analysis of the stationary solution, analogous to that in \cite{Zimdahl_Pavon_Chimento},  reveals that when $Q/3H \propto \rho$ in the vicinity of the stationary solution, the $r$ is stable for any $r<1$ .

\subsection{Interacting Chaplygin gas}

 One of the most popular models  of dark energy  is the Chaplygin gas. This model  unifies dark matter and dark energy under the same equation of state, given by
\begin{equation} \label{GrindEQSF__30_}
p=-\frac{A}{\rho },
\end{equation}
where $A$  is a positive constant. This equation of state leads to the following form of dependency of the density on the scale factor:
\begin{equation} \label{GrindEQSF__31_}
\rho =\sqrt{A+\frac{B}{a^{6} } },
\end{equation}
where $B$  is an arbitrary integration constant. Thus, for small values of the scale factor $a$ , $\rho \propto a^{-3} ,\; p\propto a^{3} $ , which implies a dust-like matter. For large values of $a$ , $\rho \sim \sqrt{A} ,\; p\sim -\sqrt{A} $ , which implies cosmological constant behavior.

    Let us find a homogeneous scalar field  $\varphi (t)$  and a self-interacting potential $V\left(\varphi \right)$ corresponding to the Chaplygin gas. Consider now the Lagrangian of the scalar field
\begin{equation} \label{GrindEQSF__32_}
L=\frac{1}{2} \dot{\varphi }^{2} -V\left(\varphi \right),
\end{equation}
The  energy density $\rho _{\varphi } $  and the pressure $p_{\varphi } $  for the scalar field are
\begin{equation} \label{GrindEQSF__33_}
\begin{array}{l} {\rho _{\varphi } =\frac{1}{2} \dot{\varphi }^{2} +V(\varphi )=\rho =\sqrt{A+\frac{B}{a^{6} } } ,} \\ {p_{\varphi } =\frac{1}{2} \dot{\varphi }^{2} -V(\varphi )=-\frac{A}{\rho } =-\frac{A}{\sqrt{A+\frac{B}{a^{6} } } } } \end{array},
\end{equation}
For a flat Universe
\begin{equation} \label{GrindEQSF__34_}
\dot{\varphi }^{2} =\frac{B}{a^{6} \sqrt{A+\frac{B}{a^{6} } } } ,\quad V\left(\varphi \right)=\frac{1}{2} \sqrt{A} \left(\cosh 3\varphi +\frac{1}{\cosh 3\varphi } \right),
\end{equation}
The Chaplygin gas model has underwent multiple generalizations, which allow us to expand this models ability to explain and correspond to observations. The simplest of these generalizations is the so-called generalized Chaplygin gas (GCG), whose equation of state has the form
\begin{equation} \label{GrindEQSF__35_}
p=-\frac{A}{\rho ^{\alpha } },
\end{equation}
The evolution of the scale factor in this model is given by
\begin{equation} \label{GrindEQSF__36_}
\rho =\left[A+\frac{B}{a^{3\left(1+\alpha \right)} } \right]^{\frac{1}{1+\alpha } }.
\end{equation}
Of couse, when $\alpha =1$ we recover the original Chaplygin gas model.

A more radical generalization of this model is the so-called new generalized Chaplygin gas (NGCG) model \cite{0411221}. The equation of state in this case is
\begin{equation} \label{GrindEQSF__36_}
p=\frac{A\left(a\right)}{\rho ^{\alpha } },
\end{equation}
where $\alpha $  is a real number and $A\left(a\right)$  is a function that depends on the scale factor of the Universe, $a$ . It can be expected that the NGCG fluid smoothly interpolates between a dust dominated phase $\rho \propto a^{-3} $ and a dark energy dominated phase $\rho \propto a^{-3(1+w_{de} )} $, where $w_{de} $  is a constant, and should be taken in such a way so as to provide for the accelerated expansion of the Universe - $w_{de} <-1/3$ . Therefore, it is natural to assume that the energy density of the NGCG should be expressed as the superposition
\begin{equation} \label{GrindEQSF__37_}
\rho =\left[Aa^{-3\left(1+w_{de} \right)(1+\alpha )} +Ba^{-3(1+\alpha )} \right]^{\frac{1}{1+\alpha } }
\end{equation}
where $A$  and $B$  are positive constants. The derivation of Eq. \eqref{GrindEQSF__37_} should be the consequence of substituting the equation of state \eqref{GrindEQSF__36_} into the energy conservation equation of the NGCG for a homogeneous and isotropic spacetime.  This requires the function $A(a)$  to be of the form
\begin{equation} \label{GrindEQSF__38_}
A(a)=-w_{de} Aa^{-3\left(1+w_{de} \right)(1+\alpha )}.
\end{equation}
We can return to the simpler Chaplygin gas models by choosing the parameters in a special way -  $\alpha =1$ and $w_{de} =-1$.

Let's show that the NGCG includes interaction in the dark sector. To do this, let's perform the following operation:
\begin{equation} \label{GrindEQSF__39_}
\rho =\rho _{de} +\rho _{dm}.
\end{equation}
Since the pressure of the NGCG fluid is provided only by the dark energy component,
\begin{equation} \label{GrindEQSF__40_}
\rho _{de} =\frac{p}{w_{de} } =\frac{Aa^{-3\left(1+w_{de} \right)(1+\alpha )} }{\left[Aa^{-3\left(1+w_{de} \right)(1+\alpha )} +Ba^{-3(1+\alpha )} \right]^{\frac{\alpha }{1+\alpha } } },
\end{equation}
and energy density of the dark matter is
\begin{equation} \label{GrindEQSF__41_}
\rho _{dm} =\frac{Ba^{-3(1+\alpha )} }{\left[Aa^{-3\left(1+w_{de} \right)(1+\alpha )} +Ba^{-3(1+\alpha )} \right]^{\frac{\alpha }{1+\alpha } } }
\end{equation}
from these expressions one obtains the scaling behavior of the ratio of energy densities
\begin{equation} \label{GrindEQSF__42_}
\frac{\rho _{dm} }{\rho _{de} } =\frac{B}{A} a^{3w_{de} \left(1+\alpha \right)}.
\end{equation}
We see explicitly from this that there must exist an energy flow between the dark matter and the dark energy, provided that $\alpha \ne 0$ . When $\alpha >0$ , the transfer direction of the energy flow is from the dark matter to the dark energy; when $\alpha <0$ , the reverse happens. Therefore, it is clear that the parameter $\alpha $ characterizes the interaction between dark energy and dark matter.

Of course, we can demonstrate the presence of interaction between the dark components in the NGCG model with the help of the traditional phenomenological approach - the ``conservation equations'' with sources \eqref{GrindEQSF__34_}. The indicator of interaction is the difference between the effective EoS parameters $w_{eff,de(dm)} $ \eqref{GrindEQSF__38_} and their initial values $w_{de(dm)} .$ In the analysed case,
\begin{equation} \label{GrindEQ__3_5_}
\begin{array}{l} {w_{eff,de} =w_{de} -\frac{\alpha w_{de} \left(1-\Omega _{0de} \right)a^{3w_{de} (1+\alpha )} }{\Omega _{0de} +\left(1-\Omega _{0de} \right)a^{3w_{de} (1+\alpha )} } ,} \\ {w_{eff,dm} =\frac{\alpha w_{de} \Omega _{0de} }{\Omega _{0de} +\left(1-\Omega _{0de} \right)a^{3w_{de} (1+\alpha )} } } \end{array}.
\end{equation}
which clearly shows that interaction is present.

\subsection{ \textit{$w=-1$} crossing and interacting models }

 In the quintessence model of dark energy, $-1<w<-1/3$ . In the phantom model with negative kinetic energy, $w<-1$ . Recent cosmological data seems to indicate that the phantom divide line was crossed in the in the near past. This means that the equation of state parameter $w_{de} $  crossed the phantom divide line $w_{de} =-1$ . This crossing to the phantom region is possible neither for an ordinary minimally coupled scalar field  nor for a phantom field. Why is this problem - the problem of crossing the phantom divide - so important \cite{0407107}? If $w<-1$, the energy density of phantom matter generally becomes infinite in a finite period of time and, hence, leads to the late-time singularity known as the "Big Rip". To avoid this singularity, one assumes that the Universe can "bounce" instead of collapsing to the singularity. Only the transition from$w\ge -1$ to $w\le -1$  just before the bounce could explain the nonsingular bouncing, without resorting to a fine-tuning of the initial energy densities of any energy form in the Universe.

    There are at least three ways to solve this problem. If dark energy behaves as quintessence at the early stage, and evolves as phantom at the later stage, a natural suggestion would be to consider a 2-field model (quintom model): a quintessence and a phantom \cite{0909.2776}. The next possibility  would be that General Relativity fails at cosmological scales. In this case, quintessence or phantom energy can cross the phantom divide line in a modified gravity theory \cite{1007.0482}.

In addition to these possibilities, the \textbf{\textit{$w=-1$ }} crossing problem can be solved by applying the model of interacting dark components \cite{1007.0482,0909.3013}.

We assume the most simple case - $Q=\delta H\rho _{dm} $ - and rewrite the expression \eqref{GrindEQ__3_25_} in terms of the redshift instead of the scale factor. Then,
\begin{equation} \label{GrindEQSF__44_}
\rho _{de} (z)=\rho _{de,0} (1+z)^{3\left(1+w_{de} \right)} +\frac{\delta \rho _{dm,0} }{\delta +3w_{de} } \left[\left(1+z\right)^{3\left(1+w_{de} \right)} -\left(1+z\right)^{3-\delta } \right].
\end{equation}
Introducing the effective EoS parameter for the dark energy \eqref{GrindEQSF__38_}, we find that
\begin{equation} \label{GrindEQSF__45_}
w_{eff,de} =\frac{p_{eff,de} }{\rho _{de} } =\frac{p_{de} +Q/3H}{\rho _{de} } =-1+\Delta ,\quad \Delta \equiv \frac{1}{3} \frac{d\ln \rho _{de} }{d\ln \left(1+z\right)}.
\end{equation}
The relation \eqref{GrindEQSF__45_}  includes the entire spectrum of scalar fields. Clearly, if $\Delta >0$ , dark energy evolves as quintessence; if $\Delta <0$ , it evolves as phantom, if  $\Delta =0$ , it is just a cosmological constant. From this it follows that if $\rho _{de} $  decreases and then increases with respect to redshift (or time), or increases and then decreases, the effective EoS parameter of dark energy crosses phantom divide. Using \eqref{GrindEQSF__44_}, we obtain
\begin{equation} \label{GrindEQSF__46_}
\frac{d\rho _{de} (z)}{dz} =3\left(1+w_{de} \right)\rho _{de,0} (1+z)^{2+3w_{de} } +\frac{\delta \rho _{dm,0} }{\delta +3w_{de} } \left[3\left(1+w_{de} \right)\left(1+z\right)^{2+3w_{de} } -\left(3-\delta \right)\left(1+z\right)^{2-\delta } \right].
\end{equation}
If  $d\rho _{de} /d\left(1+z\right)=0$ at some redshift  , the effective parameter  $w_{eff,de} $  crosses the  phantom divide. Analysis of \eqref{GrindEQSF__46_} shows \cite{0909.3013} that observations leave enough space for the parameters $\left(\delta ,w_{de} \right)$ to fulfil the condition $\frac{d\rho _{de} (z)}{dz} =0$.

\section {Structure of phase space of models with interaction}

The evolution of a Universe filled with interacting components can be effectively analysed in terms of dynamical systems theory. Let us consider the following coupled differential   equations for two variables

\begin{equation} \label{ref_4_1_}
\begin{array}{l} {\dot{x}=f(x,y,t)} \\ {\dot{y}=g(x,y,t)} \end{array}.
\end{equation}
We will be interested in so-called autonomous systems, for which the functions $f$  and $g$  do  not contain explicit time-dependent terms.

    A point $\left(x_{c} ,y_{c} \right)$  is said to be a fixed (a.k.a. critical) point of  the autonomous system if

\begin{equation} \label{ref_4_2_}
f\left(x_{c} ,y_{c} \right)=g\left(x_{c} ,y_{c} \right)=0.
\end{equation}
A critical point $\left(x_{c} ,y_{c} \right)$  is called an attractor when it satisfies  the condition
\begin{equation} \label{ref_4_3_}
\left(x(t),y(t)\right)\to \left(x_{c} ,y_{c} \right)\quad for\; t\to \infty.
\end{equation}
Let's look at the behavior of the dynamical system  \eqref{ref_4_1_} around the critical point. For this purpose, let us consider  small perturbations  around the critical point
\begin{equation} \label{ref_4_4_}
x=x_{c} +\delta x,\quad y=y_{c} +\delta y.
\end{equation}
Substituting into Eqs. \eqref{ref_4_1_} leads to the first-order differential equations:
\begin{equation} \label{ref_4_5_}
\frac{d}{dN} \left(\begin{array}{l} {\delta x} \\ {\delta y} \end{array}\right)=\hat{M}\left(\begin{array}{l} {\delta x} \\ {\delta y} \end{array}\right).
\end{equation}
Taking into account the specifics of the problem that we are solving, we made the change $\frac{d}{dt} \to \frac{d}{dN} $, where $N=\ln a$ . The matrix $\hat{M}$ is given by
\begin{equation} \label{ref_4_6_}
\hat{M}=\left(\begin{array}{cc} {\frac{\partial f}{\partial x} } & {\frac{\partial f}{\partial y} } \\ {\frac{\partial g}{\partial x} } & {\frac{\partial g}{\partial y} } \end{array}\right).
\end{equation}
The general solution for the  linear perturbations
\begin{equation} \label{ref_4_7_}
\begin{array}{l} {\delta x=C_{1} e^{\lambda _{1} N} +C_{2} e^{\lambda _{2} N} ,} \\ {\delta y=C_{3} e^{\lambda _{1} N} +C_{4} e^{\lambda _{2} N} } \end{array}.
\end{equation}
The stability around the fixed points depends on the nature of the eigenvalues.

We will look at \cite{48} the interacting dark components as a dynamical system described by the equations
\begin{equation} \label{ref_4_8_}
\begin{array}{l} {\rho '_{de} +3(1+w_{de} )\rho _{de} =-Q,} \\ {\rho '_{dm} +3(1+w_{dm} )\rho _{dm} =Q} \end{array}.
\end{equation}
Here, a prime denotes the derivative with respect to the e-folding time $N=\ln a$ . Note that although the interaction can significantly  change the cosmological evolution, the system is still autonomous. We consider the following specific interaction forms, which were already analysed before:
\begin{equation} \label{ref_4_9_}
Q_{1} =3\gamma _{m} \rho _{dm} ,\quad Q_{2} =3\gamma _{d} \rho _{de,} \quad Q_{3} =3\gamma _{tot} \rho _{tot}.
\end{equation}
Let's write the effective EoS  parameters for both dark energy and dark  matter:
\begin{equation} \label{ref_4_10_}
Q=Q_{1} ,\quad w_{eff,de} =w_{de} \left(\Omega _{de} \right)+\gamma _{m} \frac{1-\Omega _{de} }{\Omega _{de} } ,\quad w_{eff,dm} =w_{dm} -\gamma _{m}.
\end{equation}
\begin{equation} \label{ref_4_11_}
Q=Q_{2} ,\quad w_{eff,de} =w_{de} \left(\Omega _{de} \right)+\gamma _{d} ,\quad w_{eff,dm} =w_{dm} -\gamma _{d} \frac{\Omega _{de} }{1-\Omega _{de} }.
\end{equation}
\begin{equation} \label{ref_4_12_}
Q=Q_{3} ,\quad w_{eff,de} =w_{de} \left(\Omega _{de} \right)+\gamma _{m} \frac{1}{\Omega _{de} } ,\quad w_{eff,dm} =w_{dm} -\frac{\gamma _{tot} }{1-\Omega _{de} }.
\end{equation}
The syste \eqref{ref_4_8_} can be turned into a system of equations for fractional energy densities
\begin{equation} \label{ref_4_13_}
\begin{array}{l} {\Omega '_{dm} =3f_{j} \Omega _{dm} \Omega _{de} ,} \\ {\Omega '_{de} =-3f_{j} \Omega _{dm} \Omega _{de} } \\ {} \end{array},
\end{equation}
where $j=0,1,2,3$. Here,  $j=0$ corresponds to the non-interacting case $f_{0} =w_{de} -w_{dm} $

    For $j=1,2,3\; \left(Q_{1} ,Q_{2} ,Q_{3} \right)$  :
\begin{equation} \label{ref_4_14_}
\begin{array}{l} {f_{j} = w_{eff,de~j} - w_{eff,dm~j} } \\ {f_{1} =f_{0} +\frac{\gamma _{m} }{\Omega _{de} } ,} \\ {f_{2} =f_{0} +\frac{\gamma _{d} }{1-\Omega _{de} } ,} \\ {f_{3} =f_{0} +\frac{\gamma _{tot} }{\Omega _{de} \left(1-\Omega _{de} \right)} } \end{array}.
\end{equation}
Let us now obtain the critical points of the autonomous system \eqref{ref_4_12_}  by imposing the conditions $\Omega '_{dm} =\Omega '_{de} =0$ and $\Omega _{dm} +\Omega _{de} =1$. Critical points can be broken up into the following categories. The critical point $M$  is the matter dominated phase with $\Omega _{dm} =1$,  and the critical  point $E$  is the dark energy dominated phase with $\Omega _{de} =1$. If $f_{j} \propto 1/\Omega _{dm} $ or $f_{j} \propto 1/\Omega _{de}, $ these two fixed points may not exist. Besides the above two fixed points, there are other solutions with $f_{j} =0$ . Note that an attractor is one of the stable critical points of the autonomous system.

    If we analyse the linear perturbations about the critical point $\left(\bar{\Omega }_{de} ,\bar{\Omega }_{de} \right)$  of the dynamical system Eqs. \eqref{ref_4_13_} and linearize them, we get

\begin{equation} \label{ref_4_15_}
\hat{M}=\left(\begin{array}{cc} {3f\left(\bar{\Omega }_{de} \right)\bar{\Omega }_{de} } & {3\left(f\left(\bar{\Omega }_{de} \right)\bar{\Omega }_{dm} +f'\bar{\Omega }_{dm} \bar{\Omega }_{de} \right)} \\ {-3f\left(\bar{\Omega }_{de} \right)\bar{\Omega }_{de} } & {-3\left(f\left(\bar{\Omega }_{de} \right)\bar{\Omega }_{dm} +f'\bar{\Omega }_{dm} \bar{\Omega }_{de} \right)} \end{array}\right).
\end{equation}
Here, $f'\equiv df/d\Omega _{de} $ . The  two eigenvalues of the matrix $\hat{M}$ that determine the stability of the corresponding critical point are

\begin{equation} \label{ref_4_16_}
\begin{array}{l} {\lambda _{1} =0,} \\ {\lambda _{2} =3f\left(2\Omega _{de} -1\right)-3f'\Omega _{de} \left(1-\Omega _{de} \right)} \end{array}.
\end{equation}
When $\lambda _{2} $  is positive, the corresponding critical point is an unstable node. "Unstable'' means that the present phase will evolve, eventually, into other phases.  When  $\lambda _{2} $ is negative, the corresponding critical point is a stable node and the phase will last long.

    Let's analyse the structure of phase space with non-linear interactions of the type  \eqref{GrindEQ__3_44_}. For the system of equations  \eqref{GrindEQ__3_11_},  the eigenvalues of the matrix$\hat{M}$ are roots of the equation
\begin{equation} \label{ref_4_17_}
\begin{array}{l} {\lambda ^{2} +\left[2+w_{de} -w_{de} \left(1+w_{de} \right)\frac{\partial _{r} \Pi }{\Pi } \right]\lambda +\left(1+w_{de} +w_{de} \partial _{\rho } \Pi \right)=0,} \\ {\partial _{r} \Pi \equiv \frac{\partial \Pi }{\partial r} ,\quad \partial _{\rho } \Pi \equiv \frac{\partial \Pi }{\partial \rho } \quad } \end{array}.
\end{equation}
The Eq.\eqref{ref_4_17_} has the solutions
\begin{equation} \label{ref_4_18_}
\lambda _{\pm } =\frac{1}{2} \left\{\begin{array}{l} {\left[w_{de} \left(1+w_{de} \right)\frac{\partial _{r} \Pi }{\Pi } -\left(2+w_{de} \right)\right]} \\ {\pm \sqrt{\left(2+w_{de} -w_{de} \left(1+w_{de} \right)\frac{\partial _{r} \Pi }{\Pi } \right)^{2} -4\left(1+w_{de} +w_{de} \partial _{\rho } \Pi \right)} } \end{array}\right\}
\end{equation}
where we have to require $1+w_{de} +w_{de} \partial _{\rho } \Pi \ne 0$. In case these solutions are non-degenerate and real, they describe a stable critical point for $\lambda _{\pm } <0$, an unstable critical point for$\lambda _{\pm } >0$  and a saddle if $\lambda _{+} $  and $\lambda _{-} $  have different signs. For complex eigenvalues $\lambda _{\pm } =\alpha \pm i\beta $, it is
the sign of $\alpha $  that determines the character of the stationary point. For $\alpha =0$ the critical point is a center, for $\alpha <0$  it is a stable focus, and for $\alpha >0$  it is an unstable focus.

\section{Examples of realization of interaction in the dark sector}

\subsection{$\Lambda (t)$ - the simplest possibility of interaction of the dark components}

    Possibly the simplest explanation of the observed accelerated expansion of the Universe is dark energy (DE) in the form of a cosmological constant $\Lambda $ , which modifies the Einstein equations
\begin{equation} \label{ref_6_1_}
G^{\mu \nu } =8\pi GT^{\mu \nu } \to G^{\mu \nu } =8\pi GT^{\mu \nu } +\Lambda g^{\mu \nu }
\end{equation}
it is well known that flat  models with a very small cosmological term  are in good agreement with almost all sets of cosmological observations. From the theoretical viewpoint, however, at least two problems arise:  the so-called cosmological constant problem and the so-called coincidence problem. Attempts to resolve these problems on a phenomenological level are mainly tied to the introduction of interaction between the dark components. In cosmological models with interaction, $\Lambda $  is necessarily a time-dependent quantity:  the vacuum energy density is a time-dependent quantity because of its coupling with the other matter fields, the characteristics of which depend on time.

    Historically, the possibility of a time varying $\Lambda (t)$  was first  advanced by Bronstein \cite{6.1}. A summary of the evolving ideas and their state at the start of the century can  can be seen in the review papers by Peebles and Ratra \cite{6.2}, Lima \cite{6.3} and J. M. Overduin and F. I. Cooperstock \cite{6.4}. An overview of the current state of the $\Lambda (t)$ problem can be found in \cite{6.5}.

    From Eq. \eqref{ref_6_1_} the Bianchi identities imply that the coupling between a $\Lambda (t)$ term and dark matter particles must be of the type
\begin{equation} \label{ref_6_2_}
u_{\mu } T_{dm;\nu }^{\mu \nu } =-u_{\mu } \left(\frac{\Lambda }{8\pi G} g^{\mu \nu } \right)_{\, ;\nu },
\end{equation}
or, equivalently,
\begin{equation} \label{ref_6_3_}
\dot{\rho }_{dm} +3H\rho _{dm} =-\dot{\rho }_{\Lambda },
\end{equation}
where $\rho _{\Lambda } =\Lambda /8\pi G$ is the energy density of the cosmological constant. This equation requires some kind of energy exchange between matter and vacuum energy, e.g. through vacuum decay into matter, or vice versa. It must be emphasized that the equation of state of the vacuum energy density retains its usual form $p_{\Lambda } (t)=-\rho _{\Lambda } \left(t\right)$, despite the fact that $\Lambda $  evolves with time.

    It should be noted \cite{6.3} that the equation \eqref{ref_6_2_} may be rewritten to yield an expression for the rate of entropy production in the $\Lambda (t)$  model as
\begin{equation} \label{ref_6_4_}
T\frac{dS}{dt} =-\frac{\dot{\Lambda }a^{3} }{8\pi G}.
\end{equation}
From this equation, it immediately follows that $\Lambda $  must decrease over the course of time, $\dot{\Lambda }<0\, \left(dS/dt>0\right)$, while the energy is transferred from the decaying vacuum to the material component.

    Although we have been using the notation $\Lambda (t)$, the truth is that, in the majority of papers, it depends only implicitly on the cosmological time through the scale factor  $\Lambda =\Lambda (a)$ or the Hubble parameter $\Lambda =\Lambda (H)$, or even a combination of them. Phenomenological models with a variable cosmological constant are listed and reviewed in  \cite{6.4}.

    All these models have the same Achilles' heel: the expression defining $\Lambda (t)$ is obtained either using dimensional arguments or in a completely ad hoc way. The interaction between matter and dark energy cannot be derived in these models from the principle of least action in a relativistically covariant form.Essentially, we have come face to face with the previously described general problem that plagues interaction in the dark sector: in the absence of a microscopic theory of interaction, we are prevented from pointing out the exact mechanisms of energy transfer between the components.

In the field of the lagrangian description of the dynamic cosmological constant, a certain degree of progress was achieved within the framework of so-called  $\Lambda \left(T\right)$ gravity \cite{6.6}. In this theory, the cosmological constant
    is a function of the trace of the energy--momentum tensor $T$ . Within the framework of this approximation, the dynamics of the time-dependent cosmological constant can be described directly in terms of interacting components with the densities  $\rho _{\Lambda } $  and $\rho _{dm} $,
\begin{equation} \label{ref_6_5_}
\begin{array}{l} {\dot{\rho }_{dm} +3H\rho _{dm} =Q,} \\ {\dot{\rho }_{\Lambda } +3H\left(\rho _{\Lambda } +p_{\Lambda } \right)=-Q} \end{array}.
\end{equation}
where $Q$  is the rate of the energy transfer from dark energy to dark matter ,
\begin{equation} \label{ref_6_6_}
Q=3H\rho _{dm} \frac{\Lambda '+2\Lambda ''\rho _{dm} }{1+3\Lambda '+2\Lambda ''\rho _{dm} }.
\end{equation}
Here, $\Lambda '=\frac{d\Lambda }{dT} $ .We see that in this case, the interaction retains a factorized $H$ dependence, but is now a non-linear function of $\rho _{dm} $ . We return to a linear dependency when $\frac{\Lambda ''}{\Lambda '} \rho _{dm} \ll 1$.

The above-considered phenomenology of a time-depending cosmological constant can be generalized \cite{1402.3755} onto the case of bulk viscosity \cite{1307.5949} and cosmological models with entropy forces \cite{1402.3755}.

\noindent Let us consider an FLRW spatially flat Universe with the general Friedmann equations
\[\begin{array}{l} {H^{2} =\frac{1}{3} \rho +f(t),} \\ {\frac{\ddot{a}}{a} =-\frac{1}{6} \left(\rho +3p\right)+g(t)} \end{array}\]
These equations result in the generalized conservation equation (we used $\frac{\ddot{a}}{a} =\dot{H}+H^{2} $)
\[\dot{\rho }+3H\left(\rho +p\right)=6H\left(-f(t)+\frac{\dot{f}(t)}{2H} +g(t)\right)\]
For extra driving terms in the form of the cosmological constant the general conservation equation transforms into the standard conservation equation. In this case$f(t)=g(t)=\Lambda /3,\; \; \dot{f}=0$ and
\[\dot{\rho }+3H\left(\rho +p\right)=6H\left(-f(t)+\frac{\dot{f}(t)}{2H} +g(t)\right)\to \dot{\rho }+3H\left(\rho +p\right)=0\]
In case of \textbf{ $f(t)=g(t)=\Lambda /3$ }we reproduce the\textbf{ $\Lambda (t)CDM$ }model,\textbf{}
\[\dot{\rho }+3H\left(\rho +p\right)=6H\left(-f(t)+\frac{\dot{f}(t)}{2H} +g(t)\right)\to \dot{\rho }+3H\left(\rho +p\right)=\dot{\Lambda }(t)\]

\subsection{Chameleon fields as a possible realization of interaction}

In the simplest dynamical models of dark energy(quintessence, $k$-essence, phantom field), the scalar fields undergoes only self-interaction, described by the potential $V(\varphi )$. A lack of interaction with the other components of the Universe seems both unnatural and limiting. However, attempts to include interaction (a procedure that, as we've seen, is rather simple from a theoretical point of view) always face the same fundamental problem. The issue is that the available precision measurements of the local Universe (for us, the term will be synonymous with "Solar System") have been explained theoretically with the introduction of four forces: strong, weak, electromagnetic, and gravitational. The introduction of a new interaction automatically leads to the appearance of a "fifth force" that we do not observe. The fact that we do not observe it places strict limits on any possible interaction between the scalar fields and matter: either the interaction must be significantly weaker than gravity, or its quants must be very heavy - meaning that the interaction has a very short range. A natural question arises - can we build a model where dark energy is a scalar field that interacts with matter, all the while not violating the equivalence principle, which is well tested on Solar System scales? Recall that when we say "dark energy", we mean any substance that explains the accelerated expansion of the Universe.

In answer to the above question - yes, there are such models. They're called ``chameleon models'' \cite{6.7,6.8}. The chameleon scalar fields are scalar fields coupled to matter (baryonic matter too) with an intensity comparable with (and sometimes greater than) gravitational forces, and with a mass that depends on the density of the surroundings. On cosmological scales, where density is negligible, these fields are very light. However, near the Earth, where the density is significantly higher, the mass of these fields rises significantly. In other words, the characteristics of these fields, including their actual value, changes with the density of the surroundings. This is why they are called ``chameleon fields'' and ``chameleon models''. Note also that the introduction of forces that depend on density is a common practice in physics. In the 1960s, in order to calculate nuclear characteristics with the help of the Hartree-Fock method\cite{6.9}, various effective interactions between nucleons were used. As it turned out, however, none of the analysed potentials were adequate. The problem was solved by the introduction of an effective interaction that depended on density.

Let's make note of another interesting characteristic of chameleon forces, which lets us understand why forces that are responsible for the global dynamics of the Universe have only a weak impact on, say, planetary orbits. The latter are, with a good level of exactness, described by newtonian gravity (as a limit case of general relativity) and, as we've said before, the fifth force must also preserve the prescision results of the traditional dynamics. Let's look at the chameleon field that realizes the interaction between the Earth and the Sun. As it turns out, this interaction is signigicantly smaller than it appears at first. In order to calculate the field created by, for instance, Earth, let's break it up into infinitely small volumes. The input of the inner volumes will be negligible due to the high density. This means that the resulting force will be generated mainly by a thin layer near the surface of the Earth, while the input of the rest of the volume will be negligibly small. Analogous arguments are applicable to the Sun. Therefore, the introduction of an additional (chameleon) field will not lead to serious contradictions with tests of general relativity on Solar System scales.

The action of the chameleon field $\varphi $ is a sum of the Einstein-Hilbert action for gravity
\begin{equation}\label{S_EH}
    S_{\text{EH}}=\int d^4 x\sqrt{-g}\frac{1}{16\pi G}R=\int d^4 x\sqrt{-g}\frac{M_\text{pl}^2}{2}R,
\end{equation}
the action of a scalar field
\begin{equation}\label{S_FI}
S_\phi=-\int d^4 x\sqrt{-g}\left\{\frac{1}{2}(\partial\phi)^2+V(\phi)\right\},
\end{equation}
and the action of the matter fields $\psi _{m}^{(i)} $
\begin{equation} \label{1_}
S_{m} =-\int d^{4} xL_{m}  \left(\psi _{m}^{(i)} ,g_{\mu \nu }^{(i)} \right).
\end{equation}
The key characteristic of the model is the conformal relation of the chameleon field $\varphi $ with the fields $\psi _{m}^{(i)} $. This relation is chosen in such a way so as to make any perturbations (particles) of the matter fields move along the geodesics of the metric $g_{\mu \nu }^{(i)} $, which is related to the initial metric $g_{\mu \nu } $ in the following way:
\begin{equation} \label{2_}
g_{\mu \nu }^{(i)} =e^{\frac{2\beta _{i} \varphi }{M_{Pl} } } g_{\mu \nu }.
\end{equation}
where $\beta _{i} $ are dimensionless constants. From string theory, it follows that for any matter component, these constants are of the same order as 1. The full action has the form
\begin{equation} \label{3_}
S=\int d^{4} x\sqrt{-g} \left\{\frac{M_{Pl}^{2} }{2} R+\frac{1}{2} \partial _{\mu } \varphi \partial ^{\mu } \varphi -V(\varphi )-\frac{1}{\sqrt{-g} } L_{m} \left(\psi _{m}^{(i)} ,g_{\mu \nu }^{(i)} \right)\right\}.
\end{equation}

Variating by $\varphi $, we obtain the equation of motion for the field:

\begin{equation} \label{4_}
\nabla ^{2} \varphi =\frac{dV(\varphi )}{d\varphi } +\sum _{i}\frac{1}{\sqrt{-g} } \frac{\partial L_{m} \left(\psi _{m}^{(i)} ,g_{\mu \nu }^{(i)} \right)}{\partial g_{\mu \nu }^{(i)} } \frac{2\beta _{i} }{M_{Pl} }  g_{\mu \nu }^{(i)}.
\end{equation}

Using the definition, ${T_{\mu \nu }} = \frac{2}{{\sqrt { - g} }}\frac{{\delta S}}{{\delta {g^{\mu \nu }}}},$ the action, \eqref{3_} the relation, \eqref{2_}, and the assumption that all material components are an ideal fluid, we find that

\begin{equation} \label{5_}
\frac{1}{\sqrt{-g} } \frac{\partial L_{m} \left(\psi _{m}^{(i)} ,g_{\mu \nu }^{(i)} \right)}{\partial g_{\mu \nu }^{(i)} } g_{\mu \nu }^{(i)} =\frac{1}{2} \rho _{i} (1-3w_{i} )e^{(1-3w_{i} )\beta _{i} \varphi /M_{Pl} }.
\end{equation}
Putting \eqref{5_} into \eqref{4_}, we obtain the equation of motion that reconstructs the explicit dependency on the scalar field $\varphi $

\begin{equation} \label{6_}
\nabla ^{2} \varphi =\frac{dV(\varphi )}{d\varphi } +\sum _{i}\left(1-3w_{i} \right)\frac{\beta _{i} }{M_{Pl} }  \rho _{i} e^{(1-3w_{i} )\beta _{i} \varphi /M_{Pl} }.
\end{equation}
We can express the dynamics of the scalar field in terms of an effective potential:
\begin{equation} \label{7_}
\nabla ^{2} \varphi =\frac{dV_{eff} (\varphi )}{d\varphi } ;\quad V_{eff} =V(\varphi )+\sum _{i}\rho _{i} e^{(1-3w_{i} )\beta _{i} \varphi /M_{Pl} }.
\end{equation}
If matter is non-relativistic, all $w_{i} =0$, and
\begin{equation} \label{8_}
V_{eff} =V(\varphi )+\sum _{i}\rho _{i} e^{\beta _{i} \varphi /M_{Pl} }.
\end{equation}
Schematically, if we assume that there is only one matter component with the density $\rho $, the resulting effective potential can be written as
\begin{equation} \label{9_}
V_{eff} =V(\varphi )+U(\beta \varphi /M_{Pl} )\rho.
\end{equation}

The resulting potential is clearly reproduces the interaction of a scalar field (dark energy) with matter fields.

\subsubsection{Effective potentials of chameleon fields}
We wish to choose a bare potential $V\left(\phi\right)$ that can lead to accelerated expansion with the help of the mechanism of slow rolling, as in quintessence \cite{Ratra_Peebles}. The mechanism that makes $\phi$ act as a cosmological constant only today is rather large and complex, so we will assume, without loss of generality, that $\phi$ always rolls along the potential in the positive direction. For this reason, $V\left(\phi\right)$ must be a monotonically decreasing function of $\phi$.

The potential of the chameleon field must fulfil the following conditions:
\begin{enumerate}
\item$\lim_{\phi\rightarrow0}V(\phi)=\infty$;
\item$V(\phi)$ is $C^\infty$, bounded below, and strictly deceasing;
\item$V_{,\phi}(\phi)$ is strictly negative and increasing;
\item$V_{,\phi\phi}(\phi)$ is strictly positive and decreasing.
\end{enumerate}
These conditions also place restrictions on $\phi$, which only be positive.

There are two widely used types of potentials that fulfil the above conditions. The first, often found in models with quintessence (see, for instance, \cite{ZWS}), is the reverse power potential

\begin{displaymath}
V(\phi)=\frac{M^{4+n}}{\phi^n},
\end{displaymath}

where $M$ is a constant with the dimensions of mass and $n$ is a positive constant.

The second is the exponential potential

\begin{displaymath}
V(\phi)=M^4\exp\left(\frac{M^n}{\phi^n}\right),
\end{displaymath}
where, once again, $M$ is a constant with the dimensions of mass and $n$ is is a positive constant.

An important difference between these models lies in the limit $\lim_{\phi\rightarrow\infty}V(\phi)$. For the reverse power potential, it is 0, while for the exponentia potential, it is $M^4$. The differences are discussed in detail in \cite{KW} and \cite{BvdBDKW}.

\subsubsection{Chameleon fields in cosmology}
Let's use the exponential potential
\begin{displaymath}
V\left(\phi\right)=M^4\exp\left(\frac{M^n}{\phi^n}\right),
\end{displaymath}
where $M=2\times10^{-3}\,\text{eV}$.
Let's analyse a planar, uniform, isotropic Universe with the metric
\begin{displaymath}
g_{\mu\nu}=diag\left(-1,a^2,a^2,a^2\right).
\end{displaymath}
Next, assuming that $\phi$ is also uniform,
\begin{align*}
\nabla^2\phi&=g^{\mu\nu}\nabla_\mu\nabla_\nu\phi\\
&=g^{\mu\nu}\partial_\mu\partial_\nu\phi-g^{\mu\nu}\Gamma_{\nu\mu}^\rho\phi_{,\rho}\\
&=g^{00}\partial_0\partial_0\phi-\left(a^{-2}\Gamma_{11}^0+a^{-2}\Gamma_{22}^0+a^{-2}\Gamma_{33}^0\right)\phi_{,0}\\
&=-\ddot{\phi}-a^{-2}\left(3a\dot{a}\right)\dot{\phi}\\
&=-\left(\ddot{\phi}+3H\dot{\phi}\right),
\end{align*}
Therefore, the equation (\ref{7_}) takes on the form
\begin{equation}
\ddot{\phi}+3H\dot{\phi}=-V_{\text{eff},\phi}(\phi),
\label{cosmiceom}
\end{equation}
which is an ordinary result for a spatially uniform scalar field.

Let's assume that the Universe is composed of the field $\phi$, pressure-free matter with the density $\rho_\text{m}$, which interacts with the field $\phi$ through a coupling constant $\beta$, and radiation with the density $\rho_\text{r}$.

The first Friedmann equation, which is obtained from the Einstein equations $G^{\mu\nu}=8\pi G\,T^{\mu\nu}$, which can be obtained by variating the action (\ref{3_}) with respect to the Einstein-frame metric $g_{\mu\nu}$, has the following form:

\begin{equation}
3H^2M_\text{pl}^2=\frac{1}{2}\dot{\phi}^2+V\left(\phi\right)+\rho_\text{m}e^{\beta\phi/M_\text{pl}}+\rho_\text{r}.
\label{Friedmann}
\end{equation}

The critical density and the relative density of matter have the following forms:
\begin{displaymath}
\rho_\text{critical}\equiv\frac{1}{2}\dot{\phi}^2+V\left(\phi\right)+\rho_\text{m}e^{\beta\phi/M_\text{pl}}+\rho_\text{r}
\end{displaymath}
and
\begin{displaymath}
\Omega_\text{m}\equiv\frac{\rho_\text{m}e^{\beta\phi_\text{min}/M_\text{pl}}}{\rho_\text{critical}}
\end{displaymath}

\subsubsection{Chameleon forces}
The interaction between chameleon fields and matter is rooted in the conformal relation in the equation (\ref{2_}); this is analogous to how the geometry of space-time interacts with matter. Since matter fields $\psi_\text{m}^{(i)}$ couple to $g_{\mu\nu}^{(i)}$ instead of to $g_{\mu\nu}$, the worldlines of free test particles (meaning particles experiencing only gravity and the chameleon force) of the species $i$ are the geodesics of $g_{\mu\nu}^{(i)}$ rather than those of $g_{\mu\nu}$ (see also \cite{Damour}).\footnote{From this it is clear that the chameleon force violates the weak Equivalence Principle only if there exist two matter species with differing values of $\beta_i$.}

The geodesic equation for the worldline $x^\mu$ of a test mass of the species $i$ is
\begin{equation}
\ddot{x}^\rho+\tilde{\Gamma}_{\mu\nu}^{\rho}\dot{x}^\mu\dot{x}^\nu=0,
\label{geodesic}
\end{equation}
where $\tilde{\Gamma}_{\mu\nu}^{\rho}$ are Christoffel symbols and a dot denotes differentiation with respect to the proper time $\tilde{\tau}$, both in the $\tilde{g}_{\mu\nu}$ metric.

Using
\begin{displaymath}
\tilde{g}_{\mu\nu,\sigma}=\left(\frac{2\beta_i}{M_\text{pl}}\phi_{,\sigma}g_{\mu\nu}+g_{\mu\nu,\sigma}\right)e^{2\beta_i\phi/M_\text{pl}},
\end{displaymath}
the Christoffel symbols can be obtained in the following way:\footnote{The derivation of this relationship in the case of a general conformal transformation is given in \cite[pp.65--6]{FM}.}
\begin{align*}
\tilde{\Gamma}_{\mu\nu}^{\rho}&=\frac{1}{2}\tilde{g}^{\sigma\rho}\left(\tilde{g}_{\sigma\nu,\mu}+\tilde{g}_{\sigma\mu,\nu}-\tilde{g}_{\mu\nu,\sigma}\right)\\
&=\frac{1}{2}e^{-2\beta_i\phi/M_\text{pl}}g^{\sigma\rho}\left(
\begin{aligned}
\frac{2\beta_i}{M_\text{pl}}\phi_{,\mu}g_{\sigma\nu}+g_{\sigma\nu,\mu}+\frac{2\beta_i}{M_\text{pl}}\phi_{,\nu}g_{\sigma\mu}\\
\vphantom{}+g_{\sigma\mu,\nu}-\frac{2\beta_i}{M_\text{pl}}\phi_{,\sigma}g_{\mu\nu}-g_{\mu\nu,\sigma}
\end{aligned}
\right)e^{2\beta_i\phi/M_\text{pl}}\\
&=\frac{1}{2}g^{\sigma\rho}\left(g_{\sigma\nu,\mu}+g_{\sigma\mu,\nu}-g_{\mu\nu,\sigma}
\right)+\frac{\beta_i}{M_\text{pl}}g^{\sigma\rho}\left(\phi_{,\mu}g_{\sigma\nu}+\phi_{,\nu}g_{\sigma\mu}-\phi_{,\sigma}g_{\mu\nu}\right)\\
&=\Gamma_{\mu\nu}^\rho+\frac{\beta_i}{M_\text{pl}}\left(\phi_{,\mu}\delta_\nu^\rho+\phi_{,\nu}\delta_\mu^\rho-g^{\sigma\rho}\phi_{,\sigma}g_{\mu\nu}\right).
\end{align*}

Putting this into  (\ref{geodesic}), we obtain
\begin{align*}
0&=\ddot{x}^\rho+\Gamma_{\mu\nu}^{\rho}\dot{x}^\mu\dot{x}^\nu+\frac{\beta_i}{M_\text{pl}}\left(\phi_{,\mu}\delta_\nu^\rho+\phi_{,\nu}\delta_\mu^\rho-g^{\sigma\rho}\phi_{,\sigma}g_{\mu\nu}\right)\dot{x}^\mu\dot{x}^\nu\\
&=\ddot{x}^\rho+\Gamma_{\mu\nu}^{\rho}\dot{x}^\mu\dot{x}^\nu+\frac{\beta_i}{M_\text{pl}}\left(\phi_{,\mu}\dot{x}^\mu\dot{x}^\rho+\phi_{,\nu}\dot{x}^\rho\dot{x}^\nu-g^{\sigma\rho}\phi_{,\sigma}g_{\mu\nu}\dot{x}^\mu\dot{x}^\nu\right)\\
&=\ddot{x}^\rho+\Gamma_{\mu\nu}^{\rho}\dot{x}^\mu\dot{x}^\nu+\frac{\beta_i}{M_\text{pl}}\left(2\phi_{,\mu}\dot{x}^\mu\dot{x}^\rho+g^{\sigma\rho}\phi_{,\sigma}\right).
\end{align*}
The second term in the above equation is the familiar gravitational term, while the term with $\beta_i/M_\text{pl}$ is the chameleon force.

We see that in the non-relativistic limit, a test mass $m$ of the species $i$ in a static chameleon field $\phi$ experiences a force $\vec{F}_\phi$ given by
\begin{equation}
\frac{\vec{F}_\phi}{m}=-\frac{\beta_i}{M_\text{pl}}\vec{\nabla}\phi,
\label{chameleonforce}
\end{equation}
as in \cite{KW}.  Thus, $\phi$ is the potential for the chameleon force.

\subsubsection{The phantom-divide-line-crossing}

Using the chameleon cosmology model, the authors of the following paper \cite{1005.1878} described the phantom-divide-line-crossing phenomenon. This paper uses the chameleon model which was considered in \cite{1004.3508}.  The minimally-coupled-to-gravity scalar field $\phi$ with the potential $V(\phi)$, whose interaction with the perfect fluid is described by a term in the  Lagrangian, which on the Friedmann background looks like \cite{1004.3508}:
\begin{equation}
L_{scalar+matter} = w\rho f(\phi),
\label{Cham}
\end{equation}
where the coefficient $w$ relates the energy density $\rho$ and the pressure $p$ of matter:
\begin{equation}
p = w \rho,
\label{eq-of-state}
\end{equation}
and $f(\phi)$ is some function of the scalar field $\phi$.

Then the authors fixed the fundamental constants in such a way so as to give to the Friedmann equation a particularly simple form:
\begin{equation}
H^2 = \varepsilon,
\label{Friedmann1}
\end{equation}
where $\varepsilon$ is the total energy density of the scalar field and matter. On the flat Friedmann background this total energy density is
\begin{equation}
\varepsilon = \frac{\dot{\phi}^2}{2} + V(\phi) + \rho f(\phi).
\label{energy}
\end{equation}
The Klein-Gordon equation for the scalar field $\phi$ is
\begin{equation}
\ddot{\phi} + 3H\dot{\phi} + V'(\phi) + w \rho f'(\phi) = 0,
\label{KG}
\end{equation}
 where ``prime'' stands for the derivative with respect to $\phi$.
The total energy density $\varepsilon$ satisfies the energy conservation law
\begin{equation}
\dot{\varepsilon} + 3H(\varepsilon + P) = 0,
\label{conserve}
\end{equation}
where $P$ is the total pressure of the matter and of the scalar field which is equal to
\begin{equation}
P = w \rho + \frac{\dot{\phi}^2}{2} - V(\phi).
\label{pressure}
\end{equation}

After all necessary mathematical manipulations, the Friedmann and Klein-Gordon equations can be rewritten as
\begin{equation}
H^2 = \frac{\dot{\phi}^2}{2} + V + \frac{\rho_0}{f^{1-w}a^{3(1+w)}},
\label{Friedmann2}
\end{equation}
\begin{equation}
\ddot{\phi} + 3 H \dot{\phi} + V' + \frac{w\rho_0 f'}{f^{1-w}a^{3(1+w)}} = 0.
\label{KG1}
\end{equation}

The authors of the article being discussed here, \cite{1005.1878},  find the explicit expressions
for the potential $V(\phi)$ and the function $f(\phi)$:
\begin{eqnarray}
&&V(\phi) = \frac{8\cosh^4\frac{\phi}{2\phi_0}}{3(1+w)}\left(6\alpha^2(1+w)+3\phi_0^2(1-w) + 4\alpha\tanh\frac{\phi}{2\phi_0}\right),
\label{potential1}
\end{eqnarray}
\begin{eqnarray}
&&f(\phi) = \left(-\frac{16\cosh^4\frac{\phi}{2\phi_0}\exp\left(3\alpha(1+w)\frac{\phi}{\phi_0}\right)}
{3M t_R^2 (1+w)}\left(3\phi_0^2+2\alpha\tanh\frac{\phi}{2\phi_0}\right)\right)^{\frac{1}{w}}.
\label{f2}
\end{eqnarray}
where the function $f(\phi)$ describes the interaction between the chameleon scalar field and the matter.
It is found that in the case when
\begin{equation}
\phi_0 \geq \sqrt{\frac{2\alpha}{3}},
\label{ineq}
\end{equation}
\begin{equation}
w < -1.
\label{ineq1}
\end{equation}
then if the parameter $\alpha > \frac13$ and
\begin{equation}
\phi_0 > \sqrt{\frac{4\alpha - 6\alpha^2(1+w)}{3(1-w)}}
\end{equation}
then in this case, potential $V(\phi)$ is always negative. If $\alpha > \frac13$ and
\begin{equation}
\sqrt{\frac{2\alpha}{3}} < \phi_0 < \sqrt{\frac{4\alpha - 6\alpha^2(1+w)}{3(1-w)}}
\end{equation}
the potential $V(\phi)$ changes  sign at
\begin{equation}
\phi = 2\phi_0\  {\rm arctanh} \frac{5\alpha^2(1+w) + 3\phi_0^2(1-w)}{4\alpha}.
\end{equation}
If $ \alpha < \frac13$ the potential is always negative.

In the case when if at least one of two inequalities (\ref{ineq}), (\ref{ineq1}) is broken the expression for $f^{w}$ in Eq. (\ref{f2}) cannot be always nonnegative. Hence, when imposing the following condition on the factor $w$:
\begin{equation}
w = \frac{2m+1}{n},
\label{gamma}
\end{equation}
where $m$ and $n$ are integers, the expression for $f$ is well defined.
The sign of the potential depends on the interplay of three parameters $\phi_0, w$ and $\alpha$.

The Universe in this solution begins its evolution from the Big Bang singularity, undergoes a phantom divide line crossing and ends in the Big Rip singularity. The two potential-like functions of the chameleon scalar field have a rather  simple analytic form.
Note that this form is simpler than the potential functions in two-scalar model, providing the same cosmological evolution
\cite{we-PT1}.

\subsubsection{An FLRW Cosmology with a Chameleon Field}
In the article \cite{1401.2626}, the authors derive the field equations of a chameleon theory of gravitation with a general matter Lagrangian term and represent them in the framework of cosmology.
The field equations in this framework after some mathematical manipulations, take the form
\begin{equation}
\left(\frac{\dot a}{a}\right)^2=\frac{8\pi}{3\phi}\epsilon_m+\frac{\omega}{6}\left(\frac{\dot\phi}{\phi}\right)^2-\frac{\dot a}{a}\frac{\dot\phi}{\phi}\label{90}\hspace{6pt},
\end{equation}
\begin{equation}
\frac{\ddot a}{a}=-\frac{8\pi}{3\phi}\left[\rho_m\left(\frac{3+\omega}{3+2\omega}\right)+3p_m\left(\frac{\omega}{3+2\omega}\right)\right]-\frac{\omega}{3}\left(\frac{\dot\phi}{\phi}\right)^2+\frac{\dot a}{a}\frac{\dot\phi}{\phi}+\frac{4\pi}{3+2\omega}S\label{91}\hspace{6pt},
\end{equation}
\begin{equation}
\frac{\ddot\phi}{\phi}+3\frac{\dot a}{a}\frac{\dot\phi}{\phi}=\frac{8\pi}{(3+2\omega)\phi}(\rho_m-3p_m)-\frac{8\pi}{3+2\omega}S\label{101}
\end{equation}
and
\begin{equation}
\dot\rho_m+3\gamma\frac{\dot a}{a}\rho_m=\frac{1}{2}S\dot\phi\label{102}\hspace{6pt},
\end{equation}
\begin{equation}
\dot\rho_\phi+6\frac{\dot a}{a}\rho_\phi=-\frac{1}{16\pi}R\dot\phi-\frac{1}{2}S\dot\phi\label{109}\hspace{6pt},
\end{equation}
where \emph{dot} denotes derivation with respect to cosmic time $t$ and
\begin{equation}
\rho_\phi\equiv\frac{\omega\dot\phi^2}{16\pi\phi}\label{100}
\end{equation}

Then by introducing the following dynamical variables
\begin{equation}
X=\frac{\dot a}{a}\hspace{6pt},\hspace{6pt}Y=\frac{\dot\phi}{\phi}\hspace{6pt}\text{and}\hspace{6pt}Z=\frac{\rho}{\phi}\hspace{6pt},\label{200}
\end{equation}
the authors studied the dynamical behaviour of the Universe. For these variables, the field equations take the following form
\begin{eqnarray}
\dot X&=&\left(Q_3-\frac{2Q_1Q_4}{Q_2}\right)^{-1}\left[\left(\frac{\omega A}{6}-\frac{\omega}{3}+\frac{Q_4}{Q_2}-\frac{\omega B}{6\alpha^2}-\frac{\omega BQ_4}{6 \alpha^2Q_2}-\frac{\gamma\omega}{2\alpha}\frac{X}{Y}\right)Y^2\right.\nonumber\\
&+&\left(1-A+\frac{B}{\alpha^2}+\frac{3\gamma}{2\alpha}\frac{X}{Y}\right)X^2\nonumber\\
&+&\left.\left(1-A-\frac{3Q_4}{Q_2}+\frac{B}{\alpha^2}+\frac{B Q_4}{\alpha^2 Q_1}+\frac{3\gamma}{\alpha}\frac{X}{Y}\right)XY\right]\label{300}
\end{eqnarray}

\begin{eqnarray}
\dot Y&=&\left(1-2\frac{Q_1}{Q_2Q_3}\right)^{-1}\left\lbrace\left[\frac{1}{Q_2}\left(1-\frac{BC}{3}-\frac{\omega B}{6\alpha^2}-\frac{3\omega B}{6\alpha^2}\frac{X}{Y}\right)+\frac{\omega Q_1}{Q_2Q_3}\left(\frac{A}{3}-\frac{2}{3}-\frac{B}{3\alpha^2}-\frac{\gamma B}{2\alpha}\frac{X}{Y}\right)\right]Y^2\right.\nonumber\\
&+&\left[\frac{1}{Q_2}\left(\frac{2BC}{\alpha}+\frac{B}{\alpha^2}+\frac{3B}{2\alpha^2}\frac{X}{Y}\right)+\frac{Q_1}{Q_2Q_3}\left(2-2A+\frac{2B}{\alpha^2}+\frac{3(\gamma+1)B}{\alpha^2}\frac{X}{Y}\right)
\right]X^2\nonumber\\
&+&\left.\left[\frac{1}{Q_2}\left(-3+\frac{2BC}{\alpha}+\frac{B}{\alpha^2}+\frac{3B}{2\alpha^2}\frac{X}{Y}\right)+\frac{Q_1}{Q_2 Q_3}\left(-2A+\frac{2B}{\alpha^2}+\frac{3\gamma B}{\alpha^2}\frac{X}{Y}\right)\right]XY\right\rbrace\label{301}
\end{eqnarray}
where
\begin{equation}
\alpha=\frac{8\pi}{3}\hspace{6pt},\hspace{6pt}
A=\frac{3-2\omega+3\gamma}{3+2\omega}\hspace{6pt},\hspace{6pt}
B=\frac{4\pi}{3+2\omega}\hspace{6pt}\text{and}\hspace{6pt}
C=2-3\gamma\hspace{6pt}.
\end{equation}
\begin{eqnarray}
Q_1&=&\frac{B}{\alpha}\left(1-2\frac{X}{Y}\right)\hspace{6pt},\hspace{6pt}\\
Q_2&=&1+\frac{2B\omega}{3\alpha}-\frac{2B}{\alpha}\frac{X}{Y}\hspace{6pt},\hspace{6pt}\\
Q_3&=&1-\frac{B}{\alpha}-\frac{2B}{\alpha}\frac{X}{Y}\hspace{6pt}
\end{eqnarray}
and
\begin{eqnarray}
Q_4&=&\frac{B}{\alpha}\left(\frac{X}{Y}-\frac{\omega}{3}\right)\hspace{6pt}.
\end{eqnarray}

\begin{figure}
\includegraphics[scale=0.45]{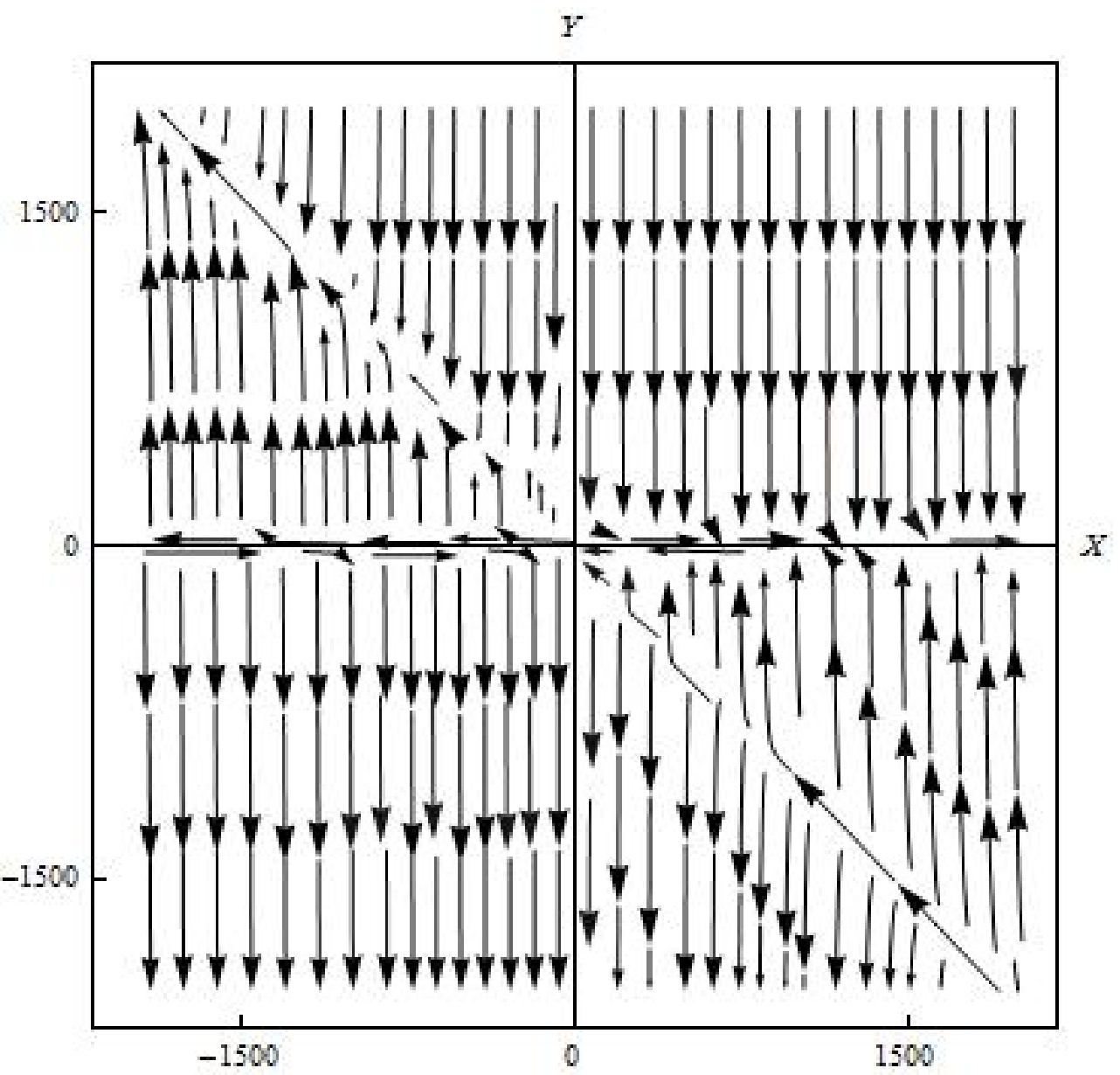}
\includegraphics[scale=0.45]{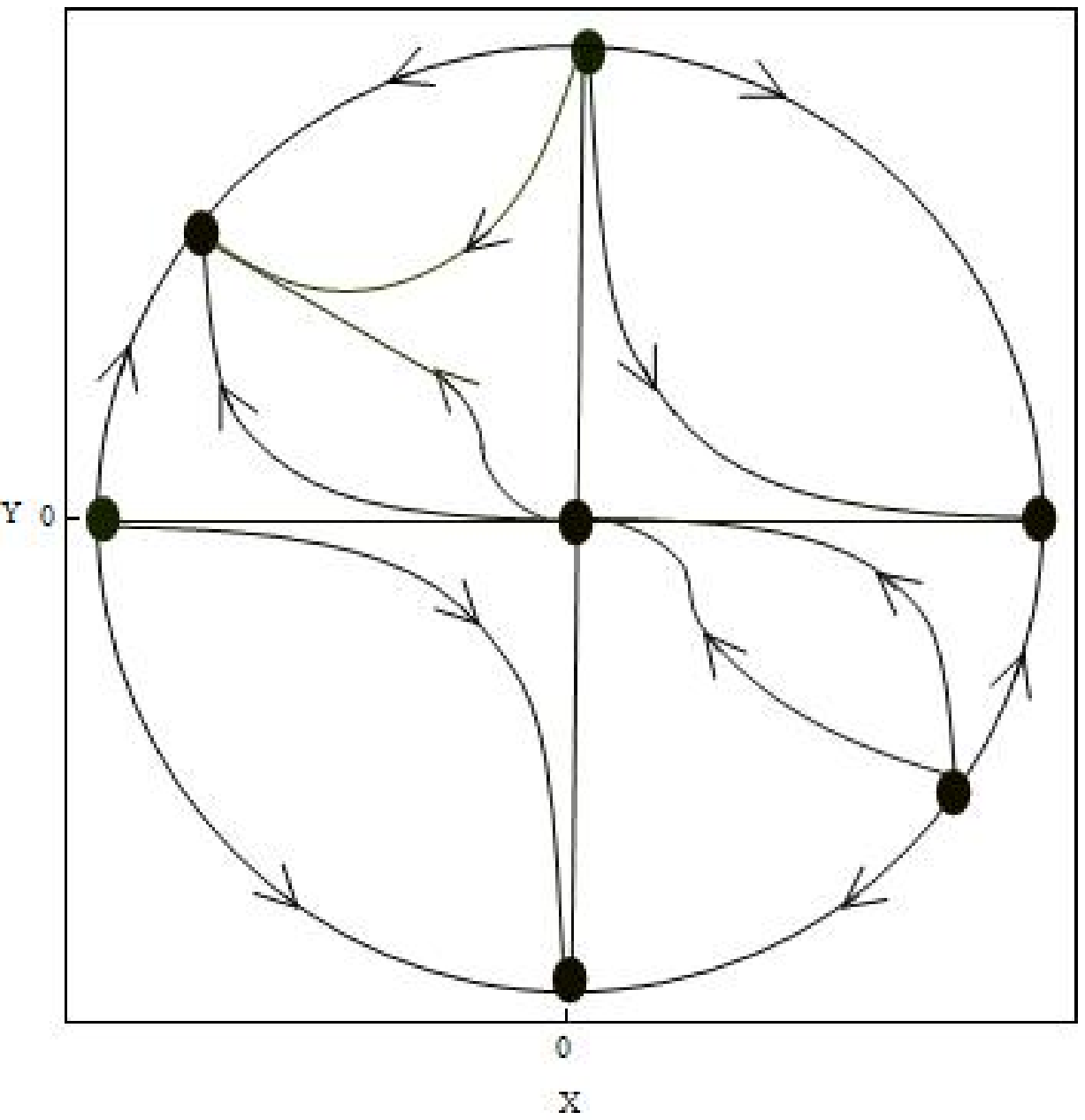}
\caption{The global phase portrait for $\omega=-1.49$ and $\gamma=1$. These diagrams show the evolution of a dust dominated Universe and include curves which can be interpreted as both the inflationary phase and the late time acceleration.}
\label{fig:fig12}
\end{figure}

\begin{figure}
\includegraphics[scale=0.45]{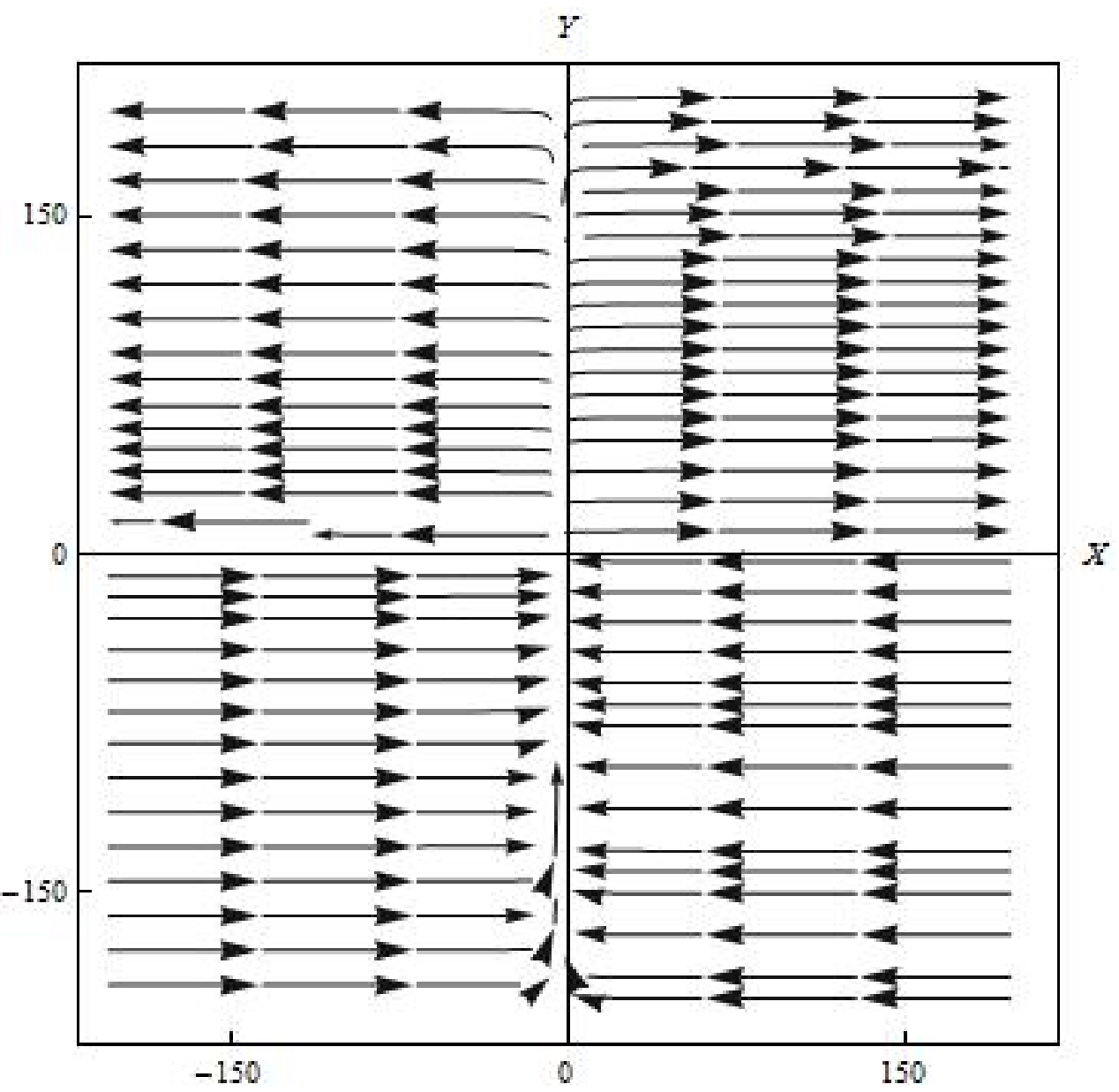}
\includegraphics[scale=0.45]{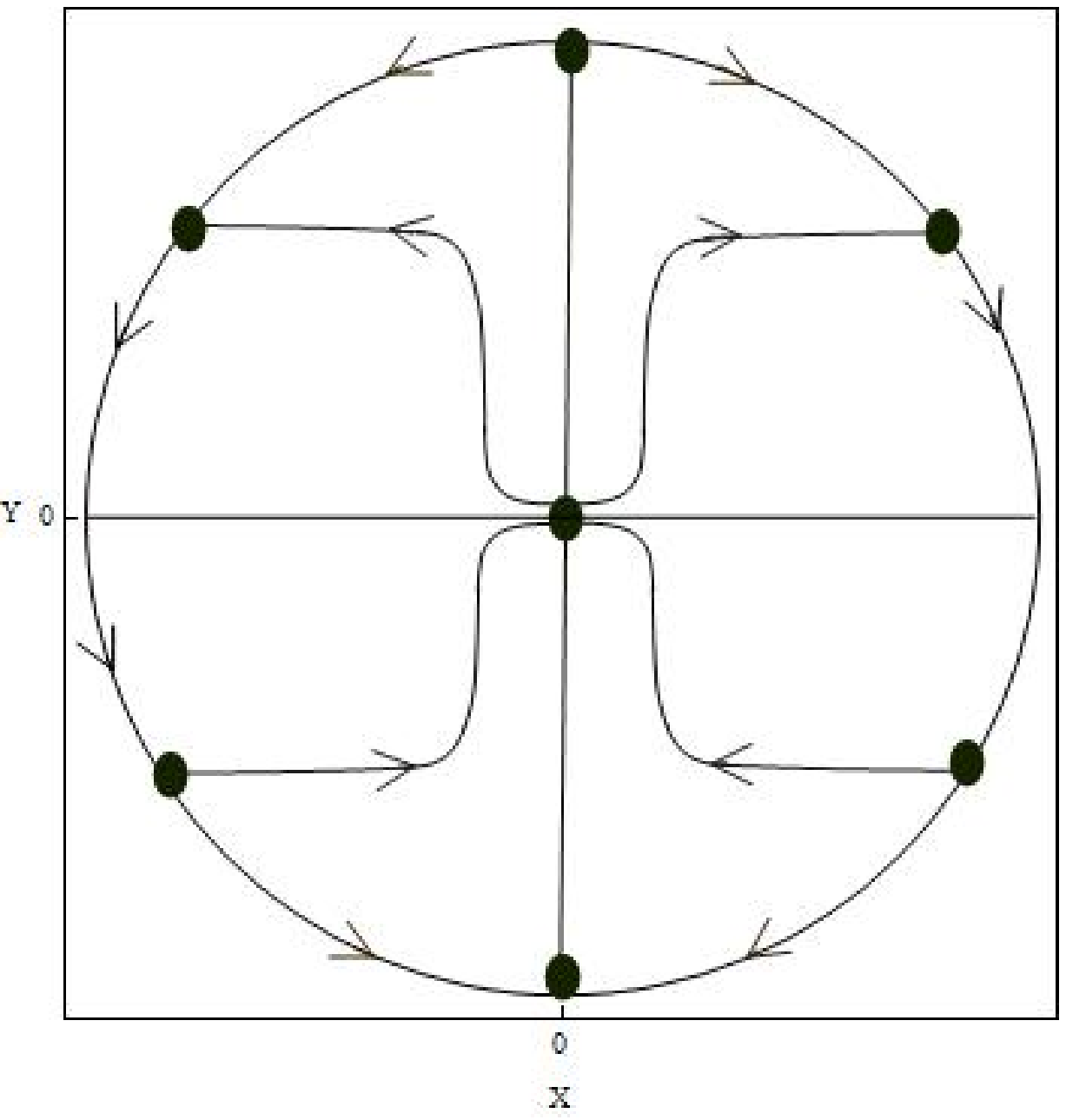}
\caption{The global phase portrait for $\omega =-1.49$ and $\gamma=0$. These diagrams show the evolution of a dark matter dominated universe and include curves which can be interpreted as the manifestations of the late time acceleration.}
\label{fig:fig13}
\end{figure}

\begin{figure}
\includegraphics[scale=0.45]{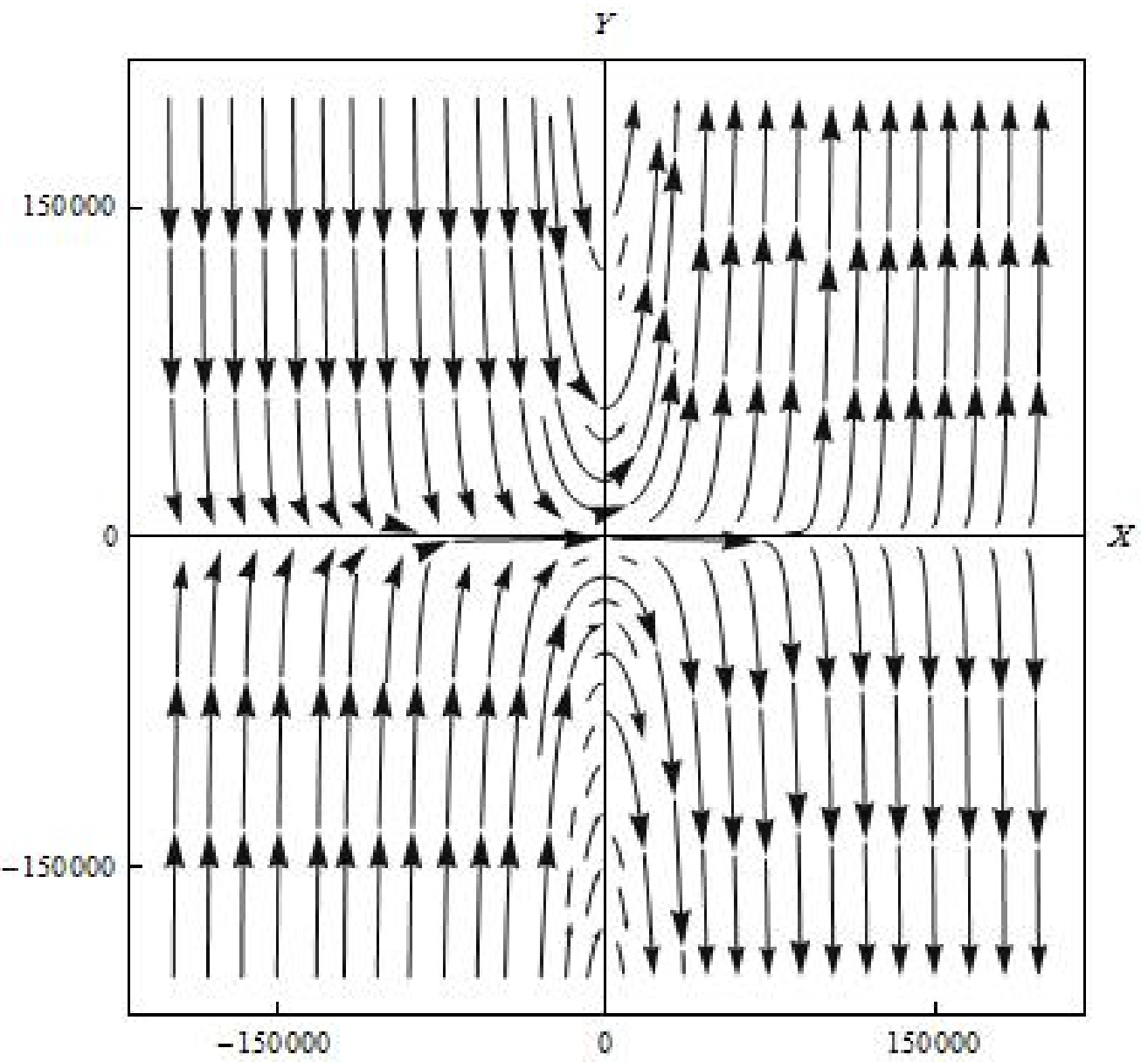}
\includegraphics[scale=0.45]{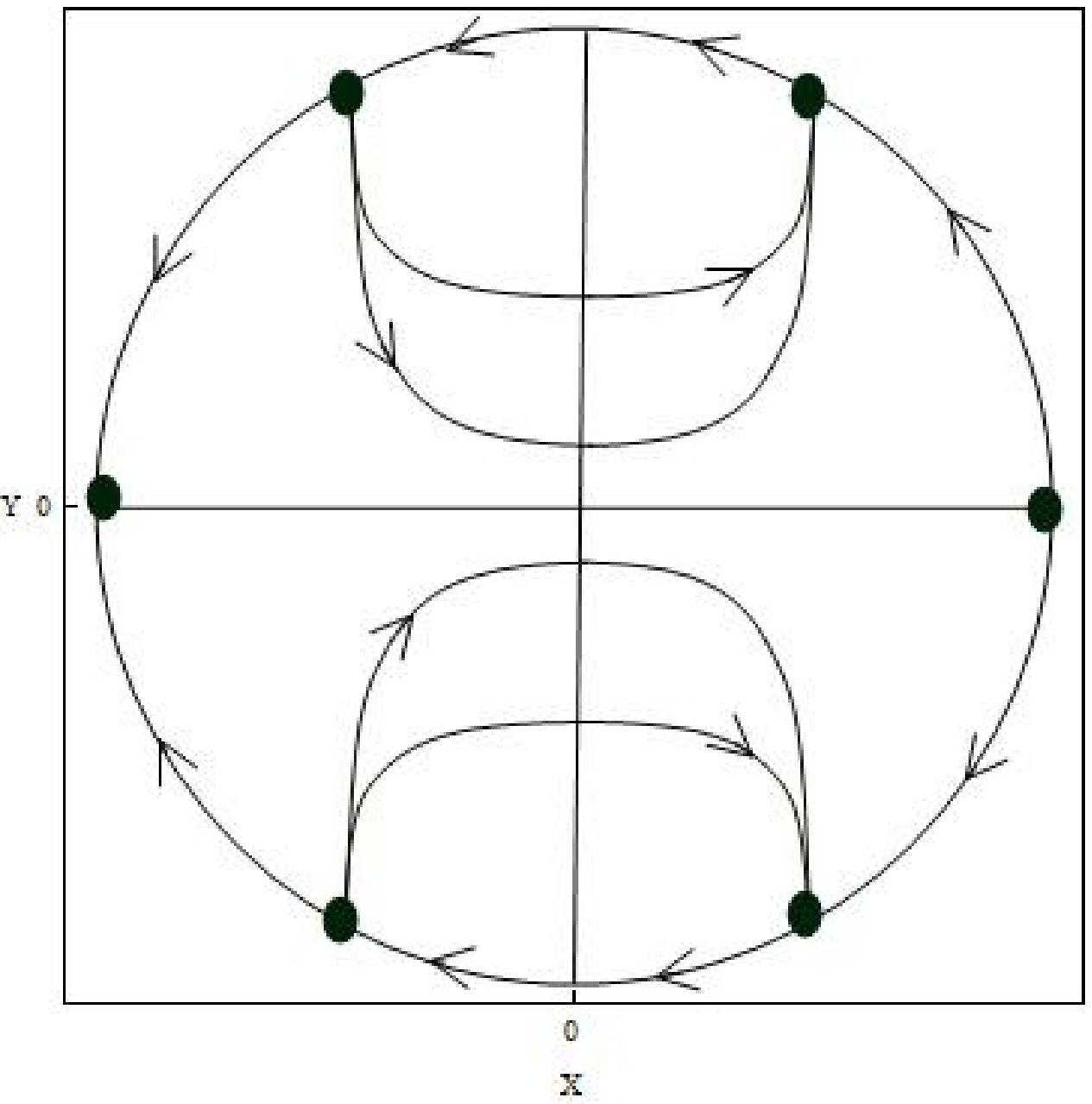}
\caption{The global phase portrait for $\omega=50000$ and $\gamma=1$. These diagrams show the evolution of a cold dark matter dominated universe and include curves which can be interpreted as the manifestations of the late time acceleration.}
\label{fig:fig14}
\end{figure}

Results of the analysis of this dynamical system are shown in Figures \ref{fig:fig12}, \ref{fig:fig13} and \ref{fig:fig14}. The case $\omega=50000$  is of special interest.

\subsection{Interacting models in $f(R)$ -gravity}

One of the possible ways to explain the acceleration of the Universe is to modify Einstein gravity by making the substitution $R \to f(R)$ .The  action in f(R) gravity in the Jordan frame is
\begin{equation} \label{ref_6_15_}
S=\frac{1}{2\kappa } \int d^{4} x\sqrt{-g} f(R)+ S_{m} \left(g^{\mu \iota } ,\psi \right),\quad S_{m} =\int d^{4} x\sqrt{-g} L_{m} \left(g^{\mu \iota } ,\psi \right),
\end{equation}
where $R$  is the Ricci scalar, $\kappa =8\pi G$ , and $L^{(m)} $ is the matter Lagrangian. and $\psi $ represents all matter fields. It is possible to transform the action \eqref{ref_6_15_}  from the original Jordan frame to the Einstein frame by using conformal transformations \cite{6.10,  6.12}. In the Einstein frame, the model contains a coupling between the canonical scalar fields (dark energy)  and  the non-relativistic matter.

    Variation of \eqref{ref_6_15_} with respect to the metric $g_{\mu \nu } $  yields the field equation
\begin{equation} \label{ref_6_16_}
f'R_{\mu \nu } -\frac{1}{2} fg_{\mu \nu } -\nabla _{\mu } \nabla _{\nu } f'+g_{\mu \nu } \square f''=kT_{\mu \nu }^{\left(m\right)} ,\quad f'\equiv \frac{df}{dR}.
\end{equation}
Here, the matter stress-energy tensor $T_{\mu \nu }^{\left(m\right)} $is
\begin{equation} \label{ref_6_17_}
T_{\mu \nu }^{\left(m\right)} =-\frac{2}{\sqrt{-g} } \frac{\delta \left(\sqrt{-g} L_{m} \right)}{\delta \left(g^{\mu \nu } \right)}
\end{equation}
$f(R)$ gravity may be written as a scalar-tensor theory, by introducing a Legendre transformation $\left\{R,f\right\}\to \left\{\phi ,U\right\}$,  defined  as
\begin{equation} \label{ref_6_18_}
\begin{array}{l} {\phi \equiv f'(R),} \\ {U\left(\phi \right)\equiv R\left(\phi \right)f'-f\left[R\left(\phi \right)\right]} \end{array}.
\end{equation}
In this representation the field equations of f(R) gravity can be derived from a Brans-Dicke type action given by
\begin{equation} \label{ref_6_19_}
S=\frac{1}{2\kappa } \int d^{4} x\sqrt{-g} \left(\phi R-U(\phi )+L_{m} \right)
\end{equation}
This is the so-called Jordan frame representation of the action. One can perform a canonical transformation and rewrite the action \eqref{ref_6_19_} in what is called the Einstein frame. Rescaling the metric as
\begin{equation} \label{ref_6_20_}
g_{\mu \nu } \to \tilde{g}_{\mu \nu } =f'\, g_{\mu \nu }.
\end{equation}
and redefining $\phi \to \tilde{\phi }$ with
\begin{equation} \label{ref_6_21_}
d\tilde{\phi }=\sqrt{\frac{3}{2k} } \frac{d\phi }{\phi }.
\end{equation}
the original theory can be mapped into the Einstein frame, in which the `new' scalar field  $\tilde{\phi }$ couples minimally to the Ricci curvature, and has canonical kinetic energy,
\begin{equation} \label{ref_6_22_}
S=\int d^{4} x\sqrt{-g} \left[\frac{\tilde{R}}{2\kappa } -\frac{1}{2} \partial ^{\mu } \tilde{\phi }\, \partial _{\mu } \tilde{\phi }-V\left(\tilde{\phi }\right)\right] +S_{m} \left(e^{-2\beta \tilde{\phi }} \tilde{g}_{\mu \nu } ,\psi \right).
\end{equation}
The self-interacting potential $V\left(\tilde{\phi }\right)$  is given by
\begin{equation} \label{ref_6_23_}
V\left(\tilde{\phi }\right)=\frac{Rf'-f}{2\kappa f'^{2} }.
\end{equation}
Clearly, a coupling of the scalar field $\tilde{\phi }$  with the matter sector is now induced. The strength of this coupling $\beta =\sqrt{1/6}$ 6 is  fixed and is same for all matter fields.

    Taking $\tilde{g}_{\mu \nu } $ and $\tilde{\phi }$  as two independent variables, the variations of the action \eqref{ref_6_22_} yield the following field equations
\begin{equation} \label{ref_6_24_}
\tilde{G}_{\mu \nu } =\kappa \left(\tilde{T}_{\mu \nu }^{\tilde{\phi }} +\tilde{T}_{\mu \nu }^{m} \right).
\end{equation}
\begin{equation} \label{ref_6_25_}
\square \tilde{\phi }-\frac{dV(\tilde{\phi })}{d\tilde{\phi }} =-\beta \sqrt{\kappa } \tilde{T}^{m},
\end{equation}
where $\tilde{T}^{m} \equiv \tilde{g}^{\mu \nu } \tilde{T}_{\mu \nu }^{m} $.  The latter equation shows that the evolution of the field $\phi $ is directly coupled to matter. Radiation, for which $\tilde{T}^{m} =0$, is an obvious exception.

    For a spatially flat, homogeneous, and isotropic Universe, the field equation \eqref{ref_6_25_} reduces to
\begin{equation} \label{ref_6_26_}
\ddot{\phi }+3H\dot{\phi }+\frac{dV}{d\phi } =-\beta \sqrt{\kappa } \rho _{m},
\end{equation}
Utilizing the usual definitions of density and pressure of a scalar field, the equation \eqref{ref_6_26_} can be transformed into
\begin{equation} \label{ref_6_27_}
\dot{\rho }_{\phi } +3H\left(1+w_{\phi } \right)\rho _{\phi } =-Q,\quad Q=\beta \sqrt{\kappa } \dot{\phi }\rho _{m}.
\end{equation}
 In the final expressions (the formulas \eqref{ref_6_27_} and \eqref{ref_6_29_} we opted not to write the tildes overhead. Let's now move to the Einstein frame for the matter conservation equation, which in the Jordan frame has the standard form $\dot{\rho }_{m} +3H\rho _{m} =0$. The transfer is realized by the transforms
\begin{equation} \label{ref_6_28_}
d\tilde{t}=\sqrt{F} dt,\quad \tilde{a}=\sqrt{F} a,\quad \tilde{H}=\frac{1}{\tilde{a}} \frac{d\tilde{a}}{d\tilde{t}} =\frac{1}{\sqrt{F} } \left(H+\frac{\dot{F}}{2F} \right),\quad F=e^{-2\beta \sqrt{\kappa } \phi }.
\end{equation}
Performing the transforms, we will obtain
\begin{equation} \label{ref_6_29_}
\dot{\rho }_{m} +3H\rho _{m} =Q,
\end{equation}
The equations \eqref{ref_6_27_} and \eqref{ref_6_29_}  represent a standard system of interacting components. It is important to note that interaction is "created" by our deviation from general relativity. Interaction  vanishes when $\phi =const$  i.e. when $f(R)$  is linear.

Since the function $ f (R) $ is given mainly phenomenologically, it is interesting to impose some limitations on it which follow from observations.

\subsubsection{Determination of the function $f(R)$ from observations}

It is known that the observed equality of the dark energy density and the matter density of in the Universe in the order of magnitude are coincidence (coincidence problem).
If one assumes that the density ratio $r$ remains constant or changes very slowly during the Universe's evolution, then one can with certainty assume that  $\dot{r}=0$ then, as was shown in, \cite{1212.1111}

\begin{equation}\label{rdot4}
(2q-1)H = -\beta\sqrt{k}\dot{\phi}.
\end{equation}

Using the relation \eqref{ref_6_26_}, in \cite{1212.1111} the equation \eqref{fr} was obtained,  which connects the model function $q$ and the phenomenological function determined from observations
\begin{equation}\label{fr}
\frac{f''}{f'}\dot{R}=-2(2q-1)H
\end{equation}
where a prime denotes a derivative with respect to $R$.\\

The article \cite{1212.1111} provides constraints on the function $f(R)$,  which could alleviate coincidence problem.
So, using the functions $H$ and $q$ derived from observations, it is possible to solve the equation (\ref{fr}) and to obtain the functional form of $f(R)$ (in the $\dot{r}=0$ regime) can be determined in principle.

Note that since there is a certain degeneracy between models with dark matter- dark energy interaction and $f(R) -$ gravity models, the observations that confirm one of the theories indirectly confirm the other theory.

In \cite{1310.0693v3}, the 579 clusters pressure profiles were considered. This analysis was based on the
Yukawa-like correction to the Newtonian potential obtained in the weak field approximation of $f(R) -$ gravity. Based on this analysis, it was shown that the dynamics of clusters at the very least does not contradict the pressure profiles observed in the foreground clean SMICA map released by the Planck Collaboration.

\subsection {Interacting models in $f(T)$ -gravity}

The theory of $f(T)$ -gravity was introduced to explain the current expansion of the universe without the need for a dark energy component. The $f(T)$ theory is a generalization of the teleparallel gravity and becomes equivalent to General Relativity in the absence of torsion. The original idea of the $f(T)$ theory is a generalization of teleparallel gravity, just like $f(R)$ gravity is a generalization of General Relativity - we replace the torsion scalar $T$ in teleparallel gravity with a certain function $f(T)$.
Nevertheless, the positive feature of the  $f(T)$ theory is that the field equations are second order as opposed to the fourth order equations of the $f(R)$
theory.
Below we consider as a model of $f(T)$ -gravity applied to the interacting dark matter and dark energy paradigm.

The action $I$ of modified teleparallel gravity in the movement of $f(T)$ gravity
has the form\cite{1401.8283v1}
\begin{equation}
I=\frac{1}{16\pi\, G}\int\,d^4x\,\sqrt{-g}\,[f(T)+L_m],
\end{equation}
here $L_m$ is linked to the Lagrangian density of the matter
inside of the Universe.

 So, in order to describe the $f(T)$ theory of gravity, we usually begin from the
field equations in a FLRW background filled with
non-relativistic  matter. The Hubble equation has the form  \cite{1401.8283v1}

\begin{equation}
H^2+\frac{k}{a^2}=\frac{1}{3}\,(\rho_m+\rho_T),\label{H_TG}
\end{equation}
the equation for the acceleration
\begin{equation}
\dot{H}-\frac{k}{a^2}=-\frac{1}{2}\,(\rho_m+\rho_T+p_T),\label{dH_TG}
\end{equation}
where the energy density and pressure contributions that are associated with the torsion take the form
\begin{equation}
\rho_T=\frac{1}{2}\,(2Tf'-f-T),\label{rhoT_TG}
\end{equation}
\begin{equation}
p_T=-\frac{1}{2}\,[-8\dot{H}T\,f''+(2T-4\dot{H})f'-f+4\dot{H}-T], \label{pT_TG}
\end{equation}

the primes denotes derivatives with respect to the torsion scalar $T$. Also note that we work in the units in which $8\pi G=1$.

For the non-flat background the torsion scalar is defined, as

\begin{equation}
T=-6\left(H^2+\frac{k}{a^2}\right).\label{T_TG}
\end{equation}

Taking into account the listed-below formulas, one can obtain \cite{1103.0824v3}

\begin{equation}
\rho_m=\frac{1}{2}\,[f-2Tf'].\label{rhom_TG}
\end{equation}

In these models, introducing interaction between dark matter and dark energy (torsion scalar) is no different from other models with interaction in the dark sector \cite{1103.0824v3,1207.2735v1,1210.4612}. So, the corresponding energy-balance equations are

\begin{equation}
\dot{\rho_m}+3H\rho_m=\,Q,\label{drm_TG}
\end{equation}
and
\begin{equation}
\dot{\rho_T}+3H(\rho_T+p_T)=\,-Q,\label{rt_TG}
\end{equation}

Rewriting the last equation in terms of the effective EoS, we obtain
\begin{equation}
\dot{\rho_T}+3H\rho_T\left(1+w_{eff}\right)=0.\label{rt2_TG}
\end{equation}
where the  effective EoS  is given by
\begin{equation}
w_{eff}=w_T+\frac{Q}{3H\rho_T}.\label{rt3_TG}
\end{equation}
Using the equations \eqref{rhoT_TG}, \eqref{pT_TG} and  \eqref{rt3_TG}, we get

\begin{equation}
w_{eff}=-1+\left(\frac{4k}{a^2}-\frac{\dot{T}}{3H}\right)\,\left(\frac{2Tf''+f'-1}{2Tf'-f-T}\right)+\frac{Q}{3H\rho_T}.\label{rt3}
\end{equation}

Following  \cite{1103.0824v3}, we find the time derivative of Eq.\eqref{rhoT_TG}
\begin{equation}
\dot{\rho_T}=\frac{\dot{T}}{2}\,\,[f'+2Tf''-1],\label{ddot_TG}
\end{equation}

thus, the equation of state for the torsion scalar has the form

\begin{equation}
w_T=-\left[1+\frac{Q}{3H\rho_T}+\frac{\dot{T}}{3H}\,\frac{(2Tf''+f'-1)}{(2Tf'-f-T)}\right].\label{wt2}
\end{equation}

Conversely, using \eqref{dH_TG} and \eqref{T_TG},  it  can be easily obtained that
\begin{equation}\label{dotT_TG}
\dot{T}=\frac{12H}{(f'+2Tf'')}\,\left[\frac{(f-2Tf')}{4}+\frac{k}{a^2}\,(f'+2Tf''-1)\right].
\end{equation}

Using the above equations,  it can be finally obtained that

\begin{equation}\label{wtt_TG}	
w_T=-\Bigl[1+\frac{Q}{3H\rho_T}+\frac{4}{(f'+2Tf'')}\,\frac{(2Tf''+f'-1)}{(2Tf'-f-T)}\left(\frac{(f-2Tf')}{4}+\frac{k}{a^2}\,(f'+2Tf''-1)\right)\Bigr]
\end{equation}

The deceleration parameter $q$ can be written as \cite{1103.0824v3}
\begin{equation}\label{q_TG}
q=\frac{1}{2}-\frac{k}{2a^2}\,\left[\frac{T}{6}+\frac{k}{a^2}\right]^{-1}
+\left[\frac{T}{6}+\frac{k}{a^2}\right]^{-1}
\Bigl[\frac{(2Tf'-f-T)}{4}+\frac{Q}{6H}+\,
\frac{(2Tf''+f'-1)}{(f'+2Tf'')}\,\times
\end{equation}
$$
\left(\frac{(f-2Tf')}{4}+\frac{k}{a^2}\,(f'+2Tf''-1)\right)\Bigr].
$$

It should be noted that in the special flat ($k=0$), non-interacting
$Q=0$ the Einstein teleparallel gravity limit in which $f(T)=T$, the
upper formula is given (becomes) $q=1/2$, which represents the matter dominated epoch.

\section{Interacting dark energy models in fractal cosmology}
The fractal properties of quantum gravity theories in $D$ dimensions have been explored in several contexts. To start off, the renormalizability of perturbative gravity at and near two topological dimensions drew much interest to $D=2+\epsilon$ models, with the hope of improving our understanding of the $D=4$ case \cite{GKT,ChD,KN,Wei79,JJ,KKN,AKNT,KPZ}.

Assuming that matter is minimally coupled with gravity, the total action is \cite{Calcagni1,Calcagni2}
\begin{equation}\label{action}
    S=S_g+S_{\rm m}\,,
\end{equation}
where $S_g$ is
\begin{equation}\label{action_grav}
    S_g=\frac{M_p^2}{2}\int {\rm d}\varrho(x)\,\sqrt{-g}\,\left(R-2\lambda-\omega\partial_\mu v\partial^\mu v\right)\,,
\end{equation}
and
\begin{equation}\label{action_matter}
S_{\rm m}=\int {\rm d}\varrho \sqrt{-g} {\cal L}_{\rm m}
\end{equation}
is the matter action. Here,  $g$ is the determinant of the dimensionless metric, $g_{\mu\nu}$, $M_p^{-2}=8\pi G$ is the reduced Planck mass, $R$ is the Ricci scalar, $\lambda$ is the bare cosmological constant, and the term proportional to $\omega$ has been added because $v$, like the other geometric field $g_{\mu\nu}$, is now dynamical. Note that $d\varrho(x)$ is Lebesgue--Stieltjes measure generalizing the  D-dimensional measure ${\rm d}^Dx$. The scaling dimension of $\varrho$ is $[\varrho]=-D\alpha\neq -D$, where $\alpha>0$ is a positive parameter.

 The derivation of the Einstein equations goes almost like it does in scalar-tensor models. Taking the variation of the action (\ref{action}) with respect to
the Friedmann-Lemaitre-Robertson-Walker (FLRW) metric $g_{\mu\nu}$, one can obtain the Friedmann equations in a fractal Universe, as was shown in \cite{Calcagni2}
\begin{equation}\label{fried111}
    \left(\frac{D}2-1\right)H^2+H\frac{\dot v}{v}-\frac{1}{2}\frac{\omega}{D-1}\dot v^2=\frac{1}{M_p^2(D-1)}\rho +\frac{\lambda}{D-1}-\frac{k}{a^2}\,,
\end{equation}
\begin{equation}\label{fried2}
    \frac{\Box v}{v}-(D-2)\left(H^2+\dot H-H\frac{\dot v}{v}+\frac{\omega}{D-1}\dot v^2\right)+\frac{2\lambda}{D-1}=\frac{1}{M_p^2(D-1)}\left[(D-3)\rho+(D-1)p\right].
\end{equation}
where $H=\dot{a}/a$ is the Hubble parameter, $\rho$ and $p$ are the total energy density and pressure of the ideal fluid composing the Universe. The parameter $k$ denotes the curvature of the Universe,
where $k=-1, 0 , +1$ for the close, flat and open Universe respectively. Clearly, when $v=const$, Eqs.(\ref{fried111}) and (\ref{fried2}) transform to the standard Friedmann equations in Einstein GR.

If $\rho+p\neq 0$, the following (purely gravitational) equation is valid (see \cite{Calcagni2} for details):
\begin{equation}\label{grav_constr}
    \dot{H}+(D-1)H^2+\frac{2k}{a^2}+\frac{\Box v}{v}+H\frac{\dot v}{v}+\omega (v\Box v-\dot v^2)=0\,.
\end{equation}
The continuity equation in fractal cosmology takes the form
\begin{equation}\label{cont_eq}
    \dot\rho+\left[(D-1)H+\frac{\dot v}{v}\right](\rho+p)=0\,,
\end{equation}
When $v=1$ and $D=4$, we recover the standard Friedmann equations in four dimensions, eqs. (\ref{fried111}) and (\ref{fried2}) (no gravitational constraint):
\begin{eqnarray}
&&H^2=\frac{1}{3M_p^2}\rho+\frac{\lambda}{3}-\frac{k}{a^2}\,,\label{fr1}\\
&&H^2+\dot{H}=-\frac{1}{6M_p^2}(3p+\rho)+\frac{\lambda}{3}\,.
\end{eqnarray}

On the other hand, for the measure weight
\begin{equation}\label{vt}
v=t^{-\beta},
\end{equation}
where $\beta$ is given by $\beta\equiv D(1-\alpha)$, the gravitational constraint is switched on.  The UV regime, in fact, describes short scales at which inhomogeneities should play some role. If these are small, the modified Friedmann equations define a background for perturbations rather than a self-consistent dynamics.

Recently \cite{Karami}   the holographic, new agegraphic and ghost dark energy models in the framework of fractal cosmology were investigated. In the next section we consider a Universe in which dark energy interacts with dark matter.

For four-dimensional space with a FLRW-metric in the fractal case, and the natural parameterization of the function as $v = t^{-\beta}$, the equations (\ref{cont_eq}) transform to:
\begin{eqnarray}
  &&\dot{\rho}_m+\left(3H-\beta t^{-1}\right)\rho_m=Q,\label{CeQFrm}\\
 &&\dot{\rho}_{x}+(1+w)\left(3H-{\beta} t^{-1}\right)\rho_{x}=-Q,\label{CeQFrm}
\end{eqnarray}
where $\rho_m$ and $\rho_{x}$ are densities of dark matter and
dark energy respectively, and $w$ is the EoS parameter for dark energy.
It is convenient to use the relative energy densities of dark energy and dark matter  in accordance with standard definitions:
\begin{equation}\label{Rel_den}
\Omega_m = \frac{\rho_m}{3M_p^2H^2},~~ \Omega_x = \frac{\rho_x}{3M_p^2H^2}.
\end{equation}
The above equation can be written in terms of these density parameters as the following:
\begin{equation}\label{OmegaCeQ_sys}
\begin{array}{c}
 \dot{\Omega}_m+\left(3H-\beta t^{-1}\right)\Omega_m+2\Omega_m\frac{\dot{H}}{H}=\frac{Q}{3M_p^2H^2},\\
 \\
\dot{\Omega}_x+(1+w_x)\left(3H-\beta t^{-1}\right)\Omega_x+2\Omega_x\frac{\dot{H}}{H}=-\frac{Q}{3M_p^2H^2},\\
\end{array}
\end{equation}

where the dot denotes a derivative with respect to the cosmic time $t.$
The differential equation for the Hubble parameter has the form
\begin{equation}\label{dot_H}
\dot H+H^2-\frac{\beta  H}{2t}+\frac{\beta (\beta +1)}{2t^2}+\frac{\omega  \beta ^2}{3t^{2(\beta +1)}}=-\frac{1}{2}((1+3w)\Omega_x +\Omega_m)H^2.
\end{equation}
In order to obtain the Friedmann equation  in terms of the relative densities, it is necessary to introduce the fictitious density in the same way as
$\Omega_k=k/(a^2H^2).$  So, we introduce the fractal relative density:
\begin{equation}\label{Rel_denFr}
\Omega_f = \frac{\omega \dot{v}^2}{6H^2}- \frac{\dot{v}}{Hv}.
\end{equation}
Taking into account the ansatz $v=t^{-\beta},$ we obtain the equation of motion for fractal relative density
\begin{equation}\label{Rel_denFr_time}
\Omega_f = \frac{\omega \beta^2}{6H^2t^{2(\beta+1)}}+ \frac{\beta}{Ht},
\end{equation}
Thus, the Friedman equation can be re-written in a very elegant form
\begin{equation}\label{Fr_omega}
\sum_{\alpha = k, f, x, m}\Omega_\alpha  \equiv 1.
\end{equation}
Note that within the framework of this definition, the values of the relative densities  $\Omega_x$ or $\Omega_m$ can exceed 1.

\subsection{Linear interaction of dark matter and dark energy}

Below, we consider the simplest form of interaction -- a linear combination of the densities of dark matter and dark energy in a flat Friedmann-Lemaitre-Robertson-Walker fractal Universe with:
\begin{equation}\label{Q lin}
Q \equiv H(\delta\rho_{x} + \gamma\rho_{m}).
\end{equation}
In this case, the equations of motion take the form
\begin{eqnarray}
&&\dot{\Omega}_m+\left(3H-\beta t^{-1}\right)\Omega_m+2\Omega_m\frac{\dot{H}}{H} = H(\delta\Omega_{x} + \gamma\Omega_{m}), \nonumber \\
&&\dot{\Omega}_x+(1+w_x)\left(3H-\beta t^{-1}\right)\Omega_x+2\Omega_x\frac{\dot{H}}{H}=-H(\delta\Omega_{x} + \gamma\Omega_{m}), \label{Omega_xsys}\\
&&\dot{\Omega}_f+\left(\frac{\dot{H}}{H}+2(1+\beta) t^{-1}\right)\Omega_f-\frac{(1+2\beta)\beta}{H t} = 0.\nonumber
\end{eqnarray}
Since the equations explicitly depend on time, it is not possible to find their analytical solution.

\subsection{Analyzable case of dark matter - dark energy interaction}

The analytical solution can be found only in the case when the Hubble parameter is inversely proportional to time, which is typical, for example, at the stage  of nonrelativistic matter dominance. Suppose that at this stage the Hubble parameter has the form $H=\sigma t^{-1}$. Then the equations \eqref{OmegaCeQ_sys} take the following form

\begin{equation}\label{Omega_m_xsys}
  \begin{array}{c}
   {\Omega}'_m=\theta\Omega_m+\sigma\delta\Omega_{x},\\
    {\Omega}'_x=-\delta\gamma\Omega_{m}+\upsilon\Omega_x,
  \end{array}
\end{equation}
where $\theta=2+\gamma\sigma+\beta-3\sigma, ~\upsilon=2-(1+w)(3\sigma-\beta)-\delta\sigma,$ and    the prime   denotes a derivative with respect to the logarithm of cosmic time $'\equiv\frac{d}{d\ln t}.$  Note also that the parameter $\theta$ is physically meaningful  under the condition $\sigma > 0,$ because we do not consider a collapsing Universe.
In this regime of evolution of the Universe, the system of equations is autonomous and can be solved exactly.
The characteristic equation of the system \eqref{Omega_m_xsys} has the form
\begin{equation}\label{Omega_m_xsys_char_eq}
\tau^2-(\theta+\upsilon)\tau+\delta^2\sigma\gamma+\theta\upsilon=0,
\end{equation}
its roots are equal to:
\begin{equation}\label{Omega_m_xsys_roots}
    \tau_{\pm}=\frac{\theta+\upsilon}{2}\left[1\pm\sqrt{1-4\frac{(\delta^2\sigma\gamma+\theta\upsilon)}{(\theta+\upsilon)^2}}\right]
\end{equation}
Let us consider possible types of solutions, and indicate the critical points that correspond to them.
As one can see, this model contains many parameters, making it cumbersome to analyze. Note that due to this feature, the system  describes all possible types of critical points typical of coarse equilibrium states.

Recall that the values of $\beta$ in the IR and UV regimes are $\beta_{\rm
IR}=0$ and $\beta_{\rm UV}=2$ respectively. The UV regime, in fact, describes short scales at which inhomogeneities should play some role. If these are small, the modified Friedmann equations define a background for perturbations rather than a self-consistent dynamics.

\begin{figure}[t]
\centering
\includegraphics[width=0.9\textwidth]{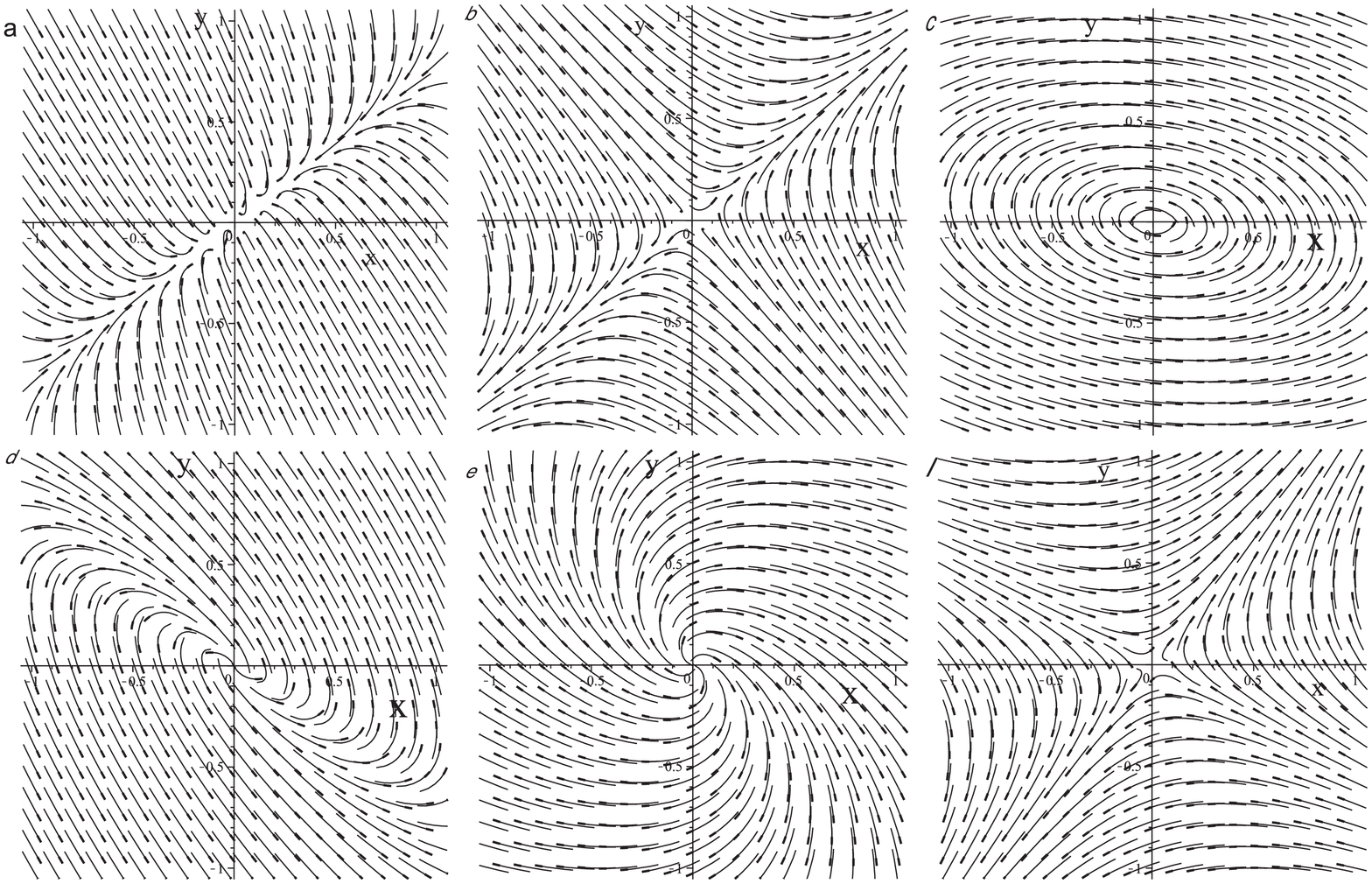}
\caption{Phase portraits for some types of critical points for $w=-1$:
a) stable node,    $\gamma =-2,\; \sigma=3,\;\beta=-1,\;\delta=3 $,
b) stable focus,   $\gamma =-3,\; \sigma=1,\;\beta=2,\;\delta=3 $,
c) center,         $\gamma =3,\; \sigma=3,\;\beta=-1,\;\delta=1 $,
d) unstable focus, $\gamma =3,\; \sigma=1,\;\beta=2,\;\delta=3 $,
e) unstable node,  $\gamma =3,\; \sigma=-3,\;\beta=2,\;\delta=3 $,
i) saddle,         $\gamma =3,\; \sigma=-3,\;\beta=1,\;\delta=-2 $.  }
\label{fig:3}
\end{figure}

There are six types of critical points:
\begin{enumerate}

 \item Stable node $\tau_{\pm}\in\Re,~\tau_{\pm}< 0, \tau_+ >\tau_- >0 $, $ \theta+\upsilon<0, ~4(\delta^2\sigma\gamma+\theta\upsilon) <(\theta+\upsilon)^2, \delta^2\sigma\gamma+\theta\upsilon>0.$

  \item  Unstable node:  $\tau_{\pm}\in\Re,~\tau_{\pm}> 0, \tau_+ >\tau_- >0 $,$ \theta+\upsilon>0, ~4(\delta^2\sigma\gamma+\theta\upsilon) <(\theta+\upsilon)^2, \delta^2\sigma\gamma+\theta\upsilon>0.$

  \item Saddle point: $\tau_{\pm}\in\Re,~ \tau_+\tau_- <0,$
 $ ~\delta^2\sigma\gamma+\theta\upsilon<0.$

  \item Stable spiral point: $\tau_{\pm}\in\mathbb{C},~ \tau\pm = \tau_1\pm i\tau_2,~\tau_1,\tau_2\in\Re~\tau_1,\tau_2>0, $ $\theta+\upsilon<0,(\theta+\upsilon)^2<4(\delta^2\sigma\gamma+\theta\upsilon).$

  \item Unstable spiral point:  $\tau_{\pm}\in\mathbb{C},~ \tau\pm = \tau_1\pm i\tau_2,~\tau_1,\tau_2\in\Re~\tau_1,\tau_2<0, $ $\theta+\upsilon>0,(\theta+\upsilon)^2<4(\delta^2\sigma\gamma+\theta\upsilon).$

 \item Elliptic fixed point $\tau_{\pm}\in\Im,~ \tau_\pm =\pm i\tau ,~\tau\in\Re, $ $\theta=\upsilon,~ \delta^2\sigma\gamma+\theta\upsilon >0.$
\end{enumerate}
     These are all the possible critical points in the system \eqref{Omega_m_xsys}.
Some types of critical points that are typical of this system are shown in figure \ref{fig:3}.

In most cases, linearized system \eqref{Omega_m_xsys} will have real eigenvalues. In these cases, it is important to identify which orbits are attracted to the singular point, and which are repelled away as the independent variable (usually t) tends to infinity.

This is not a true phase-space plot, despite the superficial similarities. One important difference is that a Universe passing through one point can pass  through the same point again but moving backwards along its trajectory, by first going to
infinity and then turning around (recollapsing).

The local dynamics of a singular point may depend on one or more parameters. When small continuous changes in the parameter result in dramatic changes in the dynamics, the singular point is said to undergo a bifurcation. The values of the parameters which result in a bifurcation at the singular point can often be located by examining the linearized system. Singular point bifurcations will only occur if one (or more) of the eigenvalues of the linearized system is a function of the parameter. The bifurcations are located at the parameter values for which the real part of an eigenvalue is zero.
The figure \ref{fig:3} actually shows such bifurcations. Different types of critical points correspond to different values of parameters, and hence different roots \eqref{Omega_m_xsys_roots} of the characteristic equation \eqref{Omega_m_xsys_char_eq}.

\section{Interacting holographic dark energy}

The cosmological constant problem consists of the enormous difference (120 orders of magnitude) between the
observed DE density in the form of the cosmological constant and its 'expected' value.
The expectations are based on rather natural assumptions concerning the cutoff parameter of
the integral that represents the density of zero-point vacuum oscillations.
The holographic principle lets us replace 'natural assumptions' with more rigorous quantitative estimates.

In any effective quantum field theory defined in a spatial region of a characteristic size $L$ and using an ultraviolet cutoff $\Lambda,$ the entropy of the system has the form $S\propto \Lambda^3L^3.$  For example, fermions situated at the nodes of a spatial lattice that has the characteristic size $L$ and the period $\Lambda^{-1}$ are in one of the $2^{(\Lambda L)^3}$ states.
 Consequently, the entropy of such a system is $S\propto \Lambda^3L^3.$ In accordance with the holographic principle, this quantity should satisfy the inequality \cite{CohenKaplanNelson}

\begin{equation}
\label{conditionQFT}
    L^3\Lambda^3\le S_{BH}\equiv \frac{1}{4}\frac{A_{BH}}{l_p^2}=\pi
    L^2M_p^2,
\end{equation}
where $S_{BH}$ is the entropy of a black hole and $A_{BH}$ is the surface area of a black hole event, which in the simplest
 case coincides with the surface of a sphere of the radius $L.$ The relation \eqref{conditionQFT} shows that the value of the infrared (IR)
 cutoff cannot be chosen independently of the value of the ultraviolet (UV) cutoff.

We have obtained an important result \cite{CohenKaplanNelson}: in the framework of holographic dynamics, the value of the IR cutoff is strictly related to the value of the UV cutoff. In other words, physics at small UV scales depends on the physical parameters at large IR scales. For instance, when inequality \eqref{conditionQFT} tends to an exact equality,
\begin{equation}
\label{LsimLambda{-3}}
    L\sim \Lambda^{-3}M_p^{2}.
\end{equation}

Effective field theories with UV cutoffs \eqref{LsimLambda{-3}} necessarily involve numerous states that have a gravitational radius that exceeds the size of the region within which the theory is defined. In other words, for any cutoff parameter, a sufficiently large volume exists in which the entropy in quantum field theory exceeds the Bekenstein limit.
To verify this, we note that the effective quantum field theory is usually required to be capable of describing the system at the temperature $T\le \Lambda .$ For $T\gg 1/L,$ this system has the thermal energy $M\sim L^3T^{4}$ and the entropy $S\sim L^3T^3.$. The condition (\ref{conditionQFT}) is satisfied for $T\le \left( M_{Pl}^2/L \right)^{1/3}$, which corresponds to the gravitational radius $r_g\sim L(LM_{Pl})\gg L.$

 To overcome this difficulty, an even stricter constant is proposed in \cite{CohenKaplanNelson} for the IR cutoff, $L\sim \Lambda^{-1},$  which excludes all states that are within the limits of their gravitational radii. Taking into account that (\ref{rho_{vac}})
\begin{equation}
    \label{rho_{vac}}
		\rho_{vac}\approx \frac{\Lambda^4}{16\pi^2},
\end{equation}
we can rewrite the condition (\ref{conditionQFT}) as
\begin{equation}
   L^3\rho_\Lambda\le L M_{Pl}^2\equiv 2M_{BH},
\end{equation}
where $M_{BH}$ is the mass of a black hole of the gravitational radius $L.$  So, by the order of magnitude, the total energy contained in a region of size $L$  does not exceed the mass of the black hole of the same size.
The quantity $\rho_\Lambda$ is conventionally called ``holographic dark energy''.

In the cosmological context we are interested in, if the total energy contained in a region of size $L$ is postulated to not exceed the mass of the black hole of the same size, i.e.,

\begin{equation}\label{rho-L}
L^3\rho_\Lambda\le M_{BH}\sim LM_{Pl}^2.
\end{equation}
we reproduce the relation between small and large scales in a natural way. If the inequality \eqref{rho-L} were violated, the Universe would only be composed of black holes. Applying this relation to the Universe as a whole, it is natural to identify the IR scale with the Hubble radius $H^{-1}.$
  \begin{equation}\label{LsimLambda{-2}}
 \rho_\Lambda\sim L^{-2}M_{Pl}^{2}\sim H^2M_{Pl}^{2}.
 \end{equation}
 Taking into account that
\[
  M_{Pl} \simeq  1.2\times {{10}^{19}} \,\mbox{GeV}; ~~~  H_0 \simeq  1.6\times 10^{-42} \,\mbox{GeV},
\]

The last quantity is in good agreement with the observed value of DE density $\rho_\Lambda \sim 3\cdot 10^{-47} GeV^4.$ Therefore, in the framework of holographic dynamics, there is no cosmological constant problem.

 We represent the holographic DE density as \cite{Hsu}
\begin{equation}\label{rho_{DE}ci}
    \rho_{L}=3c^2M_p^2L^{-2}.
\end{equation}
The coefficient $3{{c}^{2}}\,(c>0)$ is introduced for convenience,
and $M_p$ continues to stand for reduced Planck mass: $M_p^{-2}=8\pi G.$

When choosing the IR cutoff scale, we have many options, and therefore there is an equally large number of holographic DE models.

Some of these models are flawed: a problem with the equation of state arises in choosing the Hubble radius as the IR scale: in this case,
the holographic DE does not account for the accelerating expansion of the Universe \cite{Hsu}.
The first thing that suggests itself is to replace the Hubble radius with the particle horizon $R_p=a\int_{0}^{t}\frac{dt}{a}=a\int_{0}^{a}\frac{da}{Ha^2}.$
 Regretfully, such a replacement does not yield the desired result. To resolve this and other problems that arise in models with holographic dark energy, models of interacting holographic dark energy were proposed.

As we know, models featuring an interaction between matter and DE were introduced by C. Wetterich to lower the value of the cosmological term by using scalar field \cite{Wetterich_88}, and Horvat first used the holographic principle to analyse cases with decaying cosmological constants  \cite{raul1}.
The holographic dark energy model with interaction between dark energy and dark
matter was first investigated by B. Wang, Y. G. Gong and E. Abdalla in \cite{WangGongAbdalla}. As mentioned above, if  dark energy interacts with cold dark
matter,the continuity equations for them are
\begin{eqnarray}
\label{CEM}
 \dot\rho_{dm}+3H\rho_{dm} &=&Q,\\
\label{CEL}
 \dot\rho_{L}+3H(\rho_{L}+p_{L}) &=& -Q.
\end{eqnarray}
 where $Q$ represent the interaction term.
The interaction between the dark sectors in the HDE model has been
extensively studied in, e.g., \cite{SadjadiHonardoost}. It was found that the
introduction of interaction may not only alleviate the cosmic coincidence problem,
but can also help to arrive at or cross the phantom divide line \cite{Sadjadi,LepePena}.

The density parameters, meanwhile, are
\begin{eqnarray}
\Omega_L={{8\pi\rho_L}\over {3M_{Pl}^2H^2}}\;, \qquad
\Omega_m={{8\pi\rho_m}\over {3M_{Pl}^2H^2}}\;, \qquad \Omega_k ={k \over {H^2a^2}}\;.
\end{eqnarray}
for generality, we consider a Universe with an arbitrary spatial curvature.
The first Friedmann equation in this case takes the form

\begin{equation}
H^2= {{8\pi G}\over {3}}(\rho_L+\rho_m)- {{k}\over {a^2}},
\label{fried}
\end{equation}
which gives
\begin{equation}
\Omega_L +\Omega_m=1+\Omega_k.
\label{omegas}
\end{equation}
The ratio $r$ is related to the density parameters by
\begin{equation}
r={{1-\Omega_L+\Omega_k}\over {\Omega_L}}\;,
\end{equation}
and its time evolution is
\begin{equation}
\dot{r}=3Hr\left [w_L-w_m+{{1+r}\over r}{{\Gamma}\over {3H}}\right ]
=3Hr\left [w_L^{\rm eff}-w_m^{\rm eff}\right ]\;.
\label{rdot}
\end{equation}
It is obvious from Eq.~(\ref{rdot}) that when $w_m^{\rm eff}=w_L^{\rm eff}$ takes place,  the effective
equations of state give $\dot{r}=0$. When this equilibrium takes place, the ratio of dark energy and
dark matter densities is a constant.
can be

To find the time evolution of the Hubble parameter, we combine the Friedmann equation Eq.~(\ref{fried}), and the time evolution of densities:
\begin{equation}
{{1}\over {H}}{{dH}\over{dx}}=-{{3}\over{2}}-{{1}\over{2}}\Omega_k
-{{3}\over{2}}w_L\Omega_L\;,
\label{Hdot}
\end{equation}
where $x=ln(a/a_0)$ with some fixed scale factor $a_0$.
The density parameters satisfy the differential equations
\begin{eqnarray}
{{d\Omega_L}\over {dx}}&=& 3\Omega_L\left [{{1}\over {3}}\Omega_k
+w_L(\Omega_L - 1)-{{\Gamma}\over {3H}}\right ]\;,
\nonumber \\
{{d\Omega_k}\over {dx}}&=&\Omega_k(1+\Omega_k+3w_L\Omega_L)\;.
\label{prelim}
\end{eqnarray}
It was shown in Ref.~\cite{Berger:2006db} that it is enough to make
two physical assumptions in order to determine the parameters of evolution
of the Universe. For instance, we can make an assumption about the dark energy, $\rho_L$, specifically about the nature of DE and its equation of state, or we can make an assumption regarding its interaction parameter $Q$, which is equivalent to an assumption about $\Gamma$. These three values are related by
\begin{equation}
\Gamma=3H(-1-w_L)+2{{\dot L}\over{L}}.
\end{equation}
This equation demonstrates that the interaction should generally be
of the same dimension as the Hubble parameter, and likewise suggests that holographic
definitions for the interaction may be helpful.

So long as the physical meaning of the effective equation of
state \ref{GrindEQ__3_5_} is clear, one can eliminate $w_L$ and $\Gamma$ in favor of
$w_L^{\rm eff}$ and $w_m^{\rm eff}$, and therefore obtain the equations
\begin{eqnarray}
{{d\Omega_L}\over {dx}}&=&-3\Omega_L(1-\Omega_L)
(w_L^{\rm eff}-w_m^{\rm eff})+\Omega_k\Omega_L(1+3w_m^{\rm eff})\;,
\label{lambdadot}
\nonumber \\
{{d\Omega_k}\over{dx}}&=&3\Omega_k\Omega_L(w_L^{\rm eff}
-w_m^{\rm eff})+\Omega_k(1+\Omega_k)(1+3w_m^{\rm eff})\;.
\label{kdot}
\end{eqnarray}
These equations are conformable with the analysis
of Ref.~\cite{Kim:2006kk}, and with the replacement $\Omega_k=0$, one returns to the equation for the flat case
from Ref.~\cite{Berger:2006db}.
The asymptotic behavior and equilibria of these coupled differential equations is defined by its fixed points.

 The emergence of the multipliers
$w_L^{\rm eff}-w_m^{\rm eff}$ and $1+3w_m^{\rm eff}$ are easy to
comprehend from a physical point of view. The first factor only compares whether
dark energy or matter comes to dominate as the Universe expands. The
second factor compares matter to curvature
(``$w_k^{\rm eff}=-{{1}\over{3}}$''),
so it measures whether the density of matter increases or decreases as the Universe expands.

Next, we consider some models of interacting holographic dark energy,
the main difference between them consisting in the the choice of the infrared cutoff scale.
There are various choices for the forms of $Q$. The most common
choice is
\begin{equation}\label{QHDE}
Q = 3\alpha H \rho,
\end{equation}
where $\alpha$ is a dimensionless constant, and $\rho$ is taken to
be the density of dark energy, dark matter, or their sum. In this section, unless otherwise stated, we will consider the case
\begin{equation}\label{QHDE_L}
Q = 3\alpha H \rho_L.
\end{equation}

We will obtain some useful expressions, without specifying the type of holographic dark energy.
For the beginning we can differentiating in time the expression , we obtain a simple relation
To start, we differentiate both sides of the expression  \eqref{rho_{DE}ci}  with respect to time, and obtain.
\begin{equation}\label{DtrhoDE}
    \dot{\rho}_{L}=-2\rho_{L} \frac{\dot{L}}{L},
\end{equation}
substituting this expression into the conservation equations allows us to obtain obtain the effective equation of state parameter for the interacting holographic dark energy:

\begin{equation}\label{W_L}
   w_L \equiv \frac{p_L}{\rho_{L}}=\frac23\frac{\dot{L}}{LH}- \alpha - 1.
\end{equation}

\subsection{Interacting holographic dark energy with the Hubble radius as the IR cutofff}

Setting $L = H^{-1}$ in \eqref{rho_{DE}ci} and working with the equality, it becomes
\begin{equation}\label{W_m_IHDE}
  \rho_{H} = 3\, c^{2}M^{2}_pH^{2}.
\end{equation}
The effective equation of state parameter takes on the form
\begin{equation}
w_H =-\frac23\frac{\dot{H}}{H^2}-\alpha-1, \label{w_H}
\end{equation}

As we see from \eqref{QHDE_L}, the parameter responsible for interaction, $\alpha$,contributes to accelerated expansion when it is positive.
In this case, the Friedmann equation has an exact solution, and so for the Hubble parameter, we obtain
\begin{equation}
H= \frac{2(1-\alpha c^2)}{3-2\alpha c^2} \frac{A}{t},\label{H_H}
\end{equation}
where $A$ is an integration constant, and $t$ is the cosmic time. The time dependence of the scale factor $a(t)$  has the form
\begin{equation}
a(t)=a_0 t^{\frac{2(1-\alpha c^2)}{3-2\alpha c^2} },\label{a_H}
\end{equation}
where $a_0$ is an integration constant.
In conclusion, we find that for this model the deceleration parameter is $q = -1 - \frac{\dot{H}}{H^2}=\frac{1}{2}\left(1 - \frac{Q}{H \rho_{m}}\right):$
\begin{equation}
q(t)=\frac{1}{2(1-2\alpha c^2)},\label{q_H}
\end{equation}
Clearly, if we assume in equations \eqref{H_H}-\eqref{q_H} that $ \alpha = 0,$ (interaction-free case) we obtain expressions for a Universe filled with non-relativistic matter and the Einstein-de Sitter value $q=\frac{1}{2}$.
As seen in this model, the deceleration parameter is constant throughout the evolution  of the Universe and,
therefore, cannot explain the change of phases from a slow (matter dominated) expansion  to an accelerated
(dark energy dominated) expansion of the Universe.

Evidently, a change of $\rho_{H}/\rho_{m}$ needs a corresponding change of
$c^{2}$. Within the framework of this model so far, a dynamical
evolution of the energy density ratio is impossible.
In the article \cite{Pavon_Zimdahl}, the authors consider not only the $L= H^{-1}$ model, but also  studied the case of $c(t).$
In this case, the deceleration parameter is not a constant, which makes it possible (with an appropriate choice of $c(t)$) to describe transient acceleration.
As a way to resolve the problem, it has been suggested that we replace the Hubble scale with various other suitable cosmological scales.

As shown in Ref. \cite{1403.1103v1}, within the framework of this type of holographic dark energy, the transition from decelerated to accelerated expansion of the universe is possible when choosing the interaction term in the following  form  \cite{HDE,1403.1103v1}:
\begin{equation}\label{}
  \frac{Q}{3H\rho_{m}} = \mu \left(\frac{H}{H_{0}}\right)^{-n}\,,
\end{equation}
where $\mu$ is an interaction constant.
The Hubble parameter in this case takes the following form
\begin{equation}
\frac{H}{H_{0}} = \left(\frac{1}{3}\right)^{1/n}\left[1 - 2q_{0} + 2\left(1 + q_{0}\right)a^{-3n/2}\right]^{1/n}\,,
\label{Hqqqqqq}
\end{equation}
where $\mu$ is  determined by the current value $q_{0}$ of the deceleration parameter $q$ by
\begin{equation}
\mu = \frac{1}{3}\left(1 - 2q_{0}\right).
\label{muqqqqqqqqq}
\end{equation}
For $n=2$ one reproduces the $\Lambda$CDM model.

\subsection{Interacting Holographic DE density with the future event horizon as the IR Cutoff}

Although the Hubble radius is the simplest and most theoretically motivated choice for the IR cutoff, we have seen that such a choice cannot recreate the observed phenomena even in the presence of interaction between dark energy and dark matter. Furthermore,  if the interaction rate is given by $Q=9c^{2}\alpha M_{p}^{2}H^{3}~(\alpha >0),$
 the matter density $\rho_{m}$ becomes negative for $a \ll 1.$
This problem does not occur for $\alpha < 0$. Nevertheless, the case $\alpha<0$ not in agreement with observations.

In this subsection, we will consider cosmological models where the future even horizon is chosen as the IR cutoff scale .

It is worth noting that the cosmological horizons being discussed here (with the exception of horizons in de Sitter space and perturbations around it)
do not rapidly settle down to a quasiequilibrium state, but rather go on evolving for all time. The absence of a quasiequilibrium
state manifests also in the absence of a well defined  Hawking temperature for such horizons. Their thermodynamic
significance is therefore much less clear than for either black hole or de Sitter spaces \cite{Davies}.

So, consider the model of interacting holographic dark energy, with the future event horizon chosen as the infrared cutoff scale \cite{li}.
In this case,

\begin{equation}\label{L_f}
L_f=a(t)\int_t^\infty \frac{1}{a(t')}dt'.
\end{equation}
This horizon is the boundary of the volume that a stationary observer may ultimately observe.
In the presence of a big rip at $t=t_s$, the $\infty$ in (\ref{L_f}) must be replaced with $t_s$. Using
\begin{equation}\label{10}
\dot{L_f}=HL_f-1.
\end{equation}
Substituting (\ref{L_f}) into (\ref{W_L}) yields
\begin{equation}\label{w_f}
w=-\frac{1}{3}-\frac{2}{3 c}\sqrt{\Omega_f}-\alpha.
\end{equation}
Then we can calculate the deceleration parameter
\begin{eqnarray}
q& = & -\frac{\ddot{a}}{aH^2}=\frac{1}{2}+\frac{3}{2}w\Omega_{f} =\frac{1}{2}-(1+3\alpha)\Omega_{f}-\frac{1}{c}\Omega_{f}^{\frac{3}{2}}.\label{q_f}
\end{eqnarray}

\begin{figure}[hbtp]
\begin{center}
\includegraphics[width=10cm]{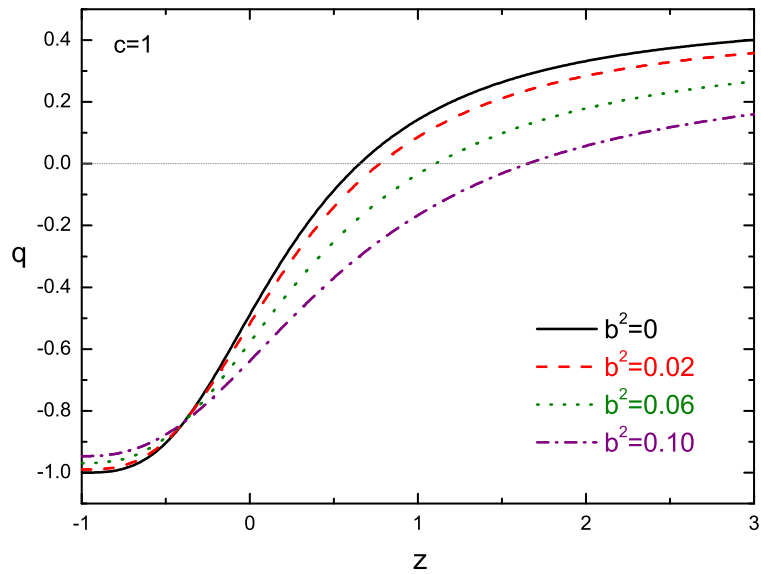}
\end{center}
\caption{Evolution of the deceleration parameter $q$ with a fixed
parameter $c$. In this plot, we take $c=1$, $\Omega_{f0}=0.73$,
and take $\alpha$ as 0, 0.02, 0.06, and 0.10, respectively\cite{ZhangZhangLiu}.}\label{qzgivenc}
\end{figure}

\begin{figure}[hbtp]
\begin{center}
\includegraphics[width=10cm]{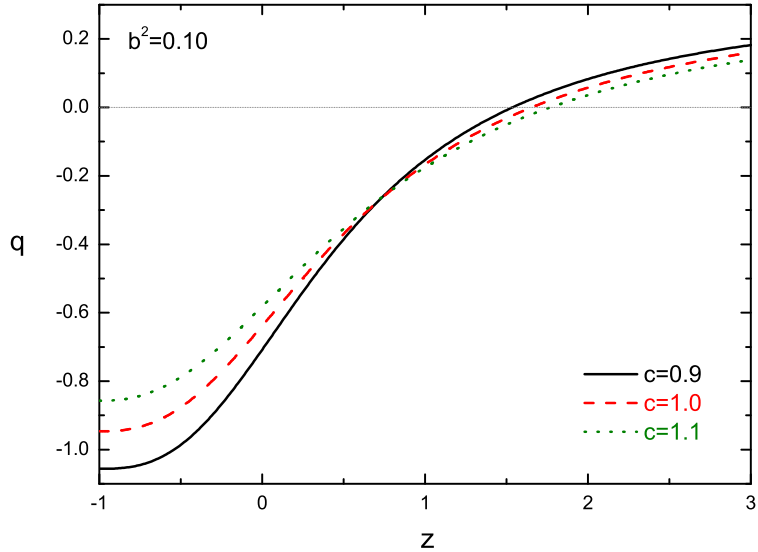}
\end{center}
\caption{Evolution of the deceleration parameter $q$ with a fixed
coupling $\alpha$. In this plot, we take $\alpha=0.10$, $\Omega_{f0}=0.73$, and take $c$ as 0.9, 1.0 and 1.1, respectively\cite{ZhangZhangLiu}.\label{qzgivenb}}
\end{figure}

In order to see how interaction acts on the evolution of the Universe, dependences of the deceleration parameter $q$ on $z$ at various values of the interaction parameter $\alpha$ are shown in Fig. \ref{qzgivenc} . In Fig. 1, we fix
$c=1$ and take the coupling constant $\alpha$ as 0, 0.02, 0.06, and
0.10. Furthermore, cases with a fixed $\alpha$ and
various values of $c$ are also interesting. In Fig. \ref{qzgivenb}, fixing the
coupling constant as $\alpha=0.10$, we plotted the evolution diagram of the
deceleration parameter $q$ with different values of $c$
(here we take the values of $c$ as 0.9, 1.0, and 1.1).
From Figs. \ref{qzgivenc} and \ref{qzgivenb} we find that the Universe underwent early
deceleration and late-time acceleration. Fig. \ref{qzgivenc} displays that, for a
constant parameter $c$, the cosmic acceleration starts earlier for the
cases with interaction than the ones without coupling (formerly, it was
discussed by Amendola in \cite{Amendola:2002kd}). Furthermore, the
stronger the coupling and dark energy and dark matter, the
earlier the start of accelerated expansion. Nevertheless, cases with
smaller coupling lead to bigger acceleration in the
 distant future. In addition, Fig.\ref{qzgivenb} shows that the acceleration starts
earlier when $c$ is larger for the same coupling $\alpha$, but
a smaller $c$ will eventually result in greater acceleration. It should be
indicated that, in the interacting holographic dark energy model,
the interaction intensity has an upper limit because of the
evolutionary behavior of the holographic dark energy, more specifically,
its tracking of dark matter. For more explicit discussions about the relationship of the coupling $\alpha$ and the
parameter $c$, see \cite{Wang:2005jx}. It is of note that, in
the presence of interaction between dark energy and dark matter, the case of
$c=1$ cannot create a de Sitter phase in the infinite future. In a word, the effect that the interaction between dark energy and dark
matter has on the evolution of the Universe is evident, as demonstrated by Figs. \ref{qzgivenc}
and \ref{qzgivenb}.


\subsection{Interacting Holographic Ricci dark energy}

The present subsection concentrates on the holographic Ricci dark energy
(RDE) model. In this model, the IR cutoff
length scale $L$ takes the form  of the absolute value of the Ricci scalar
curvature ${|\cal R|}^{-1/2}$. Therefore, in this instance, the density of the
holographic dark energy is $\rho_{_{\cal R}}\propto {\cal R}$.

The energy density of dark energy in the IRDE model is defined as~\cite{gao}
 \begin{eqnarray}
 \label{rde}
  \rho_{_{\cal R}}
       &=& 3\alpha M_{p}^{2}
       \left (
       \dot{H}+2H^{2} + \frac{k}{a^{2}}
       \right),
                \label{RDE}
 \end{eqnarray}
 where $\alpha$ is a dimensionless parameter.
 Note that $\rho_{_{\cal R}}$ is proportional to the Ricci scalar curvature
 \begin{eqnarray}
 {\cal{R}}= -6
                      \left(\dot{H}+2H^{2}+\frac{k}{a^{2}}
                      \right).
 \end{eqnarray}

This subsection, for the sake of completeness, and in order to follow the article \cite{SuwaNihei}, will also considered relativistic matter to be one of the components of the Universe.

The time evolution of the scale factor $a(t)$ is described by the Friedmann equation
 \begin{eqnarray}
  H^{2} = \frac{1}{3M_{p}^{2}}
                (\rho_{_{\cal R}}+\rho_{m}+\rho_{\gamma}+\rho_{k}),
  \label{Friedmann}
  \end{eqnarray}
where $\rho_{_{\cal R}}, \rho_{m}, \rho_{\gamma}$ and $\rho_{k}$ represent the energy densities of dark energy,
matter, radiation and curvature, respectively.

The interaction rate is given by
 \begin{eqnarray}
 \label{intrate}
   Q=\gamma H \rho_{_{\cal R}},
  \end{eqnarray}
  where $\gamma$ is a dimensionless parameter.
The energy density of radiation is given by
$\rho_{\gamma}=\rho_{\gamma 0}a^{-4}$, where $\rho_{\gamma 0}$
is the present value of radiation density.
We adopt the convention that $a(t_{0})=1$ for the present age of the Universe
$t_{0} \approx 14$ Gyr.
According to eq.~(\ref{CEM}) with $Q$ taken from eq.~(\ref{intrate}),
the interaction can be relevant as long as $\gamma \rho_{_{\cal R}}$ and $\rho_{m}$ are
comparable, whether or not the Universe is in the radiation-dominated epoch.

 Combined with eqs. (\ref{RDE}) and (\ref{CEM}), the Friedmann equation (\ref{Friedmann})
 is written as
  \begin{eqnarray}
\nonumber
    \frac{ \alpha}{2} \frac{d^{2}H^{2}}{dx^{2}}
    - \left(1-\frac{7\alpha}{2} -\frac{\alpha \gamma}{2} \right)\frac{d H^{2}}{dx}
    -(3-6\alpha-2\alpha \gamma) H^{2}
    &&\\
\label{EOH}
               - \frac{\rho_{\gamma 0}}{3M_{p}^{2}} e^{-4x}
                -\{1- \alpha(1+\gamma ) \}  k e^{-2x}
  &=&0,
 \end{eqnarray}
 where $x=\ln{a}$.
The solution to eq. (\ref{EOH}) is obtained as
 \begin{eqnarray}
 \label{hubbleparameter}
  \frac{H^{2}}{H_{0}^{2}}    &=& A_{+} e^{\sigma_{+} x} + A_{-} e^{\sigma_{-} x}
            + A_{\gamma} e^{-4 x} + A_{k} e^{-2 x} ,
\label{solution}
 \end{eqnarray}
where
 \begin{eqnarray}
 \label{sigmapm}
  \sigma_{\pm}
   &=& \frac{2-7\alpha-\alpha \gamma \pm\sqrt{(2-\alpha)^{2}
   -2\alpha (\alpha +2) \gamma+\alpha^{2}\gamma^{2}}}{2\alpha},
 \end{eqnarray}
   $\Omega_{\gamma 0}=\rho_{\gamma 0}/\rho_{c0}$,
 $\Omega_{k 0}= - k/H_{0}^{2}$ and
 $\rho_{c0}=3M_{p}^{2}H_{0}^{2}$.
 Note that $\sigma_{\pm}$ can be imaginary for
 sufficiently large $\alpha$ and $\gamma$.
 This implies that there is a parameter region where $H^{2}$
 has oscillatory behavior.
 However, this region is not phenomenologically viable.
 The constants $\Omega_{\gamma 0}$ and $\Omega_{k 0}$ are the present
 value of $\Omega_{\gamma}$ and $\Omega_{k}$, respectively.
 The constants $A_{\gamma}$, $A_{k}$ and $A_{\pm}$ are given by
  \begin{eqnarray}
  A_{\gamma} &=&  \Omega_{\gamma 0} ,
 \end{eqnarray}
 \begin{eqnarray}
  A_{k} &=&   \Omega_{k 0} ,
 \end{eqnarray}
 \begin{eqnarray}
 \label{rholambda}
 A_{\pm} &=& \pm \frac{
            \alpha(\sigma_{\mp}+3) \Omega_{k 0}
                + 2 \Omega_{\Lambda 0}
                -\alpha (1-\Omega_{\gamma 0})(\sigma_{\mp}+4)  }
               {\alpha(\sigma_{+}-\sigma_{-})}.
 \end{eqnarray}
In  the absence of interaction ($\gamma=0$), eq. (\ref{solution}), the result is reduced to the one obtained
in the article~\cite{gao}.
In this case, the constants in eq. (\ref{sigmapm})
are $\sigma_{+}=-4 +2/\alpha$ and $\sigma_{-}=-3$.

 Replacing eq. (\ref{solution}) to eq. (\ref{RDE}),
 the Ricci dark energy density is given by
 \begin{eqnarray}
 \label{rdes}
  \rho_{_{\cal R}}
    &=&   \rho_{c0}
              \sum_{i=+,-} \alpha
              \left(\frac{ \sigma_{i}}{2}+2
                                    \right)
                                    A_{i} e^{\sigma_{i}x} .
 \end{eqnarray}
Moreover , the matter density is
 \begin{eqnarray}
 \label{rms}
  \rho_{m}
    &=&  \rho_{c0}
              \sum_{i=+,-}
             \left\{1-
                       \alpha \left(\frac{ \sigma_{i}}{2} +2 \right)
             \right\}
                         A_{i} e^{\sigma_{i}x}.
  \end{eqnarray}
The equation of state of dark energy  can be found by
 substituting eq. (\ref{rdes}) into the following expression:
 \begin{eqnarray}
  w_{_{\cal R}}
    =  -1-\frac{1}{3}
             \left(
                      \gamma + \frac{1}{\rho_{_{\cal R}}}\frac{d\rho_{_{\cal R}}}{dx}
             \right).
  \end{eqnarray}

In eqs.~(\ref{rdes}) and (\ref{rms}), the term proportional to  $e^{\sigma_{-}x}$ is dominant in the past( $a$ $\ll$ 1), while
the term proportional to $e^{\sigma_{+}x}$ is dominant in the future ($a$ $\gg$ 1).
As an illustration, let us consider the case $\alpha$ $=$ 0.45 and $\gamma$
$=$ 0.15, which corresponds to $\sigma_{+}$ $\approx$ 0.25 and
$\sigma_{-}$ $\approx$ $-3.0$.
In the past ($a$ $\ll$ 1), the ratio of eq.~(\ref{rms}) to eq.~(\ref{rdes}) is
$\rho_m/\rho_{_{\cal R}}$ $\approx$ $\alpha^{-1}(2+\sigma_{-}/2)^{-1}-1$ $\approx$ 3.4,
while $\rho_m/\rho_{_{\cal R}}$ $\approx$ $\alpha^{-1}(2+\sigma_{+}/2)^{-1}-1$
$\approx$ 0.045 in the future ($a$ $\gg$ 1).

Note that the evolution of both $\rho_{_{\cal R}}$ and $\rho_{m}$ is not characteristic of these types of dark energy, and is actually caused by interaction.
 This leads to a constant ratio of $\rho_{_{\cal R}}$ to $\rho_{m}$,
 and it may help in resolving the coincidence problem.
 As shown in Ref.~\cite{gao}, the coincidence problem is less of an issue
in the original RDE model without interaction between dark matter and dark energy where
$\rho_{_{\cal R}}$ and $\rho_{m}$ were comparable with each other in the
past Universe. Due to this, $\rho_{_{\cal R}}$ starts to increase at low redshifts,
and the ratio $\rho_{_{\cal R}}/\rho_{m}$ rapidly grows in the future, since
$\rho_{m} \sim e^{-3x}$ in the absence of interaction. On the other hand, in the
IRDE model, the behavior in the past is similar to that in the original
RDE model, but the ratio $\rho_{_{\cal R}}/\rho_{m}$ is constant even in
the future \cite{SuwaNihei}.
\subsubsection{Exact solutions for various linear interactions between Ricci DE and DM}
In \cite{Chimento3}, the author considered a model  with cold dark matter
coupled to a modified holographic Ricci dark energy by means of a general interaction term linear in the energy
densities of dark matter and dark energy, the total energy density and its derivative. This parameterization is valuable because it lets us obtain analytical solutions of systems of cosmological equations of motion.

In a FLRW background, the Einstein equation for a model of cold dark matter of the energy density $\rho_c$ and modified holographic Ricci dark energy of the energy density $ \rho_x =\left(2\dot H + 3\alpha H^2\right)/\Delta$, reads
\begin{equation}
\label{1}
3H^2=\rho = \rho_c + \rho_x,
\end{equation}
where $\alpha$ and $\beta$ are constants and $\Delta=\alpha -\beta$.

In terms of the variable  $\eta = 3\ln(a/a_0)$, the  compatibility between the global conservation equation
\begin{equation}
\label{2}
\rho ' = d\rho/d\eta= -\rho_c-(1+\omega_x)\rho_x,
\end{equation}
and the equation deduced from the expression of the modified holographic Ricci dark energy
\begin{equation}
\label{3}
\rho ' = -\alpha\rho_c-\beta\rho_x,
\end{equation}
namely, $(\rho_c + \gamma_x\rho_x)=(\alpha\rho_c + \beta\rho_x)$, gives a relation between the EoS parameter of the dark energy component $\omega_x = \gamma_x - 1$ and the ratio $r= \rho_c/\rho_x$
\begin{equation}
\label{4}
\omega_x=(\alpha  - 1)r+\beta-1.
\end{equation}
Solving the system of equations (\ref{1}) and (\ref{3}), we get $\rho_c$ and $\rho_x$ in terms of $\rho$ and $\rho'$:
\begin{equation}
\label{5}
\rho_c=-(\beta\rho+\rho')/\Delta,  \qquad   \rho_x=(\alpha\rho+\rho')/\Delta.
\end{equation}
The interaction between the dark components is introduced through the term $Q$ by means of splitting the Eq.(\ref{3}) into $\rho'_c+ \alpha \rho_c = - Q $ and  $ \rho'_x+ \beta \rho_x  =   Q $. Then, differentiating $\rho_c$ or $\rho_x$ in (\ref{5}) and using the expression of  $Q$,  we obtain a second order differential equation for the total energy density $\rho$   \cite{arXiv:0911.5687}
\begin{equation}
\label{6}
\rho''+(\alpha+\beta)\rho'+\alpha\beta\rho = Q \Delta.
\end{equation}
For a given interaction $Q$, solving Eq. (\ref{6}) gives us the total energy density $\rho$ and the energy densities $\rho_{c}$ and $\rho_{x}$ after using Eq. (\ref{5}).
The general linear interaction $ Q$ \cite{arXiv:0911.5687}, linear  in $\rho_{c}$, $\rho_{x}$, $\rho$, and $\rho'$, can be written as
\begin{eqnarray}
\label{7}
Q= c_1 \frac{(\gamma_s - \alpha)(\gamma_s-\beta)}{\Delta}\,\rho + c_2 (\gamma_s-\alpha)\rho_c \\
\nonumber
- c_3 (\gamma_s -\beta)\rho_x -c_4 \frac{(\gamma_s - \alpha)(\gamma_s-\beta)}{\gamma_s\Delta}\,  \rho',
\end{eqnarray}
where $\gamma_s$ is constant and  the coefficients $c_{i}$ fulfill the condition $c_{1}+c_{2}+c_{3}+c_{4}=1$ \cite{arXiv:0911.5687}.
Now, using Eqs. (\ref{5}), we rewrite the interaction (\ref{7}) as a linear combination of $\rho$ and $\rho'$,
\begin{equation}
\label{8}
Q=\frac{u\rho+\gamma^{-1}_{s}[u-(\gamma_s-\alpha)(\gamma_s-\beta)]\rho'}{\Delta},
\end{equation}
where $u=c_1(\gamma_s -\alpha)(\gamma_s-\beta)-c_{2}\beta(\gamma_s-\alpha)-c_{3}\alpha(\gamma_s-\beta)$.
Placing the interaction (\ref{8}) into the source equation (\ref{6}), we obtain
\begin{equation}
\label{9}
\rho'' + (\gamma_s+ \gamma^{+})\rho'+ \gamma_{s}\gamma^{+}\rho=0.
\end{equation}
 where the  roots of the characteristic polynomial associated with the second order linear differential equation (\ref{9}) are  $\gamma_{s}$ and $\gamma^{+}=(\beta\alpha -u)/{\gamma_s}$.
In what follows, we adopt $\gamma^{+}=1$ in order to mimic the dust-like behavior of the Universe at early times. In that case, the general solution of (\ref{9}) is   $\rho=b_1a^{-3\gamma_s}+b_2a^{-3}$,  from which we obtain
\begin{subequations}
\label{10}
\begin{equation}
\label{10a}
\rho_c=\frac{(\gamma_s-\beta)b_1a^{-3\gamma_s}+(1-\beta)b_2a^{-3}}{\Delta},
\end{equation}
\begin{equation}
\label{10b}
\rho_x=\frac{(\alpha-\gamma_s)b_1a^{-3\gamma_s}+(\alpha-1)b_2a^{-3}}{\Delta}.
\end{equation}
\end{subequations}
Interestingly, Eqs. (\ref{10}) tell us that  the interaction (\ref{8}) seems to be a good candidate for alleviating the cosmic coincidence problem, since the ratio $\Omega_{c}/\Omega_{x}$ becomes bounded for all times.

\subsubsection{DM and Ricci-like holographic DE coupled through a quadratic interaction}

Now we consider cosmological models where the interaction $Q$ between the dark components is nonlinear and includes a set of terms  which are homogeneous of degree 1 in the total energy density and its first derivative \cite{jefe1},
\begin{equation}
\label{Q}
Q=\frac{(\alpha\beta -1)}{\Delta\gamma}\,\rho+\frac{(\alpha + \beta -\nu-2)}{\Delta\gamma}\,\rho'-\frac{\nu\rho'^{2}}{\rho\Delta\gamma},
\end{equation}
where $\nu$ is a positive constant that parameterizes the interaction term $Q$. Putting (\ref{Q}) into (\ref{6}) turns it into a nonlinear second order differential equation for the energy density: $\rho\rho''+(2+\nu)\rho\rho'+\nu\rho'^{2}+\rho^2=0$. Introducing the new variable $y=\rho^{(1+\nu)}$ into the latter equation, one gets a second order linear differential equation, $y''+(2+\nu)y'+(1+\nu)y=0$, whose solutions allow us to write the energy density as
\begin{equation}
\label{Et}
\rho=\left[\rho_{10}a^{-3}+\rho_{20}a^{-3 (1+\nu)}\right]^{1/(1+\nu)}
\end{equation}
where $\rho_{10}$ and $\rho_{20}$ are positive constants. Using Eqs. (\ref{10})-(\ref{Et}), as well as the fact that  $p =-\rho -\rho'$, we find both dark energy densities and the total pressure:
\begin{equation}
\label{cI}
\rho_c=\frac{-\rho}{\alpha-\beta}\left[\beta-1+\frac{\nu}{(1+\nu)(1+\rho_{20}a^{-3\nu}/\rho_{10})}\right],\,\,\,
\end{equation}
\begin{equation}
\label{xI}
\rho_x=\frac{\rho}{\alpha-\beta}\left[\alpha-1+\frac{\nu}{(1+\nu)(1+\rho_{20}a^{-3\nu}/\rho_{10})}\right],\,\,\,\,\,\,\,\,\,\,\,\,	
\end{equation}
\begin{equation}
\label{24}
p=-\frac{\nu\rho_{10}}{1+\nu}\,\frac{a^{-3}}{\rho^\nu}.
\end{equation}
From these equations we see that an initial model of interacting dark matter and dark energy can be associated with
an effective one-fluid description of an unified cosmological scenario where the effective one-fluid, with energy density $\rho=\rho_c+\rho_x$ and pressure (\ref{24}), obeys the equation of state of a relaxed Chaplygin gas $p=b\rho+f(a)/\rho^\nu$, where $b$ is a constant \cite{jefe1}. The effective barotropic index $\omega=p/\rho=\omega_x\rho_x/\rho$ reads
\begin{equation}
\label{eb}
\omega=-\frac{\nu\rho_{10}}{(1+\nu)(\rho_{10}+\rho_{20}a^{-3\nu})}.
\end{equation}
At early times and for $\nu>0$,  the effective energy density  behaves as $\rho\approx a^{-3}$, the effective barotropic index behaves as (\ref{eb}) $\gamma\approx 1$ and  the effective fluid describes a Universe dominated by nearly pressureless dark matter. However, a late time accelerated Universe ($\omega<-1/3$) that has positive dark energy densities requires that $\nu>1/2$, $\beta<1$ and $\alpha>1$. From now on we adopt the latter restrictions.

\subsection{Interacting agegraphic dark energy models}
From the first days of quantum mechanics, the concept of measurements [real and thought (gedankenexperiment)] has played a fundamental role in our understanding of physical
reality. GR asserts that the laws of classical physics can be verified with unlimited accuracy. The relation revealed above between the macroscopic (IR) and microscopic scales dictates the necessity of a more profound analysis of the measurement process. The uncertainty relation, together with GR, produces the fundamental space time scale—the Planck length ${{L}_{p}}\sim {{10}^{-33}}\, cm.$
The existence of a fundamental length influences the process of measurement in a critical manner \cite{PCBGFL}. We assume that a fundamental length $L_f$ exists. Because the space time coordinate system must be physically reasonable, it has to be attached to physical bodies. Therefore, postulating the fundamental length is equivalent to imposing restrictions on the realizability of precise coordinate systems. In terms of experiments with light signals, this means, for example, that the time required for a light signal to travel from body A to body  and back, measured by clocks in the system of A, is subject to uncontrollable fluctuations. Fluctuations in experiments with light signals should be considered indications of fluctuations of the metric, i.e., the gravitational field. Therefore, postulating the existence of a fundamental length is equivalent to postulating fluctuations of the gravitational field.

A direct consequence of the existence of quantum fluctuations of the metric \cite{Sasakura,Maziashvili_1,Maziashvili_2,0707.4049} is the following conclusion, related to the problem of measuring distances in Minkowski space: the distance $t$  (we recall that we use the system in which $c=\hbar =1,$ whence ${{L}_{p}}={{t}_{p}}=M_{p}^{-1}$) cannot be measured with an accuracy exceeding\cite{Karolyhazy}
\begin{equation}
\label{Karolyhazy_rel}
\delta t=\beta t_{p}^{2/3}{{t}^{1/3}},
\end{equation}
where $\beta$ is a coefficient of the order of unity. Following \cite{CohenKaplanNelson}, we can consider the result \eqref{Karolyhazy_rel} to be a relation between the UV and IR scales in the framework of the effective quantum field theory satisfying the entropic peculiarities of black holes. Indeed, rewriting the relation \eqref{LsimLambda{-2}} in terms of length and performing the substitution $\Lambda \to \delta t$, we reproduce \eqref{Karolyhazy_rel}, but in the holographic interpretation.

The relation \eqref{Karolyhazy_rel}, together with the quantum mechanical energy-time uncertainty relation, allows us to estimate the energy density of quantum fluctuations of the Minkowski space time. In accordance with \eqref{Karolyhazy_rel}, we can regard a region of volume $t^3$ as composed of cells of volume $\delta {{t}^{3}}\sim t_{p}^{2}t$. Consequently, each such cell represents a minimally detectable unit of space time for the scale t. If the age of the region chosen is t, its existence, in accordance with the time-energy uncertainty principle, cannot be realized with an energy less than $\sim {{t}^{-1}}.$ We thus arrive at the conclusion: if the lifetime (age) of a certain spatial region of the linear size $t$ is equal to $t,$ there exists a minimal cell with the volume $\delta {{t}^{3}},$ whose energy cannot be less than
 \begin{equation}
  \label{E_delta_t^3}
  E_{\delta t^3} \sim t^{-1}.
  \end{equation}
It immediately follows from \eqref{Karolyhazy_rel} and \eqref{E_delta_t^3} that in accordance with the energy time uncertainty principle, the energy density of metric (quantum!) fluctuations in Minkowski space is \cite{Sasakura,Maziashvili_2,0707.4049}
\begin{equation}\label{rho_q}
{{\rho }_{q}}\sim \frac{{{E}_{\delta {{t}^{3}}}}}{\delta {{t}^{3}}}\sim \frac{1}{t_{p}^{2}{{t}^{2}}}.
\end{equation}
It is essential that the dynamic behavior of the density of metric fluctuations \eqref{rho_q} coincides with that of the holographic
DE introduced in \eqref{LsimLambda{-2}} and \eqref{rho_{DE}ci}, although the derivations of these expressions are based on absolutely different physical principles. The holographic DE density was obtained from entropic constraints (the holographic principle), while the energy density of metric fluctuations in Minkowski space is only related to their quantum nature, namely, to the uncertainty principle.

The relation \eqref{rho_q} allows us to introduce an alternative model of holographic DE \cite{0707.4049}, in which the age of the Universe $T$ is used as the IR scale. In such a model,
\begin{equation}\label{rho_qn}
\rho_q=\frac{3{{n}^{2}}M_{p}^{2}}{{{T}^{2}}}.
\end{equation}
where $n$ is a free parameter of the model, and the numerical coefficient was introduced for convenience.
The age of the Universe $T,$ involved in \eqref{rho_qn}, is related to the scale factor as

\begin{equation}\label{AGE_U}
T=\int_{0}^{a}{\frac{d{a}'}{H{a}'}}.
\end{equation}
 It is convenient to introduce the fractional energy densities
 $\Omega_i\equiv\rho_i/(3M_p^2H^2)$ for $i=m$ and $q$. From
 Eq.~\eqref{rho_qn}, it is easy to find that
 \begin{equation}\label{eq7}
 \Omega_q=\frac{n^2}{H^2T^2}.
\end{equation}

Although agegraphic dark energy (ADE) is the quantum fluctuation of spacetime, it might decay into matter, similar to the SCM model in which the vacuum fluctuations can decay into matter. This effect could be described by the interaction term $Q$ phenomenologically. From Eq.~(\ref{eq7}),
 we get
\begin{equation}\label{eq13}
 \Omega_q^\prime=\Omega_q\left(-2\frac{\dot{H}}{H^2}
 -\frac{2}{n}\sqrt{\Omega_q}\right).
\end{equation}
 Differentiating Eq.~\eqref{eq7} it is easy to find that
\begin{equation}\label{eq14}
 -\frac{\dot{H}}{H^2}=\frac{3}{2}\left(1-\Omega_q\right)
 +\frac{\Omega_q^{3/2}}{n}-\frac{Q}{6M_p^2 H^3}.
\end{equation}
 Therefore, we obtain the equation of motion for $\Omega_q$,
\begin{equation}\label{eq15}
 \Omega_q^\prime=\Omega_q\left[\left(1-\Omega_q\right)
 \left(3-\frac{2}{n}\sqrt{\Omega_q}\right)
 -\frac{Q}{3M_p^2 H^3}\right],
\end{equation}
 where
\begin{equation}\label{eq16}
 \frac{Q}{3M_p^2 H^3}=\left\{
 \begin{array}{ll}
 3\alpha\Omega_q & {\rm ~for~~} Q=3\alpha H\rho_q \\
 3\beta\left(1-\Omega_q\right) & {\rm ~for~~} Q=3\beta H\rho_m \\
 3\gamma & {\rm ~for~~} Q=3\gamma H\rho_{tot}
 \end{array}
 \right..
\end{equation}
 From
  (\ref{AGE_U}) and~(\ref{eq7}), we get the
 EoS of the ADE, namely
\begin{equation}\label{eq17}
 w_q=-1+\frac{2}{3n}\sqrt{\Omega_q}-\frac{Q}{3H\rho_q},
\end{equation}
 where
\begin{equation}\label{eq18}
 \frac{Q}{3H\rho_q}=\left\{
 \begin{array}{ll}
 \alpha & {\rm ~for~~} Q=3\alpha H\rho_q \\ \vspace{0.75mm}
 \beta\left(\Omega_q^{-1}-1\right) & {\rm ~for~~} Q=3\beta H\rho_m\\
 \gamma\,\Omega_q^{-1} & {\rm ~for~~} Q=3\gamma H\rho_{tot}
 \end{array}
 \right..
\end{equation}
 Using Eq.~(\ref{eq14}), the deceleration parameter is given by
\begin{equation}\label{eq19}
 q\equiv -\frac{\ddot{a}a}{\dot{a}^2}=-1-\frac{\dot{H}}{H^2}
 =\frac{1}{2}-\frac{3}{2}\Omega_q
 +\frac{\Omega_q^{3/2}}{n}-\frac{Q}{6M_p^2 H^3}.
\end{equation}
 The  total EoS $w_{tot}\equiv p_{tot}/\rho_{tot}=\Omega_q w_q$,
 where $w_q$ is given in Eq.~(\ref{eq17}). On the other hand, èñïîëüçóÿ the Friedmann and  Raychaudhuri equation,
 ìîæíî íàéòè $w_{tot}=-1-\frac{2}{3}\frac{\dot{H}}{H^2}=-1/3+2q/3$.
 As mentioned above, in the case of $Q=0$, $n>1$ is necessary to drive the (present)
 accelerated expansion of our Universe. In the case of $Q\not=0$,
 the situation is changed. For example, if $Q=3\alpha H\rho_q$,
 to drive the accelerated expansion of our Universe,
 we should have $w_{tot}=\Omega_q w_q<-1/3$, which means that
 $n>2\Omega_q^{3/2}[3(1+\alpha)\Omega_q-1]^{-1}$. It is easy to see that the minimum of the right hand side of this inequality
 is $(1+\alpha)^{-3/2}$ at $\Omega_q=(1+\alpha)^{-1}$, if
 $\Omega_q>[3(1+\alpha)]^{-1}$~(nb. $\Omega_q\simeq 0.7$
 today). For $\alpha>0$, this minimum $(1+\alpha)^{-3/2}$ is
 smaller than $1$.

When obtaining the dependencies  $\Omega_q$, $w_q$,
 $q$ and $w_{tot}$, \cite{0707.4052} shows some numerical plots by using
 Eqs.~(\ref{eq15})---(\ref{eq19}) and $w_{tot}=\Omega_q w_q$.
 However, for the sake of brevity, we do not present plots for all forms
 of interaction $Q$. In what follows, we mainly focus on
 the case of $Q=3\alpha H\rho_q$ as an example. Note that
 in the numerical integration of Eq.~(\ref{eq15}) we use
 the initial condition $\Omega_{q0}=0.7$ for demonstration.

\begin{center}
 \begin{figure}[htbp]
 \centering
 \includegraphics[width=0.75\textwidth]{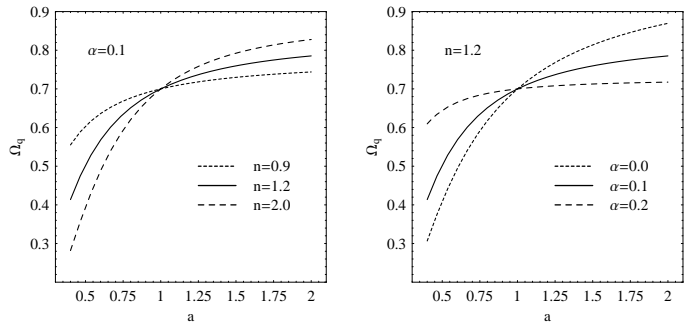}
 \caption{\label{fig1} Evolution of $\Omega_q$ for various
 model parameters $n$ and $\alpha$ in the case of
 $Q=3\alpha H\rho_q$ \cite{0707.4052}.}
 \end{figure}
 \end{center}

Fig.\ref{fig1} shows the evolution of $\Omega_q$ for
 different model parameters $n$ and $\alpha$ in the case of
 $Q=3\alpha H\rho_q$. It is easy to see that \cite{0707.4052} for a fixed
 $\alpha$, which describes the interaction between the
 agegraphic dark energy and the pressureless (dark) matter,
 the agegraphic dark energy starts to be effective earlier
 and $\Omega_q$ tends to a lower value at the late time when
 $n$ is smaller. On the other hand, for a fixed $n$, the
 agegraphic dark energy starts to be effective earlier and
 $\Omega_q$ tends to a lower value at the late time when
 $\alpha$ is larger \cite{0707.4052}. Interestingly enough, these behaviors are
 exactly opposite to the ones found in the interacting holographic
 dark energy model.

 \begin{center}
 \begin{figure}[htbp]
 \centering
 \includegraphics[width=0.75\textwidth]{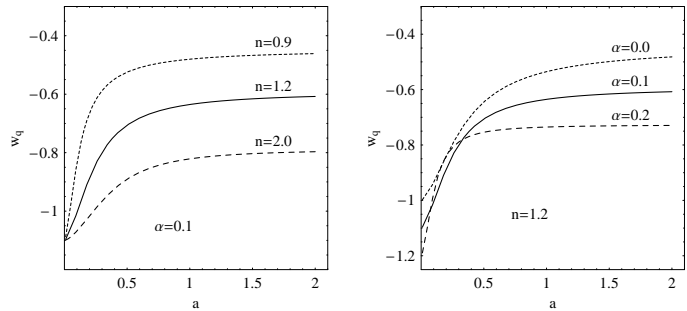}
 \caption{\label{fig2} Evolution of $w_q$ for various
 model parameters $n$ and $\alpha$ in the case of
 $Q=3\alpha H\rho_q$ \cite{0707.4052}.}
 \end{figure}
 \end{center}

Fig.~\ref{fig2}  shows \cite{0707.4052} the evolution of $w_q$ for different
  $n$ and $\alpha$ in the case of
 $Q=3\alpha H\rho_q$. It is easy to see that the EoS of the ADE $w_q$ can cross the phantom divide $w_{de}=-1$. In the
 case of $Q=0$ (i.e.~without interaction), as mentioned above,
 $w_q$ is always larger than $-1$ and cannot cross the phantom
 divide. With the help of interaction between the
 ADE and the pressureless matter, the  situation is changed. From Eq.~(\ref{eq17}), along with the first line
 of Eq.~(\ref{eq18}), it is easy to understand that $w_q$
 converges to the value $-1-\alpha$ at the early time in the case
 of $Q=3\alpha H\rho_q$. The most interesting  observation from Fig.~\ref{fig2} is that $w_q$ crosses the
 phantom divide from $w_q<-1$ to $w_q>-1$.   In the cases of negative $\alpha$, $\beta$ and
 $\gamma$, from Eq.~(\ref{eq17}) along with Eq.~(\ref{eq18}), one can
 see that $w_q$ is always larger than $-1$ and cannot cross the
 phantom divide. Obviously, the cases of positive $\alpha$,
 $\beta$ and $\gamma$ are more interesting since the $w_q$
 can cross the phantom divide from $w_q<-1$ to $w_q>-1$.

\section{Impact of interaction on cosmological dynamics}

\subsection{Transition from decelerated to accelerated expansion through interaction}

    In SCM, the transition from decelerated to accelerated expansion is related to the increase of the relative density of the cosmological constant. During analysis of interaction in the dark sector, we ask ourselves an obvious question: can we build a viable cosmological model in which this transition is the result of interaction in the dark sector \cite{8.1, 8.2}? The question appears to be a valid one, since interaction regulates the relative densities of the accelerating (DE) and decelerating (matter) components.

 We assume the dark components interact with each other according to
\begin{equation} \label{ref_8_1_}
\begin{array}{l} {\dot{\rho }_{dm} +3H\rho _{dm} =\frac{\dot{f}}{f} \rho _{dm} ,} \\ {\dot{\rho }_{de} +3H\left(1+w\right)\rho _{de} =-\frac{\dot{f}}{f} \rho _{dm} ,} \\ {} \end{array}
\end{equation}
where the interaction is described by a time dependent function $f(t)$. Let's write the Friedmann equations in the form
\begin{equation} \label{ref_8_2_}
\begin{array}{l} {3H^{2} =8\pi G\rho ,} \\ {\frac{\dot{H}}{H^{2} } =-\frac{3}{2} \left(1+\frac{p}{\rho } \right)} \end{array},
\end{equation}
where $\rho =\rho _{dm} +\rho _{de} ,\quad p=p_{de} $.  The matter energy density behaves as
\begin{equation} \label{ref_8_3_}
\rho _{dm} =\rho _{dm,0} \left(\frac{a_{0} }{a} \right)^{3} \frac{f}{f_{0} }
\end{equation}
Because the total  energy has to be conserved, the dark energy density, therefore, behaves according to
\begin{equation} \label{ref_8_4_}
\dot{\rho }_{de} +3H\left(1+w_{eff} \right)\rho _{de} =0,\quad w_{eff} \equiv w+\frac{\dot{f}}{3Hf} r
\end{equation}
where $r=\rho _{dm} /\rho _{de} $ . In case of $r=const$, we find that
\begin{equation} \label{ref_8_5_}
w_{eff} =-\frac{\dot{f}}{3Hf} ,\quad w=\left(1+r\right)w_{eff}.
\end{equation}
Under this condition, the total equation of state is
\begin{equation} \label{ref_8_6_}
\frac{p}{\rho } =\frac{p_{de} }{\rho _{de} +\rho _{dm} } =\frac{w}{1+r} =w_{eff}.
\end{equation}
From the last equation \eqref{ref_8_2_}, the deceleration parameter $q=-1-\frac{\dot{H}}{H^{2} } $ is
\begin{equation} \label{ref_8_7_}
q=\frac{1}{2} \left(\frac{3p}{\rho } +1\right)
\end{equation}
Using \ref{ref_8_5_},\ref{ref_8_6_}, we obtain
\begin{equation} \label{ref_8_8_}
q=\frac{1}{2} \left(1-\frac{\dot{f}}{Hf} \right)
\end{equation}
The sign of $q$ is defined by the ratio $\frac{\dot{f}}{Hf} $. For $\frac{\dot{f}}{Hf} <1$  we have $q>0$, i.e., decelerated expansion. For $\frac{\dot{f}}{f} >1$  we have $q<0$ - accelerated expansion. If, in particular, $f$  is such that the ratio $\frac{\dot{f}}{f} $ changes from$\frac{\dot{f}}{f} <1$ to $\frac{\dot{f}}{f} >1$, this corresponds to a transition from decelerated to accelerated expansion under the condition of a constant energy density ratio $r$ . Consequently, this transition occurs solely due to interaction.

The analysed case of $r=const$ exotic, and clearly contradicts SCM, where $r\propto a^{-3} $.  But, as we saw,  it is exactly this dependence that is found in the context of holographic dark energy models. This relation has the attractive feature that, by identifying the infrared cutoff length with the present Hubble scale, the corresponding ultraviolet cutoff energy density turns out to be of the order of the observed value of the cosmological constant parameter. However, the choice of infrared cutoff length is not consistent with the accelerated expansion of the Universe. As we see, this clear contradiction can become a positive feature \cite{8.1} if we take into account interaction in the dark sector.

\subsection{Interacting models as solutions to the cosmic coincidence problem} \label{CCP}

For constant $w_{de}$, the energy density of  DE scales as $\rho _{de} \propto a^{-3\left(1+w_{de} \right)} $. Observations constrain $w_{de} $  to be very close to $-1$. Thus, the DE density varies relatively slowly with the scale factor. The matter density, in contrast, scales as $\rho _{dm} \propto a^{-3} $. This leads to the ''well-known coincidence problem'': while the matter and DE densities today are nearly within a factor of two of each other, at early times $\rho _{dm} \gg \rho _{de} $, and in the far future we expect $\rho _{de} \gg \rho _{dm} $.  It would appear, then, that we live in a very special time. Now, the question is: why is it happening now? Is it a mere coincidence, or is there some deep underlying reason behind it? For the sake of viability, any cosmological model should give an answer to this question. Models attempting to solve the coincidence problem must take into account interactions between components. Attempts to resolve  the coincidence problem as a consequence of interaction between the matter sector and DE have a rich history \cite{8.3, 8.4, 8.5, 8.6}.

Let's analyse the main approaches we can take to resolving the coincidence problem in models with interaction in the dark sector. The key idea is exceedingly simple \cite{8.3}. A numeric ratio of the ''coincidence'' is the ratio $r\equiv \rho _{dm} /\rho _{de} $. As we saw in  Section \ref{Phenomenology_of_Interacting_Models}, if we assume that  $r\propto a^{-\xi },$ then it can be shown that  $\xi $ is related directly to the interaction between the dark components - $Q$ (see equations  \eqref{GrindEQ__2_25_}). In SCM  $\xi =3$. Therefore, for any value  $\xi <3$  the coincidence problem is less severe than for the SCM  model. Let's stop and think about that statement. Using the first Friedmann equation for a spatially flat Universe, and the conservation equation, we obtain $\left(8\pi G=1\right)$
\begin{equation} \label{ref_8_9_}
\dot{r}=3Hr\left[w_{de} +\frac{Q}{9H^{3} } \frac{\left(r+1\right)^{2} }{r} \right].
\end{equation}
For $Q>0$ (i.e., when energy transfers from DE to  DM) the ratio $r$ evolves more slowly than in the SCM model. This certainly alleviates the coincidence problem.

  Using
\begin{equation} \label{ref_8_10_}
\dot{r}=\dot{H}\frac{dr}{dH} ,\quad \dot{H}=-\frac{1}{2} \left(\rho _{dm} +\rho _{de} +p_{de} \right)=-\frac{3}{2} \frac{1+w_{de} +r}{1+r} H^{2}.
\end{equation}
Let's write  \eqref{ref_8_9_} as
\begin{equation} \label{ref_8_11_}
\frac{dr}{dH} =\frac{I}{H} ,\quad I\equiv -2r\frac{1+r}{1+w_{de} +r} \left[w_{de} +\frac{Q}{9H^{3} } \frac{\left(r+1\right)^{2} }{r} \right]
\end{equation}
The eq.\eqref{ref_8_11_} can be integrated whenever an expression for the interaction $Q$  in terms of $H$  and $r$ is given.
The three following linear coupling models were considered
\begin{equation}
Q=3\alpha H\left( {{\rho }_{dm}}+{{\rho }_{de}} \right),\quad Q=3\beta H{{\rho }_{dm}},\quad Q=3\gamma H{{\rho }_{de}},
\end{equation}
where the phenomenological parameters $\alpha $ , $\beta $ , and $\eta $  are dimensionless, positive constants. Consider, as an example, the first model. This model  fits very well with data from SN Ia, CMB, and large scale structure formation provided that $\alpha <2.3\times {{10}^{-3}}$ \cite{Olivares}.  The remarkable property of this model\cite{0812.2210} is that  the ratio $r$  tends to a stationary but unstable value at early times, $r_{s}^{+}$, and to a stationary and stable value, $r_{s}^{-}$ (an attractor), at late times. Consequently, as the Universe expands, $r\left( a \right)$  smoothly evolves from $r_{s}^{+}$ to the attractor solution $r_{s}^{-}$.

\begin{figure}[th]
\includegraphics[width=5.0in,angle=0,clip=true]{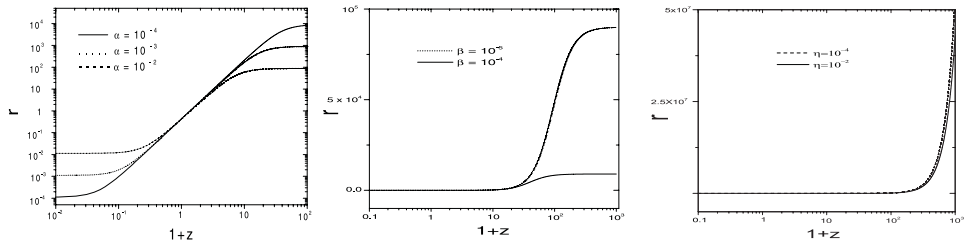}
\caption{The evolution of the ratio \cite{0812.2210}
$r=\rho_{dm}/\rho_{de}$ with redshift for  models alpha, beta,
and eta. For all of them $r$ either tends to a constant or varies
very slowly at small redshift. The initial conditions are
$r_{0}= 3/7 \,$ and $\, w= -0.9$. } \label{fig:quinteshor}
\end{figure}

We determine the critical points of Eq. (\ref{ref_8_9_}) by setting
$\dot{r}$ to zero. For $w = const$ the stationary solutions of
the resulting quadratic equation are:
\begin{equation}
r^{\pm}_{s} = -1 + 2b \pm 2 \sqrt{b(b-1)}\, ,  \qquad b = -
\frac{w}{4\, \alpha} > 1.
\label{r+-}
\end{equation}
Using the standard analysis methods of critical points, the stationary solution  $r^{+}_{s}$ proves to be
unstable while $r^{-}_{s}$ is stable \cite{plbwdl,prdladw}. The
general solution of Eq. (\ref{ref_8_10_})
\begin{equation}
r(x) = \frac{r^{-}_{s}+ x r^{+}_{s}}{1 + x}\, ,
\label{r(x)}
\end{equation}
interpolates between $r^{+}_{s}\, $ and $\, r^{-}_{s}$. Here, $x =
(a/a_{*})^{- \mu}$, with $\mu \equiv 12\, \alpha \sqrt{b(b-1)}$,
and $a_{*}$ denotes the scale factor at which $r$ takes the
arithmetic medium value $(r^{+}_{s}+ r^{-}_{s})/2$. In the range
$r^{-}_{s} < r < r^{+}_{s}$ the function $r(x)$ decreases
monotonously. Consequently, as the Universe expands, $r(x)$
smoothly evolves from $r_{s}^{+}$ to the attractor solution
$r^{-}_{s}$. The evolution from one asymptotic solution to the
other is illustrated on the graphs - see Figs.
\ref{fig:quinteshor}.

\subsection{The problem of transient acceleration}
Unlike fundamental theories, physical models only reflect the current state of our understanding of a process or phenomenon for the description of which they were developed. The efficiency of a model is to a significant extent determined by its flexibility, i.e., its ability to update when new information appears. Precisely for this reason, the evolution of any broadly applied model is accompanied by numerous generalizations aimed at resolving conceptual problems, as well as a description of the ever increasing number of observations. In the case of the SCM, these generalizations can be divided into two main classes. The first is composed of generalizations that replace the cosmological constant with more complicated dynamic forms of DE, for which the possibility of their interaction with DM must be taken into account. Generalizations pertaining to the second class are of a more radical character. The ultimate goal of these generalizations (explicit or latent) consists in the complete renunciation of dark components by means of modifying Einstein's equations. The generalizations of both the first and second classes can be demonstrated by means of a phenomenon that has been termed ``transient acceleration''.

A characteristic feature of the dependency of the deceleration parameter $q$ on the redshift $z$ in the SCM is that it monotonically tends to its limit value $q(z) = —1$ as $z \to — 1.$ Physically, this means that when DE became the dominant component (at $z \sim 1$), the Universe in the SCM was doomed to experience eternal accelerating expansion.
In what follows, we consider several cosmological models that involve dynamic forms of DE that lead to transition acceleration, and we also discuss what the observational data says about the modern rate of expansion of the Universe.

Barrow \cite{Barrow} was among the first to indicate that transient acceleration is possible in principle. He showed that within quite sound scenarios that explain the current accelerated expansion of the Universe, the possibility was not excluded of a return to the era of domination of nonrelativistic matter and, consequently, to decelerating expansion. Therefore, the transition to accelerating expansion does not necessarily mean eternal accelerating expansion. Moreover, in Barrow's article, it was shown to be neither the only possible nor the most probable course of events.

\subsubsection{Observational evidence}

\begin{figure}[t]
\centering
\psfig{figure=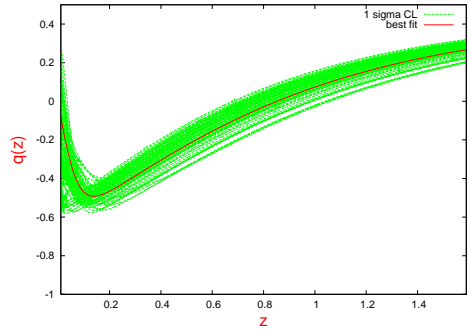,width=0.35\textwidth,angle=-90}
\caption{ The cosmological deceleration parameter $q(z)$,  reconstructed using a combination of SN Ia, BAO and CMB data and the
ansatz (\ref{eq:step}).
The solid red lines show the best fit reconstructed results,
while the dashed green lines show reconstructed results within $1\sigma$ CL\cite{Starobinsky} .}
\label{fig_Srarob}
\end{figure}

Based on independent observational data, including SNe-Ia brilliance curves, signatures of baryon acoustic oscillations (BAO) in the galaxy distribution
 and fluctuations in the cosmic microwave background (CMB), it was shown in \cite{Starobinsky} that the acceleration with which the Universe expands has reached its maximum value and is decreasing at present (Fig. \ref{fig_Srarob}). In terms of the deceleration parameter, this means that this parameter has reached its minimum value and is increasing at present. Hence, the main result of the analysis in Ref. \cite{Starobinsky} is that the SCM is not the only explanation of observational data (although it is the simplest), and the accelerated expansion of the Universe in which DE presently dominates is merely a transition phenomenon.
We note that it is also shown in Ref. \cite{Starobinsky} that using the Chevallier-Polarski-Linder (CPL) parameterization,
 \begin{equation}
\label{cplwz}
w(z)=w_0+\frac{w_a \, z}{1+z},
\end{equation}
for the  equation of state parameterdoes not allow us to unambiguously combine  data obtained from observations of close supernovae, such as SNe-Ia, and of the CMB anisotropy. A possible way to resolve this contradiction is to renounce this parameterization and adopt a different one. In Ref.\cite{Starobinsky}, a parameterization was proposed that is capable of uniting these arrays of data:
 \begin{equation}
    w(z)=- \frac{1+ \tanh\left[(z-z_t)\Delta\right]}{2}.
\label{eq:step}
\end{equation}
In this approximation, $w = — 1$ at the early stages of the evolution of the Universe, and $w$ increases to its maximum value $w\sim 0,$ at small $z.$
Figure \ref{fig_Srarob} shows the dependence of the deceleration parameter $q$ restored using the parameterization \eqref{eq:step}.

In 2010, in the framework of the Supernova Cosmology Project (SCP), the most recent array of data on bursts of supernovae was published \cite{amanullah}, which includes 557 events, making it the largest present-day body of data in this field. Moreover, the array of data on supernovae with small red shifts $(z<0.3)$ has been significantly enlarged.

\begin{figure}[t]
\centering
\includegraphics[width=0.45\textwidth]{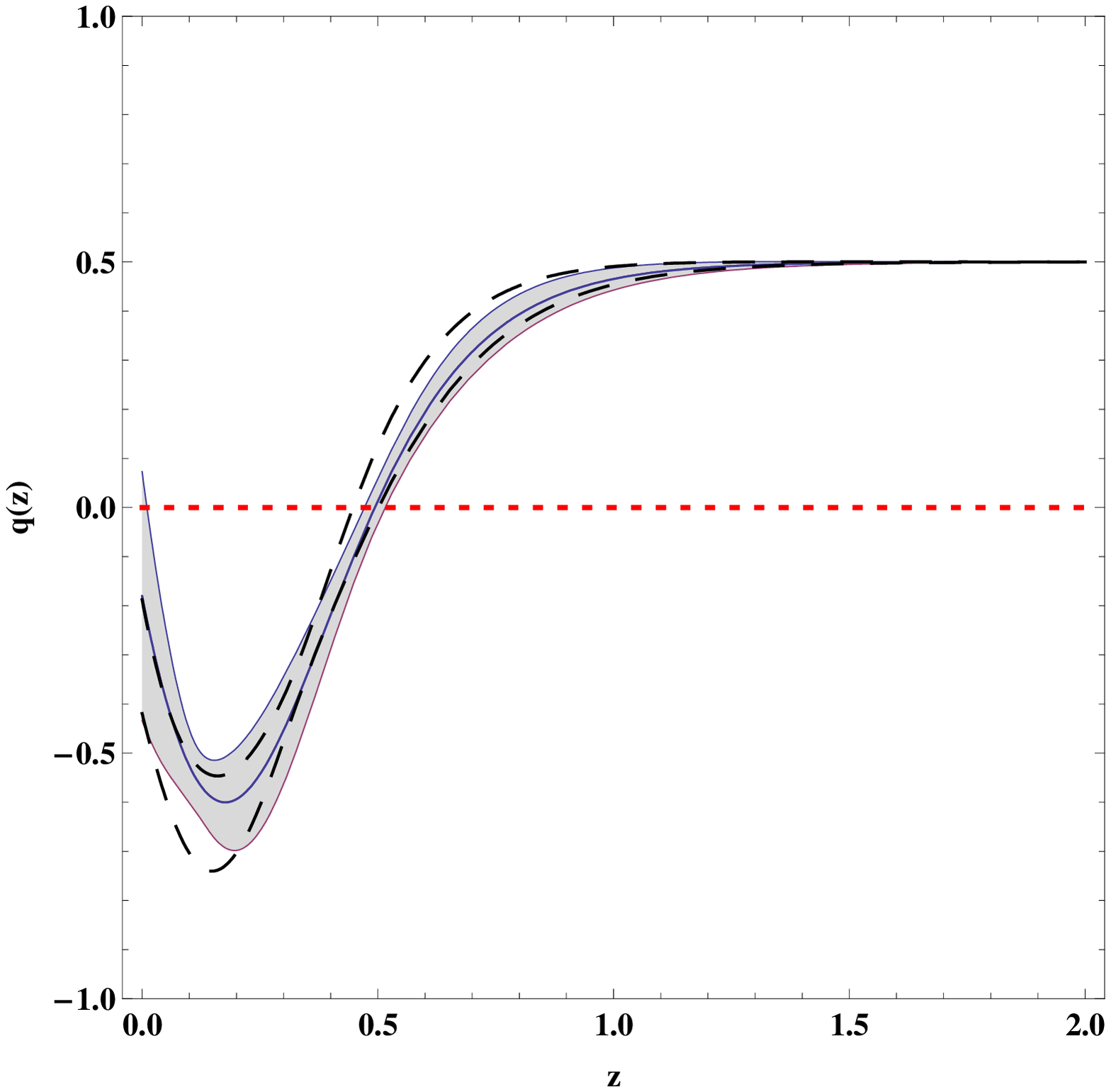}
\includegraphics[width=0.45\textwidth]{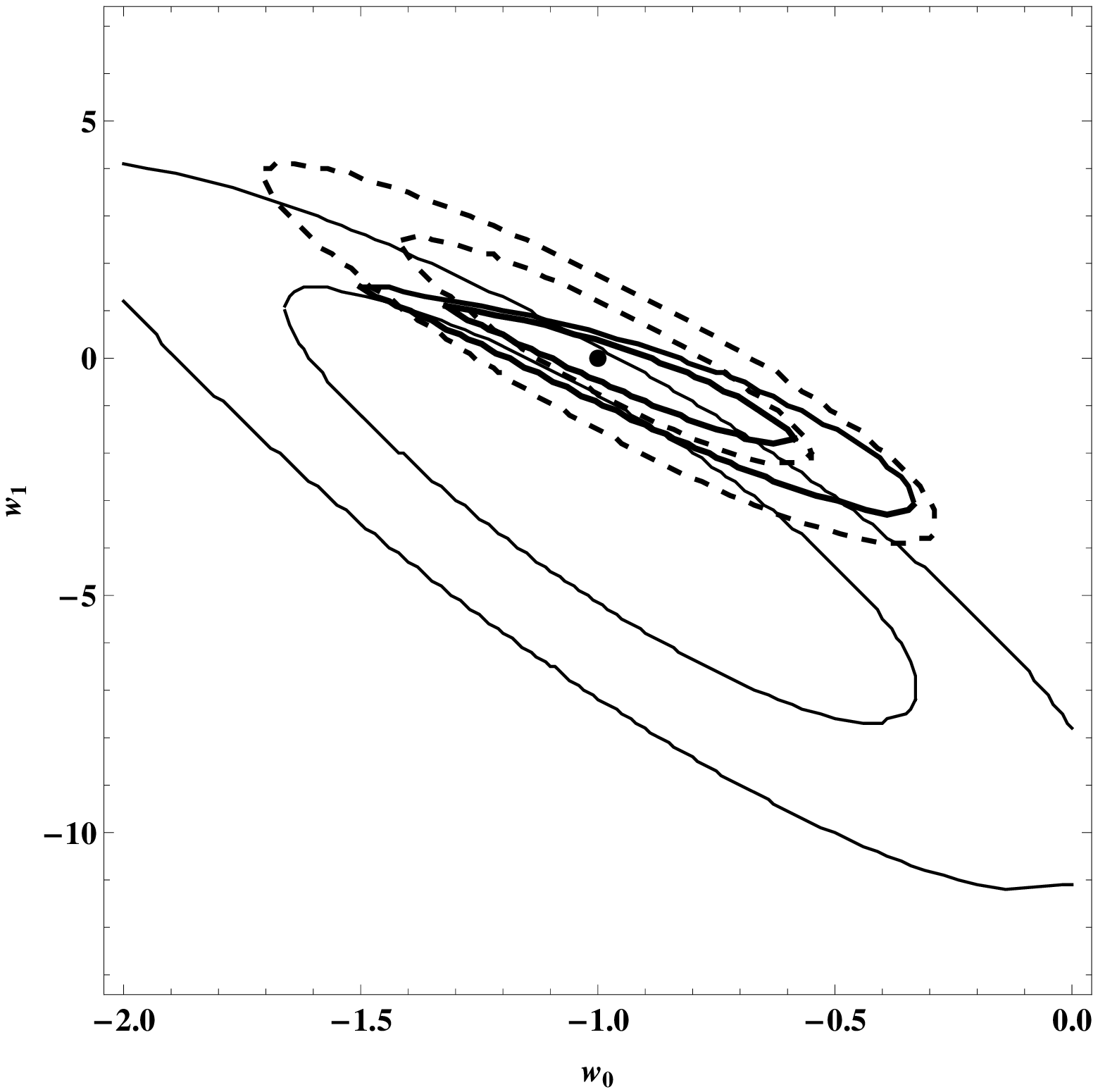}
\caption{The left panel represents the results reconstructed  from Union2+BAO, and show the evolutionary behaviors of $q(z)$  at the $68.3\%$
confidence level. The gray regions and the regions between the two long
dashed lines  show the results without and with the systematic
errors in the SNIa, respectively.
The right panel represents the $68.3\%$ and $95\%$ confidence level regions
   for $w_0$ versus $w_1$ in the CPL parameterization,  $w=w_0+w_1z/(1+z)$.  In  the right panel, the system error in the SNIa is considered.  The dashed, solid   and
   thick solid lines represent the results obtained from Union2S, Union2S+BAO and Union2S+BAO+CMB, respectively. The point at $w_0=-1$, $w_1=0$ represents the spatially flat $\Lambda$CDM model.\cite{LiWuYu1} }
\label{fig:TrAcCh}
\end{figure}

At present, there are already several studies \cite{LiWuYu,LiWuYu1} in which these observations are analyzed in order to check the hypothesis of transient acceleration.
All the authors agree that the final answer can only be given by repeated, more precise observations. Moreover, it seems that in order to obtain consistent results, the entire technique of data handling has to be corrected. For example, as shown in \cite{LiWuYu,LiWuYu1} (Fig. \ref{fig:TrAcCh}), there are contradictions between the data obtained from observations of SNe-Ia and BAO at small red shifts and CMB observations at large $z.$
The contradiction consists in the fact that the analysis of two separate series of data yields opposite results. For example, when only the SNe-Ia and BAO data are used, the probability that the acceleration rate of the expansion of the Universe has already reached its maximum at $z \sim 0.3$, and is at present starting to decrease turns out to be quite high. However, if these data are supplemented with the CMB observations, the results of the analysis change substantially and no deviations from the $\Lambda$CDM model are revealed.

Therefore, the restoration of the DE evolutionary dependence and the answer to the question of whether the expansion of our Universe will decelerate or if the accelerating expansion will go on forever (as in the SCM) depends strongly on the data obtained from observations of SNE-Ia, their quality, the technique used in the reconstruction of the cosmological parameters (such as $q(z), w(z)$  and $\Omega_{DE}$), and the actual parameterization of the dark energy equation of state. For a detailed answer to this question, we must wait for more precise observational data, and find methods of their analysis that are less model-dependent.

\subsubsection{Decaying cosmological constant and transient acceleration}
As a simple example of transient acceleration, we consider a model with a decaying cosmological constant:
\begin{equation} \label{ec}
\dot{\rho}_{m} + 3\frac{\dot{a}}{a}\rho_{m} = - \dot{\rho}_\Lambda\;,
\end{equation}
where $\rho_{m}$ and $\rho_\Lambda$ are the densities of the DM energy and of the cosmological constant $\Lambda$. At the early stages of the expansion of the Universe, when $\rho_\Lambda$ is quite small, such a decay does not influence cosmological evolution in any way. At later stages, as the DE contribution increases, its decay has an ever increasing effect on the standard dependence of the DM energy density $\rho_{m} \propto a^{-3}$ on the scale factor $a$. We consider the deviation to be described by a function  of the scale factor - $\epsilon(a)$.
\begin{equation} \label{dm}
\rho_{m} = \rho_{m, 0}a^{-3 + \epsilon(a)}\;,
\end{equation}
where $a_0 = 1$ in the present epoch.
Other fields of matter (radiation, baryons) evolve independently and are conserved. Hence, the DE density has the form
\begin{equation}\label{decayv}
\rho_{\Lambda} =  \rho_{m0} \int_{a}^{1}{\epsilon(\tilde{a}) + \tilde{a}\epsilon' \ln(\tilde{a}) \over \tilde{a}^{4 - \epsilon(a)}} d\tilde{a} + {\rm{X}}\;,
\end{equation}
where the prime denotes the derivative with respect to the scale factor, and ${\rm{X}}$ is the integration constant. If radiation is neglected, the first Friedmann equation takes the form
\begin{equation}
\label{friedmann} {{H}}= H_0\left[\Omega_{b,0}{a}^{-3} + \Omega_{m0}\varphi(a) + {\Omega}_{{\rm{X,0}}}\right]^{1/2},
\end{equation}
The function $\varphi(a)$ is then written as
\begin{equation} \label{f(a)}
\varphi(a) = a^{-3 +\epsilon(a)} + \int_{a}^{1}{\epsilon(\tilde{a}) + \tilde{a}\epsilon' \ln(\tilde{a}) \over \tilde{a}^{4 - \epsilon(a)}} d\tilde{a}\;,
\end{equation}
where ${\Omega}_{{\rm{X,0}}}$,  is the relative contribution of the constant ${\rm{X}}$ to the common relative density. To proceed, it is necessary to make some assumptions concerning the concrete form of $\epsilon(a)$. Here, we follow the original work \cite{cosmolog}, and consider the simplest case
\begin{equation}
\label{Parametrization_a}
\epsilon(a) = \epsilon_0a^\xi\ = \epsilon_0(1+z)^{-\xi},
\end{equation}
where $\epsilon_0$ and $\xi$ can take both positive and negative values. It follows from the expression (\ref{decayv}) that
\begin{equation}\label{decayv2}
\rho_{\Lambda} =  \rho_{m0}\epsilon_{0} \int_{a}^{1}{[1 + \ln(\tilde{a}^{\xi})] \over \tilde{a}^{4 - \xi - \epsilon_{0}\tilde{a}^{\xi}}} d\tilde{a} + {\rm{X}}\;.
\end{equation}

We note that the case $\epsilon_0 = 0$ corresponds to the SCM, i.e., ${\rm{X}} \equiv {\rho}_{\Lambda}$.
Using the formulas presented above, it is not difficult to also obtain the dependences for the relative densities $\Omega_b(a)$, $\Omega_{m}(a)$ and $\Omega_{\Lambda}(a)$:
\begin{subequations}
\begin{equation} \label{8a}
\Omega_{b}(a) = \frac{a^{-3}}{{\rm{A}} + a^{-3} + {\rm{B^{-1}}}\varphi(a)}\;,
\end{equation}
\begin{equation} \label{8b}
\Omega_{m}(a) = \frac{a^{-3 + \epsilon(a)}}{{\rm{D}} + {\rm{B}}a^{-3} + \varphi(a)}\;,
\end{equation}
\begin{equation} \label{8c}
\Omega_{{\rm{\Lambda}}}(a) = \frac{{\rm{D}} + \varphi(a) - a^{-3 + \epsilon(a)}}{{\rm{D}} + {\rm{B}}a^{-3}  + \varphi(a)}\;,
\end{equation}
\end{subequations}
 where ${\rm{A}} = {\Omega_{{\rm{X}},0}}/{\Omega_{b,0}}$, ${\rm{B}} = {\Omega_{b,0}}/{\Omega_{m0}}$ and  ${\rm{D}} = {\Omega_{{\rm{X}},0}}/{\Omega_{m0}}$.

\begin{figure}[t]
\centerline{\psfig{figure=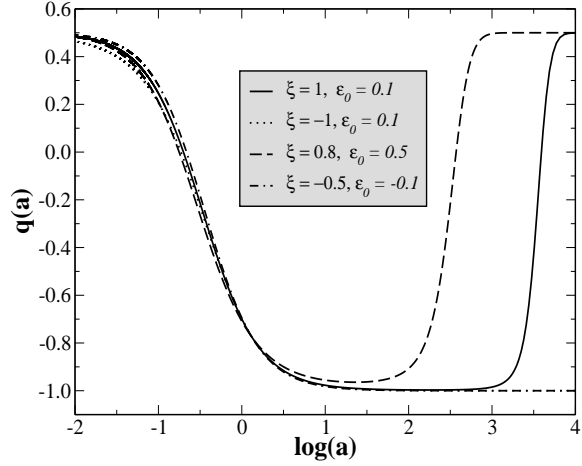,width=2.5truein,height=3.1truein,angle=-90}}
\caption{The deceleration parameter as a function of $\log(a)$ for
various values of $\epsilon_0$ and $\xi$\cite{cosmolog}. } \label{fig:qzw}
\end{figure}

Within this simple model, it is practically possible to obtain any dynamics of the Universe with the aid of an appropriate choice of the parameters $\epsilon_0$ and $\xi$. In the context
of this paper, the case of immediate interest is where $\epsilon_0 > 0$ and $\xi$ takes on large positive values ($\xi \gtrsim  0.8$). The solid curve in Fig. \ref{fig:qzw} shows the dependence of the deceleration parameter for $\xi = 1.0$ and $\epsilon_0 = 0.1.$ We note that at present, for these parameters, when $a\sim 1$, the expansion of the Universe is accelerating, but the dominance of DE is not eternal, unlike in the case of the SCM, and when $a\gg 1$ , the Universe will enter a new era of dominance of nonrelativistic matter. Such a form of dynamic behavior is unusual for most models with $\Lambda(t)$ or models with interacting quintessence discussed in literature, but it is characteristic of the so-called thawing \cite{thaw} and hybrid \cite{cqg} potentials that follow from string theory or M-theory \cite{fischler}  (also see \cite{ed}).

To better represent the phenomenon of transient acceleration, we find the explicit form of the deceleration parameter $q=-a\ddot{a}/\dot{a}^2$, in this model:
\begin{equation}
q(a) = \frac{3}{2}\frac{\Omega_{b,0}a^{-3} + \Omega_{m0}a^{\epsilon(a) - 3}}{\Omega_{b,0}{a}^{-3} + \Omega_{m0}\varphi(a) + {\Omega}_{\rm{X},0}} -1,
\end{equation}
The parameter $q$ is represented as a function of $\log(a)$ for different values of $\xi$  and $\epsilon_0$ in Fig. \ref{fig:qzw}. We note that in the distant past $(a \ll 1)$, the deceleration parameter $q(a) \rightarrow 1/2$, which corresponds to a matter-dominated Universe. However, for certain values of parameter $\xi$, a long (but finite, in contrast to the case of the SCM) era of accelerated expansion sets in. In the distant future $(a \gg 1),$ the Universe again returns to decelerated expansion $(q > 0).$

\subsubsection{Transient Acceleration in a Universe with Interacting Components}
We consider a spatially flat Universe consisting of three components: DE, DM, and baryons. The first Friedmann equation for such a Universe has the form
\begin{equation}\label{GrindEQ__32_}
    3M_{Pl}^2H^{2} =\rho _{_{DE}} +\rho _{m} +\rho _{b},
\end{equation}
where, as usual, $\rho_{_{DE}}$ is the DE density, $\rho_{m} $ is the DM energy density, $\rho_{b}$ is the baryon energy density. The equation of state for DE has the form $p_{_{DE}} =w\rho _{_{DE}} $.

The conservation equation for the baryon component is
\begin{equation}\label{GrindEQ__34_}
\dot{\rho }_{b} +3H\rho _{b} =0\, \, \, \Rightarrow \, \, \, \rho _{b} =\rho _{b0} \left(\frac{a_{0}}{a} \right)^{3}.
\end{equation}
The total density is $\rho =\rho _{m} +\rho _{b} +\rho _{_{DE}}.$
Without loss of generality, we assume that the energy density $\rho _{m}$ is expressed as
\begin{equation}\label{GrindEQ__35_}
    \rho _{m} =\tilde{\rho }_{m0} \left(\frac{a_{0}}{a} \right)^{3} f\left(a\right),
\end{equation}
where $\tilde{\rho }_{m0} $ and $a_{0}$ are constants and  $f(a)$ is an arbitrary differentiable function of the scale factor.
From conservation equations and \eqref{GrindEQ__35_}, we obtain
\begin{equation}\label{GrindEQ__36_}
    Q=\rho _{m} \frac{\dot{f}}{f} =\tilde{\rho }_{m0} \left(\frac{a_{0}}{a} \right)^{3} \dot{f}.
\end{equation}
Let's take \cite{FFZ_TCA}
\begin{equation}\label{GrindEQ__37_}
    f(a)=1+g(a).
\end{equation}
In the absence of interaction, $f(a)=1$; therefore, the function $g(a)$ is responsible for interaction. Then, taking into account that
\begin{equation}\label{GrindEQ__38_}
    \dot{f}=\dot{g}=\frac{dg}{da} \dot{a},
\end{equation}
we obtain
\begin{equation}\label{GrindEQ__39_}
    Q=\tilde{\rho }_{m0} \frac{dg}{da} \dot{a}\left(\frac{a_{0}}{a} \right)^{3} .
\end{equation}
This means that
\begin{equation}\label{GrindEQ__40_}
    \rho _{m} =\tilde{\rho }_{m0} \left(1+g\right)\left(\frac{a_{0}}{a} \right)^{3},
\end{equation}
where $\rho _{m0} =\rho _{m} (a_{0} )$ if the interaction exists, and $\tilde{\rho }_{m0} $ in the absence of interaction. The two initial values of the DM density are related as
\begin{equation}\label{GrindEQ__41_}
    \rho _{m0} =\tilde{\rho }_{m0} \left(1+g_{0} \right),
\end{equation}
where $g_{0} \equiv g(a_{0} )$.
As can be seen from \eqref{GrindEQ__36_} when $Q > 0,$ DE decays into DM, $\frac{dg}{da} >0$. When $\frac{dg}{da} <0$, the decay proceeds in the opposite direction. From  conservation equations and \eqref{GrindEQ__39_}  we obtain
\begin{equation}\label{GrindEQ__42_}
    \dot{\rho }_{_{DE}} +3H\left(1+w\right)\rho _{_{DE}} =-\tilde{\rho }_{m0} \frac{dg}{da} \dot{a}\left(\frac{a_{0}}{a} \right)^{3} .
\end{equation}
When $w=const$ the solution of \eqref{GrindEQ__42_} has the form
\begin{equation}\label{GrindEQ__43_}
\begin{gathered}
\rho _{_{DE}} =\left(\rho_{{m0}} +\tilde{\rho }_{m0} g_{0} \right)\left(\frac{a_{0}}{a} \right)^{3\left(1+w\right)} -\tilde{\rho }_{m0} \left(\frac{a_{0}}{a} \right)^{3} g+3w\tilde{\rho }_{m0} a_{0}^{3} a^{-3\left(1+w\right)} \int _{a_{0}}^{a}daga^{3w-1} .
\end{gathered}
\end{equation}
We rewrite the second Friedmann equation in terms of $g(a)$
\begin{equation}
\label{GrindEQ__44_}
   \begin{array}{l} {\frac{\ddot{a}}{a} =-\frac{1}{6} \left\{\tilde{\rho }_{m0} \left(1+g\right)\left(\frac{a_{0}}{a} \right)^{3} +\rho _{b0} \left(\frac{a_{0}}{a} \right)^{3} +\left(1+3w\right)\right. \times } \\ {\times \left[\left(\rho_{{m0}} +\tilde{\rho }_{m0} g_{0} \right)\left(\frac{a_{0}}{a} \right)^{3\left(1+w\right)} \right. \left. \left. -\tilde{\rho }_{m0} \left(\frac{a_{0}}{a} \right)^{3} g+3w\tilde{\rho }_{m0} a_{0}^{3} a^{-3\left(1+w\right)} \int _{a_{0}}^{a}daga^{3w-1}  \right]\right\}.} \end{array}
\end{equation}

To solve \eqref{GrindEQ__44_}, it is necessary to define the function $g(a)$.Since the nature of DE and DM is unknown, it is impossible to indicate the form of $g(a)$ based on first principles; therefore, we introduce the interaction in this modelin such a way so as to make the dynamics of the model be consistent with observational data

Consider the interaction for which the function $g(a)$ is represented as $g\left(a\right)=a^{n} \exp \left(-a^{2} /\sigma ^{2} \right)$,  where $n$  is a natural number and $\sigma $ is a positive real number. The existence of transient acceleration implies that the DE density starts to decrease, i.e., its decay occurs, $\frac{dg}{da} >0$. This condition requires that $n$ and $\sigma $ satisfy the inequality $n\sigma ^{2} >2.$

\begin{figure}[ht]
\includegraphics[width=.45\textwidth]{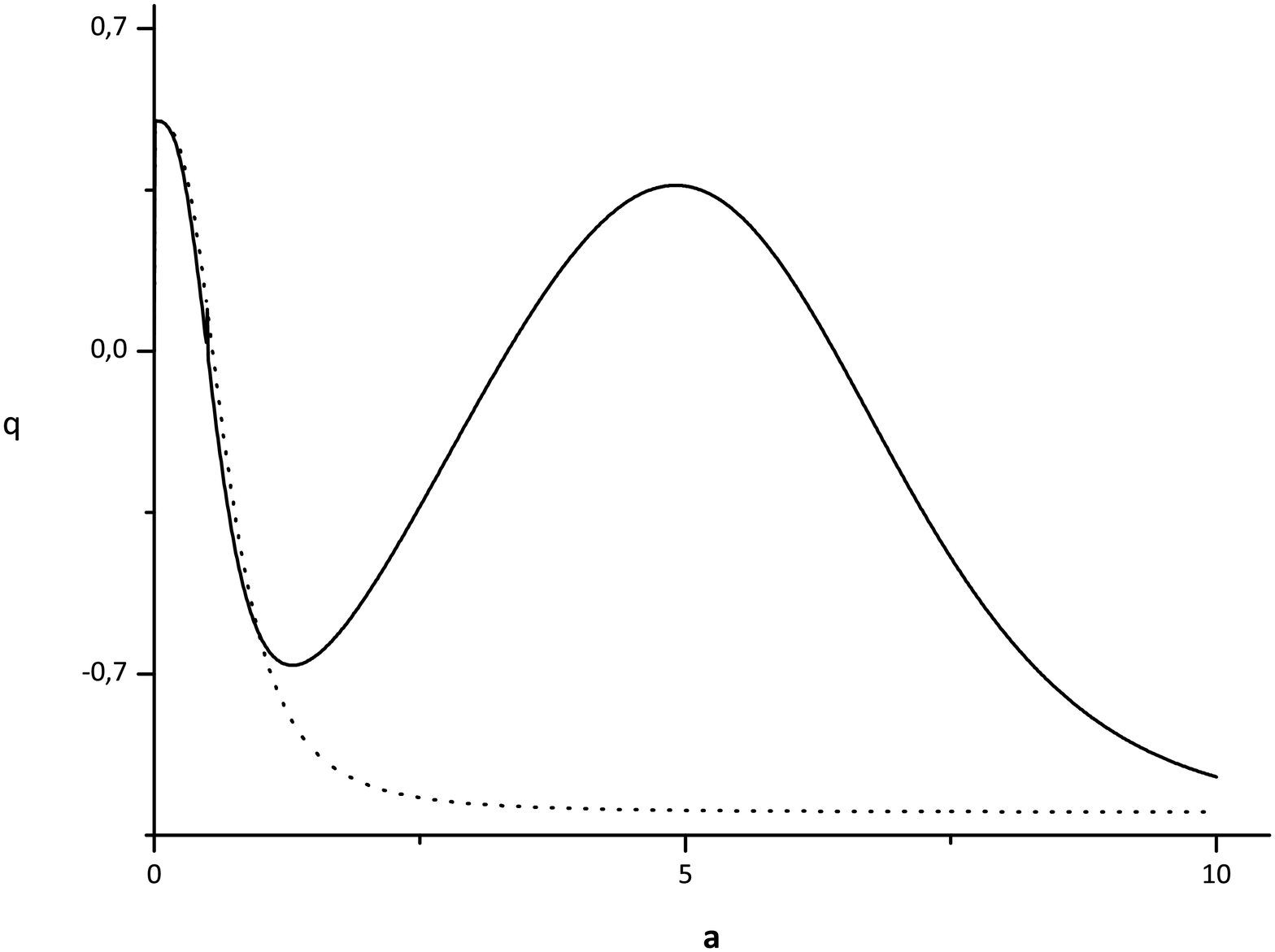}
\includegraphics[width=.45\textwidth]{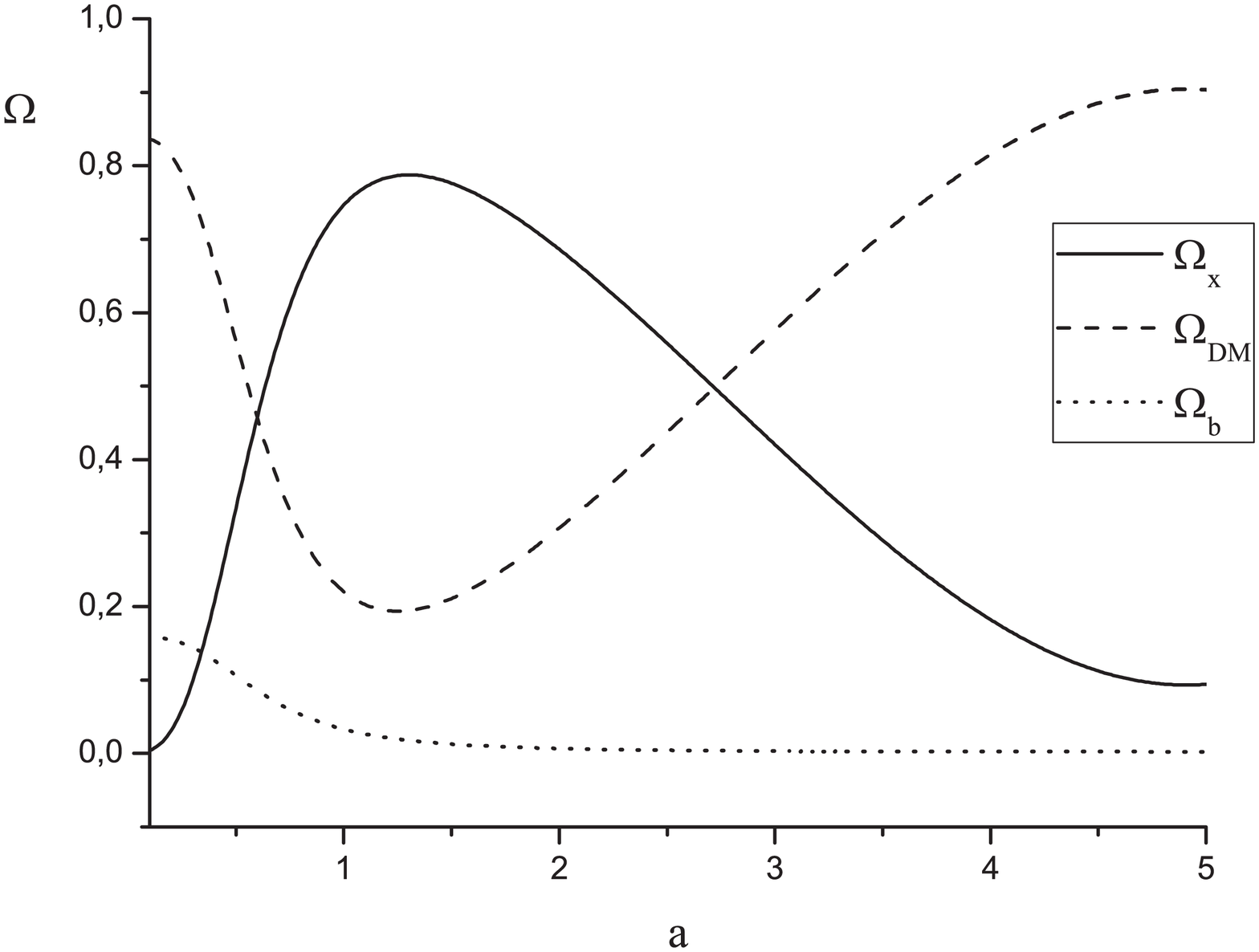}
\caption{On the right: the dependencies of the relative densities on the scale factor
for $n=7$ and $\sigma =1.5$. On the left: the
dependencies of the deceleration parameter on the scale factor in the
model with interacting dark energy and dark matter $q(a)$(solid
line) with $n=7$ and $\sigma =1.5$ in comparison to SCM (dashed
line).} \label{fig:W(a)}
\end{figure}

In Fig. \ref{fig:W(a)}, the dependencies of the relative densities on the scale factor are shown for $n=7$ and a = $\sigma =1,5.$ The model results in transient acceleration for a certain choice of the interaction parameters, but it is indistinguishable from the SCM for large (as well as small) values of the scale factor.

For a complete picture we consider the possibility of an accelerating transient regime within the interacting scalar field model. As mentioned above, Barrow \cite{Barrow} first discovered that there are many potentials of scalar fields that lead to the evolution of the Universe with a regime of transient acceleration. We will consider the recent article \cite{Cui_Zhang_Fu} as an example that the regime of transient acceleration is provided by the scalar field evolving in a specially chosen potential.
It has been shown that, in a Universe filled only by a scalar field $\phi$  \cite{Russo,Carvalho} that evolves in the potential of the form
\begin{equation}
\label{V_fransient}
V(\phi)=\rho_{\phi\,0}[1-\frac{\lambda}{6}(1+\alpha \sqrt{\sigma}\phi)^2)]
\exp{[-\lambda\sqrt{\sigma}(\phi+ \frac{\alpha \sqrt{\sigma}}{2}\phi^2)]},
\end{equation}
 where $\rho_{\phi\,0}$ is a constant energy density,
 $\sigma=8\pi G/\lambda$, and $\alpha$ and $\lambda$
are two dimensionless, positive parameters of the model, that the deceleration parameter is non-monotonically dependent on the scale factor.
In the case that is of interest to us, parameters can take values around $\alpha$,  $\lambda \sim 1$.
In the limit $\alpha\rightarrow 0$ the potential in Eq.(\ref{V_fransient})
reduces to an  exponential potential,
$V(\phi) =V_0 \exp{[- \sqrt{8\pi G \lambda }\phi]}$, a case that was examined in Ref.\cite{Russo}.

Recently the article \cite{Cui_Zhang_Fu} considered a Universe filled with a scalar field $\phi$ that interacts with dark matter $\rho_m$. After a change of variables, the Friedmann and conservation equations take the following form:
\begin{equation}
 \label{h2}
    h^{2}=\frac{U(y)+ x}{1-\frac{1}{2}(\frac{d y}{dN})^{2}},
\end{equation}
\begin{equation}
 \label{y}
\frac{d^{2}y}{dN^{2}}-\frac{3}{2}(\frac{dy}{dN})^{3}
+3\frac{dy}{dN}=(\frac{\Gamma}
{\frac{dy}{dN}}+1.5\frac{d y}{dN} x-U^{'}(y))h^{-2},
\end{equation}
\begin{equation}
\label{x}
\frac{d x}{dN}=-\Gamma-3x,
\end{equation}
where
$h\equiv H/H_0$,
$N\equiv\ln a(t)= -\ln (1+z)$,
$y\equiv\sqrt{\frac{8\pi G}{3}}\phi$,
$ x\equiv \rho_m/\rho_c$,
$\Gamma\equiv  Q/H \rho_c $,
and
$U(y)\equiv V(\phi)/\rho_c $.
It is easily seen that all of these quantities are dimensionless.
As noted in \cite{Cui_Zhang_Fu}, the parameter range of $\lambda$ and $\alpha$
for transient acceleration in our model differs from that in Refs.\cite{Russo,Carvalho}, which assumed the absence of matter.

 Fig. \ref{fig8} shows $q(z)$ for the coupling model where the scalar field transfers energy into the matter,
in which the rate is taken to be proportional to the matter density, $Q \propto -H \rho_m $, i.e.,
$\Gamma \propto -x$.
The dependency on $\Gamma$ is demonstrated in  Fig. \ref{fig8},
and larger values of  $\Gamma$ yield an earlier return of deceleration.

\begin{figure}
\centerline{\includegraphics[width=10cm]{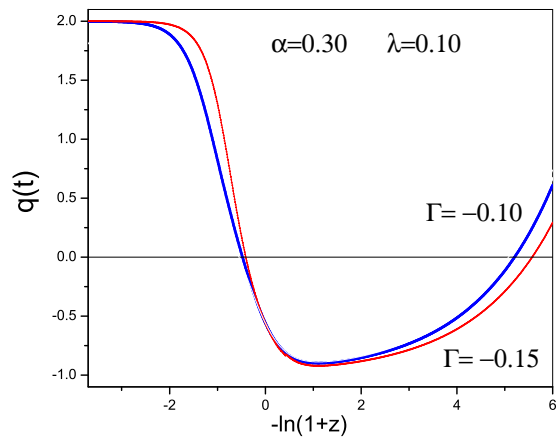}}
\caption{ \label{fig8}
$q(t)$ for various values of  $\Gamma<0$.
$q(t)$ is negative within $a \sim (0.5, 5) $.
}
\end{figure}
\begin{figure}
\centerline{\includegraphics[width=10cm]{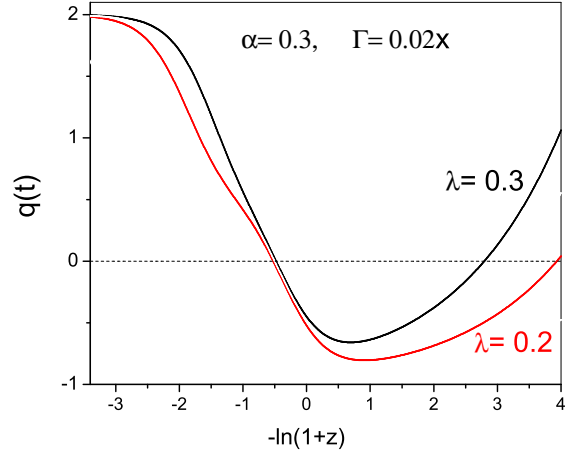}}
\caption{\label{fig11}
The $\Gamma>0$ model for various $\lambda$.
$q(t)$ is negative within $a \sim (0.5, 2.7) $ for $\lambda   =  0.4$.
A greater $\lambda $ yields a shorter duration of acceleration\cite{Cui_Zhang_Fu}.}
\end{figure}

Fig. \ref{fig11} shows the results for the coupling model in which the matter transfers energy into the scalar field with $\Gamma \propto x$.
The dependency on the parameter $\lambda $ is demonstrated in  Fig. \ref{fig11}, and larger values of $\lambda$ yield a shorter  period of transient acceleration.
It should be noted that the transient acceleration is also present in a Universe filled with a non-interacting scalar field with the potential \eqref{V_fransient} and dark matter. The interaction in this case makes the model more adaptable to observations, providing additional degrees of freedom.

\subsubsection{Simplest model of transient acceleration }
\label{simplest}

It is easy to see that the increased complexity of  the interaction parameter offers
many opportunities to obtain transient acceleration, namely the replacement of the
constant  of interaction with a function.
We have already examined similar types of interactions in the previous sections (see Section \ref{non-linear}).
 Let's now determine the form of the interaction term $Q$.
In the article \cite{Saridakis1} a simple model was considered, and it illustrates the possibility of a non-monotonic dependency of the deceleration parameter on the scale factor.
A simple parameterization has been considered:
\begin{equation}
\label{Qsimple}
Q=3\beta(a)H \rho_{de}
\end{equation}
with a simple power-law ansatz for $\beta(a)$, namely:
\begin{equation}
\label{cas32}
\beta(a)=\beta_0 a^\xi.
\end{equation}
Substituting this interaction form into  Eq. (\ref{eom1}), (\ref{eom2})
\begin{eqnarray}\label{eom1}
\dot{\rho}_{dm}+3H\rho_{dm}=Q,
\end{eqnarray}
\begin{eqnarray}\label{eom2}
\dot{\rho}_{de}+3H(\rho_{de}+p_{de})=-Q,
\end{eqnarray}
we get
 \begin{equation}
\label{rhophi2}
\rho_{de}=\rho_{de0}\, a^{-3(1+w_0)}\cdot
\exp{\left[\frac{3\beta_0(1-a^\xi)}{\xi}\right]},
\end{equation}
where the integration constant $\rho_{de0}$ is value of the dark energy at present,
and the dark energy EoS parameter $w\equiv p_{de}/\rho_{de}$ is a constant-$w_0$.
Substituting Eq. (\ref{rhophi2}) into Eq. (\ref{eom2}), we get
the dark matter energy density,
 \begin{equation}\label{rhom2}
\rho_{dm}=f(a)\rho_{dm0},
\end{equation}
where
\begin{equation}
\label{f1}
f(a)\equiv \frac{1}{a^3}\left\{1-\frac{\Omega_{de0}}{\Omega_{dm0}}\frac{3\beta_0 a^{-3w_0}e^{\frac{3\beta_0}{\xi}}}{\xi}\cdot\left[a^\xi E_{\frac{3w_0}{\xi}}\left(\frac{3\beta_0 a^\xi}{\xi}\right)-a^{3w_0} E_{\frac{3w_0}{\xi}} \left(\frac{3\beta_0}{\xi}\right)\right]\right\},
\end{equation}
where $\rho_{dm0}$ is dark matter density at present day, and $E_n(z)=\int_1^\infty t^{-n}e^{-xt}dt$ the
usual exponential integral function.
Note however that Eq. (\ref{rhom2}) is an
analytical expression, while in the corresponding expressions were left as
integrals and were calculated numerically. Obviously, in the case of
non-interaction (that is, for $\beta_0=0$), Eq. (\ref{rhom2}) recovers the standard
result $\rho_{dm}=\rho_{dm0}/a^3$. For the special case $\xi=0$, the energy densities of the dark sectors are
\begin{equation}
\label{dedens1}
\rho_{de}=\rho_{de0}a^{-3(1+w_0+\beta_0)},
\end{equation}
\begin{equation}
\label{dmdens1}
\rho_{dm}=\rho_{dm0}a^{-3}\left[1+\frac{\Omega_{de0}}{\Omega_{dm0}}\frac{\beta_0}{w_0+\beta_0}\left(1-a^{-3(w_0+\beta_0)}\right)
\right].
\end{equation}

It is now easy to use the Friedmann equation  to define
the dimensionless Hubble parameter, namely
\begin{eqnarray}
\label{hubeq2}
E^2(z)\equiv
\frac{H^{2}}{H^{2}_0}&=&\Omega_{b0}a^{-3}+\Omega_{dm0}f(a)+\Omega_
{de0}\,a^{-3(1+w_0)}\,
e^{\frac{3\beta_0(1-a^\xi)}{\xi}},\ \ \
\end{eqnarray}
where $\Omega_{i}\equiv\kappa^2\rho_{i}/3H^2_{0}$, and $\Omega_{i0}\equiv\kappa^2\rho_{i0}/3H^2_{0}$
are the present values of the energy density parameters.
Therefore, from Eqs. (\ref{rhophi2}), (\ref{rhom2}) and (\ref{hubeq2})
we can straightforwardly obtain the evolution of the density parameters
as
 \begin{eqnarray}\label{Omegab2}
\Omega_b(a)=\frac{a^{-3}}{a^{-3}+A f(a)+B
\,a^{-3(1+w_0)}\,
e^{\frac{3\beta_0(1-a^\xi)}{\xi}}}\ \
\end{eqnarray}
\begin{eqnarray}\label{Omegam2}
\Omega_{dm}(a)=\frac{f(a)}{A^{-1}a^{-3}+ f(a)+A^{-1}B
\,a^{-3(1+w_0)}\,
e^{\frac{3\beta_0(1-a^\xi)}{\xi}}}\ \
\end{eqnarray}
\begin{eqnarray}\label{Omegaphi2}
\Omega_{de}(a)=\frac{\,a^{-3(1+w_0)}\,
e^{\frac{3\beta_0(1-a^\xi)}{\xi}}}{B^{-1}a^{-3}+AB^{-1} f(a)+
\,a^{-3(1+w_0)}\,
e^{\frac{3\beta_0(1-a^\xi)}{\xi}}},\ \,
\end{eqnarray}
where
$A=\Omega_{dm0}/\Omega_{b0}$ and $B=\Omega_{de0}/\Omega_{b0}$.
Finally, we can easily analytically calculate the deceleration parameter
\begin{eqnarray}
\label{deceleration00}
q\equiv-\frac{\ddot{a}}{aH^2}=-1+\frac{3}{2}\left[\frac{
\Omega_b+\Omega_m+(1+w_0)\Omega _{de}} { \Omega_b+\Omega_m+\Omega
_{de}}\right],
\end{eqnarray}
 which leads to {\small{
\begin{equation}\label{deceleration2}
q=-1+\frac{3}{2}\left[{\frac{a^{-3}+A
f(a)+B(1+w_0)\,a^{-3(1+w_0)}\,
e^{\frac{3\beta_0(1-a^\xi)}{\xi}}}{a^{-3}+A
f(a)+B\,a^{-3(1+w_0)}\,
e^{\frac{3\beta_0(1-a^\xi)}{\xi}}}}\right].
\end{equation}}}

For the special case $\xi=0$, using Eqs. (\ref{dedens1}) and (\ref{dmdens1}), we get
\begin{equation}
\label{decqeq1}
q=\frac{1}{2}+\frac{w_0\Omega_{de0}}{w_0\Omega_{de0}/(w_0+\beta_0)+(1-w_0\Omega_{de0}/(w_0+\beta_0))a^{3(w_0+\beta_0)}}.
\end{equation}
So for $\xi=0$, when $\beta_0>-w_0-1/2$, the cosmic acceleration is transient.

Up to now we derived analytical expressions for the
evolution of the various density parameters and the
deceleration parameter\cite{Saridakis1}, with only the present density parameter values
and the dark energy equation-of-state parameter as
free parameters. It is therefore straightforwardly simple to construct their
evolution graphs, using the observational values $\Omega_{de0}\approx
0.72$, $\Omega_{dm0}\approx 0.24$, $\Omega_{b0}\approx 0.04$
, and setting the present scale factor value to 1.

\begin{figure}[htp]
    \label{fig1:subfig:a}
\includegraphics[width=0.45\textwidth]{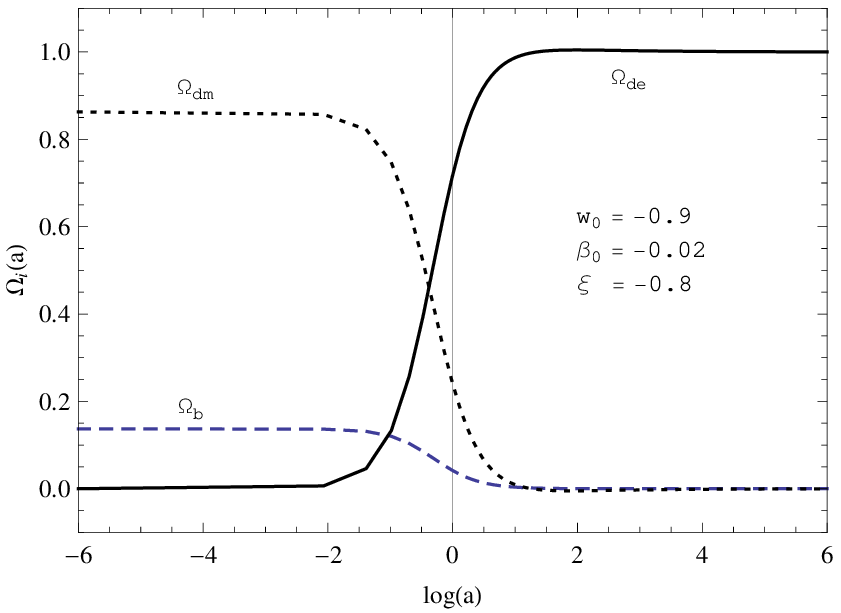}
    \label{fig1:subfig:b}
\includegraphics[width=0.45\textwidth]{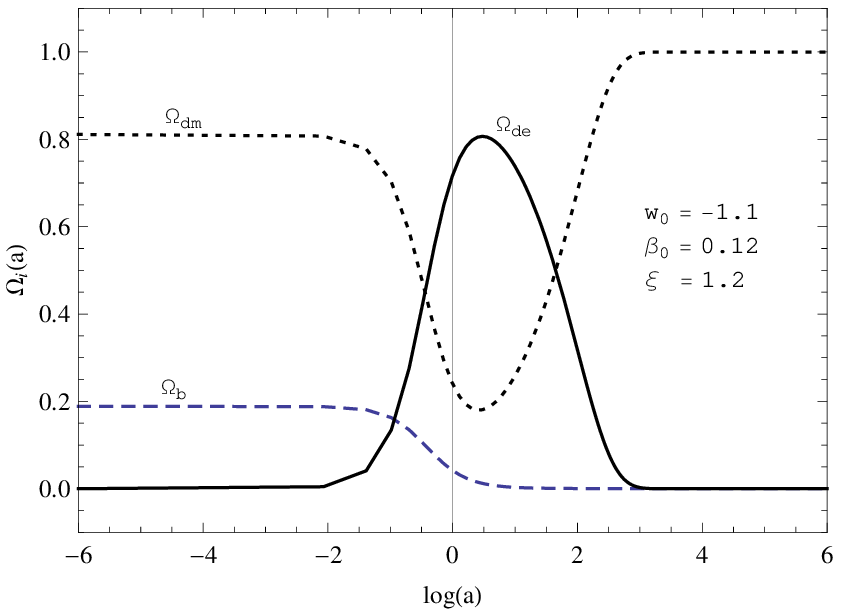}
    \label{fig1:subfig:c}
\includegraphics[width=0.45\textwidth]{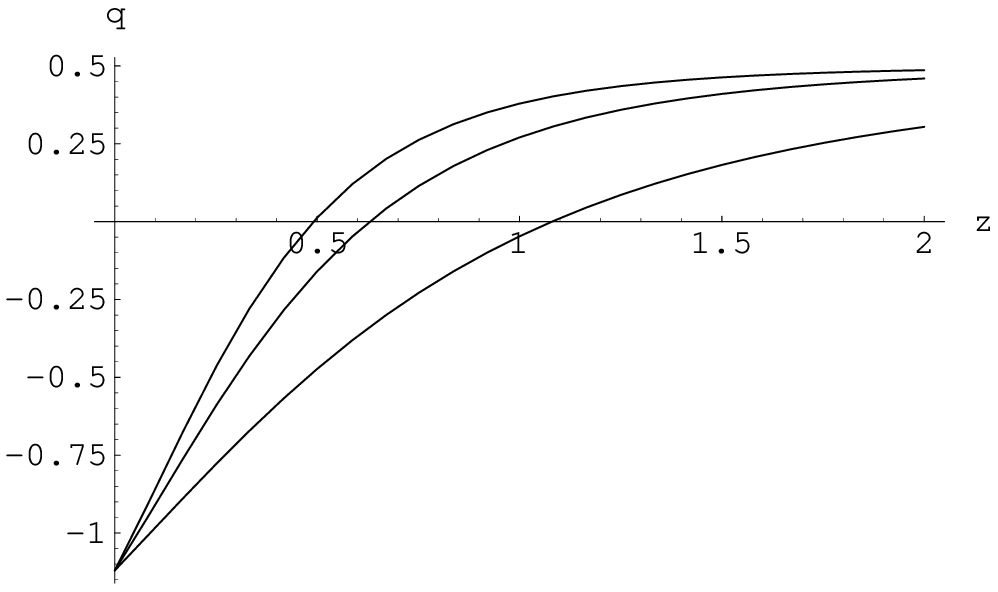}
    \label{fig1:subfig:d}
\includegraphics[width=0.45\textwidth]{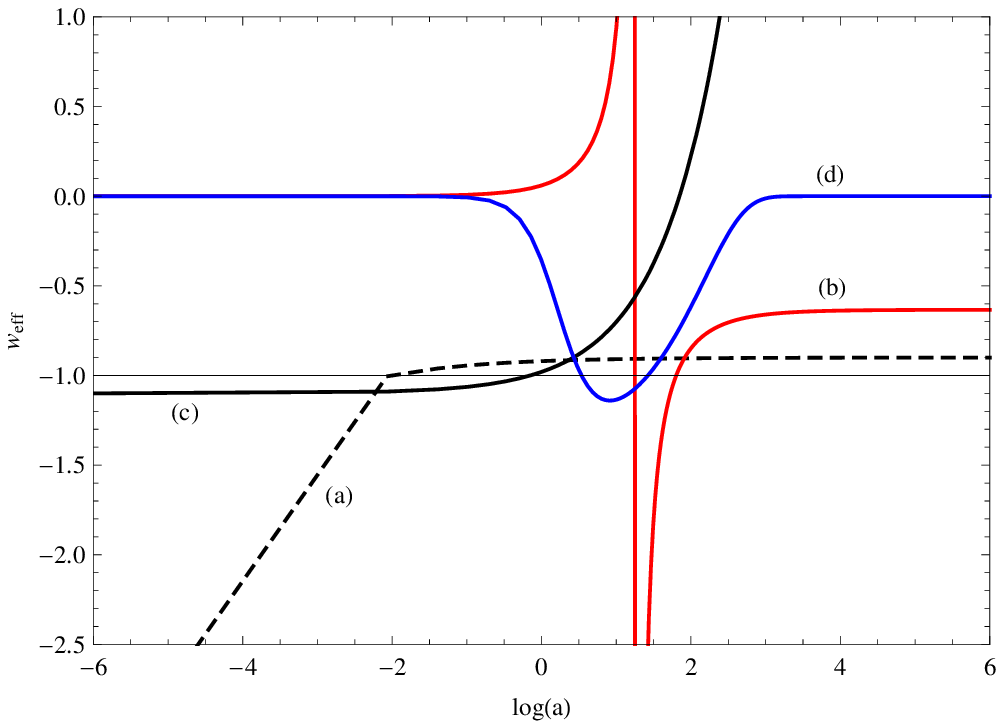}
\caption{The results for the simplest interacting model $Q=3\beta_0 a^\xi H \rho_{de}$.
Upper left panel (a): The evolution of the various density parameters
for $\beta_0=-0.02$, $\xi=-0.8$ and $w_0=-0.9$.
Upper right panel (b): The evolution of the various density parameters
for $\beta_0=0.12$, $\xi=1.2$ and $w_0=-1.1$.
Lower left panel (c): The corresponding
evolution of the deceleration parameter $q$. Line (a) is for the parameters
$\beta_0=-0.02$, $\xi=-0.8$ and $w_0=-0.9$ and line (b) is for the parameters
$\beta_0=0.12$, $\xi=1.2$ and $w_0=-1.1$. Lower right panel (d): the evolution
of the effective equation of state for dark energy (lines (a) and (c)) and dark matter (lines (b) and (d)).
Lines (a) and (b) are for the parameters
$\beta_0=-0.02$, $\xi=-0.8$ and $w_0=-0.9$ and lines (c) and (d) are for the parameters
$\beta_0=0.12$, $\xi=1.2$ and $w_0=-1.1$\cite{Saridakis1}.
\label{mod2}}
\end{figure}

In the upper left panel of Fig.~\ref{mod2} we plot the evolution of the
various density parameters with $\beta_0=-0.02$, $w_0=-0.9$ and $\xi=-0.8$,
corresponding to energy transfer from dark matter to dark energy.
Due to the energy transfer from dark matter to dark energy, despite the
fact that the energy transfer decreases as time passes ($\xi$ is negative),  we obtain the expected
result of complete dark energy domination in the future. This result is
independent of the values of $\xi$ and $w_0$, and a positive $\xi$ would just make
the dark energy domination occur earlier. In the lower left panel of Fig.~\ref{mod2} we depict the
corresponding evolution of the deceleration parameter. We can clearly see
that in this scenario, the late-time cosmic acceleration is permanent.

The upper right panel of Fig.~\ref{mod2} depicts the evolution of the various density
parameters with $\beta_0=0.12$, $w_0=-1.1$ and $\xi=1.2$. It is clear that the
cosmic acceleration is transient. Because a positive $\beta_0$ corresponds to energy
transfer from dark energy to dark matter and a positive $\xi$ means increasing energy transfer
as the Universe evolves, dark matter will finally become the dominant component.

In the phantom case ($w_0<-1$), we find that the interaction can not only save the
Universe from a Big Rip, but can also lead to dark matter domination.
Additionally, in the lower left panel of Fig.~\ref{mod2} we
plot the evolution of the deceleration parameter. From these plots we can
clearly see that the present acceleration of the Universe is transient
when both $\beta_0$ and $\xi$ are positive. This is a very interesting result
from the phenomenological point of view, and one of the main results of the
present work. The result of transient acceleration is quite general for interacting models in which more and more energy transfers from dark energy to dark matter.

In the lower right panel of Fig.~\ref{mod2}, we show the evolution of the effective equation of state parameters $w_{eff}$
for both dark energy and dark matter. We see that the effective equation of state parameter of dark energy becomes positive in
the future due to the energy transfer from dark energy to dark matter in the case of transient acceleration, while dark matter behaves like dark energy in the future
due to the energy transfer from dark matter to dark energy in the case of eternal acceleration.

\section{Constraints on coupled dark energy models}

\subsection{ Reconstruction of interacting dark energy models from parameterizations}

Interacting models, on a fundamental level, are specified  by choosing a functional form for the scalar potential and for the interaction term. However, in order to compare to observational data it is usually more convenient to use parameterizations of the dark energy equation of state and the evolution of the dark matter energy density. Once the relevant parameters are fitted  it is important to obtain the shape of the fundamental functions. In this section we show how to reconstruct the scalar potential and the scalar interaction with dark matter using such parameterizations \cite{9.1}.

Let us consider a spatially flat  Universe composed of three perfect fluids, namely dark energy, non-baryonic dark matter and baryons. The dark matter and baryons are nonrelativistic pressureless  fluids, and Einstein's equations result in

\begin{equation} \label{ref_9_1_}
\begin{array}{l} {H^{2} =\frac{8\pi G}{3} \left(\rho _{\varphi } +\rho _{dm} +\rho _{b} \right),} \\ {\dot{H}+H^{2} =-\frac{4\pi G}{3} \left(\rho _{\varphi } +\rho _{dm} +\rho _{b} +3\rho _{\varphi } \right)} \end{array}
\end{equation}
Introducing the coupling function $\delta $(a) between dark energy and dark matter  as
\begin{equation} \label{ref_9_2_}
\delta (a)=\frac{d\ln m_{\psi } (a)}{d\ln a}.
\end{equation}
(see Section \ref{Delta_Q_a}) results in the following equation for the evolution of the DM energy density $\rho _{dm} $
\begin{equation} \label{ref_9_3_}
\dot{\rho }_{dm} +3H\rho _{dm} -\delta \left(a\right)H\rho _{dm} =0.
\end{equation}
Conservation of baryon number and the total energy density implies  that the dark energy density  should obey
\begin{equation} \label{ref_9_4_}
\dot{\rho }_{\varphi } +3H\left(\rho _{\varphi } +p_{\varphi } \right)+\delta (a)H\rho _{dm} =0.
\end{equation}
Notice that the parameterization \eqref{ref_9_2_}  implies
\begin{equation} \label{ref_9_5_}
W\left(\varphi (a)\right)=\exp \left(-\int _{a}^{1}\delta \left(a'\right)d\ln a' \right).
\end{equation}
normalized in such a way that $W\left(\varphi \left(a=1\right)\right)=1$. Remember that the function $W(\varphi )$ determines the coupling of the scalar field $\varphi $ to fermionic dark matter. From a lagrangian point of view this coupling is $W(\varphi )m_{0} \bar{\psi }\psi $ .

  Combining Eqs. \eqref{ref_9_3_}-\eqref{ref_9_5_}, one obtains a modified Klein-Gordon equation for the scalar field:
\begin{equation} \label{ref_9_6_}
\ddot{\varphi }+3H\dot{\varphi }+\left(\frac{dV}{d\varphi } +\frac{\rho _{dm}^{\left(0\right)} }{a^{3} } \frac{dW}{d\varphi } \right)=0.
\end{equation}
One can now proceed to reconstruct the potential and the interaction for a given parameterization of the equation of state $W(a)$  and the interaction $\delta (a)$ . The first step is to find the time variation of the dark matter energy density:
\begin{equation} \label{ref_9_7_}
\rho _{dm} (a)=\rho _{dm}^{(0)} a^{-3} \exp \left(-\int _{a}^{1}\delta (a')d\ln a' \right).
\end{equation}
where $\rho _{dm}^{(0)} $  is the non-baryonic DM energy density today. It is more useful to work with the variable $u=\ln a$ , and one can write
\begin{equation} \label{ref_9_8_}
\rho _{dm} (u)=\rho _{dm}^{(0)} e^{-3u} \exp \left(-\int _{a}^{1}\delta (u')du' \right).
\end{equation}
The second step is to substitute $\rho _{dm} \left(u\right)$ into eq. \eqref{ref_9_4_}, which in terms of $u$  reads:
\begin{equation} \label{ref_9_9_}
\rho '_{\varphi } \left(u\right)+3\left(1+w_{\varphi } (u)\right)\rho _{\varphi } (u)+\delta (u)\rho _{dm} (u)=0.
\end{equation}
where $'=d/du$, and find  a solution $\rho _{\varphi } (u)$  with the initial condition $\rho _{\varphi } \left(u=0\right)=\rho _{\varphi }^{\left(0\right)} $, with $\rho _{\varphi }^{\left(0\right)} $  being the dark energy density today.

  In the third step, one constructs the Hubble parameter:
\begin{equation} \label{ref_9_10_}
\frac{H^{2} \left(u\right)}{H_{0}^{2} } =\Omega _{b} e^{-3u} +\Omega _{dm} e^{-3u} \exp \left(-\int _{u}^{0}\delta \left(u'\right)du' \right)+\Omega _{\varphi } f\left(u\right).
\end{equation}
where $\Omega _{X} =\rho _{X}^{\left(0\right)} /\rho _{c}^{\left(0\right)} $, the critical density today is $\rho _{c}^{\left(0\right)} =3H_{0}^{2} /8\pi G$ and $H_{0} $  is the Hubble constant. The function $f(u)$  that determines the evolution of the dark energy density is, in general, obtained numerically.

  Having obtained the Hubble parameter, the fourth step consists of solving the evolution equation for the scalar field obtained from  \eqref{ref_9_1_}:
\begin{equation} \label{ref_9_11_}
\left(\frac{d\tilde{\varphi }}{du} \right)^{2} =-\frac{1}{4\pi } \left(\frac{d\ln H(u)}{du} +\frac{3}{2} \left(\Omega _{dm} (u)+\Omega _{b} (u)\right)\right),
\end{equation}
where $\tilde{\varphi }=\varphi /M_{Pl} $  is the scalar field in units of the Planck mass $M_{Pl} =1/\sqrt{G} $ and
\begin{equation} \label{ref_9_12_}
\Omega _{dm,b} (u)=\frac{\rho _{dm,b} (u)}{\rho _{\varphi } (u)+\rho _{dm} (u)+\rho _{b} (u)}
\end{equation}
In the fifth step, one numerically inverts the solution $\tilde{\varphi }(u)$ in order to determine $u\left(\tilde{\varphi }\right)$  and  finally obtain
\[\tilde{V}\left(\tilde{\varphi }\right)\equiv \frac{V\left(u\left(\tilde{\varphi }\right)\right)}{\rho _{c}^{\left(0\right)} }=\]
$$ =\frac{1}{3} \frac{H(u)}{H_{0} } \frac{d\left(H/H_{0} \right)}{du} +\frac{H^{2} (u)}{H_{0}^{2} } -\frac{1}{2} \Omega _{b} e^{-3u} -\frac{1}{2} \Omega _{dm} e^{-3u} \exp \left(-\int _{u}^{0}\delta \left(u'\right)du' \right)$$

and
\begin{equation} \label{ref_9_14_}
W\left(u\left(\tilde{\varphi }\right)\right)=\exp \left(-\int _{u}^{0}\delta \left(u'\right)du' \right).
\end{equation}
This completes the reconstruction procedure.

  Let us consider now  the simple example of a constant EoS parameter $w_{\varphi } $  and a constant coupling  $\delta $ . In this case, one has
\begin{equation} \label{ref_9_15_}
\rho _{dm} =\rho _{dm}^{\left(0\right)} a^{-3+\delta }
\end{equation}
and the solution of  eq \eqref{ref_9_4_}  is
\begin{equation} \label{ref_9_16_}
\rho _{\varphi } (a)=\rho _{\varphi }^{\left(0\right)} a^{-3\left(1+w_{\varphi } \right)} +\frac{\delta }{\delta +3w_{\varphi } } \rho _{dm}^{\left(0\right)} \left(\right)\left(a^{-3\left(1+w_{\varphi } \right)} -a^{-3+\delta } \right).
\end{equation}

  The first  term of the solution is the usual evolution of  DE without the coupling to DM. From this solution it is easy to see that one must require a positive value of the coupling $\delta >0$  in order to have a consistently positive value of $\rho _{\varphi } $  for earlier epochs of the Universe. One can also easily reconstruct the interaction $W$  in this simple case:
\begin{equation} \label{ref_9_17_}
W(\tilde{\varphi }(u))=e^{\delta u}.
\end{equation}

\subsection{Cosmography as a way of testing models with interaction}

The method used in this section for testing interaction between dark components is fully based on the cosmological principle and, has been termed `cosmography' \cite{9.2}. The cosmological principle allows us to construct the metric of the Universe and take the first steps toward the interpretation of cosmological observations. Like kinematics, that is, the part of mechanics that describes the motion of bodies regardless of the forces causing this motion, cosmography only represents the kinematics of cosmological expansion.

  The rate at which the Universe expands is determined by how  the Hubble parameter $H(t)$ depends on time. A measure of this dependence is the deceleration parameter $q(t)$ . For a more complete description of the kinematics of cosmological expansion, it is useful to consider an extended set of parameters \cite{9.3,9.4,9.5}:
\begin{equation} \label{ref_9_18_}
\begin{array}{l} {H(t)=\frac{1}{a} \frac{da}{dt} ,} \\ {q(t)=-\frac{1}{a} \frac{d^{2} a}{dt^{2} } \left(\frac{1}{a} \frac{da}{dt} \right)^{-2} ,} \\ {j(t)=\frac{1}{a} \frac{d^{3} a}{dt^{3} } \left(\frac{1}{a} \frac{da}{dt} \right)^{-3} ,} \\ {s(t)=\frac{1}{a} \frac{d^{4} a}{dt^{4} } \left(\frac{1}{a} \frac{da}{dt} \right)^{-4} ,} \\ {l(t)=\frac{1}{a} \frac{d^{5} a}{dt^{5} } \left(\frac{1}{a} \frac{da}{dt} \right)^{-5} } .\end{array}
\end{equation}

We will not make any phenomenological assumptions about the dynamics of the dark components. Based solely on kinematics (cosmography), we will show \cite{9.6} that the observation of distant SNIa offer the possibility of testing the energy transport from the vacuum sector to the nonrelativistic matter sector which includes DM. We show that the measurements of the third order term in the expansion of the luminosity distance relation with respect redshift $z$ (jerk) allows us to detect the energy transport. Higher order terms in the expansions (snap, crackle, etc.) control the velocity, acceleration, etc... of energy transport.

To start off, and to demonstrate the main ideas behind this method, we analyse a two-component fluid with effective pressure and energy

\begin{equation} \label{ref_9_19_}
p=p_{de} ,\quad \rho =\rho _{de} +\rho _{dm}.
\end{equation}
The conservation condition can be rewritten to the form
\begin{equation} \label{ref_9_20_}
\frac{1}{a^{3} } \frac{d}{dt} \left(\rho _{dm} a^{3} \right)+\frac{1}{a^{3\left(1+w_{de} \right)} } \frac{d}{dt} \left(\rho _{de} a^{3\left(1+w_{de} \right)} \right)=0.
\end{equation}
The first term describes the net rate of absorption of energy per unit time in a unit of comoving volume transfered  out of the decaying dark energy to the nonrelativistic dark matter. The relation \eqref{ref_9_20_} can be written as  \cite{9.6}, \cite{9.7}

\begin{equation} \label{ref_9_21_}
\frac{1}{a^{3} } \frac{d}{dt} \left(\rho _{dm} a^{3} \right)=\gamma (t),\quad \frac{1}{a^{3\left(1+w_{de} \right)} } \frac{d}{dt} \left(\rho _{de} a^{3\left(1+w_{de} \right)} \right)=-\gamma (t).
\end{equation}
The function $\gamma (t)$   describes the  interaction between the  two dark components. Integration of \eqref{ref_9_21_} gives
\begin{equation} \label{ref_9_22_}
\begin{array}{l} {\rho _{dm} a^{3} =\rho _{dm0} a_{0}^{3} +\int _{t_{0} }^{t}\gamma (t)a^{3} dt, } \\ {\rho _{de} a^{3(1+w_{de} )} =\rho _{de0} a_{0}^{3(1+w_{de} )} +\int _{t_{0} }^{t}\gamma (t)a^{3(1+w_{de} )} dt } \\ {} \end{array}.
\end{equation}
Using \eqref{ref_9_22_} find
\begin{equation} \label{ref_9_23_}
\begin{array}{l}
{\ddot{a}=\frac{1}{2} \left[-\frac{A(a)}{a^{2} } -\frac{\left(1+3w_{de} \right)B(a)}{a^{2+3w_{de} } } \right],} \\
 {A(a)\equiv \frac{1}{3} \rho _{dm} a^{3} ,\quad B(a)\equiv \frac{1}{3} a^{3\left(1+w_{de} \right)} }.
 \end{array}
\end{equation}
Let's represent   \eqref{ref_9_23_} in the form
\begin{equation} \label{ref_9_24_}
qH^{2} =\frac{1}{2} \left[\frac{A(a)}{a^{3} } +\frac{\left(1+3w_{de} \right)B(a)}{a^{3(1+w_{de} )} } \right].
\end{equation}
To describe  higher derivatives of the scale factor we use the cosmographical  parameters \eqref{ref_9_18_} and to describe  the interaction we introduce the dimensionless transfer parameter
\begin{equation} \label{ref_9_25_}
\nu (t)\equiv \frac{\gamma (t)}{3H^{3} }
\end{equation}
Deriving by time both sides of   \eqref{ref_9_23_}, we obtain the basic relations connecting  the jerk  $\left(j(t)\right)$ to the transfer density parameter $\left(\nu (t)\right)$
\begin{equation} \label{ref_9_26_}
\begin{array}{l} {j-\frac{3}{2} w_{de} \nu =\Omega _{dm} +\frac{1}{2} \left(1+3w_{de} \right)\left(2+3w_{de} \right)\Omega _{de} ,} \\ {j-\frac{3}{2} w_{de} \nu -1=\frac{9}{2} w_{de} \left(1+w_{de} \right)\Omega _{de} -\Omega _{c} ,\quad \Omega _{c} \equiv -\frac{k}{a^{2} H^{2} } } \end{array}
\end{equation}
Since
\begin{equation} \label{ref_9_27_}
q=\frac{1}{2} \Omega _{dm} +\frac{1+3w_{de} }{2} \Omega _{de}.
\end{equation}
for any $\Omega _{c} $  we obtain
\begin{equation} \label{ref_9_28_}
j-\frac{3}{2} w_{de} \nu +q=\frac{3}{2} \Omega _{dm} +\frac{1}{2} \left(1+3w_{de} \right)\left(1+w_{de} \right)\Omega _{de},
\end{equation}
In the special case of the flat  model $\left(\Omega _{dm} +\Omega _{de} =1\right)$ the formula \eqref{ref_9_28_} reduces to
\begin{equation} \label{ref_9_29_}
j-\frac{3}{2} w_{de} \nu +q=-\frac{3}{2} \Omega _{dm} \left(4+3w_{de} \right)w_{de} +\frac{3}{2} \left(1+3w_{de} \right)\left(1+w_{de} \right),
\end{equation}
Therefore, interaction between nonrelativistic matter and DE is described by the third (and higher) derivate of the scale factor - a cosmographic parameter.

The methods described above can be applied to more complicated forms of dark energy \cite{9.5}, but the main principles remain the same: using the series expansions of the scale factor. Aside from supernovae, Hubble parameter measurements, Gamma Ray Bursts and Baryonic Acoustic Oscillations can be used in cosmography.

 Questions arise regarding the truncation and convergence \cite{9.3,9.5} of the series, as well as the choice of which redshift to use \cite{9.3}. Indeed, the traditional redshift z has built-in divergence for all redshifts >1. Mathematically, this is seen in the fact that the following series has divergence around $z=-1:$

\begin{equation}
{1\over 1+z} =  {a(t)\over a_0} = 1 + H_0 \; (t-t_0) -  {q_0 \; H_0^2\over 2!}  \;(t-t_0)^2
+{j_0\; H_0^3\over 3!} \;(t-t_0)^3
+ O([t-t_0]^4).
\end{equation}

Due to this, when we revert the series \cite{9.3} to obtain the lookback time as a function $T(z)$ of $z,$ the series will also diverge for $z>1,$ since by standard complex variable theory, the radius of convergence in this case is at most 1. This can be seen in picture \ref{convergence_radius_z}. On a physical level, this divergence is caused by the fact that $z=-1$ corresponds to an infinite scale factor $a=\infty,$ and one cannot physically expect to extrapolate beyond that. Because of this physical fact, the conclusions drawn for the lookback time can be extended onto all observable quantities expanded in terms of the redshift $z$ the same parameters (Hubble, deceleration, jerk, etc...) - any series (including photometric distance) will diverge for $z>1.$ This all poses a problem since many of the supernovae being discovered today are in the $z>1$ range (recall that the Big Bang corresponds to $z=\infty$).

 \begin{center}
 \begin{figure}[tbp]
 \centering
 \includegraphics[width=0.7\textwidth]{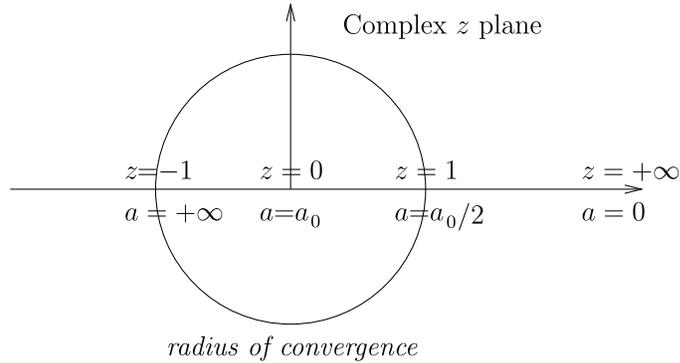}
 \caption{\label{Fconvergence_radius_z}
Qualitative sketch of the behaviour of the scale factor $a$ and the radius of convergence of the Taylor series in $z$-redshift\cite{9.3}.}
 \end{figure}
 \end{center}

It bears mentioning that this problem can be partially mitigated with a technique known as ``pivoting'' – expanding the Taylor series not around zero, but around a certain non-zero ``pivot'' value. While this technique can certainly help, it does not quite address the root causes of the divergence.

There are, however, ways of attacking the divergence problem head-on by introducing alternative redshifts. Visser \cite{9.3} proposed the so-called y-redshift, which is related to the old z-redshift in the following way:

\begin{equation}
y = {z\over1+z}; \qquad z={y\over1-y}.
\end{equation}

Like the z-redshift, the new y-redshift also has a simple physical interpretation:
\begin{equation}
y = {\lambda_0-\lambda_e\over\lambda_0} = {\Delta\lambda\over \lambda_0}.
\end{equation}

\begin{center}
 \begin{figure}[tbp]
 \centering
 \includegraphics[width=0.7\textwidth]{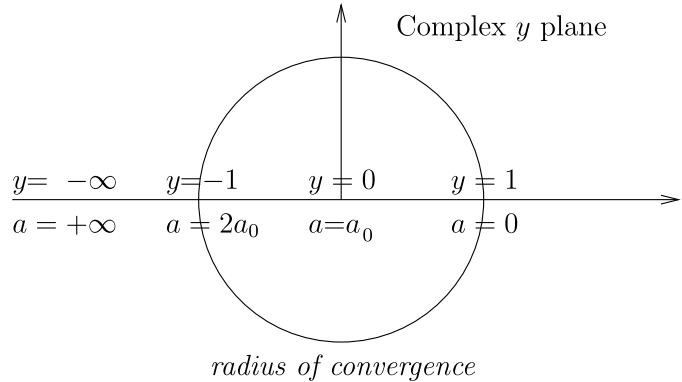}
\caption{\label{convergence_radius_y}
Qualitative sketch of the behaviour of the scale factor $a$ and the radius of convergence of the Taylor series in $y$-redshift\cite{9.3}.}
 \end{figure}
 \end{center}

More importantly, however, is that this parameterization, the entire past of the Universe all the way up to the Big Bang is located in the small limit (0,1), where 1 corresponds to the Big Bang. Physically, we assume that we cannot, like we could not before, interpolate past the Big Bang. For this reason, the y-redshift parameterization also has a convergence radius of 1, only now it's not a problem, since the entire past of the Universe lies within this radius. A visual demonstration of the advantages of the y-redshift can be seen in picture \ref{convergence_radius_y}. The formulas based on y-redshifts are no more complicated than those based on z-redshifts, which clearly shows the wide array of advantages that y-redshifts have when analyzing supernovae.

It must be said, however, that the z-redshift is more useful when interpolating into the future, as  it converges all the way up to $a=\infty,$ while y-redshift encounters problems for Universes twice as large as today's.

Aside from the y-redshift, many have proposed other redshifts, notably the "y4" redshift proposed in \cite{9.5} - $y_4= \arctan z. $ The article outlines various criteria for testing redshift models, and shows that while the y-redshift is perfectly serviceable, there are certain advantages (such as constraints of various parameters) offered by the "y4" redshift. This opens the door to the use of other redshifts in specific cases that call for it.

\subsection{Statefinder diagnostic for interacting models }

The lack of a microscopic theory of dark components, as well as our inability to properly interpret the results of observations, has led to the creation of many phenomenological models. To start off, we pick out the models that do not explicitly contradict fundamental theories and observations. This process can be divided into two phases. First, we test how well the model corresponds to certain fundamental physical principles, as well as how well it corresponds ``well studied'' areas of parameters. Second, models must be in agreement with the massive amount of data that has been obtained by modern cosmology. Obviously, the second step should come after the first. It must be said, however, that on the fundamental level, most of today's popular models stand their ground, which means that we are forced to test them using observations. Among the most popular testing methods are the so-called ${\it {\rm O} }m$-diagnostic \cite{9.8} and the use of a method based on the introduction of so-called statefinder parameters \cite{9.9}.

At the heart of ${\it {\rm O} }m$-diagnostics is a construct that depends only on the Hubble parameter
\begin{equation} \label{ref_9_301_}
{\it {\rm O} }m(x)\equiv \frac{h^{2} (x)-1}{x^{3} -1} ,\quad x=1+z,\quad h(x)=\frac{H(x)}{H_{0} }
\end{equation}
For a planar Universe composed of DE with an EoS parameter $w=const$ and non-relativistic matter
\begin{equation} \label{ref_9_31_}
h^{2} (x)=\Omega _{m0} x^{3} +\left(1-\Omega _{m0} \right)x^{\alpha } ,\quad \alpha =3(1+w)
\end{equation}
Therefore,
\begin{equation} \label{ref_9_32_}
{\it {\rm O} }m(x)=\Omega _{m0} +\left(1-\Omega _{m0} \right)\frac{x^{\alpha } -1}{x^{3} -1},
\end{equation}
From this relation, it follows that ${\it {\rm O} }m(x)=\Omega _{m0} $ for when DE is the cosmological constant $\left(\alpha =0\right)$, ${\it {\rm O} }m(x)>\Omega _{m0} $ for the quintessence case $\left(\alpha >0\right)$, and ${\it {\rm O} }m(x)<\Omega _{m0} $ for phantom energy $\left(\alpha <0\right)$. Therefore, measurements of ${\it {\rm O} }m(x)$, which are equivalent to measurements of the Hubble parameter at two different redshifts, provide us with a possible test, and help us choose an adequate DE model.

As it turns out, ${\it {\rm O} }m$-diagnostics proved to be ineffective when analysing models with interaction.  The reason is simple. The derivative $\dot{H}$ is related to the deceleration parameter
\[q=-1-\left(\dot{H}/H\right)^{2} =1/2\left(1+3w_{de} \Omega _{de} \right)\],
and does not depend on whether or not the components are interacting. On the other hand,

\begin{equation} \label{ref_9_33_}
\frac{\ddot{H}}{H^{3} } =\frac{9}{2} \left(1+\frac{p_{de} }{\rho } \right)+\frac{9}{2} \left[w_{de} \left(1+w_{de} \right)\frac{\rho _{de} }{\rho } -w_{de} \frac{\Pi }{\rho } -\frac{\dot{w}_{de} }{3H} \frac{\rho _{de} }{\rho } \right],
\end{equation}
Unlike with $H$  and $\dot{H}$ , the second derivative $\ddot{H}$  does depend on the interaction between the components. Consequently, in order to discriminate between models with different interactions, or between interacting and non-interacting models, it is desirable to additionally characterize cosmological dynamics additionally by parameters that depend on $\ddot{H}$. This role is played by the statefinder parameters
\begin{equation} \label{ref_9_34_}
r\equiv \frac{\mathop{a}\limits^{...} }{aH^{3} } ,\quad s\equiv \frac{r-1}{3(q-1/2)},
\end{equation}

  The parameters are dimensionless, and are constructed from the scale factor and its derivatives. The parameter $r$ is the next (after the Hubble parameter and the deceleration parameter) member of the set of kinematic characteristics that describe the Universe's expansion. The parameter $s$ is a combination of $q$ and $r$, âûáðàííàÿ òàêèì îáðàçîì,chosen in such a way so as not to depend on the density of dark energy. What are the reasons behind this choice? The characteristics chosen to describe dark energy can be either geometric, if they are derived directly from the space-time metric, or physical, if they depend on the characteristics of the fields that represent dark energy. Physical characteristics are, obviously, model-dependent, while geometric characteristics are more universal. Moreover, the latter are free from the uncertainties that arise during measurements of physical values like density of energy. For this very reason, geometric characteristics are more reliable during analysis of DE models. The values of the geometric parameters, with a good degree of prescision, are reconstructed from cosmological data. After this, statefinder parameters can be successfully used to identify various DE models.

For a planar Universe filled with a two-component liquid, composed of non-relativistic matter (dark matter + baryons) and dark energy with the relative density $\Omega _{de} $ , the statefinder parameters take on the form

\begin{equation} \label{ref_9_36_}
\begin{array}{l} {r=1+\frac{9}{2} \Omega _{de} w_{de} (1+w_{de} )-\frac{3}{2} \Omega _{de} \frac{\dot{w}_{de} }{H} ;} \\ {s=1+w_{de} -\frac{1}{3} \frac{\dot{w}_{de} }{w_{de} H} ;\quad w_{de} \equiv \frac{p_{de} }{\rho _{de} } } \end{array},
\end{equation}
Let's write the statefinder parameters $\left\{r,s\right\}$ for a) the cosmological constant; b) for time-independent  $w_{de} $; c) quintessence:
\begin{equation} \label{ref_9_37_}
\begin{array}{l} {a)\left\{r,s\right\}=\left\{1,0\right\};} \\ {b)\left\{r,s\right\}=\left\{1+\frac{9}{2} \Omega _{DE} (1+w_{de} ),1+w_{de} \right\};} \\ {c)\left\{r,s\right\}=\left\{1+\frac{12\pi G\dot{\varphi }^{2} }{H^{2} } +\frac{8\pi G\dot{V}}{H^{3} } ,\frac{2\left(\dot{\varphi }^{2} +\frac{2\dot{V}}{H} \right)}{\dot{\varphi }^{2} -2V} \right\}} \end{array},
\end{equation}
Much like with ${\it {\rm O} }m(x)$ - diagnostics, the statefinder parameters demonstrate the clear difference between the cosmological constant and dynamical forms of DE.

  For interacting $\left(Q=-3\Pi H\right)$ two-component fluids $(de,dm)$ in a flat Universe, the statefinder parameters take the form \cite{9.10}
\begin{equation} \label{ref_9_38_}
r=1+\frac{9}{2} \frac{w_{de} }{1+R} \left[1+w_{de} -\frac{\Pi }{\rho _{de} } -\frac{\dot{w}_{de} }{3w_{de} H} \right],\quad R\equiv \frac{\rho _{dm} }{\rho _{de} }
\end{equation}
\begin{equation} \label{ref_9_39_}
s=1+w-\frac{\Pi }{\rho _{de} } -\frac{\dot{w}_{de} }{3Hw_{de} },
\end{equation}
For non-interacting models i.e., for $\Pi =0$ , these parameters reduce to \eqref{ref_9_36_}.

Previously, we saw that the scaling solution of the form $R\propto a^{-\xi } $, where $\xi $ is a constant parameter in the range [0, 3], can be obtained when the dark energy component decays into the pressureless matter fluid. If $w_{de} =const$, it can be shown \cite{9.10}  that the interactions that produce the scaling solutions are given by
\begin{equation} \label{ref_9_40_}
\Pi =\rho _{de} \left(w_{de} +\frac{\xi }{3} \right)\frac{R_{0} \left(1+z\right)^{\xi } }{1+R_{0} \left(1+z\right)^{\xi } }
\end{equation}
Inserting this expression into Eqs. \eqref{ref_9_38_} and \eqref{ref_9_39_} yields the following expressions for the statefinder parameters

\begin{equation} \label{ref_9_41_}
r=1+\frac{9}{2} \frac{w_{de} }{1+R_{0} \left(1+z\right)^{\xi } } \left[1+w_{de} -\left(w_{de} +\frac{\xi }{3} \right)\frac{R_{0} \left(1+z\right)^{\xi } }{1+R_{0} \left(1+z\right)^{\xi } } \right]
\end{equation}
\begin{equation} \label{ref_9_42_}
s=1+w_{de} -\left(w_{de} +\frac{\xi }{3} \right)\frac{R_{0} \left(1+z\right)^{\xi } }{1+R_{0} \left(1+z\right)^{\xi } }
\end{equation}

\subsection{Statefinder parameters for some interaction models}

\subsubsection{Statefinder parameters for  Ricci dark energy}

Recently, the Ricci dark energy ìîäåëü (RDE model) was expanded in the following way  \cite{Granda:2008plb275}
\begin{equation}\label{ghde}
\rho_{de}=3M_p^2(\alpha H^2+\beta\dot{H}),
\end{equation}
where $\alpha$ and $\beta$ are constants to be determined.
Obviously, this extended model can be reduced to the
RDE model ~\cite{Gao:2009prd043511} for the case of $\alpha=2\beta$.

In order to determine the statefinder parameters, let's briefly describe the extended RDE model. The conservation equations for this model have the form \eqref{GrindEQ__3_1_} where $Q$  has the
form $Q=3bH(\rho_{de}+\rho_m)$ with $b$ as the coupling constant.  When introducing the parameter $r_\rho=\rho_m/\rho_{de}$ as the
density ratio of matter to dark energy, $Q$ can be rewritten in the
form $Q=3b(1+r_\rho)H\rho_{de}$. Making use of the "conservation
equations", we can get
\begin{equation}\label{dotratio}
\dot{r}_\rho=3H\left[wr_\rho+b(1+r_\rho)^2\right].
\end{equation}

Moreover, the Friedmann equation is
\begin{equation}\label{Friedmann}
3M_p^2 H^2=\rho_{de}+\rho_m,
\end{equation}
and the derivative of $H$ with respect to time can be given:
\begin{equation}\label{dotH}
\dot{H}=-\frac{3}{2}H^2\left(1+\frac{w}{1+r_\rho}\right).
\end{equation}
Defining the fractional energy densities as
$\Omega_{de}\equiv\rho_{de}/(3M_p^2 H^2)$ and
$\Omega_m\equiv\rho_m/(3M_p^2 H^2)$, the Friedmann equation reads
$\Omega_{de}+\Omega_m=1$. Therefore, $r_\rho$ also has the form
$r_\rho=\rho_m/\rho_{de}=\Omega_m/\Omega_{de}$, which leads to
\begin{equation}
\Omega_{de}=\frac{1}{1+r_\rho}.
\end{equation}
Substituting Eqs. (\ref{ghde}) and (\ref{dotH}) into Eq.
(\ref{Friedmann}), we get the relationship between $w$ and $r_\rho$,
\begin{equation}\label{w}
w=\left(\frac{2\alpha}{3\beta}-1\right)(1+r_{\rho})-\frac{2}{3\beta}.
\end{equation}

Let's now get the statefinder parameters for the given model. According to one of the basic dynamical equations of
cosmology,
\begin{equation}
\frac{\ddot{a}}{a}=-\frac{4\pi G}{3}(\rho+3p),
\end{equation}
where $\rho$ and $p$ denote, respectively, the total energy density and pressure of the Universe, the statefinder parameters \eqref{ref_9_34_} have the following
form in terms of $\rho$ and $p$:
\begin{equation}
r=1+\frac{9(\rho+p)}{2\rho}\frac{\dot{p}}{\dot{\rho}}, \ \ \
s=\frac{(\rho+p)}{p}\frac{\dot{p}}{\dot{\rho}},
\end{equation}
where the deceleration parameter is
\begin{equation}
q=-\frac{\ddot{a}}{aH^2}=\frac{1}{2}+\frac{3p}{2\rho}.
\end{equation}
Further, in view of $\rho=\rho_m+\rho_{de}$ and $p=p_m+p_{de}=p_{de}=w\rho_{de}$, $\rho$ is conserved and
satisfies $\dot{\rho}=-3H(\rho+p)$, while $\dot{p}=\dot{w}\rho_{de}+w\dot{\rho}_{de}$. Note that the conservation equation of dark energy (\ref{GrindEQ__3_1_}) is a little more complicated, so we introduce the effective EoS parameter of dark energy as
\begin{equation}
w^\textrm{eff}=w+b(1+r_\rho),
\end{equation}
then, Eq. (\ref{GrindEQ__3_1_}) recovers the standard form
\begin{equation}
\dot{\rho}_{de}+3H(1+w^\textrm{eff})\rho_{de}=0.
\end{equation}
So, the statefinder and deceleration parameters can be expressed as
\begin{eqnarray}
r &=& 1-\frac{3}{2}\Omega_{de}\left[w'-3w(1+w^\textrm{eff})\right], \\
s &=& 1+w^\textrm{eff}-\frac{w'}{3w}, \\
q &=& \frac{1}{2}+\frac{3}{2}w\Omega_{de},
\end{eqnarray}
where `` $'$ '' denotes the derivative with respect to $x=\ln a$, and
$H=dx/dt$. When there is no interaction, i.e., $b=0$, we have
$w^{\rm eff}=w$. Therefore, the LCDM model with $w=-1$ leads to the
constant statefinder parameters below:
\begin{equation}
\{r,s\}|_{\rm LCDM}=\{1,0\}.
\end{equation}
This means that the LCDM model corresponds to a fixed point $(s=0,r=1)$ in the statefinder $r-s$ plane. Thus, because of this feature, other models of dark energy can be measured in terms of the distance between them and the LCDM point in order to study their behavior.

\begin{figure*}
\centering
\includegraphics[width=2.1 in]{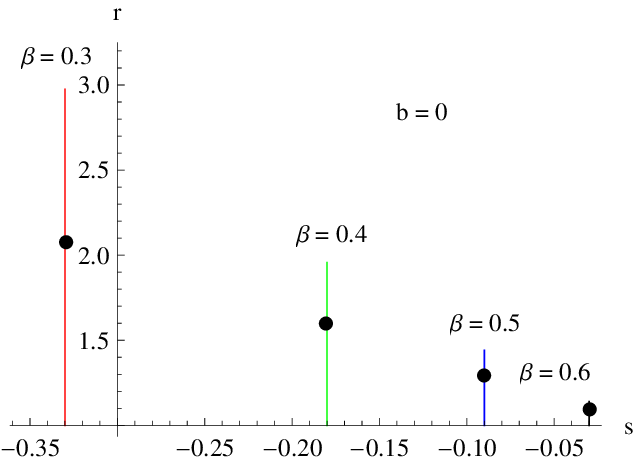}
\includegraphics[width=2.1 in]{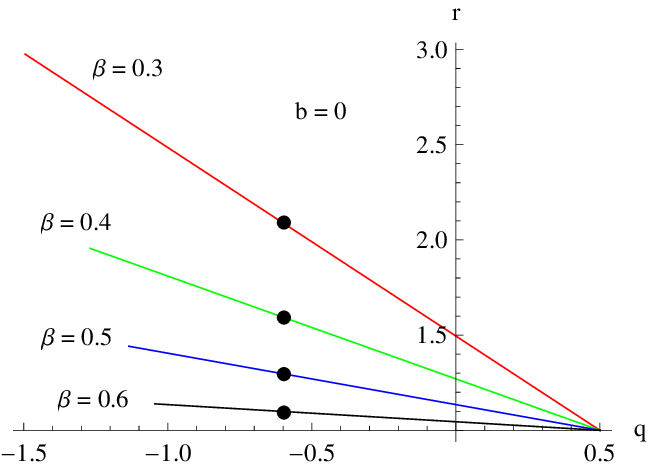}
\includegraphics[width=2.1 in]{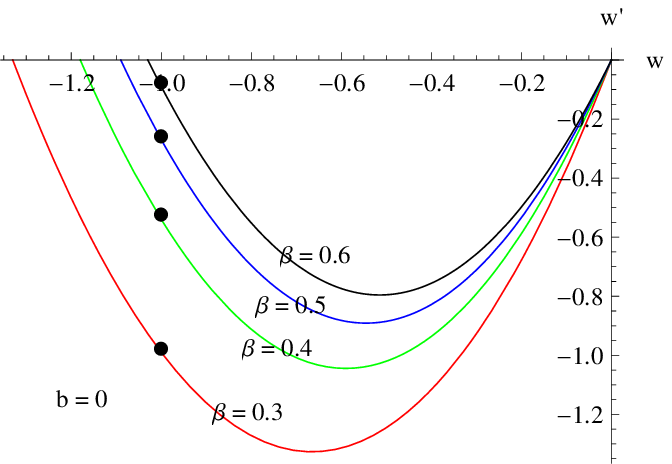}
\caption{(Color online) $r(s)$, $r(q)$ and $w'(w)$ in the extended RDE model without interaction, respectively, in the $r-s$, $r-q$
and $w'-w$ planes for the variable $\beta$. The dots denote today's values of these parameters, and $q_0=-0.595$ and $w_0=-1$ for all the
cases \cite{1305.2792}.}\label{nin}
\end{figure*}

\begin{figure*}
\centering
\includegraphics[width=2.1 in]{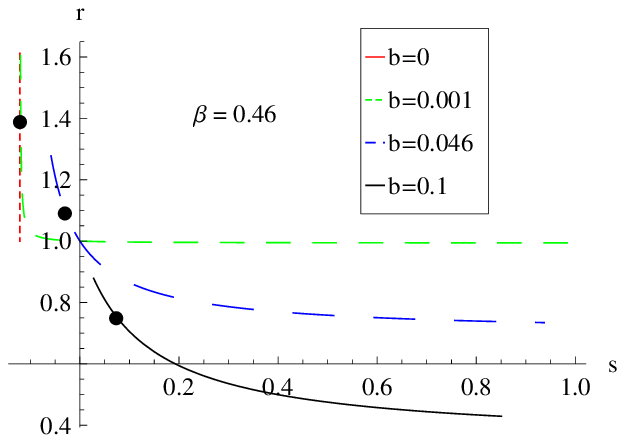}
\includegraphics[width=2.1 in]{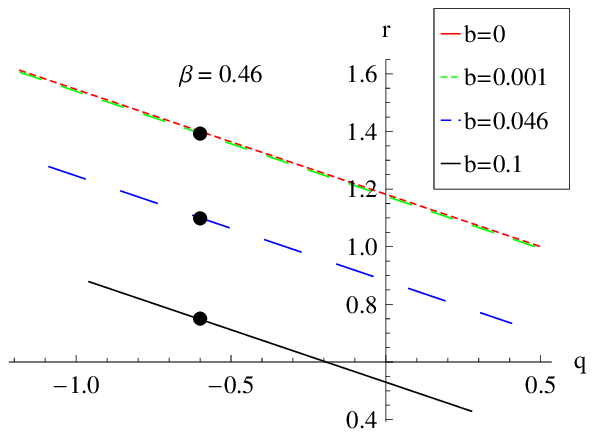}
\includegraphics[width=2.1 in]{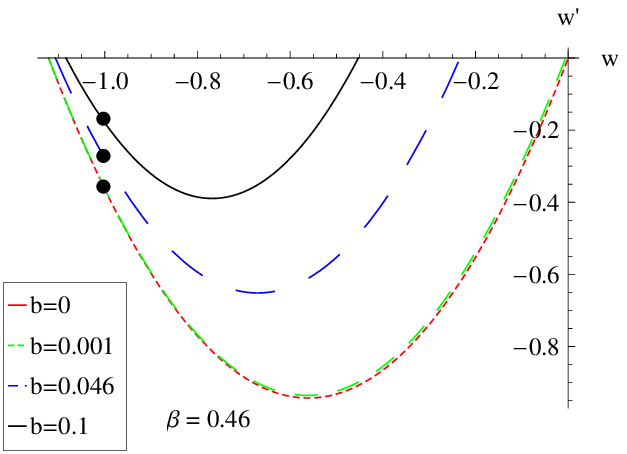}
\caption{(Color online)$r(s)$, $r(q)$ and $w'(w)$ in the extended RDE model with interaction, respectively, in the $r-s$, $r-q$ and
$w'-w$ planes for the variable $b$ with the best-fit $\beta=0.46$. The dots denote today's values of these parameters, and $q_0=-0.595$ and
$w_0=-1$ for all the cases \cite{1305.2792}.}\label{bf}
\end{figure*}

Fig.~\ref{nin} and Fig.~\ref{bf} show $r(s)$, $r(q)$ and $w'(w)$ in the ERDE model, respectively, in the $r-s$, $r-q$ and $w'-w$ planes. Fig.~\ref{nin} lacks interaction, while Fig.~\ref{bf} includes it.

\subsubsection{Statefinder parameters for the interacting ghost model of dark energy.}

The Friedmann equation for the interacting ghost model of dark energy has the form
\begin{equation}\label{fridt}
H^{2}=\frac{1}{3M_{p}^{2}}(\rho _{m}+\rho _{\Lambda})
\end{equation}%
where $H$ and $M_p$ are the Hubble parameter and the reduced Planck mass, respectively. The density of ghost dark energy is given by \cite{ghost1}
\begin{equation}\label{statet}
\rho_{\Lambda}=\alpha H
\end{equation}
where $\alpha$ is a constant of the model. The dimensionless energy densities are defined as
\begin{equation}\label{denergyt}
\Omega_{m}=\frac{\rho_m}{\rho_c}=\frac{\rho_m}{3M_p^2H^2}, ~~~\\
\Omega_{\Lambda}=\frac{\rho_{\Lambda}}{\rho_c}=\frac{\rho_{\Lambda}}{3M_p^2H^2}~~\\
\end{equation}

From the definition of $q$ and $H$, the parameter $r$  can be written as
\begin{equation}\label{stateft}
r=\frac{\ddot{H}}{H^3}-3q-2.
\end{equation}
For the given model, it is easy to find
\begin{equation}\label{stater1t}
r=1+\frac{9}{4}w_{\Lambda}\Omega_{\Lambda}(w_{\Lambda}\Omega_{\Lambda}+1)-
\frac{3}{2}\Omega_{\Lambda}w_{\Lambda}^{\prime}
\end{equation}
The parameter $s$ is obtained as
\begin{equation}\label{st6}
s=\frac{1}{2}(1+w_{\Lambda}\Omega_{\Lambda})-\frac{w_{\Lambda}^{\prime}}{3w_{\Lambda}}
\end{equation}

\begin{center}
\begin{figure}[!htbp]
\includegraphics[width=6cm]{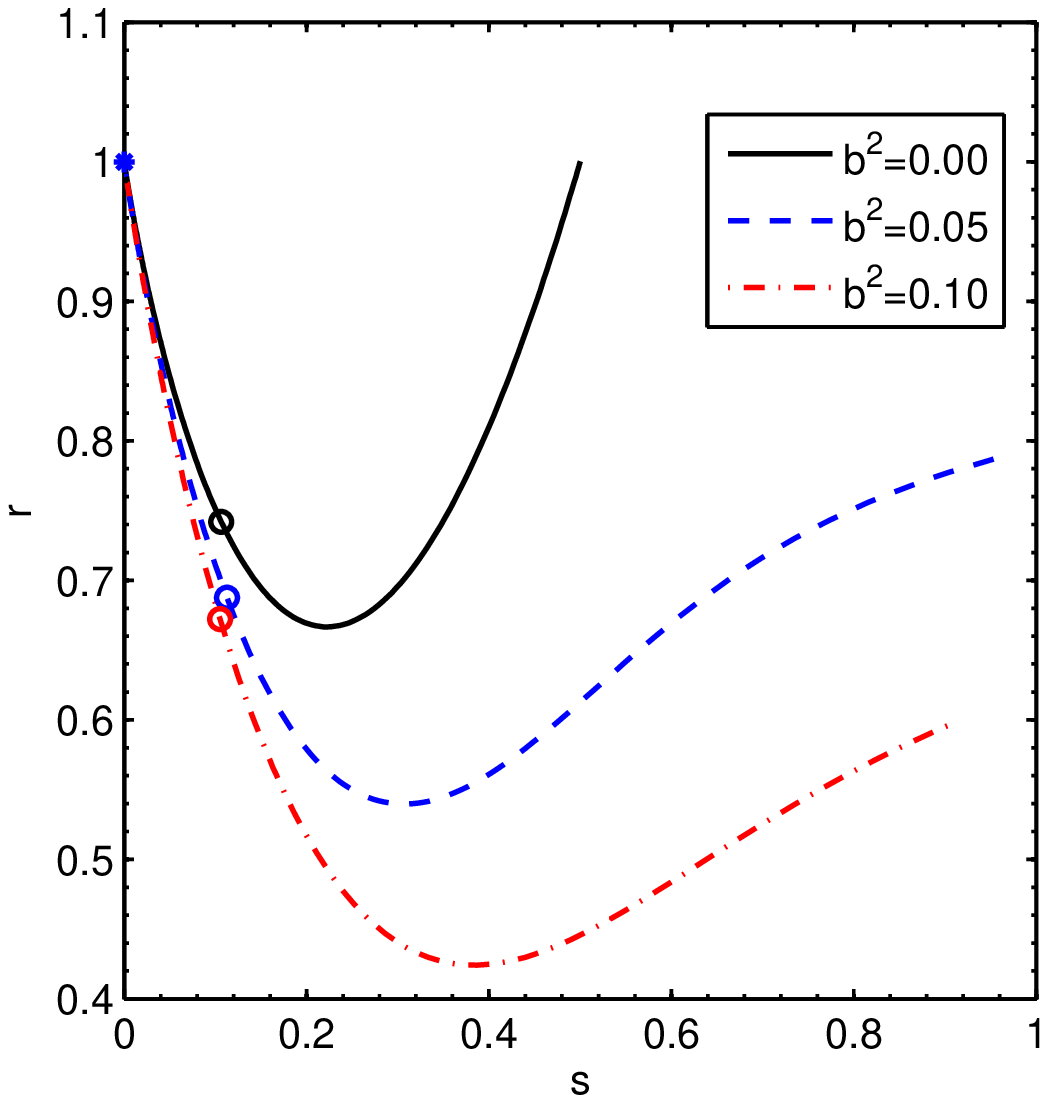}
\includegraphics[width=6cm]{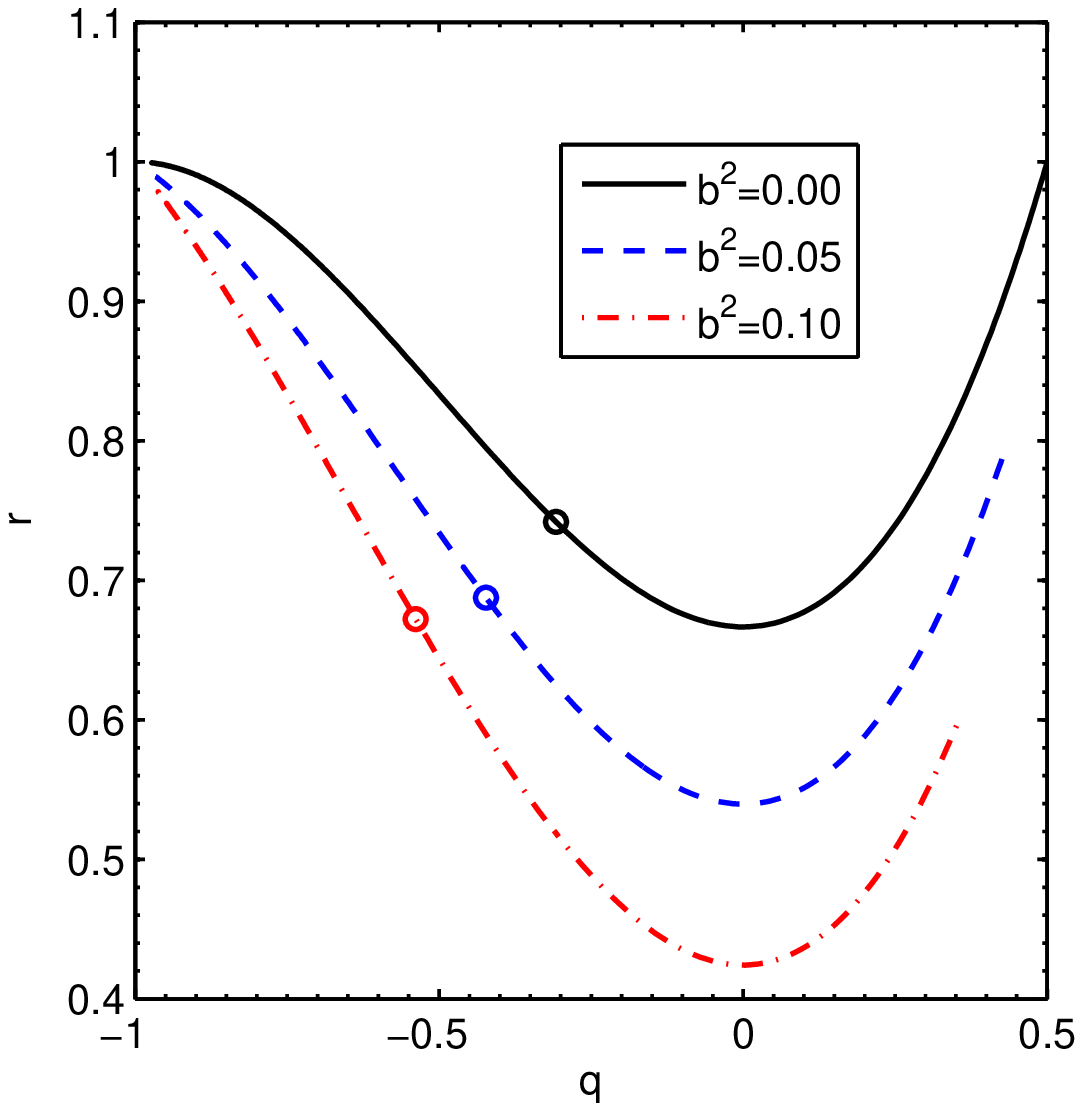}
~~~~~~ ~~~~~~\caption{Left: The evolutionary trajectories in the $s-r$ plane
for the interacting ghost dark energy model in a flat Universe with
the cosmological parameters $\Omega_{m0}=0.3$ and $\Omega_{\Lambda
0}=0.7$. The location of the standard $\Lambda$CDM fixed point is
indicated by a star symbol. The colored circle points are the locations
of the present values of the statefinder pair $\{s_0, r_0\}$ for different values of the
interaction parameter, as described in the legend.
Right: The evolutionary trajectories in the $q-r$ plane
for the interacting ghost dark energy model in a flat Universe with
the cosmological parameters $\Omega_{m0}=0.3$ and $\Omega_{\Lambda
0}=0.7$. The colored circle points are the locations of the present
values of the statefinder pair $\{q_0, r_0\}$ for different values of the interaction
parameter, as described in the legend (see \cite{IGHOST_DE}).}\label{pic_g}
 \end{figure}
 \end{center}

Fig. \ref{pic_g} illustrates the evolutionary trajectories of the ghost dark energy model in a flat Universe in the $s-r$ plane for different illustrative values of the interaction parameter $b$. Here we adopted the current values of the cosmological parameters $\Omega_{\Lambda}$ and $\Omega_m$ as $0.7$ and $0.3$, respectively. The standard $\Lambda CDM$ fixed point $\{r=1,s=0\}$ is indicated by a star symbol in this diagram. The colored circles on the curves show the present values of the statefinder pair $\{s_0, r_0\}$. By expanding the Universe, the trajectories in the $s-r$ plane move from right to left. The parameter $r$ decreases, then increases to the constant value $r=1$ at late times, while the parameter $s$ deceases from a positive value at early times to the constant value $s=0$ at late times.

In the right side of Fig. \ref{pic_g},  the evolutionary trajectories of ghost dark energy in a flat Universe are plotted for different values of the interaction parameter $b$ in the $q-r$ plane. Same as statefinder the analysis, the $q-r$ analysis can discriminate between different dark energy models. By expanding the Universe, the trajectories move from right to left. The parameter $r$ decreases, then increases to the constant value $r=1$ at late times, while the parameter $q$ decreases from a positive value (indicating decelerated expansion) at early times to a negative value (representing accelerated expansion) at the late times. Here we see the different evolutionary trajectories for different interaction parameters $b$. The current value $\{q_0, r_0\}$ can also be affected by the interaction parameter. Increasing the interaction parameter $b$ makes the parameters $r$ and $q$ smaller.

\subsubsection{Statefinder analysis for the interacting polytropic gas dark energy model}

The equation of state (EoS) of a polytropic gas is
given by (for more details and discussions, see \cite{malek1} and \cite{1201.0589})
\begin{equation}\label{poly}
p_{\Lambda}=K\rho_{\Lambda}^{1+\frac{1}{n}},
\end{equation}
where $K$ and $n$ are the polytropic constant and polytropic index, respectively \cite{c19}.
Using Eq.(\ref{poly}), the integration of the continuity equation for the interacting dark energy component gives
 \begin{equation}\label{rho1}
 \rho_{\Lambda}=\left(\frac{1}{Ba^{\frac{3(1+\alpha)}{n}}-\widetilde{K}}\right)^n,
 \end{equation}
 where $B$ is the integration constant,
$\widetilde{K}=\frac{K}{1+\alpha}$ and $a$ is the scale factor.

Substituting $Q=3\alpha H\rho_{\Lambda}$ into the EoS for DE yields
\begin{equation}\label{contd2}
\dot{\rho _{\Lambda}}+3H(1+\alpha+w_{\Lambda})\rho_{\Lambda}=0,
\end{equation}
Taking the derivative of Eq.(\ref{rho1}) with respect to time, one
obtains
\begin{equation}\label{dotrho}
\dot{\rho_{\Lambda}}=-3BH(1+\alpha)a^{\frac{3(1+\alpha)}{n}}\rho_{\Lambda}^{1+\frac{1}{n}}
\end{equation}

After obtaining the expressions for $w_{\Lambda}$, $\frac{\dot{H}}{H^2}$ and $q$, we find that
 \begin{eqnarray}\label{ddoth1}
&&\frac{\ddot{H}}{H^3}=-\frac{9}{2}\Omega_{\Lambda}(1+\alpha)(\alpha+w_{\Lambda})[(1+\alpha)(-w_{\Lambda}+\Omega_{\Lambda}\alpha+\Omega_{\Lambda}w_{\Lambda})-\alpha(\alpha+2)]\nonumber \\
&&-\frac{3}{2}\Omega_{\Lambda}(1+\alpha)w_{\Lambda}^{\prime}+\frac{9}{2}[\Omega_{\Lambda}(1+\alpha)(\alpha+w_{\Lambda})+1]^2
\end{eqnarray}
Let's now get the statefinder parameters for the given model (${s,r}$). Using the definition of the statefinder parameters, one can obtain
\begin{equation}\label{r1}
r=\frac{\dddot{a}}{aH^3}=\frac{\ddot{H}}{H^3}-3q-2
\end{equation}
Putting the expression for $q$ and (\ref{ddoth1}) into (\ref{r1}), and using the expression for $\Omega_{\Lambda}^{\prime}$, we find that
\begin{equation}\label{r2}
r=1+\frac{3}{2}\Omega_{\Lambda}(1+\alpha)[3(1+\alpha)(\alpha+w_{\Lambda})(1+\alpha+w_{\Lambda})-w_{\Lambda}^{\prime}]
\end{equation}
The parameter $s$ for the interacting polytropic gas is obtained as
\begin{equation}\label{s3}
s=\frac{2}{3}\frac{3\alpha(\alpha+1)^2+3\alpha
w_{\Lambda}(2\alpha+w_{\Lambda}+3)+3w_{\Lambda}(1+w_{\Lambda})-w_{\Lambda}^{\prime}}{\alpha+w_{\Lambda}}
\end{equation}

\begin{center}
\begin{figure}[!htb]
\includegraphics[width=10cm]{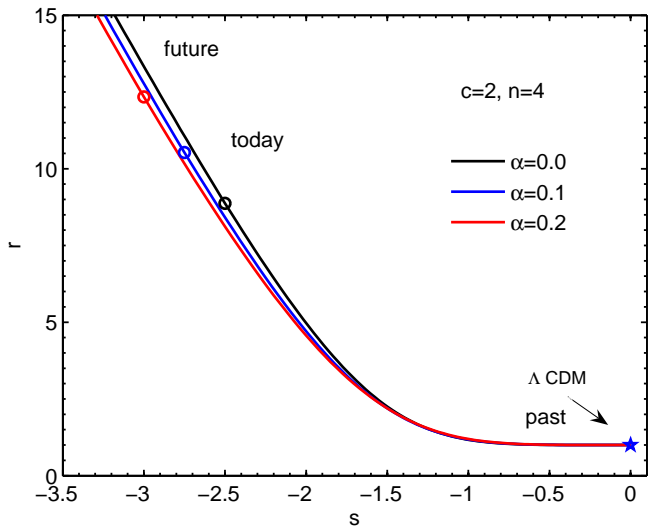}
\caption{The evolutionary trajectories for the interacting polytropic
gas model in the $s-r$ plane for different values of the interaction
parameter $\alpha$. The black curve indicates the non-interacting
case and the blue and red curves represent $\alpha=0.1$ and $\alpha=0.2$
respectively. The circles on the curves show the present values of
the statefinder pair $\{s_0, r_0\}$. The star symbol is related to
the location of the standard $\Lambda CDM$ model in the $s-r$ plane. The
parameters of the model are chosen as $c=2, n=4$ \cite{1201.0589}.}\label{fig1pol}
\end{figure}
\end{center}

In Fig. \ref{fig1pol}, the evolutionary trajectories of the interacting polytropic gas model are plotted for different values of the interaction parameter
$\alpha$. Here, we fixed the parameters of the model as $c=2$ and $n=4$. The standard $\Lambda CDM$ fixed point is indicated by a star symbol in
this diagram. The colored circles on the curves show the present values of the statefinder pair $\{s_0, r_0\}$. Different values of $\alpha$ result in different evolutionary trajectories in the $s-r$ plane. Hence, the interaction parameter can influence the evolutionary
trajectory of the polytropic gas model in the $s-r$ plane. For larger values of $\alpha$, the present value $s_0$ decreases, and the present value
$r_0$ increases. The distance of the point ($s_0, r_0$) to the $\Lambda CDM$ fixed point (i.e. $s=0,r=1$) becomes larger as the interaction parameter $\alpha$ increases. While the Universe expands, the evolutionary trajectory of the interacting polytropic gas
dark energy model evolves from the $\Lambda CDM$ at the early time, then $r$ increases and $s$ decreases. The present values of $\{s_0,
r_0\}$ are valuable, if they can be extracted from the future data of SNAP (SuperNova Acceleration Probe) experiments. Therefore, the
statefinder diagnostic tool with future SNAP observations is useful when discriminating between various dark energy models.

\begin{center}
\begin{figure}[!htb]
\includegraphics[width=0.495\textwidth]{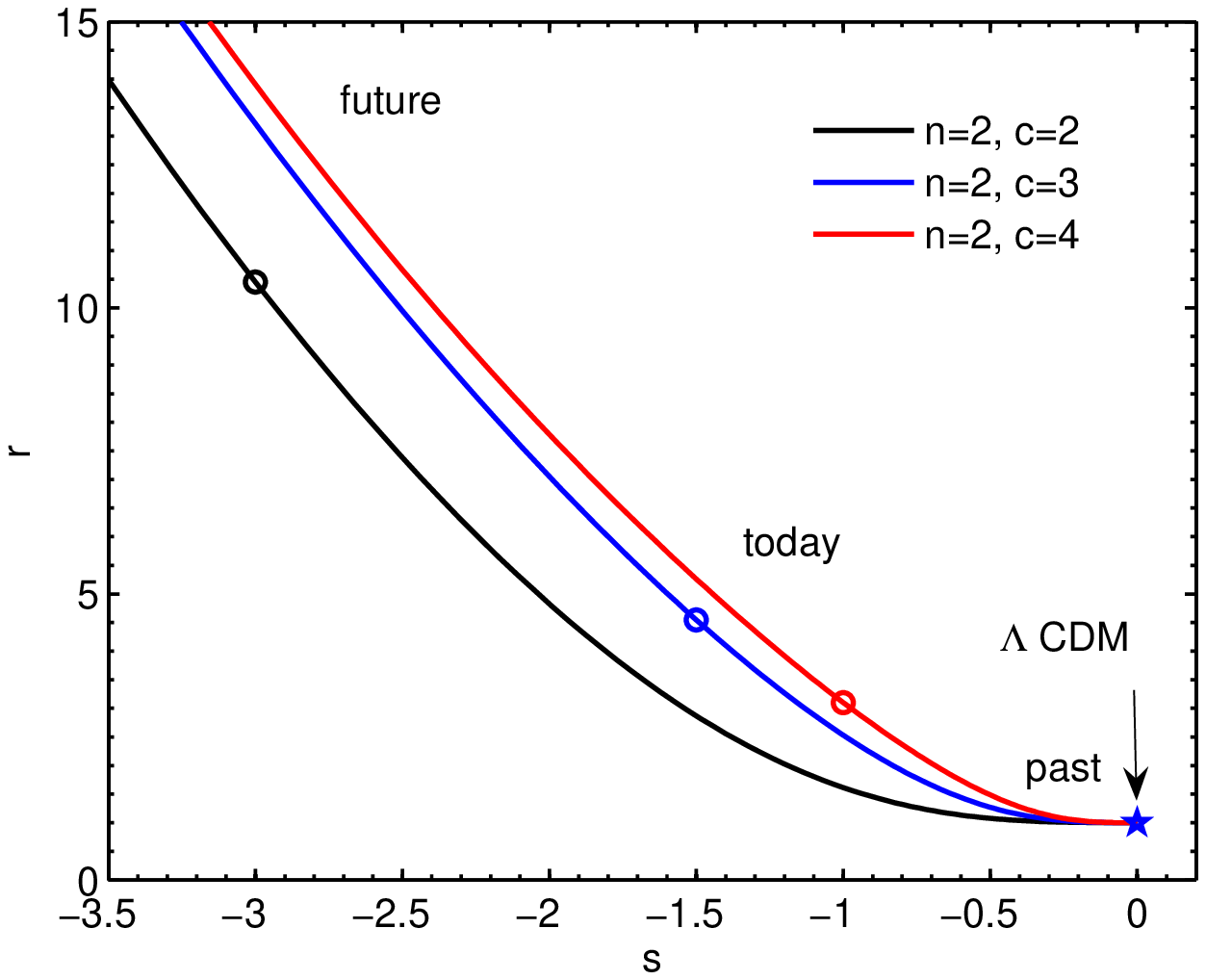}
\includegraphics[width=0.495\textwidth]{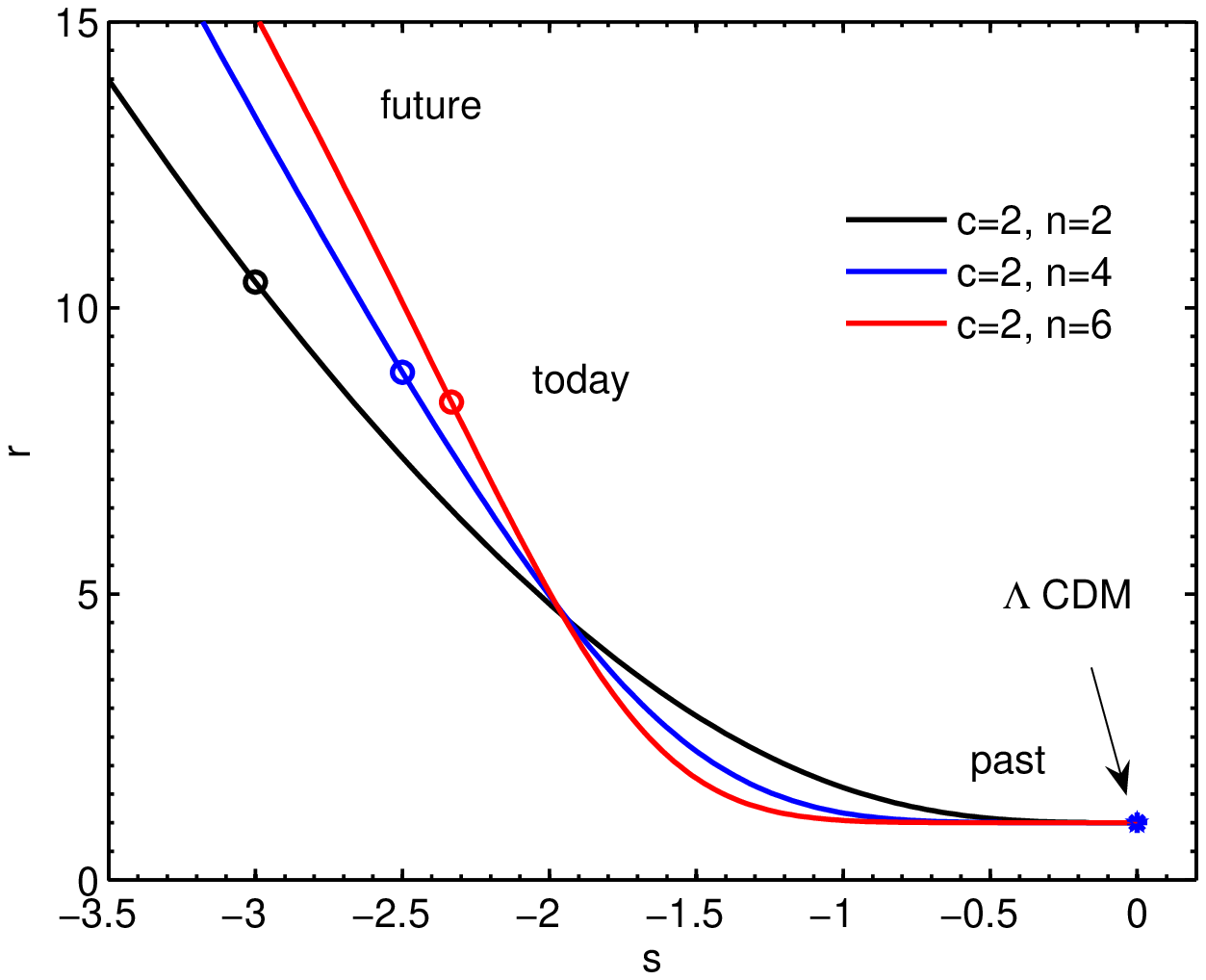}
\caption{The evolutionary trajectories for the polytropic gas model in the
$s-r$ plane for different illustrative values of the parameters $c$ and
$n$. Here we choose the interaction parameter as $\alpha=0$. In the left
panel the parameter $n$ is fixed and the parameter $c$ is varied as
$c=2$( black curve ), $c=3$ ( blue curve ), $c=4$ ( red curve ). In the
right panel the parameter $c$ is fixed and the parameter $n$ is
varied as $n=2$ ( black curve ), $n=4$ ( blue curve ) and $n=6$ (
red curve ). The circles on the curves show the present values of the
statefinder pair $\{s_0, r_0\}$. The star symbol is related to the
location of the standard $\Lambda CDM$ model in the $s-r$ plane\cite{1201.0589}.\label{fig2pol}}
\end{figure}
\end{center}

In Fig.\ref{fig2pol}, the evolutionary trajectories for the interacting polytropic
gas are plotted for different values of the parameters of the model.
The interaction parameter was fixed as $\alpha=0$. In the left panel,
the parameter $n$ is fixed and the parameter $c$ is varied.
Different values of $c$ give different evolutionary
trajectories in the $s-r$ plane. Therefore the parameter $c$ of the
model can affect the evolutionary trajectories in the $s-r$ plane.
Like Fig. \ref{fig1pol}, the present value of the statefinder pair, i.e.
$\{s_0,r_0\}$, is indicated by colored circles on the curves. For
larger values of $c$, $r_0$ decreases and $s_0$ increases. The
distance of the point ($s_0,r_0$) to the standard
$\Lambda$CDM fixed point becomes shorter for larger values of $c$. In the
right panel, the parameter $c$ is fixed and the parameter $n$ is
varied. Same as the left panel, the interaction parameter is fixed to
$\alpha=0$. Here we also see that different values of $n$ give
different evolutionary trajectories in the $s-r$ plane. For larger
values of $n$, we see that $r_0$ decreases and $s_0$ increases. Here we
see that, same as for the parameter $c$, the distance from the point
($s_0,r_0$) to the standard $\Lambda$CDM fixed point
becomes shorter for larger values of $n$.

\subsection{Observational data}

Among observational data types, the observational Hubble parameter data $H(z)$ has become an effective probe both in cosmology and astrophysics compared to the SNe Ia data, the CMB data and the baryonic acoustic oscillation (BAO) data. It is more rewarding to investigate the observational $H(z)$ data directly. The reason is quite simple: it is obvious that these probes all use the distance scale (e.g., the luminosity distance $d_L$, the shift parameter $R$, or the distance parameter $A$) measurement to determine cosmological parameters, which necessitates the integration of the Hubble parameter, and therefore destroys the fine structure of $H(z)$, as well as some more important information \cite{Lin09}. The Hubble parameter depends on the differential age as a function of the redshift $z$ of the form
 \begin{equation}
 H(z)=-\frac{1}{1+z}\frac{dz}{dt}\,.
 \end{equation}
which provides a direct measurement of $H(z)$ through a determination of $dz/dt$.

In order to obtain constraints on cosmological parameters, we use Pearson's chi-squared test. This test, sometimes called the $\chi^2$ - test, is the test most commonly used when testing hypotheses about distribution laws. In many practical problems, the exact dispersion law is unknown, and is therefore a hypothesis that demands statistical verification $\chi^2$ for $H(z)$ can be defined as
\begin{equation}
\label{chi2H}
\chi^2_H=\sum_{i=1}^{13}\frac{[H(z_i)-H_{obs}(z_i)]^2}{\sigma_{hi}^2},
\end{equation}
where $\sigma_{hi}$ is the $1\sigma$ uncertainty in the $H(z)$ data.

As it is known, the baryonic oscillations at recombination are expected to leave baryonic acoustic oscillations (BAO) in the power spectrum of galaxies. The expected BAO scale depends on the scale of the sound horizon at recombination, and on the transverse and radial scales at the mean redshift $z_{BAO}=0.35$ of galaxies in the survey. \cite{Eisenstein05} measured the quantity
\begin{equation}
A=\frac{\sqrt{\Omega_m}}{E(z_{\rm
BAO})^{1/3}}\bigg[\frac{1}{z_{BAO}}\int_0^{z_{BAO}}\frac{dz{\prime
}}{E(z^{\prime })}\bigg]^{2/3}\;,
\end{equation}
The SDSS BAO measurement \cite{Eisenstein05} gives $A_{obs}=0.469(n_s/0.98)-0.35 \pm 0.017$ where the scalar spectral index is taken to be $n_s = 0.963$, as measured by WMAP7 \cite{Komatsu10}. In
this case, $\chi^2$ can be defined as
\begin{equation}
\label{chi2BAO} \chi_{BAO}^{2} = \frac {(A-A_{\rm
obs})^2}{\sigma^2_A}.
\end{equation}

Meanwhile, the locations of the peaks in the CMB temperature power spectrum in l-space depend on the comoving scale of the sound horizon at recombination, and on the angular distance to recombination.
This is summarized by the so-called CMB shift parameter $R$ \cite{Bond97,Wang06}, which is related to cosmology by
 \begin{equation}
 R =\sqrt{\Omega_{m0}} \int^{z_{rec}}_0 \frac{dz^{'}}{E(z^{'})}
\end{equation}
where $z_{rec}\approx 1091.3$ \cite{Komatsu10} is the redshift of recombination. The 7-year WMAP data gives a shift parameter of $R =1.725\pm 0.018$ \cite{Komatsu10}. In this case, $\chi^2$ can be defined as
\begin{equation}
\label{chi2CMB} \chi_{\rm CMB}^{2} = \frac {(R-R_{\rm
obs})^2}{\sigma^2_R}
\end{equation}
Notice that both $A$ and $R$ are independent of $H_0$. Thus, these quantities can provide robust constraints on DE models in addition to the constraints provided by $H(z)$.

It is commonly believed that SNe Ia all have the same intrinsic luminosity, and thus can be used as ``standard candles''. Recently, the Supernova Cosmology Project (SCP) collaboration have released their Union2 compilation, which consists of 557 SNe Ia \cite{Amanullah}. The Union2 compilation is the largest published and spectroscopically confirmed SNe Ia sample to date. Theoretically, the distance modulus can be calculated as
\begin{equation}\label{eq.mu}
\mu=5\log\frac{d_L}{Mpc}+25=5\log_{10}H_0d_L-\mu_0,
\end{equation}
where $\mu_0=5\log_{10}[H_0/(100km/s/Mpc)]+42\cdot38$, and the luminosity distance $d_L$ can be calculated using $d_L=\frac{(1+z)}{H_0}\int_0^z\frac{dz'}{E(z')}$.
Then, $\chi^2$ from SNe Ia data is:
\begin{equation}
\label{chi2SN}
\chi^2_{SN}=A-\frac{B^2}{C}+\ln\left(\frac{C}{2\pi}\right),
\end{equation}
where $A=\sum_i^{557}{(\mu^{\rm data}-\mu^{\rm th})^2}/{\sigma^2_i}~, B=\sum_i^{557}{\mu^{\rm data}-\mu^{\rm th}}/{\sigma^2_i}~, C=\sum_i^{557}{1}/{\sigma^2_i}$, $\mu^{\rm data}$ is the distance modulus obtained from observations and $\sigma_i$ is the total uncertainty of SNe Ia data.

\subsection{Comparison of cosmological parameters in different models}
\begin{figure}
\centering
\includegraphics[width=0.4\textwidth]{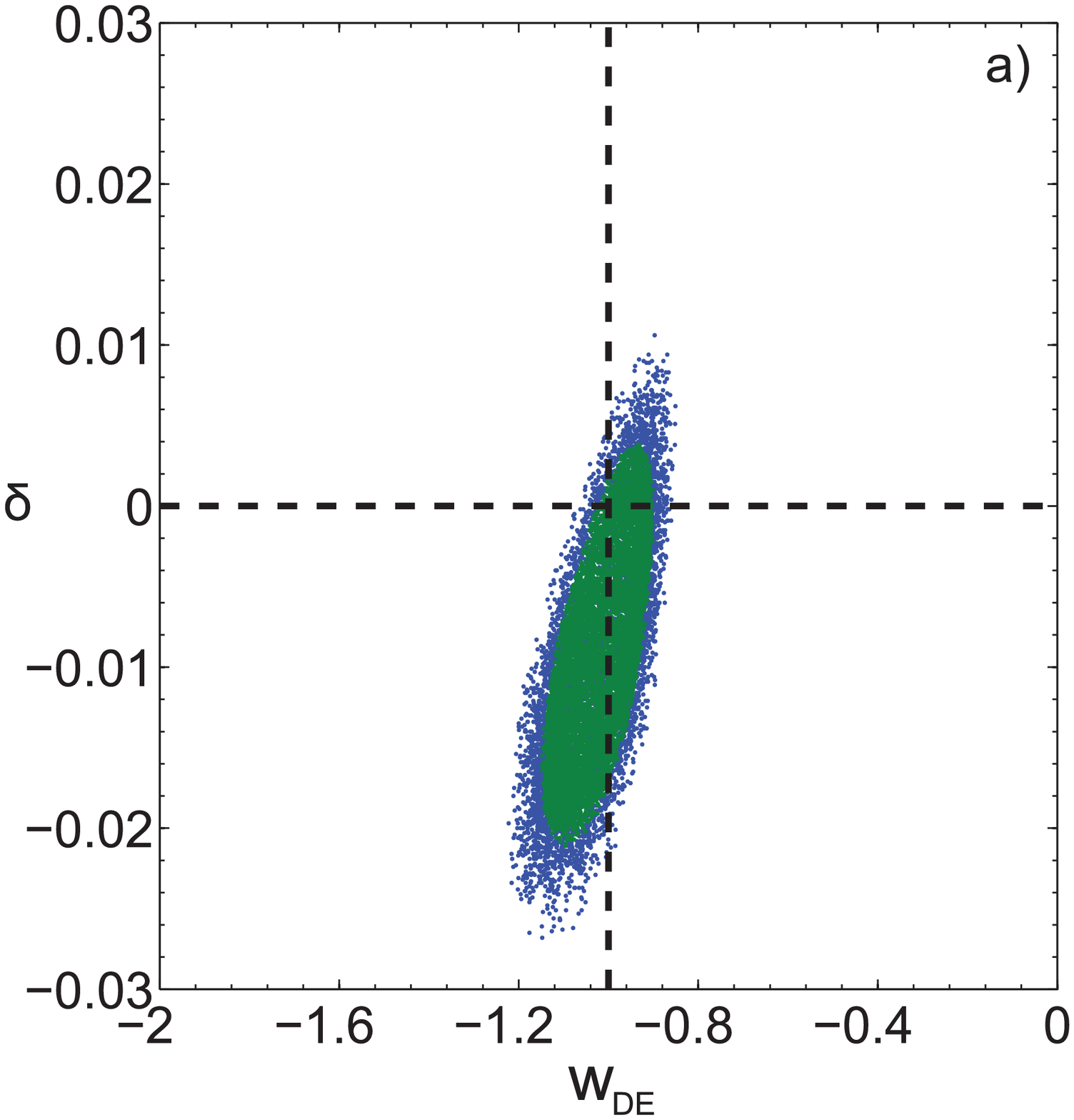}
\includegraphics[width=0.29\textwidth]{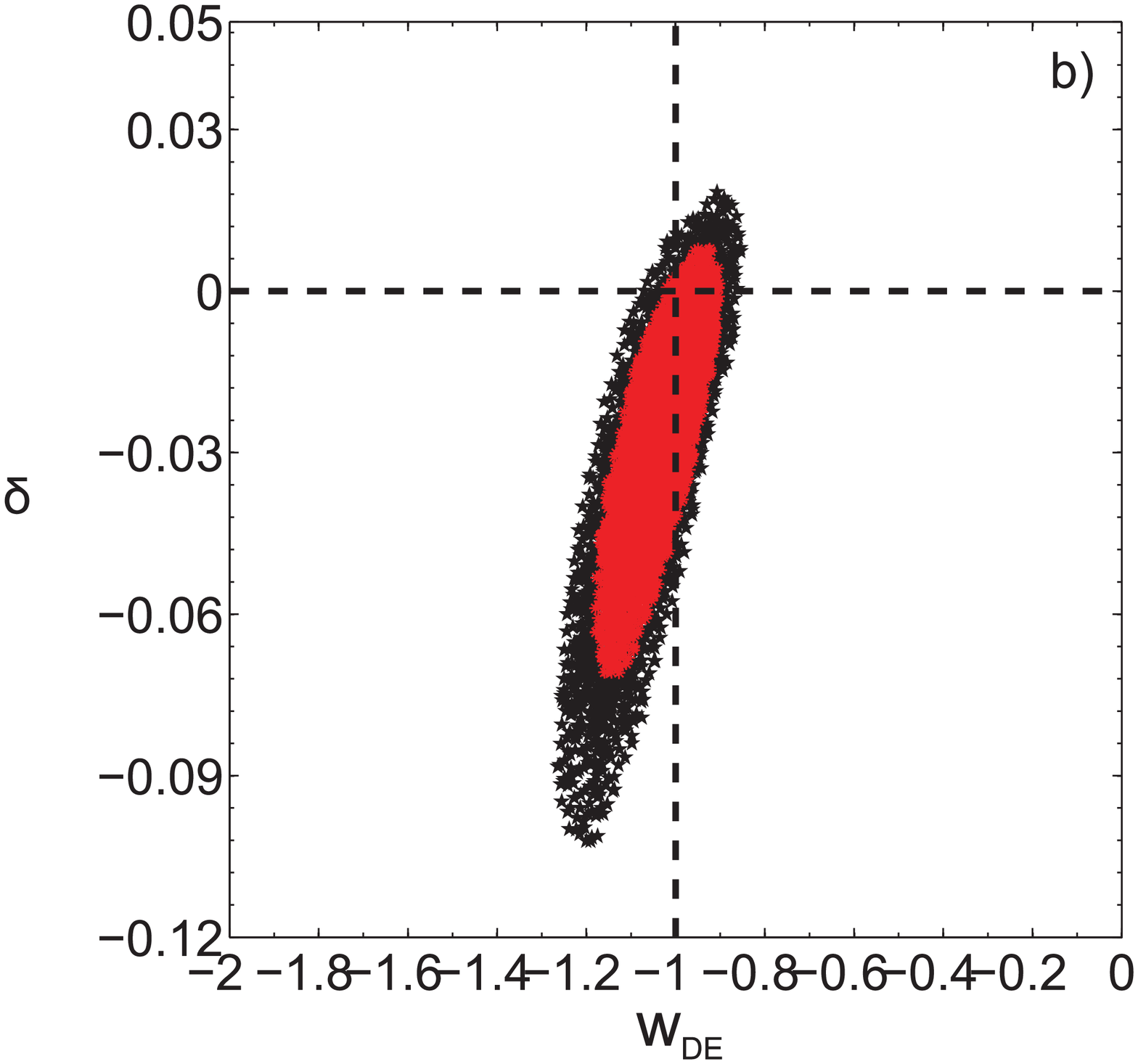}
\includegraphics[width=0.265\textwidth]{conall.eps}
\caption{ The 68.3\% and 95.4\% confidence level contours for $w_{_{DE}}$ versus $\delta$  with  SNeIa+BAO+CMB in different models.  $Q=3\delta H\rho_m$ (\textit{a})\cite{arXiv:1105.6274} ; $Q=3\delta H\rho_{_{DE}}$ (\textit{b})\cite{arXiv:1105.6274} ;  $\rho_m=\rho_{m0}a^{-3+\delta}$ (\textit{c})\cite{0702015};
  The dashed lines represent $\delta=0$ and $w_{_{DE}}=-1$.
  }
\label{wwe}
\end{figure}

In this section, we will compare cosmological parameters with various models. Table \ref{table1}contains the best-fit values of parameters for three different models with interactions in the dark sector. On Fig. \ref{wwe}, you can find the probability contours for $w_{_{DE}}$ versus $\delta$ for different models. The interaction term $\delta$ is near zero. Note, however, that even such a small value of interaction can facilitate the solution coincidence problem.
\begin{table*}\label{table1}
 \begin{center}
 \begin{tabular}{|c|c|c|c|} \hline\hline
Model &    $\Omega_{m,0}$             &      $w_{_{DE}}$                     &     $\delta$         \\ \hline

 $Q=3\delta H\rho_m$      \ \ & \ \ $0.274_{-0.029}^{+0.029}$\ \  & \ \ $-1.02_{-0.13}^{+0.12}$ \ \   & \ \ $-0.009_{-0.012}^{+0.013}$ \ \ \\
 $Q=3\delta H\rho_{_{DE}}$ \ \ & \ \ $0.272_{-0.030}^{+0.030}$\ \  & \ \ $-1.02_{-0.09}^{+0.09}$\ \  & \ \ $-0.023_{-0.040}^{+0.039}$\ \ \\
 $\rho_m=\rho_{m0}a^{-3+\delta}$\ \ & \ \ $0.270_{-0.050}^{+0.040}$\ \  & \ \ $-1.03_{-0.15}^{+0.12}$\ \  & \ \ $-0.03_{-0.05}^{+0.06}$\ \ \\
 $\Lambda CDM$\ \ & \ \ $0.270_{-0.019}^{+0.019}$\ \  & \ \ $-1.0710_{-0.0775}^{+0.0775}$\ \  & \ \ $0$\ \ \\ \hline
\hline
 \end{tabular}
 \end{center}
 \caption{\label{tab2} The best-fit values of the parameters \{$\Omega_{m0}$, $w_{_{DE}}$, $\delta$ \} for $Q=3\delta H\rho_m$ \cite{arXiv:1105.6274}, $Q=3\delta H\rho_{_{DE}}$ \cite{arXiv:1105.6274} and $\rho_m=\rho_{m0}a^{-3+\delta}$ \cite{r-m} with 1-$\sigma$ and 2-$\sigma$ uncertainty for SNe Ia+BAO+CMB data}
 \end{table*}

Let us now consider a situation in which the ratio of dark energy and dark matter has the following relation \cite{dal01}:
\begin{equation}
\label{ratio}
\frac{\rho_{_{DE}}}{\rho_m} =
\frac{\rho_{_{DE0}}}{\rho_{m0}} a^{\xi},
\end{equation}
where $\xi$ is a constant which quantifies the severity of the
coincidence problem.
In the absence of the coupling $\delta$, with a constant $w_{_{DE}}$, the energy density of dark energy scales as $\rho_X \propto a^{-3(1+w_X)}$.
Here, the ratio $\rho_{_{DE}}/\rho_m$ is proportional to $a^{-3w_{_{DE}}}$, namely, the $\xi=-3w_{_{DE}}$ case in Eq.~(\ref{ratio}).
Note that the standard $\Lambda$CDM model corresponds to $w_{_{DE}}=-1$ and $\xi=3$.

\begin{figure}
\begin{center}
\includegraphics[angle=0,scale=0.30]{vawxxi.eps}
\includegraphics[angle=0,scale=0.85]{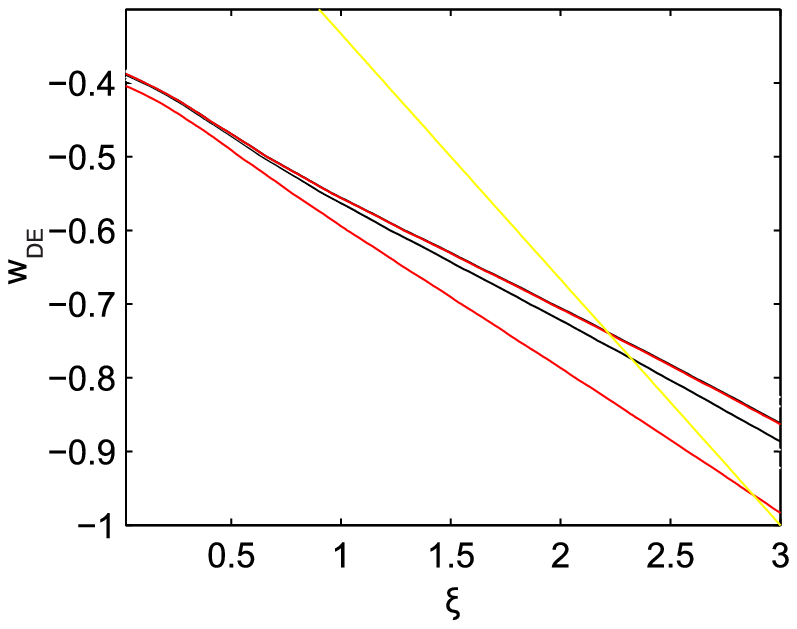}
\caption{Left: probability contours in the varying coupling models on the $(w_{_{DE}}, \xi)$ plane marginalized over $\Omega_{_{DE0}}$. The line $\xi=-3w_{_{DE}}$ corresponds to the uncoupled models. In this case we have the constraint $2.66 < \xi < 4.05$ (95\% CL)\cite{0702015}. Right: contour Plots of the first three Doppler peaks and the first trough location in the $(\xi, w_{_{DE}})$ plane with $\Omega_{m0}=0.2$ and $h=0.71$. Black, red, blue and green lines correspond to the observational bounds on the first, second, third peaks and the first trough, respectively.  The upper line corresponds to the non-interacting case $(\xi+3w_{_{DE}}=0)$ \cite{0402067}.}\label{CMB}
\end{center}
\end{figure}

As we see from Fig.~\ref{CMB}(left), the $\Lambda$CDM model, which corresponds to the point $(w_{_{DE}},\xi)=(-1, 3)$, is within the 1$\sigma$ contour bound. Recall that the uncoupled models are characterized by the line $\xi=-3w_X$. Thus, provided that the points are not on the line $\xi=-3w_{_{DE}}$, the coupled models are observationally allowed in the parameter regions $2.66 < \xi < 4.05$ (95\% CL). {}From Fig.~\ref{CMB}(left), it is obvious that the scaling models with $\xi=0$ are excluded from the data.

In figures Fig.~\ref{CMB}(right) the noninteraction line (solid yellow) stays well beyond the reach of the parameter space allowed by the CMB data. This includes the concordance $\Lambda$CDM model as well. Thus, the scaling model is more consistent with the CMB data, and is compatible with a larger parameter space than  the noninteracting standard model.

\subsection{New constraints on Coupled Dark Energy from Planck}

A truly monumental discovery was made by Salvatelli et. al based on the analysis of data obtained from the Planck satellite mission, rooted in the differences of the values of the Hubble parameter measured from the Hubble Space Telescope (HST) and the values based on the Planck mission. HST gives $H_0=73.8\pm 2.4 km/s/Mpc $, while the Planck satellite gives decidedly different results: $H_0=67.3\pm1.2 km/s/Mpc.$

Now, one must consider the different nature of these measurements: HST measures the Hubble parameter more or less directly, based on the analysis of approximately 600 Cepheid variables, while the Planck satellite analysis uses an assumption of an underlying theoretical model to obtain its results from analysis of Cosmic Microwave Background Anisotropies \cite{riess2011}. This means that, on one hand, the problem of different values can be waved away with the assumption of the presence of underlying systematic errors in both HST and Planck measurements, since neither method is ideal – HST could have certain underlying biases \cite{marra}, while Planck measurements are not direct by their very nature.

This latter point, however, may also serve as the key to resolving tensions between HST and Planck in a more physically proper manner: the Planck results change significantly when the underlying model is changed. The case analysed by Salvatelli et. al \cite{Salvatelli} included the possibility of a simple form of coupling in the dark sector, where the interaction rate is proportional to the density of the dark energy:

\begin{equation}
  \begin{array}{c}
 \nabla_\mu T_{(dm)\nu}^\mu=Qu_\nu^{(dm)}/a\\
 T_{(de)\nu}^\mu=-Qu_\nu^{(dm)}/a\\
Q=\xi H \rho_{de}    \\
  \end{array}
\end{equation}
where $\xi$ is a dimensionless parameter and $H=\dot{a}/a$ (where the dot indicates a derivative with respect to conformal time $d\tau=dt/a$.  This model is in agreement with cosmological constraints if the coupling is negative and  the dark energy EOS parameter is larger than -1. The background evolution equations here have the form \cite{Gavela:2010tm}
\begin{eqnarray}
  \label{eq:backDM}
\dot{{\rho}}_{dm}+3{\mathcal H}{\rho}_{dm}= \xi{\mathcal H}{\rho}_{de}~, \\
  \label{eq:backDE}
\dot{{\rho}}_{de}+3{\mathcal H}(1+w){\rho}_{de}= -\xi{\mathcal
  H}{\rho}_{de}~.
\end{eqnarray}

The results were the following: coupled cosmologies are not only completely compatible with the data set, but actually provided better fits than the $\Lambda$CDM model. It must also be noted that there is a strong degeneracy between the value of $\xi$ and the cold dark matter density. Negative values of the coupling $\xi$ translate into a larger matter density in the past which means that, since the dataset is sensitive to the amount of cold dark matter at recombination, the value of cold dark matter density is small today. Indeed – the more negative $\xi$ is, the larger its contribution to the value of the ''effective'' matter content - a contribution proportional to $\xi$ and $(1-a).$ The larger this contribution, the smaller the value of ''intrinsic'' dark matter density \cite{Gavela:2010tm}, which is the only part of the "effective" matter content that remains today. Therefore, by making $\xi$ more negative, the value of cold dark matter density drops, and can even drop to nearly zero and still be compatible with data.

\begin{figure}[htb!]
\centering
\includegraphics[width=0.7\textwidth]{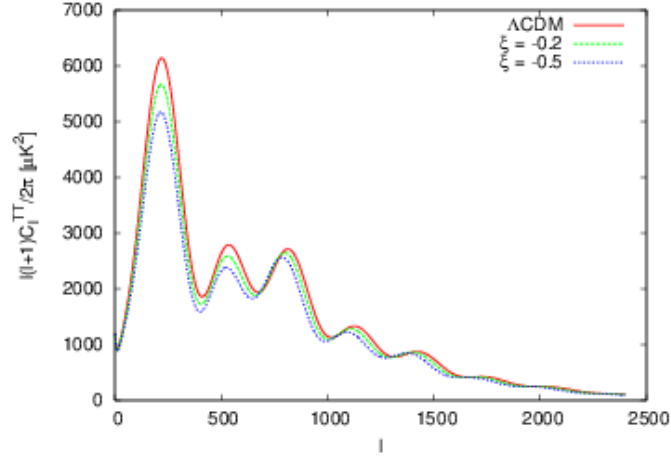}
\caption{CMB temperature power spectrum in the $\Lambda$CDM case and in the coupled cases for $\xi=-0.2, -0.5$, $\Omega_c h^2= 0.1186$, $H_0=67.9$ km/s/Mpc. The main effects of the coupling are the shifting of the positions of the acoustic peaks and changes in their amplitude \cite{1304.7119}.}
\label{figSpectra}
\end{figure}

It is this degeneracy that is the cause of the resolution of the aforementioned tensions between HST and Planck: the introduction of coupling causes the value of the Hubble parameter to rise from $H_0=67.3\pm1.2 km/s/Mpc.$ (at 68\% c.l.) to $H_0=73.3_{-1.6}^{+2.6} km/s/Mpc $ (at 68\% c.l.). Going further and including the HST prior (since the value of the Hubble parameter is compatible with the HST value), the results are fine tuned: $H_0=73.3_{-1.6}^{+2.6} km/s/Mpc$ (at 68\% c.l.) and, somewhat more importantly,  $-0.90 <\xi<-0.22$ (at 95\% c.l.).

In the synchronous gauge, the evolution of the dark matter and dark energy perturbations in the linear regime reads~\cite{Gavela:2010tm}

 \begin{eqnarray}
\label{eq:deltambe}
\dot\delta_{dm}  & = & -(k v_{dm}+\frac12 \dot h) +\xi {\mathcal H}\frac{\rho_{de}}{\rho_{dm}} \left(\delta_{de}-\delta_{dm}\right)
\\ \nonumber
&& +\xi \frac{\rho_{de}}{\rho_{dm}} \left(\frac{k v_T}{3}+\frac{\dot h}{6}\right)\,, \\
\label{eq:deltaees}
\dot\delta_{de}  & = & -(1+w)(k v_{de}+\frac12 \dot h)-3 {\mathcal H}\left(1 -w\right)
   \\ \nonumber
&& \left[ \delta_{de} +{\mathcal H} \left( 3(1+w) + \xi\right)\frac{v_{de}}{k} \right]-\xi \left(\frac{k v_T}{3}+\frac{\dot h}{6}\right) \,,\\
\label{eq:thetames}
 \dot v_{dm}  & = & -{\mathcal H} v_{dm} \,,\\
 \label{eq:thetaees}
\dot v_{de}  & = & 2 {\mathcal H}\left(1 +\frac{\xi}{1+w} \right)
    v_{de}+\frac{k}{1+w}\delta_{de}-\xi{\cal H}\frac{v_{dm}}{1+w}\,,
\end{eqnarray}
where $\delta_{dm,de}$ and $v_{dm,de}$ are the density perturbations
and velocities of the dark matter and dark energy fluids,
respectively, $v_T$ is the center of mass velocity for the total
fluid and $h$ is the usual synchronous gauge metric
perturbation.  Equations~(\ref{eq:deltambe})-(\ref{eq:thetaees})
include the contributions of the perturbation in the expansion rate
$H={\mathcal H}/a + \delta H$, the dark energy
speed of sound has been fixed to 1, i.e. $\hat c_{s\,de}^2=1$, and the
equation of state for dark energy $w$ has been taken to be
constant.

The interaction discussed in the model affects the CMB temperature spectrum
in several ways. In Fig.~\ref{figSpectra} illustrate the impact of $\xi$ up to multipole $l=2500$ for $\xi=-0.2, -0.5$ assuming a cold
dark matter density $\Omega_c h^2= 0.1186$ and $H_0=67.9$ km/s/Mpc. Notice that the presence of
a coupling among the dark matter and the dark energy fluids shifts the
position of the peaks to larger multipoles. At low multipoles, a value of
$\xi$ different from zero contributes to the late integrated Sachs-Wolfe
(ISW) effect, while at high multipoles changes the amplitude of the gravitational lensing.

In \cite{Salvatelli} it was found that considered above model with interaction in the dark sector, is in agreement with Planck data and that
even though the coupling parameter $\xi$ is weakly constrained by Planck measurements ($\xi=-0.49^{+0.19}_{-0.31}$ $68 \%$ c.l.) it induces interesting degeneracies among cosmological parameters. With such a dark interaction a
lower matter density $\Omega_{\rm m}=0.155^{+0.050}_{-0.11}$ and a
larger Hubble parameter $H_0=72.1^{+3.2}_{-2.3}\mathrm{km}/\mathrm{s}/\mathrm{Mpc}$
are favoured.

Since the value of the Hubble constant is compatible with the Hubble
Space Telescope (HST) value, in \cite{Salvatelli} authors combined the Planck and HST data sets, finding that, in this case, a non-zero value of the dark coupling is suggested by the data, with $-0.90< \xi <-0.22$ at $95\%$ c.l..

The analysis presented here points out that an interaction in the dark sector is not only allowed by current CMB data but can even resolve
the tension between the Planck and the HST measurements of the Hubble parameter.  The results we have found are in agreement with the
results obtained in former analyses for similar models using previous cosmological data
\cite{Honorez:2010rr,Valiviita:2009nu,Clemson:2011an}.

\begin{figure}[htb!]
\includegraphics[width=1.0\textwidth]{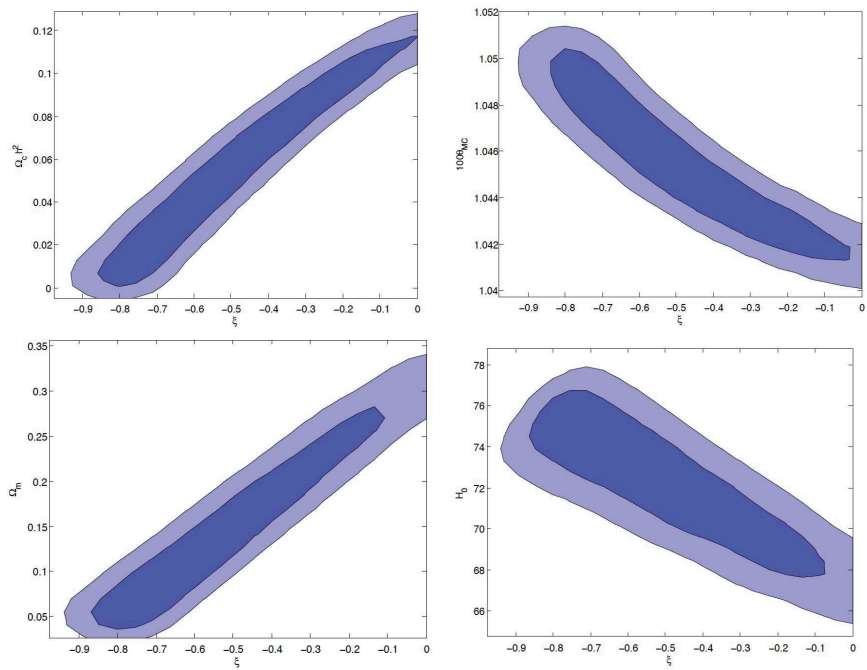}
\caption{2-D posterior distributions of parameters most degenerate with the coupling $\xi$. A strong correlation is evident with the cold dark matter density parameter. A larger absolute value of the coupling $\xi$ implies a decrement of the dark cold matter and a consequent decrease of the dark matter density. Since the assumption of a flat Universe, it also implies a larger dark energy amount that brings to an increment of $H_0$ and consequently an increase of $\theta$ \cite{1304.7119}.}
\label{fig.2Dcontour}
\end{figure}
Summing up, the presence of a coupling in the dark sector is not only possible, but is even favored by observations in comparison to the standard $\Lambda$CDM scenario, resolving tensions with the HST measurements of the Hubble parameter.


\subsection{N-body simulations}
For over 15 years now, numerical N-body simulations have been successfully used to analyse characteristics and forming processes of collapsed systems in the Universe. In addition to this, N-body simulations played a big role in the establishment of the Cold Dark Matter (CDM) paradigm as the standard scenario
for structure formation \cite{Aarseth_1963,Peebles_1970,White_1976b,Frenk_White_Davis_1983,Davis_etal_1985,White_etal_1987,Navarro_Frenk_White_1995,NFW,Klypin_etal_1999,Moore_etal_1999,Millennium,Aquarius,Angulo_etal_2012}.
Aside from this, cosmological simulations also play an important role in the analysis and understanding of the DE phenomenon. In fact, despite the undoubtable importance of the direct detection of the cosmic acceleration
by Perlmutter, Riess and Schmidt (recently recognized by the 2011 Nobel Prize in Physics),
it is worth noting that the first observational claim of a DE-dominated Universe
came about ten years before, from the comparison of the large-scale correlation of galaxies in the Automated Plate Measurement(APM) galaxy survey
with the predictions of N-body simulations \cite{Maddox_etal_1990,Efstathiou_etal_1990}.

Specifically, \cite{Maddox_etal_1990} comparing the correlation function extracted from the simulations of a CDM dominated Universe performed by \cite{White_etal_1987} with the APM observational correlation function, we find a stark discrepancy between the two for large correlation angles, with the latter showing a higher level of clustering at large scales when compared to the numerical predictions. Shortly after, \cite{Efstathiou_etal_1990} showed this such large discrepancy was removed when
comparing the data with simulations of a flat low-density Universe with $\Omega _{M}\approx 0.2$, where the missing energy, for closure, was given by a cosmological constant $\Lambda$.
Therefore, it seems appropriate to state that the first observational evidence of a DE-dominated Universe was actually derived from the outcomes of cosmological N-body simulations.

In truth, N-body simulations have only recently started to be used for DE analysis. Prior to this, most of the efforts in numerical cosmology have been devoted to improving the efficiency and the scalability of standard N-body algorithms for the  $\Lambda$CDM scenario. These attempts were mainly made in order to achieve a higher level of detailization in the description of characteristics of nonlinear structure formation, as well as to include in the integration scheme the effects of baryonic physics \cite{Teyssier_2001,Springel_Hernquist_2002,Duffy_etal_2010}, as well as a wide range of astrophysical processes such as gas cooling, star formation, feedback mechanisms from supernovae explosions and active galactic nucleus activity \cite{Springel_Hernquist_2003a,Springel_Hernquist_2003b,Kay_etal_2002,Schaye_2004,Sijacki_etal_2007, DallaVecchia_Schaye_2008}. Alternatively, large N-body simulations of the standard $\Lambda $CDM scenario have also been used to develop and calibrate semi-analytical methods to populate simulated CDM halo catalogs with realistic galaxy samples \cite{White_Frenk_1991, Lacey_Cole_1993,Kauffmann_White_Guiderdoni_1993, Cole_etal_1994,Kauffmann_etal_1999, Somerville_Primack_1999,Springel_etal_2001,DeLucia_etal_2006}.

Therefore, N-body simulations let us move further in our understanding of galaxy formation and evolution, as well as structure formation and other processes. By comparing data obtained from cosmological observations with results of N-body simulations, we can come to conclusions regarding the validity of various cosmological models.

\subsubsection{N-body simulations: general considerations and a simple model}

In order to illustrate the effect of interaction, we follow \cite{1012.0002} in considering interacting DE models where the role of DE is played by a scalar field $\phi $ evolving in
a self-interaction potential $V(\phi )$, and where the interaction with CDM particles is represented by
a source term in the respective continuity equations of the two fluids:
\begin{eqnarray}
\label{cons_c} \dot{\rho }_{dm} +3H \rho_{dm} &=& - \beta (\phi )\dot{\phi } \rho_{dm} \\
\label{cons_phi} \dot{\rho }_{\phi} +3H \rho_{\phi} &=& + \beta (\phi )\dot{\phi } \rho_{dm}
\end{eqnarray}
Clearly, the function $\beta (\phi )$ defines the intensity of the DE-CDM interaction, and, together with the
scalar potential $V(\phi )$, is fully defined by the model.

As a consequence of the interaction, and the assumption of the conservation of the CDM particle number, Eq.~\eqref{cons_c} implies the following time evolution of the CDM particle  mass, caused by the dynamic nature of the DE scalar field:
\begin{equation}
\label{mass_variation}
m_{c}(a) = m_{c}(a_{0})e^{-\int \beta(\phi )d\phi} \,,
\end{equation}
where $a_{0}$ is the cosmic scale factor at the present time.

In this subsection, unless otherwise specified, we will  assume $\beta (\phi ) = \beta.$
We will also always assume an exponential form for the potential $V(\phi )\propto e^{-\alpha \phi}$,
with $\alpha =0.1$  \cite{1012.0002}.

As we have already said (\ref{ISFMs}), the background evolution of constant coupling models is characterized by a scaling regime during matter domination
where the two interacting fluids (DE and CDM in our case) share a constant ratio of the total energy budget of the Universe,
therefore allowing for large amounts of DE in the early Universe (EDE hereafter), which will change the dynamics of the Universe from the standard
expansion history with $\Lambda $CDM model, and will also change the present values of the set of cosmological parameters. This scaling regime is sustained (for the case of positive couplings on which we focus)
by the energy transfer from the CDM particles to the DE scalar field, which determines, in turn, a decrease, in time, of the mass
of CDM particles, according to the modified continuity equation (\ref{cons_c}).
Therefore, the modified dynamics of the Universe at $z > 0$ and the dependency of the mass of the dark matter particles on time are two common features of any interacting DE
model that can affect the growth rate of density perturbations.

When taking into account the latter
effect, the normalization of the CDM masses is clearly important:
when comparing models at fixed present ($z=0$) values of cosmological parameters, it is necessary to take into account the fact that interaction causes the mass of the DM particles to rise in the past, which corresponds to an effectively larger value of the CDM density $\rho _{dm}$.

Note that there is a substantial distinction between the mass variation (that is, the fact that $\dot{m}_{dm}\neq 0$), and the mass normalization (
that is, the effective $\rho _{dm}(z)$ during the expansion history of the Universe). The former effect is found to have significant implications for the nonlinear regime of structure formation and for the internal dynamics of collapsed objects, while the latter primarily affects the linear evolution of density perturbations.

Based on these two peculiarities, we briefly look at the linear perturbation evolution in interacting DE models.
Based on the dynamic equation for CDM density perturbations in interacting DE scenarios,
\begin{equation}
\label{linear_evolution}
\ddot{\delta} _{dm} + \left( 2H - \beta \dot{\phi }\right)\dot{\delta} _{dm} - \frac{3}{2}H^2\left[\left( 1+2\beta ^{2}\right) \Omega _{dm}\delta _{dm} + \Omega _{b}\delta _{b}\right] = 0\,,
\end{equation}
shows, in fact, also the presence of an additional friction term directly proportional to the coupling
\begin{equation}
\label{friction_term}
-\beta\dot{\phi }\dot{\delta} _{dm} \,,
\end{equation}
and of an effective enhancement of the gravitational pull for CDM fluctuations by a factor of $(1 + 2\beta ^{2})$, which is known as the
``fifth-force" (see subsection \ref{PHYSICAL_MECHANISM}). Both the extra friction term and the fifth-force accelerate the growth of CDM density perturbations in the linear regime,
as clearly shown by Eqn.~(\ref{linear_evolution}).

In order to analyse the evolution of density perturbations beyond the linear regime, and to have the ability to predict the features that
interacting DE imprints on the highly nonlinear objects that we can directly observe in the sky, we need to rely on numerical integrations, as the equation (\ref{linear_evolution})
is no longer sufficient. In order to do this, it is necessary to understand how the
interaction between DE and CDM affects the laws of newtonian dynamics that govern the evolution of structure formation
in the newtonian limit of General Relativity, and apply these effects to N-body algorithms.

The article \cite{1012.0002} has shown that the acceleration equation for a CDM
particle in interacting DE cosmology for the case of a light scalar field (that is, a scalar field model for which $m_{\phi }\equiv {\rm d}^{2}V/{\rm d}\phi ^{2} \ll H$) takes the form \cite{1012.0002}
\begin{equation}
\label{acceleration_equation}
\dot{\vec{v}}_{i} = \beta \dot{\phi } \vec{v}_{i} + \sum_{j \ne i}\frac{G(1+2\beta ^{2})m_{j}\vec{r}_{ij}}{|\vec{r}_{ij}|^{3}} \,,
\end{equation}
where $\vec{v}_{i}$ is the velocity of the $i$-th particle, $\vec{r}_{ij}$ is the vector distance between the $i$-th and the $j$-th particles, and the sum extends to all the CDM particles in the Universe.

The equation (\ref{acceleration_equation}) clearly identifies the same coupling-dependent terms
already encountered in the linear perturbation equation (\ref{linear_evolution}). The friction term of Eq.~(\ref{friction_term})
now appears as a ``velocity-dependent" acceleration (see also section \ref{PHYSICAL_MECHANISM})
\begin{equation}
\vec{a}_{v} = \beta \dot{\phi }\vec{v}
\label{velocity_acc}
\end{equation}
which depends on the velocity of CDM particles, while the ``fifth-force''
term appears in the same form as in Eq.~(\ref{linear_evolution}), which is essentially equivalent to the rescaling of the gravitational constant $G$ between CDM particle pairs by a factor of $(1+2\beta^{2})$.

It is important to notice the key difference between the linear and nonlinear regimes: whereas in the linear regime the friction always accelerated structure growth, the non-linear case complicates matters by making it dependent on the relative orientation of speed and gravitational acceleration of each CDM particle. Due to this, a particle moving towards the local potential minimum will experience an effectively larger potential gradient, while a particle moving away from the local potential minimum will conversely feel an effectively smaller potential gradient . For the realistic situation of nonlinear virialized objects,
where tangential velocities are non-negligible with respect to radial velocities, the velocity-dependent acceleration will therefore have a completely different
effect than in the linear regime \cite{1012.0002}.

For this reason, when comparing the properties of nonlinear structures, it is necessary to avoid considering the linear friction term and the nonlinear velocity-dependent acceleration
as a single phenomenon. One must always distinguish between its linear and nonlinear behavior. Failing to do so can cause some further confusion when determinig which effects of interacting DE are the most relevant to the
 nonlinear dynamics of CDM particles.

As we have said before, the study of the nonlinear effects of interacting DE models with appropriately modified N-body algorithms has become popular recently.

 The first hydrodynamical high-resolution N-body simulations of interacting DE models have been performed with a modification
of the parallel TreePM code {\small GADGET-2} \cite{gadget-2} and presented in \cite{Baldi_etal_2010}.
Other studies have then been carried out by means of mesh or Tree based N-body algorithms, but without
hydrodynamics \cite{Li_Barrow_2010,Hellwing_etal_2010}, whose results are in good agreement with results from \cite{Baldi_etal_2010}.

All of the above can be summarized in the following way:
\begin{itemize}
\item The interaction between dark matter and dark energy can lead to quicker growth of linear density perturbations when compared to $\Lambda $CDM;
\item Interaction that includes only DE and CDM, while leaving baryons completely uncoupled, leads to a difference in the rates of evolution of baryon and CDM
density fluctuations in interacting DE models; this leads to a significant reduction of the relative role played by baryons in the galactic halo, as well as in collapsed objects at $z=0$;

\item For the case of constant couplings (see formulas  \eqref{cons_c}-\eqref{cons_phi}), the CDM density profiles of massive halos at $z=0$ are always less concentrated in interacting DE scenarios
when compared to $\Lambda $CDM; this does not necessarily hold for the more general case of time dependent couplings \cite{1012.0002}.
\end{itemize}

\subsubsection{Simulations of the large scale structure of the Universe}
As an example of the simulation of large scale structure of the Universe that takes dark sector coupling into account, we will look at the Coupled Dark Energy  Cosmological Simulations project (\texttt{CoDECS}). Below, briefly, we will present the  N-body simulations  for Coupled Dark Energy cosmologies in terms of simulated volume, numerical resolution, and
range of models covered in the numerical sample. These include both collisionless runs at large scales and adiabatic hydrodynamical simulations at small scales for five different
Coupled Dark Energy scenarios, besides the standard fiducial $\Lambda $CDM cosmology. The various Coupled Dark Energy models include constant coupling models \cite{Amendola1},
variable coupling models \cite{Baldi}, and the recently proposed Bouncing Coupled Dark Energy scenario \cite{Baldi_2011c}.
All the models share the same set of cosmological parameters at the present time, and
the same amplitude of density perturbations at the redshift of the last scattering surface ($z_{\rm CMB}\approx 1100$), both consistent with the latest results from the WMAP satellite \cite{Komatsu10}.

The \texttt{CoDECS} project is aimed at providing publicly available data from large N-body simulations
for a significant number of interacting Dark Energy (DE) cosmological models.

The background dynamic equations for the different cosmological components are given by
\begin{eqnarray}
\label{klein_gordon}
\ddot{\phi } + 3H\dot{\phi } +\frac{dV}{d\phi } &=& \sqrt{\frac{2}{3}}\beta _{c}(\phi ) \frac{\rho _{c}}{M_{{\rm Pl}}} \,, \\
\label{continuity_cdm}
\dot{\rho }_{c} + 3H\rho _{c} &=& -\sqrt{\frac{2}{3}}\beta _{c}(\phi )\frac{\rho _{c}\dot{\phi }}{M_{{\rm Pl}}} \,, \\
\label{continuity_baryons}
\dot{\rho }_{b} + 3H\rho _{b} &=& 0 \,, \\
\label{continuity_radiation}
\dot{\rho }_{r} + 4H\rho _{r} &=& 0\,, \\
\label{friedmann}
3H^{2} &=& \frac{1}{M_{{\rm Pl}}^{2}}\left( \rho _{r} + \rho _{c} + \rho _{b} + \rho _{\phi} \right)\,,
\end{eqnarray}
where the source terms on the right hand side of Eqs.~(\ref{klein_gordon},\ref{continuity_cdm}) represent the
interaction between the DE scalar field $\phi $ and the CDM particles.

The coupling function $\beta _{c}(\phi )$
sets the strength of the interaction, while the sign of the quantity $\dot{\phi }\beta _{c}(\phi )$
determines the direction of the energy-momentum flux between
the two components. With the convention assumed in Eqs.~(\ref{klein_gordon},\ref{continuity_cdm}), a positive
combination $\dot{\phi }\beta _{c}(\phi ) > 0$ corresponds to a transfer of energy-momentum from CDM to DE, while the opposite
trend occurs for negative values of $\dot{\phi }\beta _{c}(\phi )$.

In the range of models included in the {\small CoDECS} project, we will consider two possible candidates for the scalar field self-interaction potential $V(\phi )$, namely
an exponential potential \cite{Wetterich}:
\begin{equation}
\label{exponential}
V(\phi ) = Ae^{-\alpha \phi }
\end{equation}
and a SUGRA potential \cite{Brax_Martin_1999}:
\begin{equation}
\label{SUGRA}
V(\phi ) = A\phi ^{-\alpha }e^{\phi ^{2}/2} \,.
\end{equation}

To complete the definition of the range of models considered in the \texttt{CoDECS} project, we need to specify
the coupling function $\beta _{c}(\phi )$, for which we will assume the exponential form proposed by \cite{Amendola4,Baldi}:
\begin{equation}
\beta _{c}(\phi ) \equiv \beta _{0}e^{\beta _{1}\phi }
\end{equation}
We wil also consider in our analysis both the case of a constant coupling ($\beta _{1}=0$) and of an exponentially growing coupling ($\beta _{1}>0$).

All the cosmological models analysed in the \texttt{CoDECS} project, as well as the parameters of these models, are summarized in Table~\ref{tab:models}.

This whole parameter set assumes that the Universe is planar, and therefore that $\Omega _{c}=1-\Omega _{r}-\Omega _{b}-\Omega _{\phi }$. Specifically, all of the parameters are in agreement with the ''WMAP7 only Maximum Likelihood''
results of \cite{Komatsu10}, which are listed in Table~\ref{tab:parameters}.
\begin{table}
\begin{center}
\begin{tabular}{cc}
\hline
Parameter & Value\\
\hline
$H_{0}$ & 70.3 km s$^{-1}$ Mpc$^{-1}$\\
$\Omega _{\rm CDM} $ & 0.226 \\
$\Omega _{\rm DE} $ & 0.729 \\
${\cal A}_{s}$ & $2.42 \times 10^{-9}$\\
$ \Omega _{b} $ & 0.0451 \\
$n_{s}$ & 0.966\\
\hline
\end{tabular}
\end{center}
\caption{The set of cosmological parameters at $z=0$ assumed for all the models included in the \texttt{CoDECS} project, consistent with the latest
results of the WMAP collaboration for CMB data alone \cite{Komatsu10}}
\label{tab:parameters}
\end{table}

\begin{table}
{\tiny
\renewcommand{\tabcolsep}{0.1cm}
\begin{tabular}{l|lcccccccc}
Model & Potential  &
$\alpha $&
$\beta _{0}$&
$\beta _{1}$&
\begin{minipage}{35pt}
Scalar field \\ normalization\\
\end{minipage}&
\begin{minipage}{35pt}
Potential \\ normalization\\
\end{minipage}&
$w_{\phi }(z=0)$&
${\cal A}_{s}(z_{\rm CMB})$ &
$\sigma _{8}(z=0)$\\
\hline
$\Lambda $CDM & $V(\phi ) = A$ & -- & -- & -- & -- & $A = 0.0219$ & $-1.0$ & $2.42 \times 10^{-9}$ & $0.809$ \\
EXP001 & $V(\phi ) = Ae^{-\alpha \phi }$  & 0.08 & 0.05 & 0 & $\phi (z=0) = 0$ & $A=0.0218$ & $-0.997$ & $2.42 \times 10^{-9}$ & $0.825$ \\
EXP002 & $V(\phi ) = Ae^{-\alpha \phi }$  & 0.08 & 0.1 & 0 &$\phi (z=0) = 0$ & $A=0.0218$ & $-0.995$ & $2.42 \times 10^{-9}$ & $0.875$ \\
EXP003 & $V(\phi ) = Ae^{-\alpha \phi }$  & 0.08 & 0.15 & 0 & $\phi (z=0) = 0$ & $A=0.0218$ & $-0.992$ & $2.42 \times 10^{-9}$ & $0.967$\\
EXP008e3 & $V(\phi ) = Ae^{-\alpha \phi }$  & 0.08 & 0.4 & 3 & $\phi (z=0) = 0$ & $A=0.0217$ & $-0.982$ & $2.42 \times 10^{-9}$ & $0.895$ \\
SUGRA003 & $V(\phi ) = A\phi ^{-\alpha }e^{\phi ^{2}/2}$  & 2.15 & -0.15 & 0 & $\phi (z\rightarrow \infty ) = \sqrt{\alpha }$ & $A=0.0202$ & $-0.901$ & $2.42 \times 10^{-9}$ & $0.806$ \\
\end{tabular}
\caption{The list of cosmological models considered in the \texttt{CoDECS} project and their specific parameters. The scalar field is normalized to be zero
at the present time for all the models except the bouncing CDE scenario, for which the normalization is set in the very early Universe by placing the field
at rest in its potential minimum \cite{Baldi_2011c}. All the models have the same amplitude of scalar perturbations at $z_{\rm CMB}\approx 1100$,
as shown by the common value of the amplitude ${\cal A}_{s}$, but have very different values of $\sigma _{8}$ at $z=0$, again with the sole exception of the bouncing CDE
model SUGRA003.}}
\label{tab:models}
\end{table}

The {\small CoDECS} suite includes, at the present time, the six different cosmological models listed in Table~\ref{tab:models} \cite{CoDECS}. For all these models, two different N-body simulations (with different parameter sets) have so far been performed. They are called{\small L-CoDECS} and {\small H-CoDECS}. Both sets of simulations consist of a cosmological volume with periodic boundary conditions
filled with an equal number of CDM and baryonic particles, but differ from each other in scale and in the physical processes included
in the runs. All simulations have been carried out with the modified version (by \cite{Baldi_etal_2010}) of the widely used parallel Tree-PM
N-body code {\small GADGET} \cite{gadget-2}, specifically developed to include all the additional physical effects that characterize
CDE models \cite{Baldi_etal_2010}.

The {\small L-CoDECS} simulations have a box size of $1$ comoving Gpc$/h$ aside and include $1024^{3}$ CDM and baryon
particles for a total particle number of $2\times 1024^{3}\approx 2\times 10^{9}$. The mass resolution at $z=0$ for this
set of simulations is $m_{c}=5.84\times 10^{10}$ M$_{\odot }/h$ for CDM and $m_{b}=1.17\times 10^{10}$ M$_{\odot }/h$ for
baryons. Despite the presence of baryonic particles, these simulations do not include hydrodynamics, and are therefore purely collisionless
N-body runs. The inclusion of baryonic particles is necessary in order to realistically follow the growth of structures in the context
of specific cosmological scenarios (as the CDE models under discussion here) where baryons and CDM do not obey the same dynamical equations.

\begin{figure}[htb!]
\includegraphics[width=1.0\textwidth]{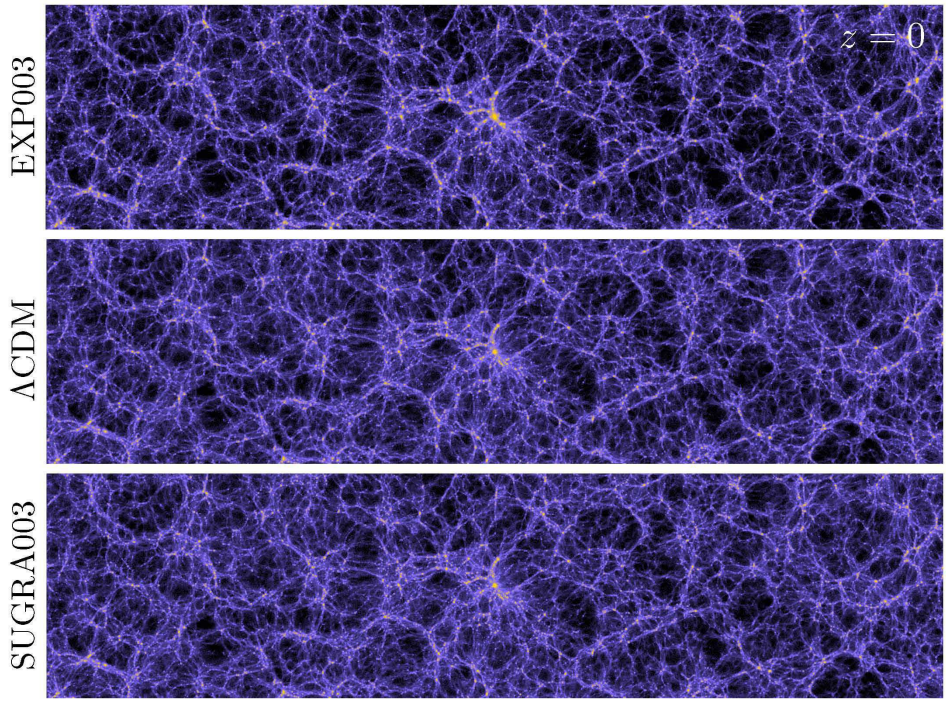}
\caption{The CDM density distribution in a slice with size $1000\times 250$ Mpc$/h$ and thickness $30$ Mpc$/h$, as extracted from the {\small L-CoDECS}
simulations of a few selected models. The middle slice shows the case of the standard $\Lambda $CDM cosmology, while the top slice is taken from the EXP003 simulation
and the bottom slice from the bouncing CDE model SUGRA003. While the latter model shows basically no difference with respect to $\Lambda $CDM at $z=0$, due to the very similar value of $\sigma _{8}$
for the two models, clear differences in the overall density contrast and in the distribution of individual halos can be easily identified for the EXP003 cosmology\cite{CoDECS}. }
\label{fig:slice}
\end{figure}
\normalsize
\begin{figure}[htb!]
\includegraphics[width=1.0\textwidth]{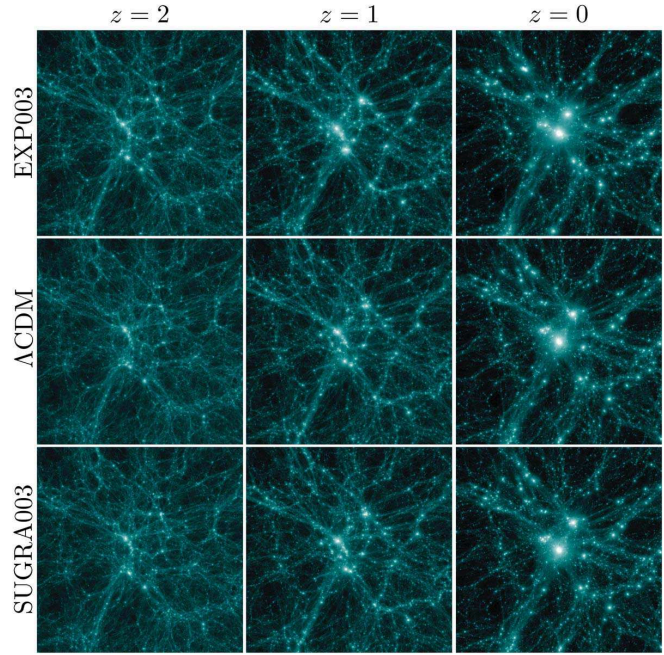}
\caption{The gas density distribution during the formation process of a massive galaxy cluster, as extracted from the {\small H-CoDECS} runs for the same three
models shown in Fig.~\ref{fig:slice}. Also, in this case, differences in the overall density contrast and in the distribution of the individual lumps are clearly visible
 when comparing the standard $\Lambda $CDM cosmology and the EXP003 CDE model at $z=0$. However, in this case, the redshift evolution shown in the figure also lets us identify differences
 between $\Lambda $CDM and the bouncing CDE model SUGRA003 at higher redshifts, where the latter model appears more evolved and shows a more pronounced density contrast
when compared to the standard cosmology\cite{CoDECS}. }
\label{fig:halo}
\end{figure}

\section*{Acknowledgments}
The research was supported in part by the Joint DFFD-RFBR Grant \# F53.2/012 .

\end{document}